\documentclass[twocolumn,floatfix,showpacs,aps,prd,nofootinbib]{revtex4}

\usepackage{amsmath}
\usepackage{amssymb}
\usepackage{graphicx}
\usepackage{dcolumn}
\usepackage{bm}
\usepackage{enumitem}

\newcommand{\BABARPubYear}    {12}
\newcommand{\BABARPubNumber}  {003}

\newcommand{\SLACPubNumber} {14996}

\newcommand{\beq}{\begin{equation}}
\newcommand{\eeq}{\end{equation}}
\newcommand{\beqn}{\begin{eqnarray}}
\newcommand{\eeqn}{\end{eqnarray}}
\newcommand{\beqns}{\begin{eqnarray*}}
\newcommand{\eeqns}{\end{eqnarray*}}

\newcommand{\intl}{\int\limits}

\newcommand{\mumu}{\ensuremath{\mu^+\mu^-}\xspace}
\newcommand{\pipi}{\ensuremath{\pi^+\pi^-}\xspace}

\newcommand{\ppg}{\ensuremath{\pi^+\pi^-(\gamma)}\xspace}

\newcommand{\mee}{e^+e^-}

\def\rq{\mathchar"0027}
 
\input babarsym
\def\sbabar{\mbox{\slshape{\small B\kern-0.1em{\smaller A}\kern-0.1em
    B\kern-0.1em{\smaller A\kern-0.2em R}}}}

\long\def\inst#1{\par\nobreak\kern 4pt\nobreak
    {\it #1}\par\vskip 10pt plus 3pt minus 3pt}

\begin{document}

\begin{flushleft}
\babar-PUB-\BABARPubYear/\BABARPubNumber\\
SLAC-PUB-\SLACPubNumber\\
~ \\
\end{flushleft}

\title{Precise Measurement of the $e^+e^-\to\pi^+\pi^-(\gamma)$ 
Cross Section with the Initial-State Radiation Method at \babar}

%
\author{J.~P.~Lees}
\author{V.~Poireau}
\author{V.~Tisserand}
\affiliation{Laboratoire d'Annecy-le-Vieux de Physique des Particules (LAPP), Universit\'e de Savoie, CNRS/IN2P3,  F-74941 Annecy-Le-Vieux, France}
\author{J.~Garra~Tico}
\author{E.~Grauges}
\affiliation{Universitat de Barcelona, Facultat de Fisica, Departament ECM, E-08028 Barcelona, Spain }
\author{A.~Palano$^{ab}$ }
\affiliation{INFN Sezione di Bari$^{a}$; Dipartimento di Fisica, Universit\`a di Bari$^{b}$, I-70126 Bari, Italy }
\author{G.~Eigen}
\author{B.~Stugu}
\affiliation{University of Bergen, Institute of Physics, N-5007 Bergen, Norway }
\author{D.~N.~Brown}
\author{L.~T.~Kerth}
\author{Yu.~G.~Kolomensky}
\author{G.~Lynch}
\affiliation{Lawrence Berkeley National Laboratory and University of California, Berkeley, California 94720, USA }
% ...
\collaboration{The \babar\ Collaboration}
\noaffiliation


\begin{abstract}
A precise measurement of the cross section of the process 
$e^+e^-\to\pi^+\pi^-(\gamma)$ from threshold to an energy of $3\gev$ 
is obtained with the initial-state radiation (ISR) method
using $232\invfb$ of data collected with the \babar\ detector at
$e^+e^-$ center-of-mass energies near $10.6\gev$. 
The ISR luminosity is determined from a study of the 
leptonic process $e^+e^-\to\mu^+\mu^-(\gamma)\gamma_{\rm ISR}$, which is found
to agree with the next-to-leading-order QED prediction to within 1.1\%. The cross section
for the process $e^+e^-\to\pi^+\pi^-(\gamma)$ is obtained with a systematic 
uncertainty of 0.5\% in the dominant $\rho$ resonance region.
The leading-order hadronic contribution to the muon magnetic anomaly
calculated using the measured $\pi\pi$ cross section from threshold 
to $1.8\gev$ is
$(514.1 \pm 2.2({\rm stat}) \pm 3.1({\rm syst}))\times 10^{-10}$.
\end{abstract}


\maketitle


\section{Introduction}

\subsection{The physics context}

The theoretical precision of observables like the running of the 
Quantum Electrodynamic (QED) fine structure constant $\alpha(s)$ or the anomalous 
magnetic moment of the muon is limited by second-order loop
effects from hadronic vacuum polarization (VP). Theoretical calculations are 
related to hadronic production rates in $e^+e^-$ annihilation via dispersion 
relations. As perturbative Quantum Chromodynamic theory 
fails in the energy regions where resonances occur, 
measurements of the $e^+e^-\to{\rm hadrons}$ cross section 
are necessary to evaluate the dispersion integrals. 
Of particular interest is the
contribution $a_\mu^{\rm had}$ to the muon magnetic moment anomaly
$a_\mu$, which requires data in a region 
dominated by the process $e^+e^-\to\pi^+\pi^-(\gamma)$.
The accuracy of the theoretical prediction for $a_\mu$ is linked to the 
advances in $e^+e^-$ measurements. 
A discrepancy of roughly 3 standard deviations ($\sigma$) including
systematic uncertainties between the
measured~\cite{bnl} and predicted~\cite{dehz03,md-pisa,hmnt2007} values of $a_\mu$ 
persisted for years before the results of this analysis became available~\cite{babar-prl}, 
possibly hinting at new physics. 
An independent approach using $\tau$ decay data leads to a smaller 
difference of $1.8\sigma$~\cite{newtau} in the same direction,
with enlarged systematic uncertainties due to isospin-breaking corrections. 

The kernel in the integrals involved in vacuum polarization 
calculations strongly emphasizes the low-energy part of the 
spectrum. 73\% of the lowest-order hadronic contribution 
is provided by the $\ppg$ final state, and about 60\% of its total 
uncertainty stems from that mode~\cite{dhmz09}. To improve on present 
calculations, the precision on the VP dispersion integrals is required to be 
better than 1\%. More precise experimental data in the $\ppg$ channel are 
needed, such that systematic uncertainties on the cross sections that are 
correlated over the relevant mass range are kept well below the percent level.
 
In this paper an analysis of the process 
$e^+e^- \rightarrow \pi^+\pi^-(\gamma)\gamma$ 
based on data collected with
the \babar\ experiment is presented. 
In addition, as a cross-check of the analysis, we measure the 
$e^+e^- \rightarrow \mu^+\mu^-(\gamma)\gamma$ cross section on the same data
and compare it to the QED prediction. The reported results and their application
to the $\pi\pi$ contribution to the muon magnetic anomaly have been already 
published in shorter form~\cite{babar-prl}.

\subsection{The ISR approach}

The initial-state radiation (ISR) method has been 
proposed~\cite{isr1,isr2,isr3,isr4} as a 
novel way to study $e^+e^-$ annihilation processes instead of the standard 
point-by-point energy-scan measurements. The main advantage of the ISR approach
is that the final-state mass spectrum is obtained in a single configuration of 
the $e^+e^-$ storage rings and of the detection apparatus, thus providing a 
cross section measurement over a wide mass range starting at threshold. Consequently, a 
better control of the systematic errors can be achieved compared to the 
energy-scan method, which necessitates different experiments and colliders to cover 
the same range. The disadvantage is the reduction of the measured cross 
section, which is suppressed by one order of $\alpha$. This is offset by the availability
of high-luminosity $e^+e^-$ storage rings, primarily designed as $B$ and $K$ 
factories in order to study \CP violation.

In the ISR method, the cross section for $e^+e^-\to X$ at the reduced energy 
$\sqrt{s'}=m_X$, where $X$ can be any final state, 
is deduced from a measurement of the radiative process 
$e^+e^-\to X\gamma$, where the photon is emitted by the initial 
$e^+$ or $e^-$ particle. 
The reduced energy is related to the energy $E_\gamma^*$ 
of the ISR photon in the $e^+e^-$ center-of-mass (c.m.) frame
by $s'=s(1-2E_\gamma^*/\sqrt{s})$, 
where $s$ is the square of the $e^+e^-$ c.m.\ energy. 
In this analysis, $s\sim (10.58\gev)^2$ and $\sqrt{s'}$ ranges from 
the two-pion production threshold to 3\gev. Two-body ISR processes 
$e^+e^-\to X\gamma$ with $X=\pi^+\pi^-(\gamma)$ and
$X=\mu^+\mu^-(\gamma)$ are measured, where the ISR photon is detected
at large angle to the beams, and the charged particle pair can be
accompanied by a final-state radiation (FSR) photon.

\begin{figure}[htbp] 
\centering
  \includegraphics[width=4.cm]{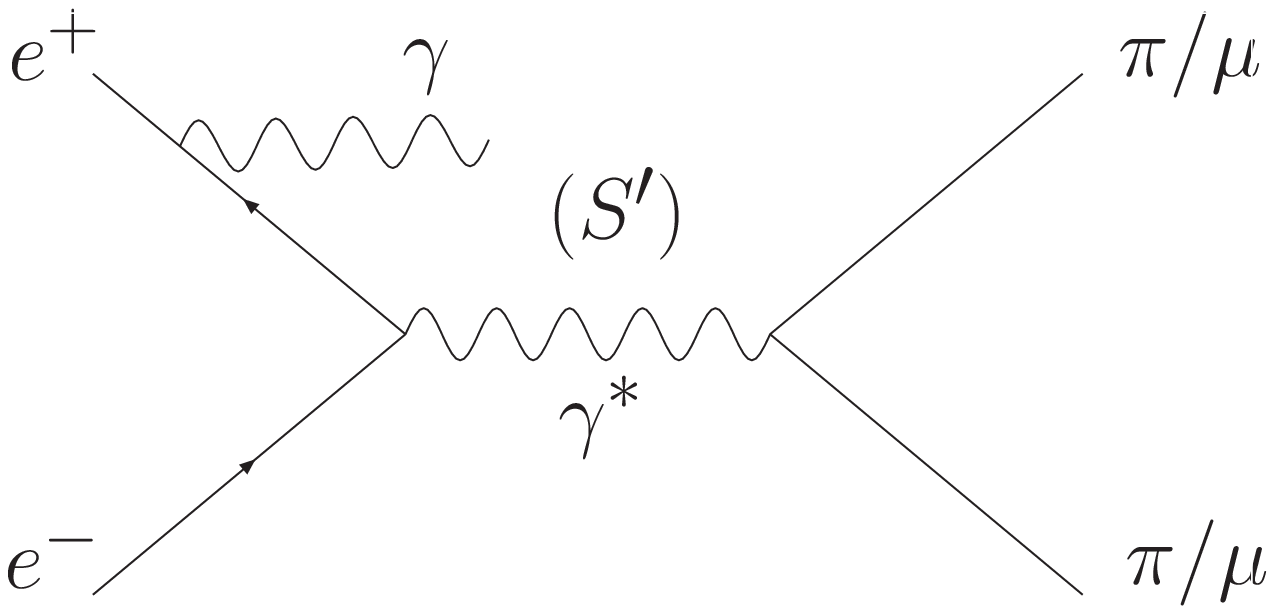}
  \hspace{5pt}
  \includegraphics[width=4.cm]{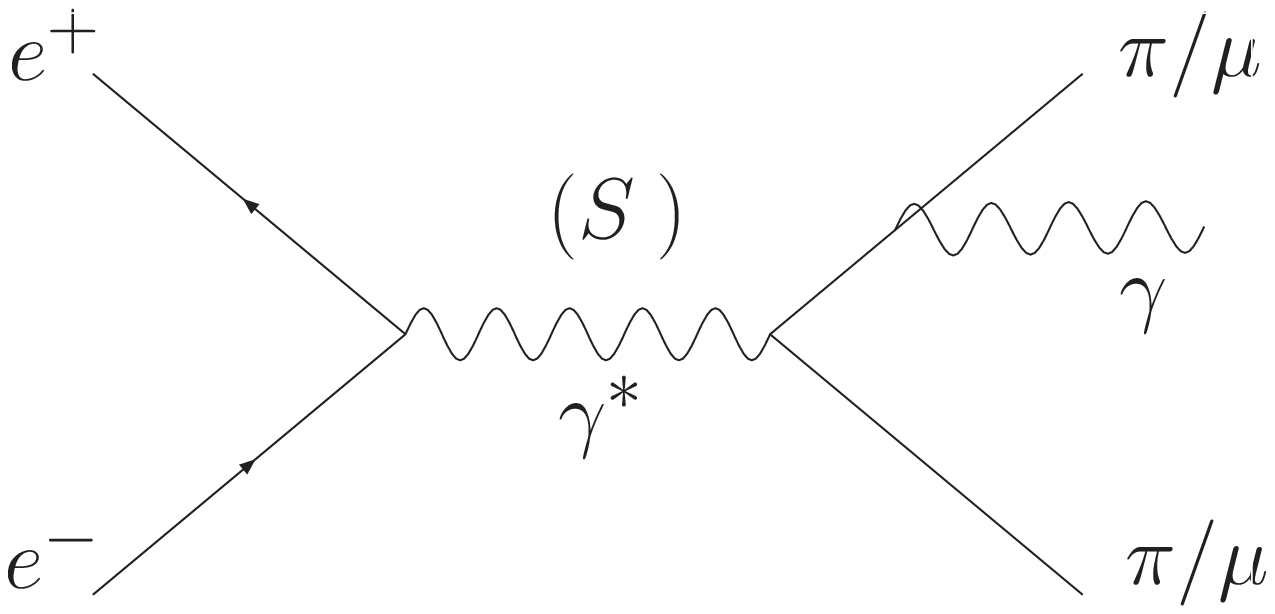} \\
  \vspace{10pt}
  \includegraphics[width=4.cm]{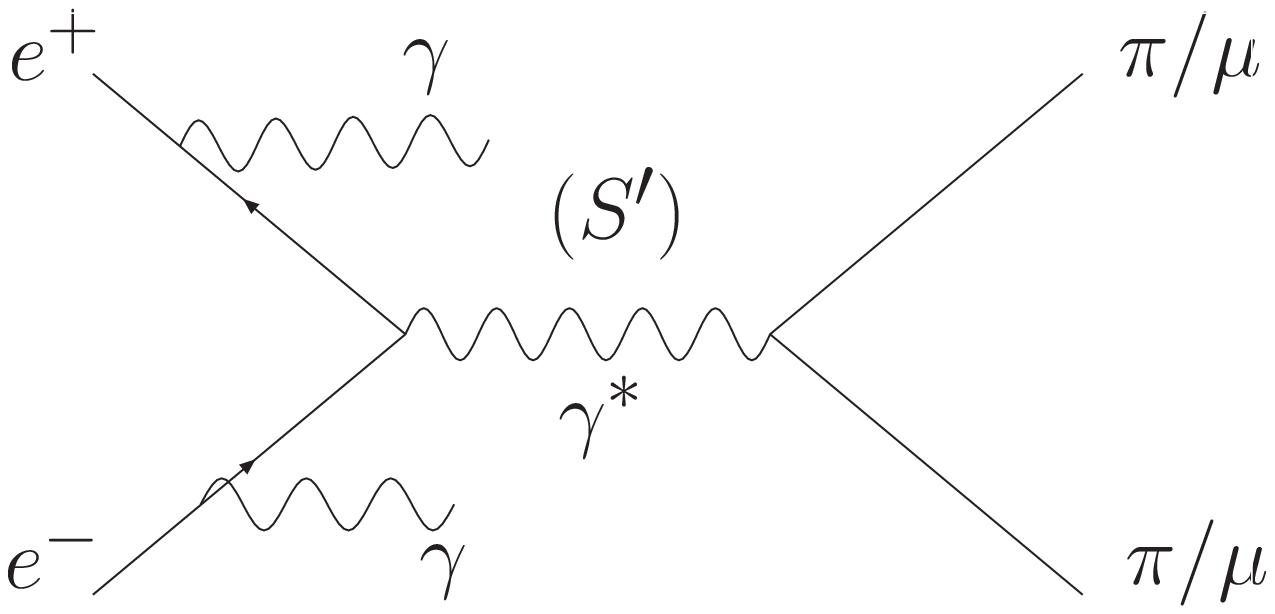}
  \hspace{5pt}
  \includegraphics[width=4.cm]{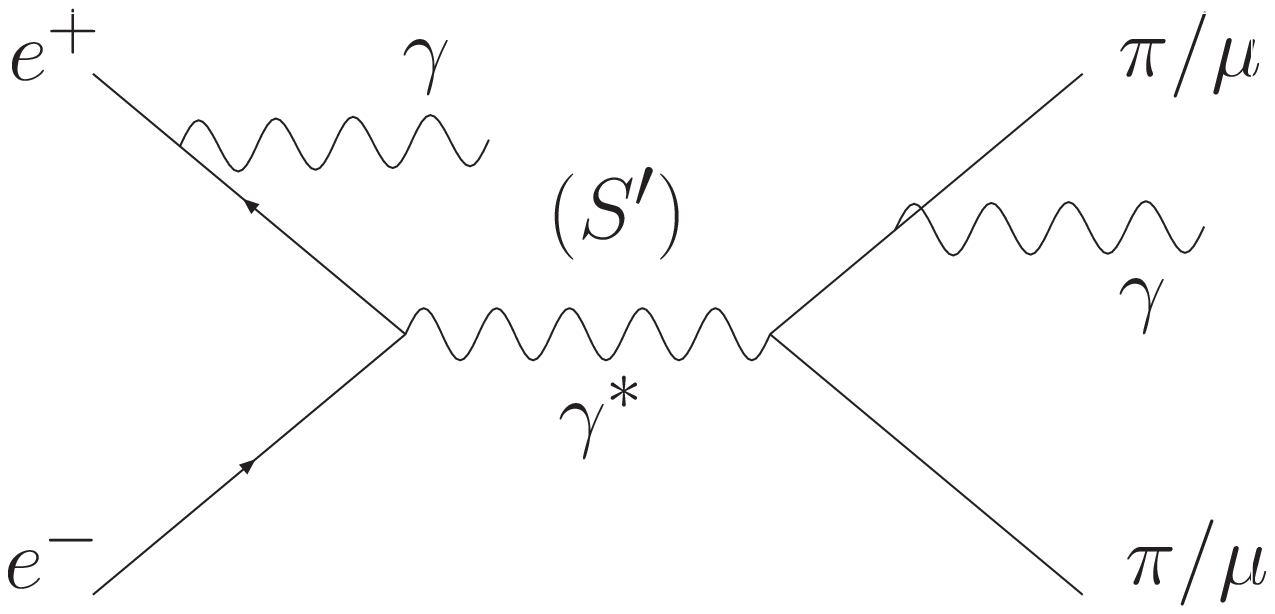}
  \caption{\label{diagrams} \small The generic Feynman diagrams for the processes
relevant to this study with one or two real photons: lowest-order (LO) ISR (top left),
LO FSR (top right), 
next-to-leading order (NLO) ISR with 
additional ISR (bottom left), NLO with additional FSR (bottom right).}
\end{figure} 

Figure~\ref{diagrams} shows the Feynman diagrams relevant to this study. 
The lowest-order (LO) radiated photon can be either from ISR or FSR. In the muon 
channel, ISR is dominant in the measurement range, but the LO FSR contribution
needs to be subtracted using QED. In the pion channel, the LO FSR 
calculation is model-dependent, but the contribution is strongly suppressed 
due to the large $s$ value. 
In both channels, interference between ISR and FSR 
amplitudes vanishes for a charge-symmetric detector. 

In order to control the overall efficiency to high precision, it is 
necessary to consider higher-order radiation. 
The next-to-leading-order (NLO) correction in $\alpha$ amounts to about 4\%~\cite{fsr-czyz} with 
the selection used for this analysis, 
while the next-to-next-to-leading-order (NNLO) correction is expected to be at least one order of 
magnitude smaller than NLO.  
Most of the higher-order contributions come from ISR and hence are independent of
the final state. As the cross section is measured through the $\pi\pi/\mu\mu$ ratio,
as explained below, most higher-order radiation effects cancel and  
NLO is sufficient to reach precisions of $10^{-3}$. 
As a result, the selection keeps $\pi\pi\gamma\gamma$ ($\mu\mu\gamma\gamma$) as well as $\pi\pi\gamma$
($\mu\mu\gamma$) final states, where the additional photon can be either ISR or FSR.

\subsection{Cross section measurement through the $\pi\pi/\mu\mu$ ratio}

The cross section for the process $e^+e^-\to X$ 
is related to the $\sqrt{s'}$ spectrum of $e^+e^-\to X\gamma_{\rm ISR}$ events through
\beqn    
\label{def-lumi}
  \frac {dN_{X\gamma_{\rm ISR}}}{d\sqrt{s'}}~=
   ~\frac {dL_{\rm ISR}^{\rm eff}}{d\sqrt{s'}}~
    \varepsilon_{X\gamma}(\sqrt{s'})~\sigma_{X}^0(\sqrt{s'})~,
\eeqn   
where $dL_{\rm ISR}^{\rm eff}/d\sqrt{s'}$ is the effective ISR luminosity, 
$\varepsilon_{X\gamma}$ is the full acceptance for the event sample,
and $\sigma_X^0$ is the `bare' cross section for the process 
$e^+e^- \to X$ (including additional FSR photons), in which the leptonic and 
hadronic vacuum polarization effects are removed.

Eq.~(\ref{def-lumi}) applies equally to $X=\pi\pi(\gamma)$ and $X=\mu\mu(\gamma)$ final states, so
that the ratio of cross sections is directly related to the ratio of the pion to muon spectra
as a function of $\sqrt{s'}$. 
Specifically, the ratio $R_{\rm exp}(\sqrt{s'})$ 
of the produced $\pi\pi(\gamma)\gamma_{\rm ISR}$ and $\mu\mu(\gamma)\gamma_{\rm ISR}$ spectra, 
obtained from the measured spectra corrected for full acceptance,
can be expressed as:
\beqn
   R_{\rm exp}(\sqrt{s'}) &=& 
\frac {\frac {\displaystyle dN^{\rm prod}_{\pi\pi(\gamma)\gamma_{\rm ISR}}}{d\sqrt{s'}}}
      {\frac {\displaystyle dN^{\rm prod}_{\mu\mu(\gamma)\gamma_{\rm ISR}}}{d\sqrt{s'}}}\\
\label{eq-3}
                      &=& 
\frac {\sigma^0_{\pi\pi(\gamma)}(\sqrt{s'})}
      {(1+\delta^{\mu\mu}_{\rm FSR})\sigma^0_{\mu\mu(\gamma)}(\sqrt{s'})}\\
                      &=& 
\frac {R^0(\sqrt{s'})}{(1+\delta^{\mu\mu}_{\rm FSR})(1+\delta^{\mu\mu}_{\rm add.FSR})}~.
\eeqn

The `bare' ratio $R^0$ (no vacuum polarization, but additional FSR included),
which enters the VP dispersion integrals, is given by
\beqn    
 R^0(\sqrt{s'})=\frac {\sigma^0_{\pi\pi(\gamma)}(\sqrt{s'})}{\sigma_{\rm pt}(\sqrt{s'})}~,
\eeqn    
where $\sigma_{\rm pt}=4\pi\alpha^2/3s'$ is the cross section for pointlike
charged fermions. The factor $(1+\delta^{\mu\mu}_{\rm FSR})$ corrects for the lowest-order 
FSR contribution, including possibly additional soft photons,
to the $e^+e^-\to\mu^+\mu^-\gamma$ final state,
as is explicitly given in Eq.~(\ref{fsr-delta}).
No such factor is included for pions because of the negligible LO FSR
contribution (see Sect.~\ref{lofsr-pi}).
The factor $(1+\delta^{\mu\mu}_{\rm add.FSR})$ corrects for additional FSR in the 
$e^+e^-\to\mu^+\mu^-$ process
at $\sqrt{s'}$,
as is explicitly given in Eq.~(\ref{born-fsr-sig}).

In this analysis, we use a procedure strictly equivalent to taking the ratio $R_{\rm exp}(\sqrt{s'})$,
namely we measure the $\sigma_{\pi\pi(\gamma)}^0(\sqrt{s'})$ cross section
using Eq.~(\ref{def-lumi}) in which the effective ISR luminosity is obtained from the 
mass spectrum of produced $\mu\mu(\gamma)\gamma_{\rm ISR}$ events divided by the 
$\sigma_{\mu\mu(\gamma)}^0(\sqrt{s'})$
cross section computed with QED. 
The ISR luminosity measurement is described in detail in Section\ref{isr-lumi}.

This way of proceeding considerably reduces the uncertainties related to the 
effective ISR luminosity function when determined through
\beqn
\label{lumi-eff}
 \frac {dL_{\rm ISR}^{\rm eff}}{d\sqrt{s'}}~=~L_{ee}~\frac {dW}{d\sqrt{s'}}~
   \left(\frac {\alpha(s')}{\alpha(0)}\right)^2~ 
   \frac {\varepsilon_{\gamma_{\rm ISR}}(\sqrt{s'})} {\varepsilon_{\gamma_{\rm ISR}}^{\rm MC}(\sqrt{s'})}~.
\eeqn
Eq.~(\ref{lumi-eff}) relies on the $e^+e^-$ luminosity measurement ($L_{ee}$) and 
on the theoretical radiator function $dW/d\sqrt{s'}$. The latter describes the probability
to radiate an ISR photon (with possibly additional ISR photons) 
so that the produced final state (excluding ISR photons) has a mass $\sqrt{s'}$. 
It depends on $\sqrt{s}$, on $\sqrt{s'}$, and on the angular range 
($\theta^*_{\rm min}$, $\theta^*_{\rm max}$) of the ISR photon in the $\mee$ 
c.m.\ system. 
For convenience, two factors that are common to the
muon and pion channels are included in the effective luminosity definition 
of Eq.~(\ref{lumi-eff}):
i) the ratio of $\varepsilon_{\gamma_{\rm ISR}}$, the 
efficiency to detect the main ISR photon, to the same quantity 
$\varepsilon_{\gamma_{\rm ISR}}^{\rm MC}$ in simulation, and ii) the vacuum polarization 
correction $(\alpha(s')/\alpha(0))^2$. The latter factor is implicitly included 
in the effective luminosity deduced from $\mu\mu(\gamma)\gamma_{\rm ISR}$ data using 
Eq.~(\ref{def-lumi}), while the former, which 
cancels out in the $\pi\pi$ to $\mu\mu$ ratio, is ignored in Eq.~(\ref{def-lumi}).
As an important cross-check of the analysis, hereafter called the QED test, 
we use Eq.~(\ref{lumi-eff}), together with Eq.~(\ref{def-lumi}), 
to measure the muon cross section and compare it to the QED prediction. 

Many advantages follow from taking the $R_{\rm exp}(\sqrt{s'})$ ratio:
\begin{itemize}
\item the result is independent of the \babar\ luminosity $L_{ee}$ measurement;
\item the determination of the ISR luminosity comes from the muon data,
independently of the number of additional ISR photons, and thus does not depend
on a theoretical calculation;
\item the ISR photon efficiency cancels out;
\item the vacuum polarization also cancels out.
\end{itemize}
Furthermore the Monte Carlo generator and the detector simulation are only used to 
compute the acceptance of the studied $X\gamma_{\rm ISR}$ processes, with 
$X=\pi\pi(\gamma),\mu\mu(\gamma)$.
The overall systematic uncertainty on the $\pi\pi$ cross section is 
reduced, because some individual uncertainties cancel between pions and muons.

\section{Analysis Outline}

\subsection{The \babar\ detector and data samples}

The analysis is based on  $232\invfb$ of data collected with the \babar\ 
detector at the SLAC PEP-II asymmetric-energy $e^+e^-$ storage 
rings operated at the $\FourS$ resonance.
The \babar\ detector is described in detail elsewhere~\cite{detector}.
Charged-particle tracks are measured with a five-layer double-sided 
silicon vertex tracker (SVT) together with a 40-layer drift chamber (DCH) 
inside a 1.5~T superconducting solenoid magnet. Photons are assumed to 
originate from the primary vertex defined by the charged tracks of the event 
and their energy is measured in a CsI(Tl) electromagnetic calorimeter (EMC).
Charged-particle identification (PID) uses the ionization losses $\dedx$
in the SVT and DCH, the Cherenkov radiation detected in a ring-imaging
device (DIRC), the shower energy deposit in the EMC ($E_{\rm cal}$) and the
shower shape in the instrumented flux return (IFR) of the magnet. The IFR 
system is constructed from modules of resistive plate chambers interspaced 
with iron slabs, arranged in a configuration with a barrel and two endcaps.

\subsection{Monte Carlo generators and simulation}

Signal and background ISR processes $e^+e^- \to X\gamma_{\rm ISR}$ are simulated with a Monte Carlo (MC) event
generator called AfkQed, which is based on the formalism of Ref.~\cite{eva}. 
The main ISR photon, $\gamma_{\rm ISR}$, is generated within the angular range 
[$\theta^*_{\rm min}=20^0$, $\theta^*_{\rm max}=160^0$]
in the c.m.\ system, bracketing the photon detection range with a 
margin to account for finite resolution. 
Additional ISR photons are generated with
the structure function method~\cite{struct-fct}, and additional FSR photons 
with {\small PHOTOS}~\cite{photos}. Additional ISR photons are emitted along the
$e^+$ or $e^-$ beam particle direction. A minimum mass $m_{X\gamma_{\rm ISR}}>8\gevcc$ is imposed at 
generation, which places an upper bound on the additional ISR photon energy. 
Samples corresponding to 5 to 10 times the number of data events are generated 
for the signal channels.
The more accurate Phokhara generator~\cite{phokhara} is used at the 4-vector level
to study some effects related to additional ISR photons.  
Background processes $e^+e^-\to\qqbar$ 
($q=u,d,s,c$) are generated with {\small JETSET}~\cite{jetset}, and $e^+e^-\to\tau^+\tau^-$ 
with {\small KORALB}~\cite{koralb}. The response 
of the \babar\ detector is simulated with {\small GEANT4}~\cite{geant}.

\subsection{Analysis method}

The $\pi\pi(\gamma)\gamma_{\rm ISR}$ and 
$\mu\mu(\gamma)\gamma_{\rm ISR}$ processes are
measured independently with full internal checks and the ratio $R_{\rm exp}(\sqrt{s'})$, 
which yields the measured $\sigma_{\pi\pi(\gamma)}^0(\sqrt{s'})$ cross section,
is only examined after these checks are successfully passed. One of the most demanding tests
is the absolute comparison of the $\mu\mu(\gamma)\gamma_{\rm ISR}$ 
cross section, which uses the \babar\ $L_{ee}$ luminosity, with the NLO QED prediction
(QED test).

After preliminary results were presented from the blind analysis~\cite{md-tau08},
a few aspects of the analysis were revisited to refine some
effects that had been initially overlooked, mostly affecting the correlated loss 
of muon identification for both tracks.  
While the final measurement is not a strictly blind analysis, 
all studies are again made independently for muons and pions and combined at the very
end. 

The selected events correspond to a final state with two tracks and the ISR
candidate, all within the detector acceptance, as described in Section~\ref{evt-sel}. 
Kinematic fits provide discrimination of the channels under study 
from other processes. However the separation
between the different two-prong final states (including $K^+K^-(\gamma)\gamma_{\rm ISR}$)
relies exclusively on the identification
of the charged particles. Thus particle identification plays a major role
in the analysis. This is the subject of Section~\ref{PID}. Background reduction and
control of the remaining background contributions are another challenge of
the analysis, in particular in the pion channel away from the $\rho$ resonance. 
This is discussed in Section~\ref{background}. 

The determination of the $\sqrt{s'}$ spectrum is described in Section~\ref{mass-spectrum}.
The relevant final-state mass is 
$m_{\pi\pi}$ ($m_{\mu\mu}$) when there is additional ISR or no additional radiation,
or $m_{\pi\pi\gamma}$ ($m_{\mu\mu\gamma}$) in the case of additional FSR.  
The $\sqrt{s'}$ spectrum is obtained from the observed $m_{\pi\pi}$ ($m_{\mu\mu}$) 
distributions through unfolding (Sect.~\ref{unfold}). 

Although selection of the final state of two-body ISR processes is rather simple,
the main difficulty of the analysis resides in the full control of all involved
efficiencies. Relying on the simulation alone cannot provide the required precision.
The simulation
is used in a first step in order to incorporate in a consistent way all 
effects entering the final event acceptance. Corrections for data-to-MC 
differences are obtained for each efficiency using dedicated studies performed
on the data and simulation samples. The main contributions for these 
corrections originate from trigger, tracking, particle 
identification, and the $\chi^2$ selection of the kinematic fits, 
so that the corrected efficiency is
\beqn
\label{eff-corr}
   \varepsilon = \varepsilon_{\rm MC}~\left (
   \frac {\varepsilon_{\rm trig}^{\rm data}}{\varepsilon_{\rm trig}^{\rm MC}}
      \right ) \left (
   \frac {\varepsilon_{\rm track}^{\rm data}}{\varepsilon_{\rm track}^{\rm MC}}
      \right ) \left (
   \frac {\varepsilon_{\rm PID}^{\rm data}}{\varepsilon_{\rm PID}^{\rm MC}}
      \right ) \left (
   \frac {\varepsilon_{\chi^2}^{\rm data}}{\varepsilon_{\chi^2}^{\rm MC}}
      \right )\,.
\eeqn
The corrections $C_i=\left (
   \frac {\varepsilon_i^{\rm data}}{\varepsilon_i^{\rm MC}}
      \right )$ are reviewed in turn in the following sections
(Sect.~\ref{eff} and \ref{kin-eff}). 
They are applied as mass-dependent corrections to the MC efficiency. 
They amount to at most a few percent and are known to a few permil level 
or better. Efficiency measurements are designed to avoid correlations between the
$C_i$. Further data-to-MC corrections deal with second-order 
effects related to 
the description of additional ISR in the generator, which was found inadequate at 
the level of precision required for this analysis.

\section{Event selection}
\label{evt-sel}

\subsection{Topological selection}

Two-body ISR events are selected by requiring a photon candidate with $E_\gamma^*>3\gev$
and laboratory polar angle in the range $0.35-2.4\rad$, 
and exactly two tracks of opposite charge, each with 
momentum $p>1\gevc$~\footnote{Unless otherwise stated, starred quantities are measured in the
$\mee$ c.m.\ and un-starred quantities in the laboratory.} and within the angular range $0.40-2.45\rad$.
A photon candidate is defined as a cluster in the EMC, with energy larger than $0.02\gev$,
not associated to a charged track. 
If several photons are detected, the main ISR photon is assumed to be that with the 
highest $E_\gamma^*$; this results in an incorrectly assigned ISR photon in 
less than $\sim10^{-4}$ of the events, mostly due to the ISR photon loss in 
inactive areas of the EMC.  
The track momentum requirement is dictated by the fall-off of the muon-identification 
efficiency at low momenta.
The tracks are required to have at least 
15 hits in the DCH, and originate within
$5\mm$ of the collision axis (distance of closest approach $doca_{\rm xy}<5\mm$) 
and within $6\cm$ from the beam spot along the beam direction
($|\Delta_z|<6\cm$).
They are required to extrapolate to the DIRC active area, whose length  
further restricts the minimum track polar angle to $\sim0.45\rad$.
Tracks are also required to extrapolate to the IFR active areas that exclude low-efficiency regions.
An additional veto based on a combination of 
$E_{\rm cal}$ and $\dedx$, $((E_{\rm cal}/p-1)/0.15)^2~+~((\dedx_{\rm DCH}-690)/150)^2~<~1$),
reduces electron contamination.
Events can be accompanied by any number of `bad' tracks, not satisfying the above criteria, 
and any number of additional photons. To ensure a rough momentum balance
at the preselection level (hereafter called `preselection cut'), the ISR photon is 
required to lie within $0.3\rad$ of the missing momentum of the tracks 
(or of tracks plus other photons).

\subsection{Kinematic Fit description and $\chi^2$ selection}
\label{kin-fit}

\begin{figure*}
  \begin{minipage}[ht]{0.45\textwidth}
  \centering
  \includegraphics[width=\textwidth]{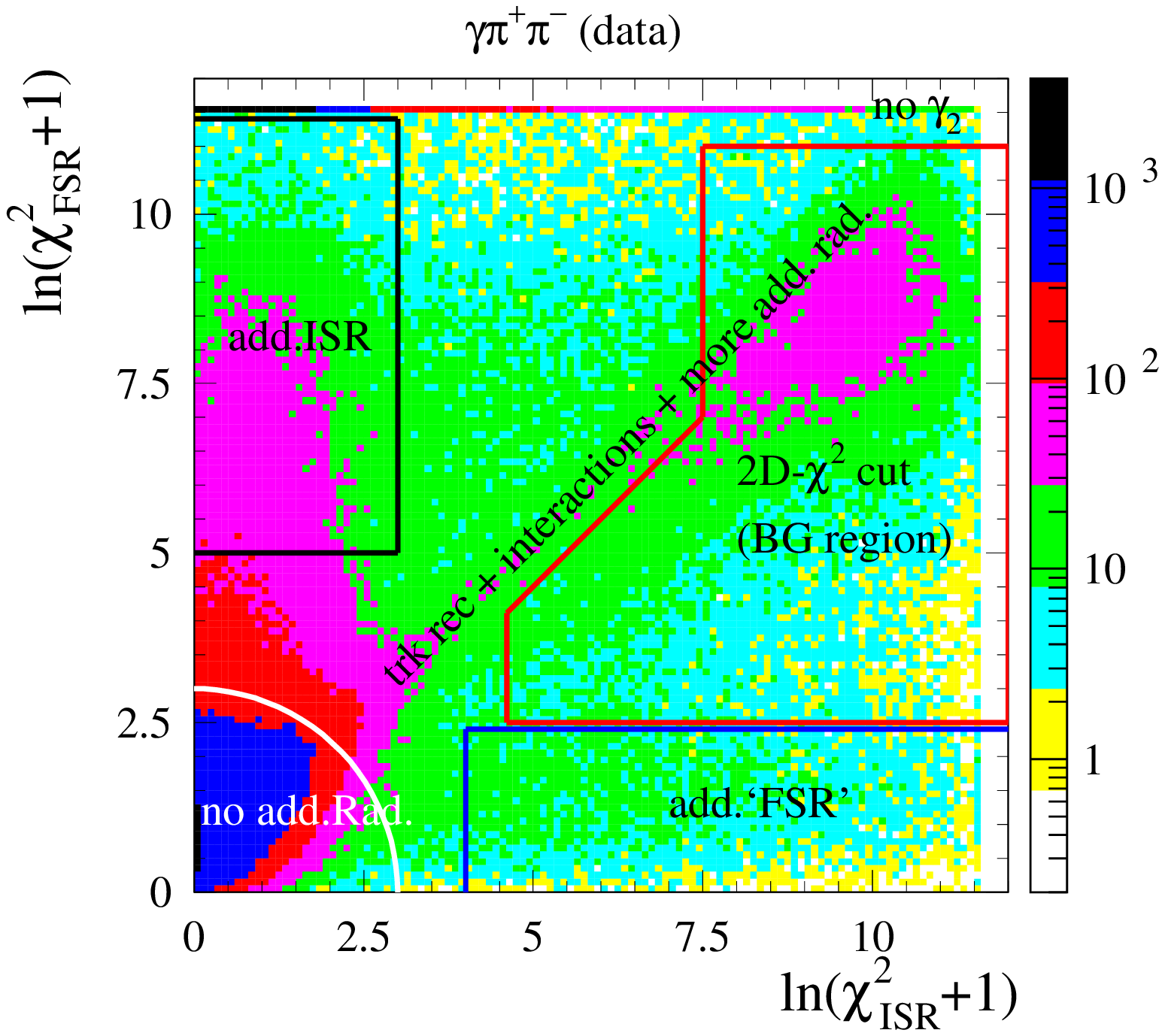}
  \caption{\small(color online). The 2D-$\chi^2$ distribution for $\pi\pi(\gamma)\gamma_{\rm ISR}$
 (data) for $0.5<m_{\pi\pi}<1.0\gevcc$, where
  different interesting regions are defined.}
  \label{pi-2d-chi2}
  \end{minipage}\hfill
  \begin{minipage}[ht]{0.45\textwidth}
  \centering
  \includegraphics[width=\textwidth]{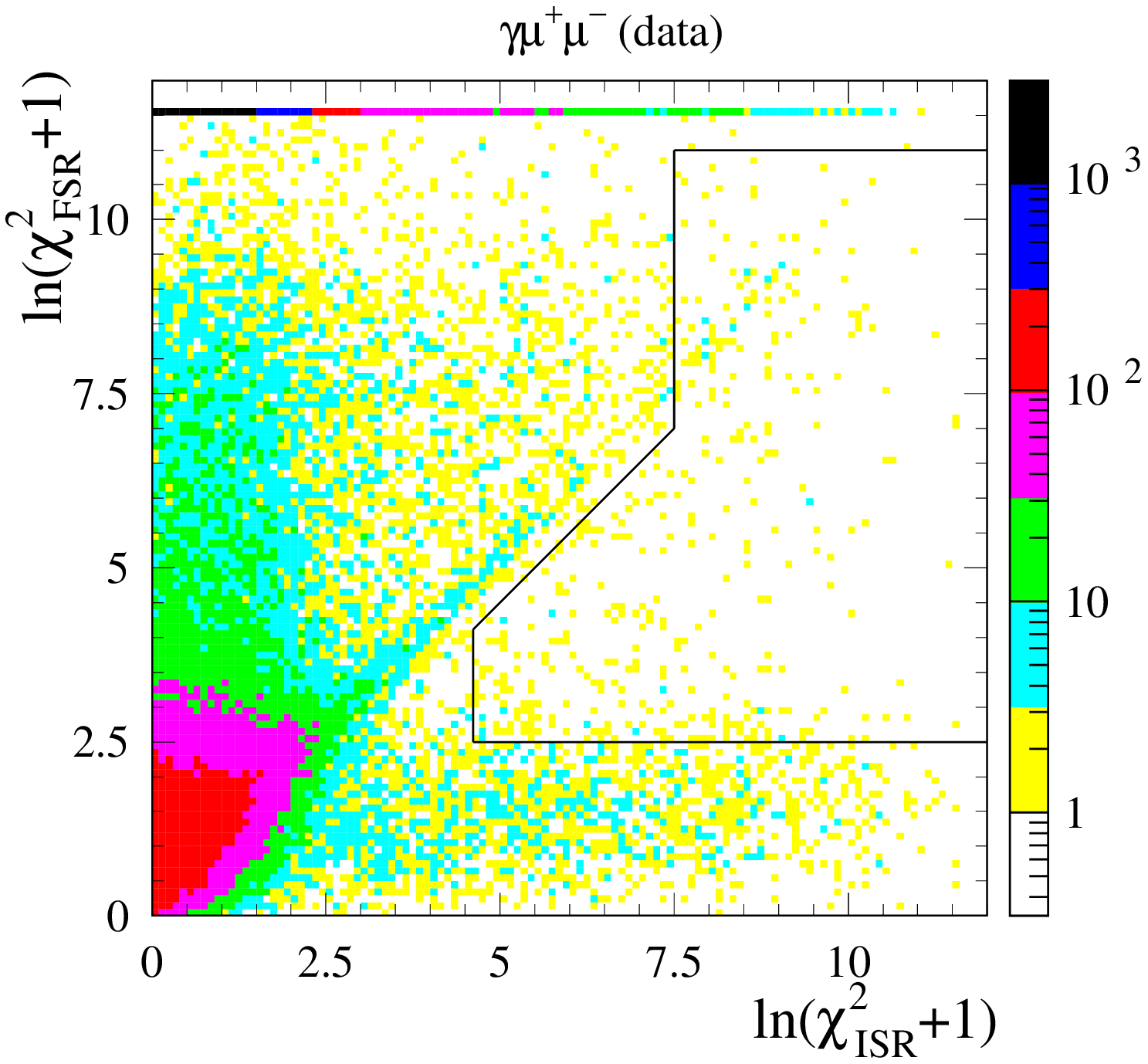}
  \caption{\small(color online). The 2D-$\chi^2$ distribution for $\mu\mu(\gamma)\gamma_{\rm ISR}$
(data) for $0.5<m_{\mu\mu}<1.0\gevcc$, where the signal and background regions 
are indicated.}
  \label{mu-2d-chi2}
  \end{minipage}
\end{figure*}

For both the $\mu\mu\gamma$ and $\pi\pi\gamma$ processes, the event definition is 
enlarged to include the radiation of one photon in addition to the already-required 
ISR photon. Two types of fits are considered, according to the 
following situations:
\begin{itemize}
\item The additional photon is detected in the EMC, in which case its energy 
and angles can be readily used in the fit: we call this a 3-constraint (3C) FSR fit, although 
the extra photon can be either from FSR or from ISR at large angle to the 
beams. The threshold for
the additional photon is kept low (20\mev). This can introduce some background,
but with little effect as the fit in that case would not be different in 
practice from a standard fit to the $\mu\mu(\gamma)\gamma_{\rm ISR}$ 
($\pi\pi(\gamma)\gamma_{\rm ISR}$) hypothesis.
\item The additional photon is assumed to be from ISR at a small angle to the 
beams. Since further information~\footnote{This is not strictly true as the 
missing photon could be completely reconstructed if the ISR photon energy were 
used in the kinematic fit. However tests have shown that the relative quality 
of this new information does not permit a significant improvement for the 
fitted direction of the additional ISR photon over the collinear assumption.}
is not available, it is presumed that the extra photon is perfectly aligned with 
either the $e^+$ or the $e^-$ beam. The corresponding so-called
2C ISR fit ignores additional photons measured in the EMC and determines the
energy of the fitted collinear ISR photon.
\end{itemize}

In both cases the constrained fit procedure uses the ISR photon direction and 
the measured momenta and 
angles of the two tracks with their covariance matrix in order to solve the
four energy-momentum conservation equations. 
The measured energy of the primary ISR photon is not used in either fit, as it adds 
little information for the relatively low masses involved. 

Each event is characterized by two $\chi^2$ values, $\chi^2_{\rm FSR}$ and
$\chi^2_{\rm ISR}$ from the FSR and ISR fits, respectively, which are examined 
on a 2D-plot.
In practice the quantities $\ln{(\chi^2+1)}$ are used so that the long tails
can be properly visualized 
(Figs.~\ref{pi-2d-chi2}, \ref{mu-2d-chi2}). 
Events without any extra measured
photons have only the $\chi^2_{\rm ISR}$ value and they are plotted separately
on a line above the $\chi^2_{\rm FSR}$ overflow.
In case several extra photons are detected, FSR fits are performed using
each photon in turn and the fit with the smallest $\chi^2_{\rm FSR}$ is retained.
The muon (pion) mass is assumed for the two charged particles, according to 
the selected channel, and in the following studies and final distributions, 
the $\mu\mu$ ($\pi\pi$) mass is obtained using the fitted parameters of the 
two charged particles from the ISR fit if $\chi^2_{\rm ISR}<\chi^2_{\rm FSR}$ and 
from the FSR fit in the reverse case.

It is easy to visualize the different interesting regions in the 2D-$\chi^2$
plane, as illustrated in Fig.~\ref{pi-2d-chi2} for $\pi\pi(\gamma)\gamma_{\rm ISR}$ data. 
Most of the events peak at small values of both $\chi^2$, but the tails 
along the axes clearly indicate events with additional radiation: 
small-angle ISR along the $\chi^2_{\rm FSR}$ axis (with large ISR energies at large
values of $\chi^2_{\rm FSR}$), or FSR or large-angle ISR along the $\chi^2_{\rm ISR}$ 
axis (with large additional radiation energies at large values of $\chi^2_{\rm ISR}$). 
Events along the diagonal do not satisfy either hypothesis and result from resolution
effects for the pion tracks (also secondary interactions) or the primary ISR 
photon, or possibly additional radiation of more than one photon.
Multibody background populates the region where both $\chi^2$ 
are large and consequently a background region is defined in the 2D-$\chi^2$
plane. This region is optimized as a compromise between efficiency 
and background contamination in the signal sample, aiming at best control of the 
corresponding systematic uncertainties.

The $\chi^2$ criteria used in the pion analysis depend on the $\pi\pi$ mass 
region considered. 
The $m_{\pi\pi}$ region between 0.5 and 1\gevcc is dominated by the $\rho$ resonance.
The corresponding large cross section provides a dominant contribution 
to vacuum-polarization dispersion integrals, so it has to be known with small
systematic uncertainties. Also background is expected to be at a small level in this
region. These two considerations argue for large efficiencies, in order
to keep systematic uncertainties sufficiently low. Therefore a loose $\chi^2$ criterion is
used, where the physical (accepted) region corresponds to the left of
the contour outlined in Fig.~\ref{pi-2d-chi2}, excluding the BG-labeled region.
The same loose $\chi^2$ criterion is applied for the $\mu\mu(\gamma)\gamma_{\rm ISR}$ analysis
(Fig.~\ref{mu-2d-chi2}).

\begin{figure*}[htp]
  \centering
  \includegraphics[width=0.45\textwidth]{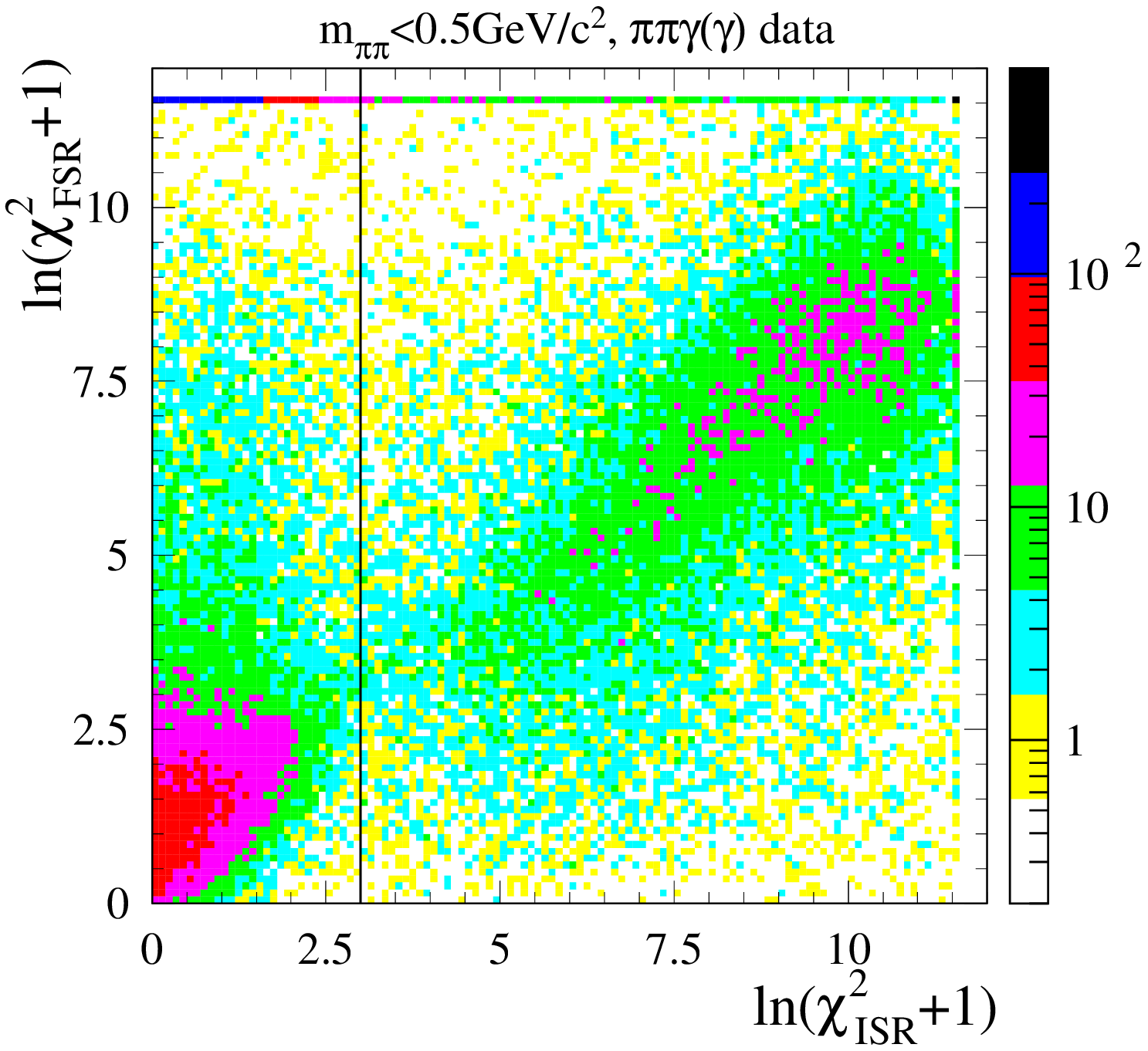}
  \includegraphics[width=0.45\textwidth]{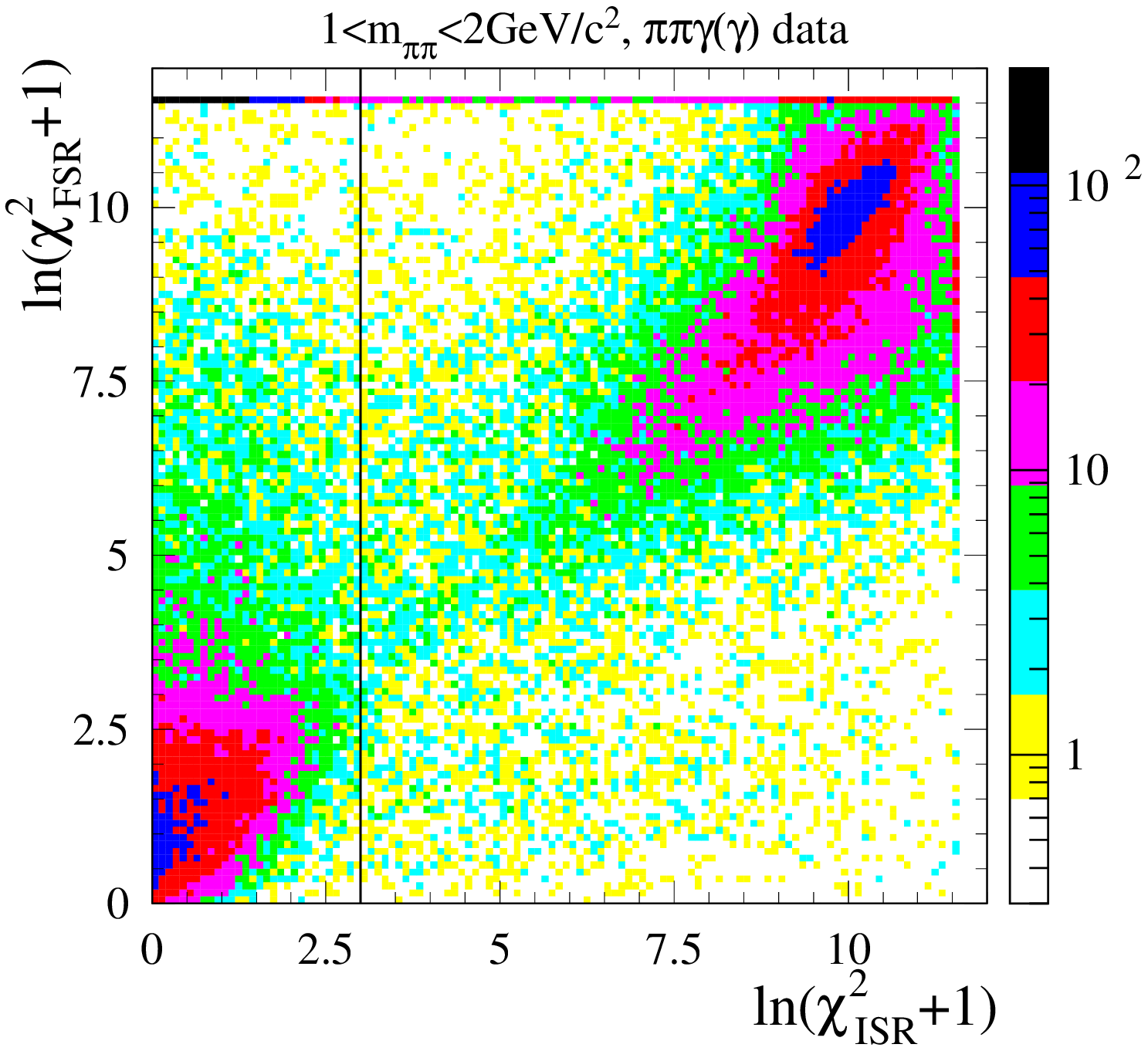}
  \caption{\small(color online). The 2D-$\chi^2$ distributions in $\pi\pi(\gamma)\gamma_{\rm ISR}$ data: 
  (left) below the central $\rho$ region ($m_{\pi\pi}<0.5\gevcc$); 
  (right) above the central $\rho$ region ($1.<m_{\pi\pi}<2.\gevcc$). 
  The line indicates the boundary for the tight $\chi^2$ selection.}
  \label{chiISR-lt-3}
\end{figure*}

The pion form factor decreases rapidly away from the $\rho$ peak, while the
backgrounds vary slowly with the $\pi\pi$ mass. The
multihadronic background in the physical sample becomes excessively large 
if the $\chi^2$ criterion as used in the $\rho$ region is applied, and 
it is necessary to tighten the selection of $\pi\pi(\gamma)\gamma_{\rm ISR}$ events.
Figure~\ref{chiISR-lt-3} shows the tight $\chi^2$ selection boundary 
$\ln(\chi^2_{\rm ISR}+1)<3$ chosen to reduce multihadronic background, and the
2D-$\chi^2$ distributions for masses below and above the central $\rho$ region.
The tight $\chi^2$ criterion retains events with additional ISR since this region in the 
$\chi^2$ plane is free of multihadronic background. The reduced 
efficiency on signal from the tight selection results in a larger relative 
uncertainty, but this is still acceptable considering the
much smaller contribution from the $\rho$ tails to the dispersion integral.

Efficiencies and systematic uncertainties resulting from the loose and tight 
$\chi^2$ selection criteria are discussed in Section~\ref{eff-chi2}.

\section{Efficiency Studies (I)}
\label{eff}
To achieve the required precision for the cross section measurement, efficiencies are validated with data
at every step of the event processing, and mass-dependent data/MC corrections are determined. 
This necessitates specific studies on data control samples whose selection criteria are designed 
to minimize biases on efficiency measurements. Residual effects are estimated and included
in the systematic errors.

\subsection{Efficiency-dedicated event selection and kinematic fit}

For trigger and tracking efficiency studies, a dedicated selection of 
$\mu^+\mu^-\gamma_{\rm ISR}$ and $\pi^+\pi^-\gamma_{\rm ISR}$ events is devised that
only requires one reconstructed track (called `primary'), identified as a muon or pion,
and the ISR photon.
A 1C kinematic fit is performed and the momentum vector of the second 
muon(pion) is predicted from 4-momentum conservation. Standard
track selection is applied to the primary track and the predicted track is 
required to be in the acceptance.

\subsection{Trigger and filtering}

A number of trigger conditions are imposed at the hardware (L1) and online software (L3) levels, 
as well as in a final filtering, before an event is fully reconstructed and 
stored in the \babar\ data sample. They are common to all \babar\ analyses, 
and hence are not specifically designed to select ISR events.
Since individual trigger and filter line responses are stored for 
every recorded event, efficiencies can be computed by comparing the response of trigger 
lines, after choosing lines that are as orthogonal and as efficient as possible. 
Trigger efficiencies are determined on data and simulation samples, after 
applying identical event selections and measurement methods, and data/MC 
corrections $C_{\rm trig}$ are computed from the comparison of measured efficiencies on
background-subtracted data and signal MC. Once the physics origins of inefficiencies are 
identified, uncertainties are estimated through studies of biases 
and data-to-MC comparison of distributions of relevant quantities.
Efficiencies and data/MC corrections are measured separately for the pion and muon 
channels. 

Trigger efficiencies are determined on samples 
unbiased with respect to the number of tracks actually reconstructed, to avoid
correlations between trigger and tracking efficiency measurements. 
In practice, one- and two-track samples are sufficient and consequently the trigger control 
samples are selected through the dedicated 1C kinematic fit described above. 
Because of the loose 
requirement with respect to tracking, the data samples contain backgrounds
with potentially different trigger efficiencies to that of the 
signal. These backgrounds are studied with simulation and are then subtracted.
To obtain data samples that are as pure as possible, criteria tighter than the 
standard track selection are applied to the primary track, including
tight PID identification.
Possible biases resulting from the tighter 
selection are studied and accounted for in the systematic errors. 
Background contributions are subtracted from the data spectra using properly-normalized 
simulated samples, and, if necessary, with data/MC correction of the trigger efficiencies 
in an iterative procedure. 

\begin{figure}[htb]
  \begin{center}
  \includegraphics[width=7.5cm]{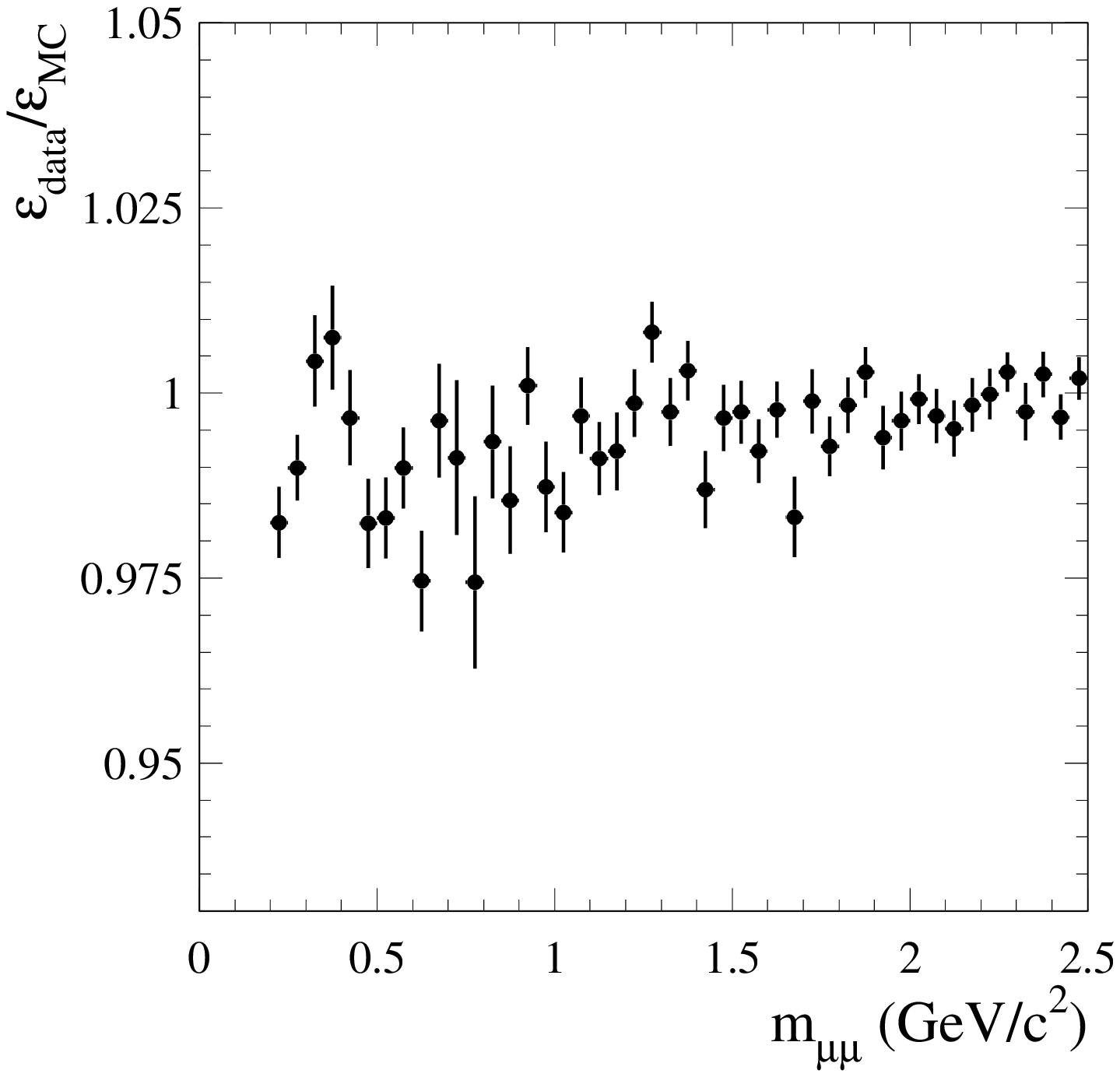}
  \includegraphics[width=7.5cm]{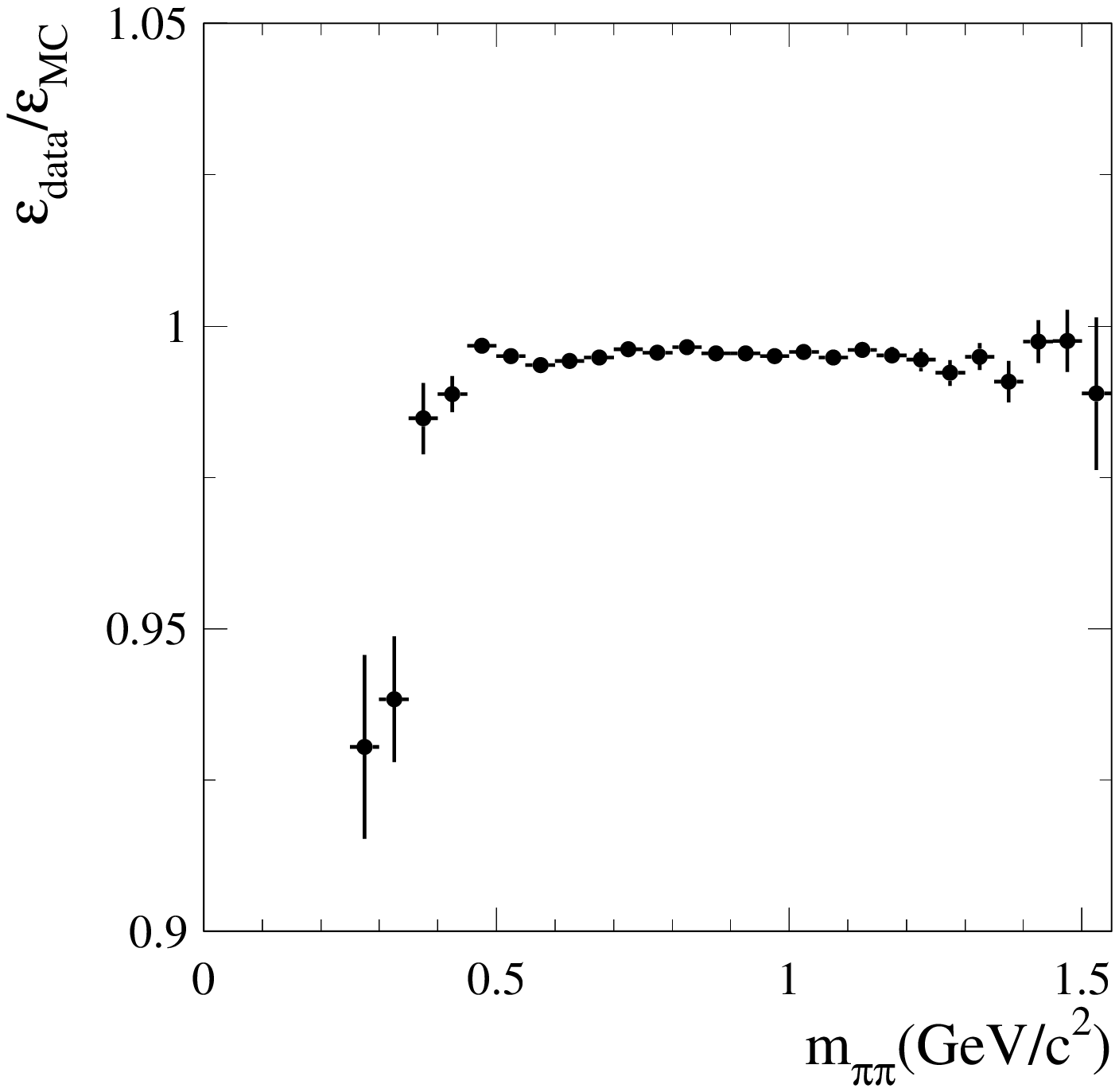}
  \caption{\small
The data/MC event trigger and filter correction $C_{\rm trig}$ for the $\mu\mu(\gamma)\gamma_{\rm ISR}$
(top) and $\pi\pi(\gamma)\gamma_{\rm ISR}$ (bottom) cross sections as a function of the $\mu\mu$
and $\pi\pi$ masses, respectively. Statistical errors only.}
  \label{c-trig}
  \end{center}
\end{figure}

The data/MC corrections for the L1 trigger are found to be
at a few $\times 10^{-4}$ level for muon and pion events. 
The L3 level involves a track trigger (at least one
track is required) and a calorimetric trigger (demanding at least one high-energy 
cluster and one low-energy cluster). 
Both of them are efficient for $\pi\pi\gamma_{\rm ISR}$ events. 
For $\mu\mu\gamma_{\rm ISR}$ events, the small efficiency of the calorimetric trigger
limits the statistical precision of the track-trigger and overall efficiency measurements.
Furthermore, a correlated
change of the two trigger line responses for close-by tracks induces both a non-uniformity
in the efficiency and a bias in the efficiency measurement. This
originates from the overlap of tracks in the drift chamber and of showers in the EMC,
which induces a simultaneous decrease in the track-trigger efficiency and an increase in the
calorimetric-trigger efficiency.  
Overlap is a major source of overall inefficiency and 
difference between data and simulation, necessitating specific studies.
The correction to the MC L3 trigger efficiency is small for pions, about 
$2\times 10^{-3}$ at the $\rho$ peak, and known to a precision better than $10^{-3}$.
The data/MC correction $C_{\rm trig}$ is larger in the $\mu\mu(\gamma)\gamma_{\rm ISR}$ channel,
due to the dominant role of the track trigger, about 1\% at a $\mu\mu$ mass of 0.7\gevcc,
and known to a precision of $3\times 10^{-3}$ (Fig.~\ref{c-trig}\,top). Uncertainties, 
which increase to $5\times 10^{-3}$ 
at the maximum overlap ($m_{\mu\mu}\sim0.4\gevcc$), are mostly statistical in nature.

The offline event filtering 
involves a large number of specific selections, including
dedicated $\mu\mu\gamma$ filters. Whereas the inefficiency and 
its correction are negligible for muons, some inefficiency at the filtering stage is observed
for pion events, mostly at low $m_{\pi\pi}$ mass. This originates again from 
the overlap of tracks in the DCH and hadronic showers in the EMC. 
The correction $C_{\rm trig}$ to the $\pi\pi(\gamma)\gamma_{\rm ISR}$ cross section 
(Fig.~\ref{c-trig}, bottom) is found to be
$(1.0\pm0.3\pm0.3)$\% at $m_{\pi\pi}\sim0.4\gevcc$ and $(2.9\pm0.1\pm1.0)\times 10^{-3}$ at the $\rho$ peak,
where the first error is statistical and the second systematic. 
Beyond $1.5\gevcc$, the background level precludes a significant measurement of the efficiency
with data and no correction is applied; a systematic error of  $0.4\times 10^{-3}$ is assigned
in the high mass range, equal to the inefficiency observed in MC.
Due to imperfect simulation of hadronic showers, filtering is the major source of trigger
systematic uncertainties in the pion channel. 

Systematic errors due to the trigger and filter are reported in Tables~\ref{mu-syst-err}
and~\ref{pi-syst-err}, for muon and pion channels, respectively.

\subsection{Tracking}

The tracking control samples of $\mu\mu\gamma_{\rm ISR}$ ($\pi\pi\gamma_{\rm ISR}$) events 
are selected through the efficiency-dedicated 1C fit described above. The rate of
predicted tracks that are actually reconstructed in the tracking 
system, with a charge opposite to that of the primary track, yields the 
tracking efficiency.

To ensure the validity of the measurement, further criteria are applied to the
tracking sample in addition to the kinematic fit. To enhance purity, a $\pi^0$ 
veto is applied if a pair of additional photons in the event can form a $\pi^0$ 
candidate with mass within $15\mevcc$ of the nominal mass.
This $\pi^0$ veto is not applied to the pion tracking sample, because of the bias 
it would introduce on the inefficiency related to secondary interactions.
The events are required to pass the triggers and online filter and are 
selected without specific requirements on the second reconstructed track, 
if any. Biases affecting the tracking-efficiency
measurement introduced by the primary-track selection or 
event-level background-rejection criteria are identified and studied with simulation,
and evaluated with data.

The predicted track is required to lie within the tracking 
acceptance, taking into account the effect of angular and 
momentum resolution. 
The method therefore determines the efficiency to
reconstruct a given track in the SVT+DCH system within a specified
geometrical acceptance, no matter how close or distant this track is with
respect to the expected one, due for instance to decays or secondary interactions. 
However, the possible mismatch in momentum and/or 
angles affects the full kinematic reconstruction of the event, and its 
effect is included in the efficiency of the corresponding $\chi^2$ 
selection applied to the physics sample (Sect.~\ref{sec-int-study}). 
Likewise, effects from pion decays 
in the detector volume are included in the pion-identification efficiency. 

The individual track efficiency is determined assuming
that the efficiencies of the two tracks are uncorrelated. However,
the tracking efficiency is observed to be sharply reduced for
overlapping tracks in the DCH,
as measured by the two-track opening angle ($\Delta{\phi}$) in the plane transverse to the beams.
Not only is the individual track efficiency  
locally reduced, but a correlated loss of the two tracks is observed.
In addition, as the final physics sample is required to have two and only two good tracks with 
opposite charge, the understanding of the tracking involves not only track 
losses, but also the probability to reconstruct extra tracks as a result of 
secondary interactions with the detector material or the presence of 
beam-background tracks.
The full tracking efficiency is then the product of the square of the 
single-track efficiencies, the probability for not losing the two tracks in a 
correlated way (loss probability $=f_0$), and the probability for not having 
an extra reconstructed track (loss probability $=f_3$). 
The event correction $C_{\rm track}$ to be applied to the MC is the
corresponding product of the data/MC ratios of each term. 
The mass-dependent quantities $f_0$
and $f_3$ are in the $(0.5-2.5)\times 10^{-3}$ range.

For muons the single-track inefficiency and the data/MC correction are driven
by the DCH overlap effect. At the maximum overlap 
($m_{\mu\mu}\sim0.4\gevcc$) the inefficiency reaches 1.7\% in simulation, but
2.5\% in data, while the intrinsic reconstruction, measured
for non-overlapping tracks, accounts for an inefficiency of $2.5\times 10^{-3}$ in data,
and $5\times 10^{-4}$ in simulation. 

Because of backgrounds, the pion tracking efficiency can be obtained 
directly from data only in the $\rho$ peak region, from $0.6$ to $0.9\gevcc$. 
The main sources of track loss are identified: the 
track overlap in the DCH and the secondary interactions. The two effects
are separated using the $\Delta{\phi}$ distribution. 
This two-component model is used to extrapolate the inefficiency to mass regions 
outside the $\rho$ peak. Results for pions are qualitatively similar to those for muons,
with inefficiencies driven by the track overlap effects.
The intrinsic track inefficiency is dominated by secondary interactions 
(2.2\% in data and 1.7\% in simulation) and is thus much larger than for muons. 
Near 0.4\gevcc the track inefficiency is determined to be 6.2\% in data and 
4.7\% in simulation. Above 1.2\gevcc for pions, where the data/MC correction is not 
expected to vary significantly, a systematic uncertainty of about 0.3\% is assigned. 

\begin{figure}[htp]
  \centering
  \includegraphics[width=7.5cm]{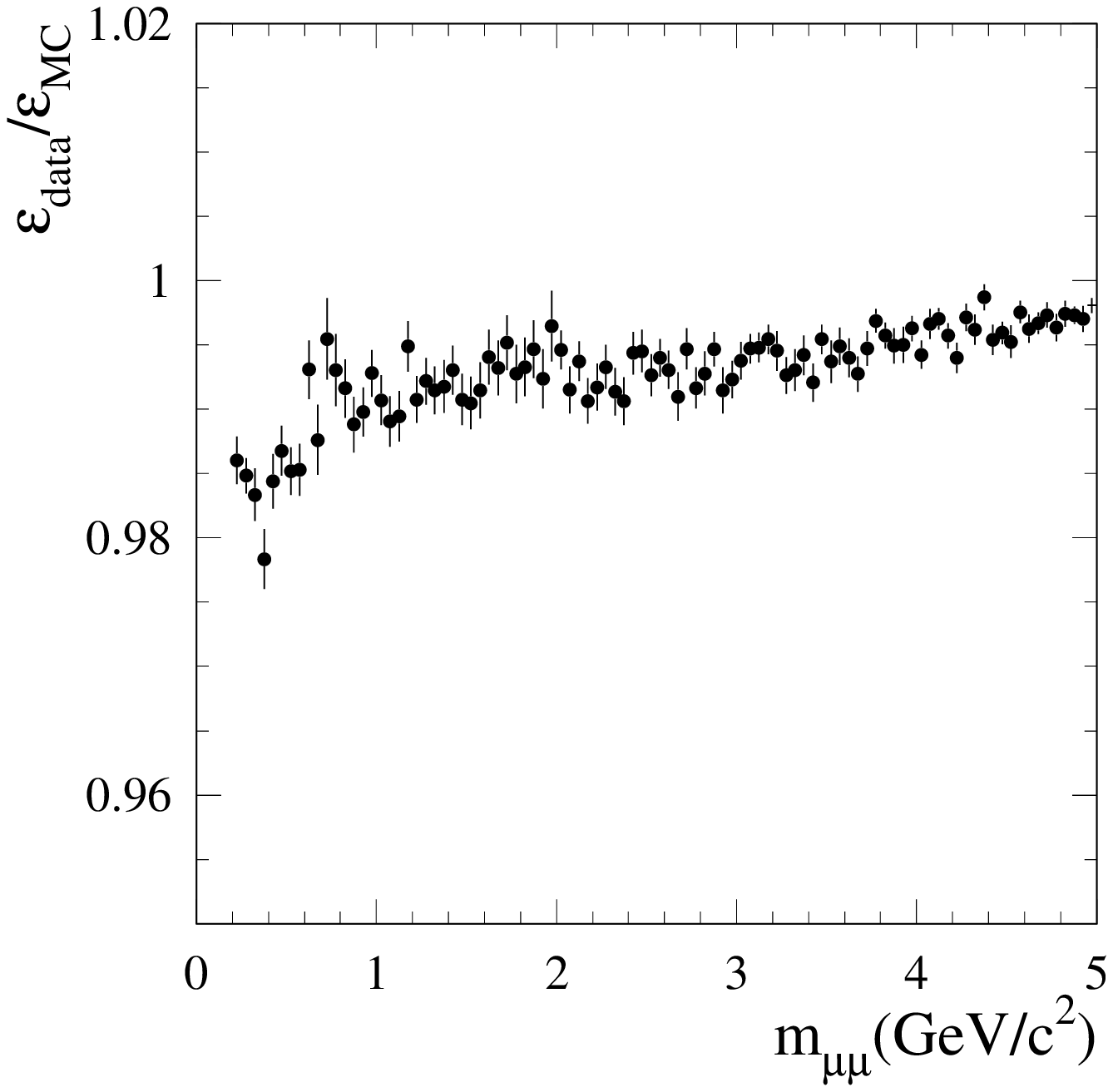}
  \includegraphics[width=7.5cm]{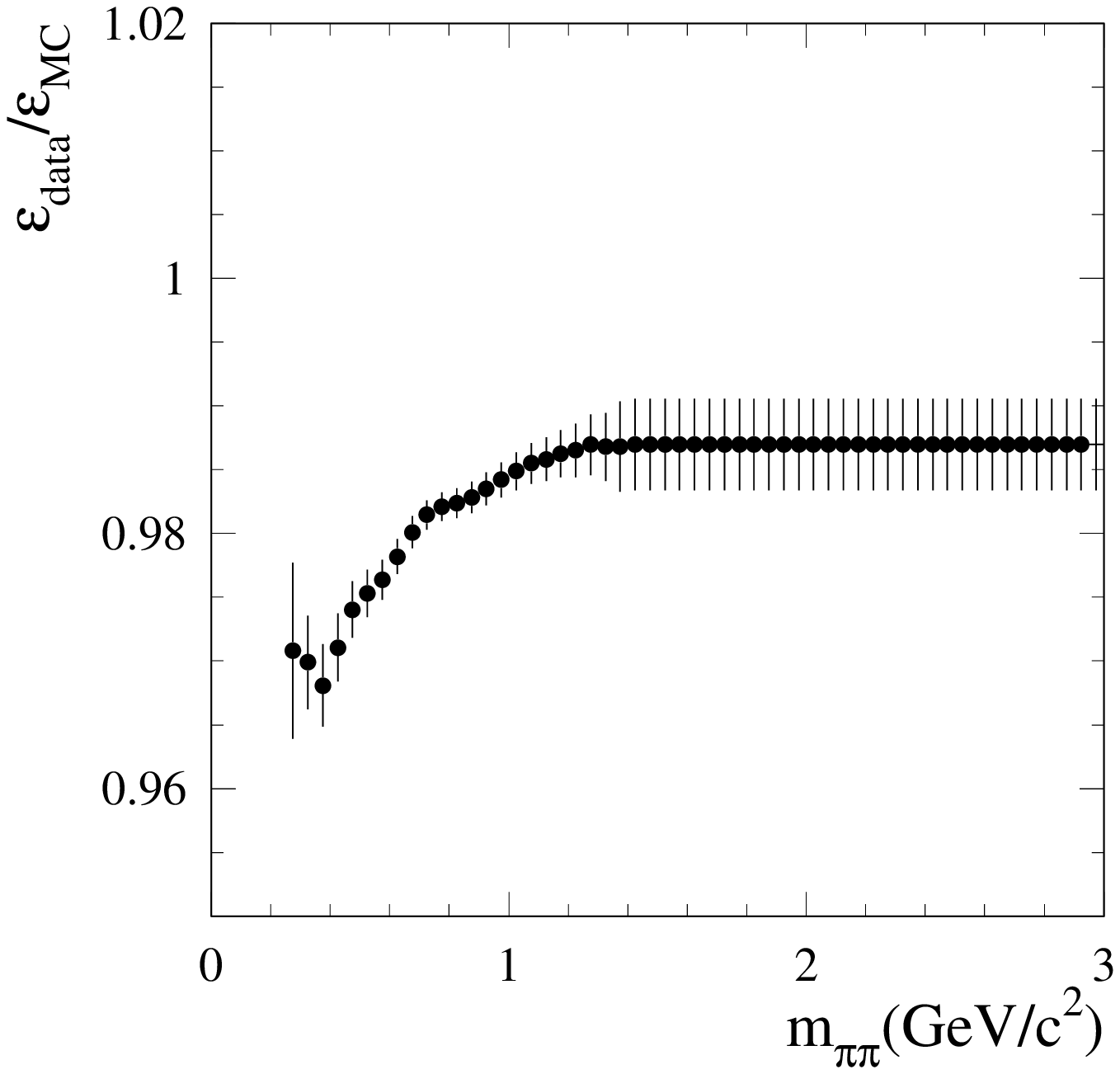}
  \caption{\small
The data/MC event tracking correction $C_{\rm track}$ for the $\mu\mu(\gamma)\gamma_{\rm ISR}$
(top) and $\pi\pi(\gamma)\gamma_{\rm ISR}$ (bottom) cross sections as a function of the $\mu\mu$
and $\pi\pi$ masses, respectively.}
  \label{tracking}
\end{figure}

The final corrections $C_{\rm track}$ to the $\mu\mu(\gamma)\gamma_{\rm ISR}$ and 
$\pi\pi(\gamma)\gamma_{\rm ISR}$ 
cross sections are presented in Fig.~\ref{tracking}.
$C_{\rm track}$ differs from unity by about 1.6\% (3.0\%) at $0.4\gevcc$, and by 0.8\% (1.5\%) 
at $1\gevcc$ for muons (pions).
Statistical uncertainties from the efficiency measurements are indicated by 
point-to-point errors. Systematic uncertainties are estimated from the study of
biases in the method, determined using the simulation and calibrated 
with data-to-MC comparison of distributions characteristic of the physics source of the bias.
Systematic uncertainties amount to $0.8\times 10^{-3}$
for muons in the mass range from 0.4 to $1.0\gevcc$, and are about a factor of two 
larger outside. For pions, the systematic uncertainty of the correction is
$1.1\times 10^{-3}$ in the $0.6-0.9\gevcc$ mass range, increasing to $2.1\times 10^{-3}$ 
($0.4-0.6\gevcc$), $3.8\times 10^{-3}$ (below 0.4\gevcc), $1.7\times 10^{-3}$ ($0.9-1.2\gevcc$),
and $3.1\times 10^{-3}$ (above 1.2\gevcc).

\subsection{Particle identification}
\label{PID}

The method to determine the PID efficiencies makes use of the $x^+x^-\gamma_{\rm ISR}$
sample itself, where one of the produced charged $x$ particles ($x=\mu,\pi,K$) 
is tagged using
strict identification criteria, and the second (`opposite') track identification is probed
(`tag-and-probe' method). 
The events are selected through a 1C kinematic fit that uses only the two 
tracks, with assumed mass $m_x$. The requirement $\chi^2_{xx}<15$ is applied to reduce 
multihadronic background. In this way the ensemble of opposite tracks constitutes a pure 
$x$ sample to be subjected to the identification process. 
The residual small impurity in the data samples is measured and corrected
in the efficiency determination. The same analysis is performed on MC
samples of pure $x^+x^-\gamma_{\rm ISR}$ events, and data/MC corrections 
$C_{\rm PID}$ are determined for each $x$ type, as explained below. 
Since the PID efficiency measurement relies on two-track events that have passed the triggers,
$C_{\rm PID}$ is not correlated with $C_{\rm trig}$ or $C_{\rm track}$, as required by Eq.~(\ref{eff-corr}).

\subsubsection{Particle identification classes}

Particle ID measurements in this analysis aim to obtain from data the
values for all the elements $\epsilon_{i\to`j\rq}$ of the 
efficiency matrix, where $i$ is the true $e$, 
$\mu$, $\pi$, or $K$ identity and `$j$' is the assigned ID from the PID procedure
 (Table~\ref{PID_def}). 
Protons (antiprotons) are not included in the particle
hypotheses because the $p\overline{p}\gamma$ final state occurs only at a
very small rate~\cite{ppb}. This contribution is estimated from simulation, normalized to data,
and subtracted statistically from the mass spectra.

We identify muon candidates by applying criteria on several
discriminant variables related to the track, such as the energy deposition 
$E_{\rm cal}$ in
the EMC, and the track length, hit multiplicity, matching between hits and extrapolated track 
in the IFR. This defines the $\mu_{\rm ID}$ selector. 
The $K_{\rm ID}$ selector is constructed from a likelihood function using the distributions
of $\dedx$ in the DCH and of the Cherenkov angle in the DIRC.
The electron identification relies on a simple $E_{\rm cal}/p>0.8$ requirement. 
As most of the electrons are vetoed at the preselection level, their fraction
in the pion sample is generally small. Their 
contribution is completely negligible in the muon sample.
 
In addition to physical particle types, we assign an ID type of `$0$' 
if the number of DIRC photons associated with the track
($N_{\rm DIRC}$) is insufficient 
to define a Cherenkov ring, thus preventing $\pi$-$K$ separation.
The ID classes defined in Table~\ref{PID_def} constitute a complete
and orthogonal set that is convenient for studying cross-feed between
different two-prong ISR final states.

The $\pi$-ID selector is a set of negative conditions since no set of
positive pion-ID criteria was found that provided both sufficient efficiency and purity. 
Pion candidates are tracks that do not satisfy any of the other ID class requirements. 
In this sense the
pion-ID is sensitive to the problems affecting the identification of all 
the other particle types.

\begin{table} \centering 
\caption{\small
Definition of particle ID types (first column) using combinations of 
experimental conditions (first row): ``+'' means ``condition satisfied'', 
``$-$'' means ``condition not satisfied'', an empty box means 
``condition not applied''. The conditions $\mu_{\rm ID}$ and $K_{\rm ID}$ 
correspond to cut-based and likelihood-based selectors, respectively. 
The variables $N_{\rm DIRC}$ and $E_{\rm cal}$ correspond to the
number of photons in the DIRC and the energy deposit in the EMC associated 
to the track, respectively.}
\vspace{0.1cm}
\label{PID_def}
\begin{tabular}{|c|c|c|c|c|} \hline\hline\noalign{\vskip2pt}
  & $\mu_{\rm ID}$ & $E_{\rm cal}/p>0.8$ & $N_{\rm DIRC}\leq 2$ & $K_{\rm ID}$ \\ \hline\noalign{\vskip2pt}
`$\mu$' & $+$ &   &  &          \\ \hline\noalign{\vskip2pt}
 `$e$'  & $-$ & $+$ &  &        \\ \hline\noalign{\vskip2pt}
 `$0$'  & $-$ & $-$ & $+$ &     \\ \hline\noalign{\vskip2pt}
 `$K$'  & $-$ & $-$ & $-$ & +   \\ \hline\noalign{\vskip2pt}
`$\pi$' & $-$ & $-$ & $-$ & $-$ \\ \hline\hline
\end{tabular}
\end{table}

\subsubsection{`Hard' pion-ID definition}
\label{pi-hard}

The standard $\pi$-ID definition in Table~\ref{PID_def} is part of a 
complete set of exclusive PID conditions that is convenient when 
backgrounds in the pion candidate sample are expected to be manageable. 
However, in some cases the standard $\pi$-ID algorithm does not deliver sufficient pion purity. 
One such case concerns the purity of the tagged pion in the tag-and-probe pion pair, 
which is crucial in the determination of $\pi$-ID efficiencies (Sect.~\ref{pi-ID}).
Improved pion purity is also necessary to reduce backgrounds 
for the $\pi\pi$ cross section measurement in mass
regions in the tails of the $\rho$ peak. 

A tighter pion-ID selector is thus developed, which we call the `hard pion' ($\pi_h$) selector, 
to improve the rejection of muons and electrons that are misidentified as pions 
with the standard definition. The $\pi_h$-ID is based on two likelihood functions 
$P_{\pi/\mu}$ and $P_{\pi/e}$: 
$P_{\pi/\mu}$ uses the EMC deposit 
$E_{\rm cal}$ associated with the track and the penetration of the track into the IFR, while 
$P_{\pi/e}$ uses $E_{\rm cal}$ and the measurements of $(\dedx)_{\rm DCH}$ and $(\dedx)_{\rm SVT}$ 
as a function of momentum. Tracks with likelihoods close to 0 correspond to pions 
while $P_{\pi/\mu}\sim 1$ ($P_{\pi/e}\sim 1$) are muon-like (electron-like).
Reference distributions used in the likelihoods are obtained from
simulation, with corrections determined from data control samples
in mass ranges that ensure backgrounds are negligible.
The pure muon sample used for PID efficiency (see below), and the sample identified 
as `$e\pi$' in the very high mass range ($m_{\pi\pi}>5\gevcc$), provide the reference 
distributions for muons and electrons, respectively. 
The $\pi\pi(\gamma)\gamma_{\rm ISR}$ sample, with $m_{\pi\pi}$ restricted to the $\rho$ peak
and with both pions satisfying the standard $\pi$-ID, provides the reference distributions 
for pions.

\subsubsection{PID measurements with the muon samples}

The method used to determine the muon ID efficiency utilizes of the $\mu\mu(\gamma)\gamma_{\rm ISR}$
sample itself, where one of the tracks is tagged as a muon
using the $\mu_{\rm ID}$ selector defined above, and the opposite track is probed. 
The sample is restricted to $m_{\mu\mu}>2.5\gevcc$ to reduce the non-$\mu$ 
background to the $(1.1\pm0.1)\times 10^{-3}$ level, so that the ensemble of 
opposite tracks constitutes a pure muon sample.

The IFR performance at the time the data for this paper were collected was non-uniform across the detector and 
deteriorated with time~\footnote{This problem was remedied for data collected subsequently, through IFR 
detector upgrades.}. In order to map the PID efficiency, the track to 
be probed is extrapolated to the IFR. 
Local coordinates ($v_1,v_2$) of the impact point are defined depending on the 
IFR geometry (barrel or endcaps). Efficiency maps are obtained for each of the four 
data-taking periods used in the analysis. The granularity of the 3-dimensional (3D) maps 
is optimized as a function of momentum and local coordinates ($p,v_1,v_2$),
so that local variations of efficiencies are described with significant
statistical precision. 

The low-efficiency regions in the IFR are removed
in order to keep as active areas only the regions where the $\mu_{\rm ID}$
efficiency was reasonably homogeneous. Removed portions include the crack 
areas between modules and some parts of the nominal active region where the
IFR performance was strongly degraded. The definition of removed 
regions is run-dependent: in the first running period only cracks are removed
(about 13\% of the IFR solid angle), while in the fourth period an additional 
$\sim15$\% is eliminated. 

Due to the mass restriction applied to the muon control sample, the 3D-maps provide 
the identification efficiency for isolated muon tracks.
They parametrize the local
performance of the IFR at the track impact point. However, 
at $\mu\mu$ masses less than 2.5\gevcc, tracks can become geometrically close 
to each other within the IFR
and their respective ID efficiencies can be significantly affected.
First, the efficiency is reduced with respect to the isolated 
track efficiency because the combination of the two sets of hits causes
some of the criteria that enter the $\mu_{\rm ID}$ selector to fail. Also the 
single-hit readout of the two-dimensional strip structure of the IFR 
chambers leads to losses. 
Second, track overlap leads to a correlated loss of PID for both tracks, 
not accounted for by the product of their uncorrelated 
single-track inefficiencies registered in the maps. 

The loss of efficiency and the correlated loss effects are studied 
and evaluated in data and in simulation using the two-track physics sample.
Since the pion background in the data sample is large in the 
$\rho$ mass range, the efficiencies are measured directly only in the mass 
regions in the resonance tails, and are then extrapolated to the $\rho$ peak region 
($0.6-0.9\gevcc$). Possible bias from this procedure is studied with
simulation and a systematic uncertainty of $2.2\times 10^{-3}$ is assigned. 
The efficiency loss (compared to the isolated muon efficiency) is determined 
using a muon-ID tagged track as for the high-mass sample. Results are stored
in mass-dependent 2D-maps as a function of the differences
($\Delta v_1, \Delta v_2$) between the impact 
points of the two tracks in the local IFR coordinate system. 
Background from pions and kaons is subtracted using simulation with data/MC 
corrections for the mis-ID probabilities. The efficiency loss is maximal for 
$m_{\mu\mu}\sim 0.7\gevcc$ with a reduction of 8.4\% of the single-track 
efficiency in data and 4.8\% in simulation. The resulting muon-ID inefficiencies 
($1-\epsilon_{\mu\to`\mu\rq})$ measured in data and simulation at 
0.4 (1.0)$\gevcc$ are 7.7\% (6.6\%) and 4.2\% (3.5\%), respectively. 

The two-track sample with no
identified muon is used to measure the correlated efficiency loss. Since the
pion background is overwhelming large in that sample, the small di-muon component is extracted
by applying the likelihood estimator described in Section~\ref{pi-hard} to both tracks. 
The correlated muon-ID loss is found to be 1.4\% in data and 0.3\% in simulation, at
$m_{\mu\mu}\sim 0.7\gevcc$. A systematic uncertainty of $1.5\times 10^{-3}$ is
assigned to the data/MC correction. Both efficiency loss and correlated loss 
decrease for higher masses and vanish at 2.5\gevcc.

\begin{figure}[thp]
  \centering
  \includegraphics[width=7.5cm]{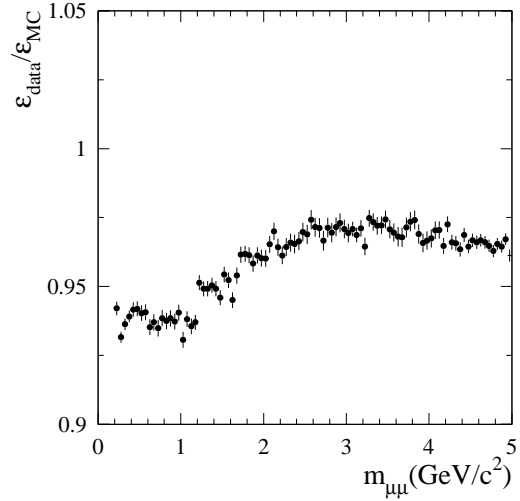}
  \caption{\small
The data/MC event correction $C_{\rm PID}$ for muon-ID efficiency as a function of $m_{\mu\mu}$.}
  \label{muid-corr}
\end{figure}

The event data/MC corrections resulting from requiring muon-ID for both 
tracks are obtained separately for the different running periods. The
overall correction is given in Fig.~\ref{muid-corr}. The plotted errors are 
statistical only.

Systematic errors are estimated for the different data-taking periods.
The overall systematic error from muon-ID on the $\mu\mu(\gamma)\gamma_{\rm ISR}$ cross section 
amounts to $3.3\times 10^{-3}$. Muon-ID is the largest source of uncertainty in the 
muon analysis. The dominant contribution arises from the procedures used to 
estimate the efficiency loss and the correlated loss of the two muons.

The pure muon sample of opposite tracks is used to measure the mis-ID probabilities. 
The largest one is $\epsilon_{\mu\to`\pi\rq}$, very close to 
$1-\epsilon_{\mu\to`\mu\rq}$. Therefore, the preceding results for muon-ID
efficiency and data/MC corrections, including those for close tracks, translate
to the muon mis-ID to `$\pi$'. The small mis-ID probabilities into particle 
types other than `$\pi$' are estimated from additional studies. 
The $\epsilon_{\mu\to`K\rq}$ mis-ID is smaller than 0.1\% for momenta below
3\gevc, with a steep increase for larger momenta, reaching about 1\% at 5\gevc, with a data/MC 
correction of 12\%. 
Mis-ID probabilities to $\epsilon_{\mu\to`0\rq}$ and $\epsilon_{\mu\to`e\rq}$ 
are 0.4\% and less than 0.1\%, respectively.

\subsubsection{PID measurements with the kaon samples}

For the kaon efficiency and mis-ID measurement, the same tag-and-probe method is used
as described for the muons, this time with a primary track satisfying
the $K_{\rm ID}$ condition. 
In addition to the restriction $\chi^2_{KK}<15$ applied to reduce multihadronic 
background, a requirement $\chi^2_{KK} < \chi^2_{\pi\pi}$ is applied to 
reduce the pion contamination. The purity of the kaon sample is further enhanced by a restriction 
on the fitted $m_{KK}$ mass, which must be in the $\phi$ window $1.01-1.04\gevcc$.
The electron background from photon conversions in the process $e^+e^-\to\gamma\gamma$, 
which populate the $m_{KK}$ threshold region,
is eliminated by a requirement on the distance 
(in the transverse plane) between the vertex of the two tracks and the 
beam axis.  
The purity achieved is ($99.0\pm0.1$)\%, determined from a fit of the $m_{KK}$ 
distribution in data, with $\phi$ signal and background shapes taken from MC.

The data/MC corrections for the $K_{\rm ID}$ efficiency are obtained as a function of 
track momentum. The restriction to the $\phi$ sample imposes kinematic
restrictions on the kaon momentum, with a lack of statistics below
1.5\gevc and above 5\gevc. This necessitates an extrapolation, which is achieved 
through a fit of the kinematically available data. A sampling of the momentum-dependent correction 
is performed
using the $KK\gamma_{\rm ISR}$ MC simulation, in order to determine the event correction 
as a function of the $m_{KK}$ invariant mass. 

The mis-ID probabilities depend on momentum, especially for the $K\to`\pi$' mis-ID, 
which increases strongly for large momenta, where the $K_{\rm ID}$ 
selector becomes inefficient. At 4\gevc\ the values in data are 5.8\% for 
$K\to`\mu$', 16.1\% for $K\to`\pi$', and 0.7\% for $K\to`e$'. The corresponding
data/MC corrections are $0.61 \pm 0.05$, $0.87 \pm 0.04$, and $2.7 \pm 0.8$.

\subsubsection{PID measurements with the pion samples}
\label{pi-ID}

The tag-and-probe method is again applied to construct a pure pion sample used 
to measure the pion ID efficiency and misidentification probabilities. 
To reduce the backgrounds, the mass range of selected 
$\pi\pi(\gamma)\gamma_{\rm ISR}$ candidates is 
restricted to the $\rho$ peak, $0.6<m_{\pi\pi}<0.9\gevcc$.
To ensure the validity of the pion ID efficiency measurement, the purity of the 
pion ID sample is further increased by requiring the primary track to satisfy 
the `hard pion' tag. In the restricted mass range, the sum of $\mu$, $K$ and $e$ 
backgrounds is reduced to the $(3.7\pm0.5)\times 10^{-3}$ level.

While for muons it is possible to measure the ID efficiencies for 
isolated tracks using events with a large $m_{\mu\mu}$ mass, for pions we use 
events in the $\rho$ region. Tracks in this region may overlap in one 
detector or another. Thus the $\pi$-ID efficiencies and mis-ID probabilities 
contain some average
of overlap effects, which are not possible to sort out in detail. However these
effects are much reduced for pions compared to muons, since showering 
in the IFR is sufficient to distinguish hadrons from muons and the overlap of
pion showers does not degrade the pion-ID efficiency.

All $\pi$-ID efficiencies and mis-ID probabilities are stored in 2D-maps 
as a function of the
momentum and local $z$ coordinate of the track extrapolated to the most relevant 
detector (IFR or DIRC). 
Biases from primary pion tagging and correlated two-track pion-ID loss are 
studied with simulation, and verified with data in the most critical cases. Both effects 
are at the $10^{-3}$ level.
The $\pi$-ID maps are sampled to build the full event efficiency distributions as 
a function of the $m_{\pi\pi}$ mass, in data and MC.
The event $\pi$-ID efficiency is weakly mass-dependent with typical 
values in data of 77.8\%, 75.3\%, and 77.0\%, at masses of $0.35\gevcc$, 
$0.6\gevcc$, and $1\gevcc$, respectively. 

\begin{figure}[thp] 
  \centering
  \includegraphics[width=7.5cm]{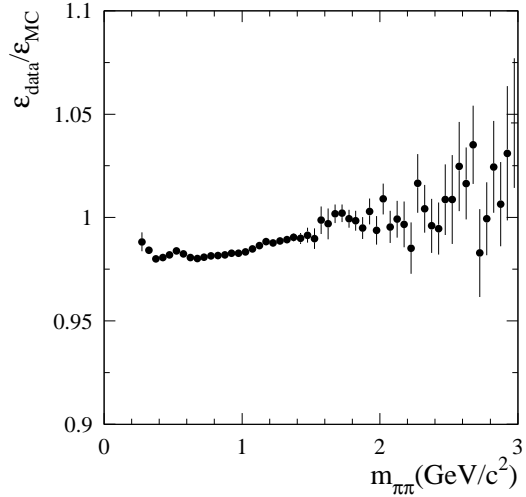}
  \caption{\label{piid-corr} \small 
    The data/MC event correction $C_{\rm PID}$ for $\pi_{\rm ID}$ efficiency as a function 
    of $m_{\pi\pi}$.}
 \end{figure}

The data/MC correction to the full event $\pi$-ID efficiency  
is shown in Fig.~\ref{piid-corr}. The correction is smaller than the corresponding 
factor for muons, which reflects a lesser sensitivity of the $\pi$-ID 
efficiency to the IFR conditions. Although it has been obtained using 
maps determined in the $\rho$ region, it shows only a few percent variation  
with mass, consistent with the fact that correlated ID losses are small.

The systematic errors on the efficiencies come from the limited granularity 
of the mis-ID maps, the biases caused by the `hard'-$\pi$ selection of the 
tagged pion, and the application of maps determined in the $\rho$ region 
(0.6--0.9\gevcc) to other mass ranges. These effects are studied with simulated
$\pi\pi(\gamma)\gamma_{\rm ISR}$ signal samples, by comparing the mass spectra of produced 
events when $\pi$-ID is either applied or not. 
The former spectrum is obtained by applying the PID process, then correcting 
the `$\pi\pi$' spectrum by the $\pi$-ID efficiency determined as in data. The latter 
spectrum is obtained by not applying any PID requirement. As expected, the agreement 
is excellent in the $\rho$ region with a
variation of at most 2 permil, while some bias is observed in the lower and 
higher mass regions: 1\% for $m_{\pi\pi}<0.4\gevcc$, $6\times 10^{-3}$ for
$0.4<m_{\pi\pi}<0.6\gevcc$, $4\times 10^{-3}$ for $0.9<m_{\pi\pi}<1.2\gevcc$, 
and 1\% for $m_{\pi\pi}>1.2\gevcc$. The full bias determined in simulation is
taken as a systematic uncertainty. The global PID test in data described 
below supports this estimate.

In this analysis a good control of the $\pi \rightarrow `\mu$' mis-ID 
probability is crucial to the determination of the $\mu\mu(\gamma)\gamma_{\rm ISR}$ cross section 
in the $\rho$ region. At track level this probability in data is found to 
vary with momentum between 4.5\% at $5\gevc$ and 7\% at $1\gevc$, and the 
data/MC correction is determined to be $0.95\pm0.05$ and $1.23\pm0.03$, 
respectively. 

\subsubsection{Global PID test with data}
\label{consistency-data}

Since the PID classes form an exclusive and complete set, every event in the 
full sample before PID, $N_{xx}$, is assigned to a $N_{`ij\rq}$ category
($`i$',$`j$'$=\mu,\pi,K,e,0$). The observed $N_{`ij\rq}$ spectra and the
measured PID efficiencies are used in global consistency checks over the full 
$N_{xx}$ data sample.

The $N_{xx}$ sample is actually composed of $N_{ii}$ pairs of particles of identical 
true types $i=\mu, \pi, K$, with small background contributions from other processes 
that are taken into account as follows.
The contribution to $N_{xx}$ from electrons stems from $ee\gamma$ and $\gamma\gamma$ 
followed by a pair conversion. It occurs mainly in the `$\pi e$'  and `$\pi\pi$' 
topologies, while being negligible in `$ee$' due to the strong rejection
of electrons at the preselection and track definition levels. 
The small electron component of the `$\pi\pi$' sample is subtracted out, after proper 
normalization (Sect.~\ref{pipi-ee}). In the PID process, protons are mainly identified 
as `$\pi$' and in the global test below, their very small contribution is included in 
$N_{\pi\pi}$. Because the $N_{xx}$ sample is selected with a
tight $\chi^2_{\pi\pi}<15$ requirement applied to the 1C fit, multihadronic 
background is reduced to a negligible level. Contributions from events with two tracks of 
different true types, from $\tau$-pair decays for instance, are found to be negligible.

Each observed `$ii$' spectrum with `diagonal' ID, {\it i.e.}, `$\pi\pi$', `$\mu\mu$', 
`$KK$', receives contributions from the true ($ii$) channel degraded by the
$\epsilon_{i\to`i\rq}$ efficiencies and from the two other channels through 
$\epsilon_{j\to`i\rq}$ mis-ID. The spectra of produced events in each channel 
are thus obtained by solving a system of three linear equations.
In each mass bin (computed with the $\pi\pi$ mass hypothesis) of the spectra for
identified pairs of type `$i$', $N_{`ii\rq}$, the following equations
\begin{widetext}
\beqn
\label{sep-pid}
 N_{`\pi\pi\rq}&=&N_{\mu\mu}\varepsilon_{\mu\mu\to`\pi\pi\rq} + 
                N_{\pi\pi}\varepsilon_{\pi\pi\to`\pi\pi\rq} +
                N_{KK}\varepsilon_{KK\to`\pi\pi\rq}\\\nonumber
 N_{`\mu\mu\rq}&=&N_{\mu\mu}\varepsilon_{\mu\mu\to`\mu\mu\rq} + 
                N_{\pi\pi}\varepsilon_{\pi\pi\to`\mu\mu\rq} +
                N_{KK}\varepsilon_{KK\to`\mu\mu\rq}\\\nonumber
 N_{`KK\rq}    &=&N_{\mu\mu}\varepsilon_{\mu\mu\to`KK\rq} + 
                N_{\pi\pi}\varepsilon_{\pi\pi\to`KK\rq} +
                N_{KK}\varepsilon_{KK\to`KK\rq}~, 
\eeqn
\end{widetext}
are solved for the produced numbers of particle pairs 
of each type, $N_{\pi\pi}$, $N_{\mu\mu}$, and $N_{KK}$. 
In Eqs.~(\ref{sep-pid}), the quantities $\varepsilon_{jj\to`ii\rq}$ represent
the product of the ID-efficiencies $\varepsilon_{j\to`i\rq}$ and possibly correlation factors
that have been established in each PID study. 

From the inferred spectra $dN_{ii}/dm_{\pi\pi}$ of particle pairs of true type $i$,
any `$ij$' spectrum, $dN_{`ij\rq\rm pred}/dm_{\pi\pi}$, is derived, using the measured efficiencies 
and mis-ID probabilities, 
and is compared to the directly observed `$ij$' distribution. A relative 
difference is computed, normalized to the spectrum $dN_{xx}/dm_{\pi\pi}$ of 
the full sample before PID assignment:
\beqn
\label{delta-ij-data}
\delta_{ij}^{\rm data}=\frac {dN_{`ij\rq\rm pred}/dm_{\pi\pi}-dN_{`ij\rq}/dm_{\pi\pi}}
                  {dN_{xx}/dm_{\pi\pi}}~,
\eeqn
All differences $\delta_{ij}^{\rm data}$ are within a few permil. 

\begin{figure}[thbp] \centering
  \includegraphics[width=7.cm]{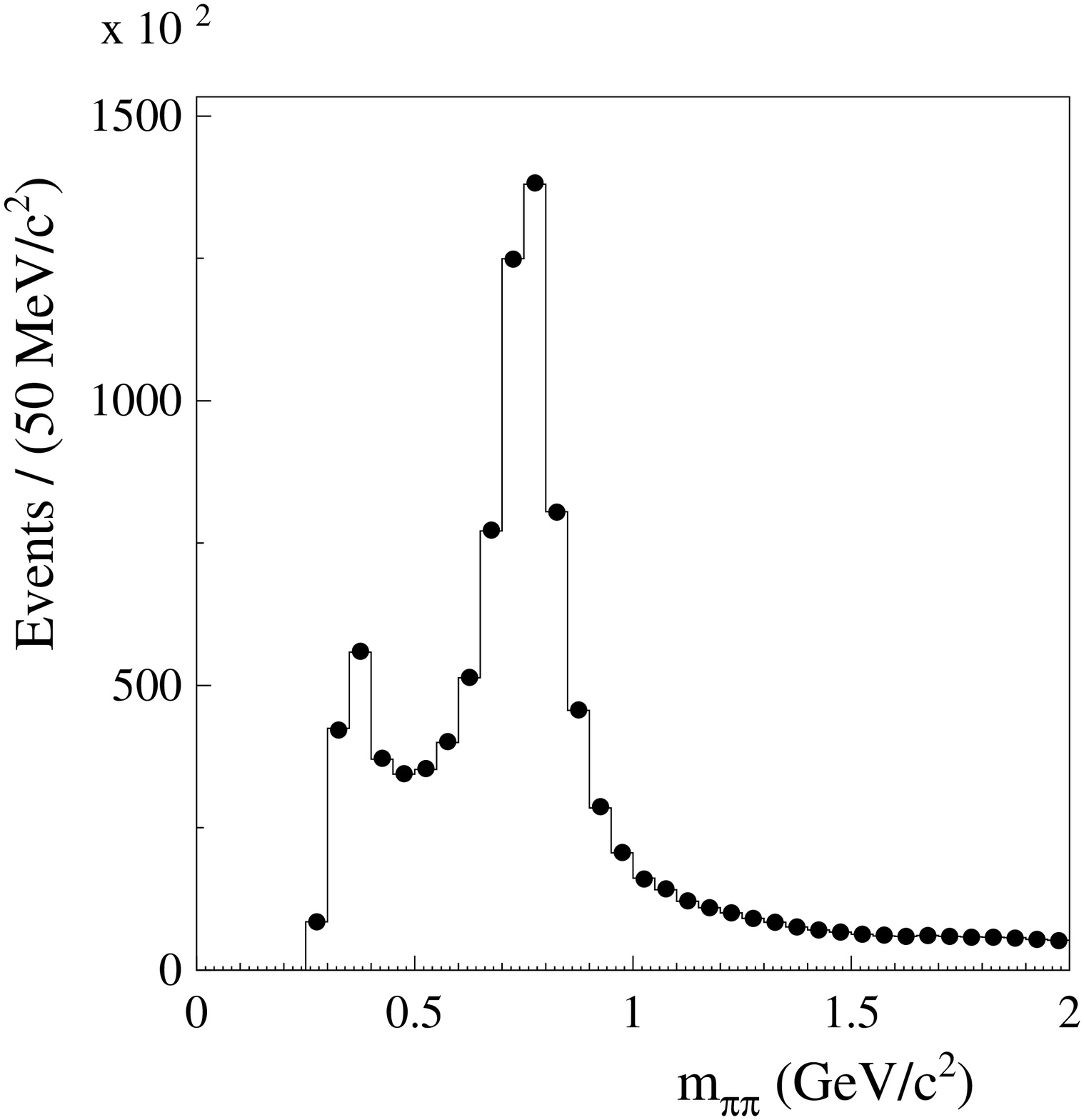}
  \includegraphics[width=7.cm]{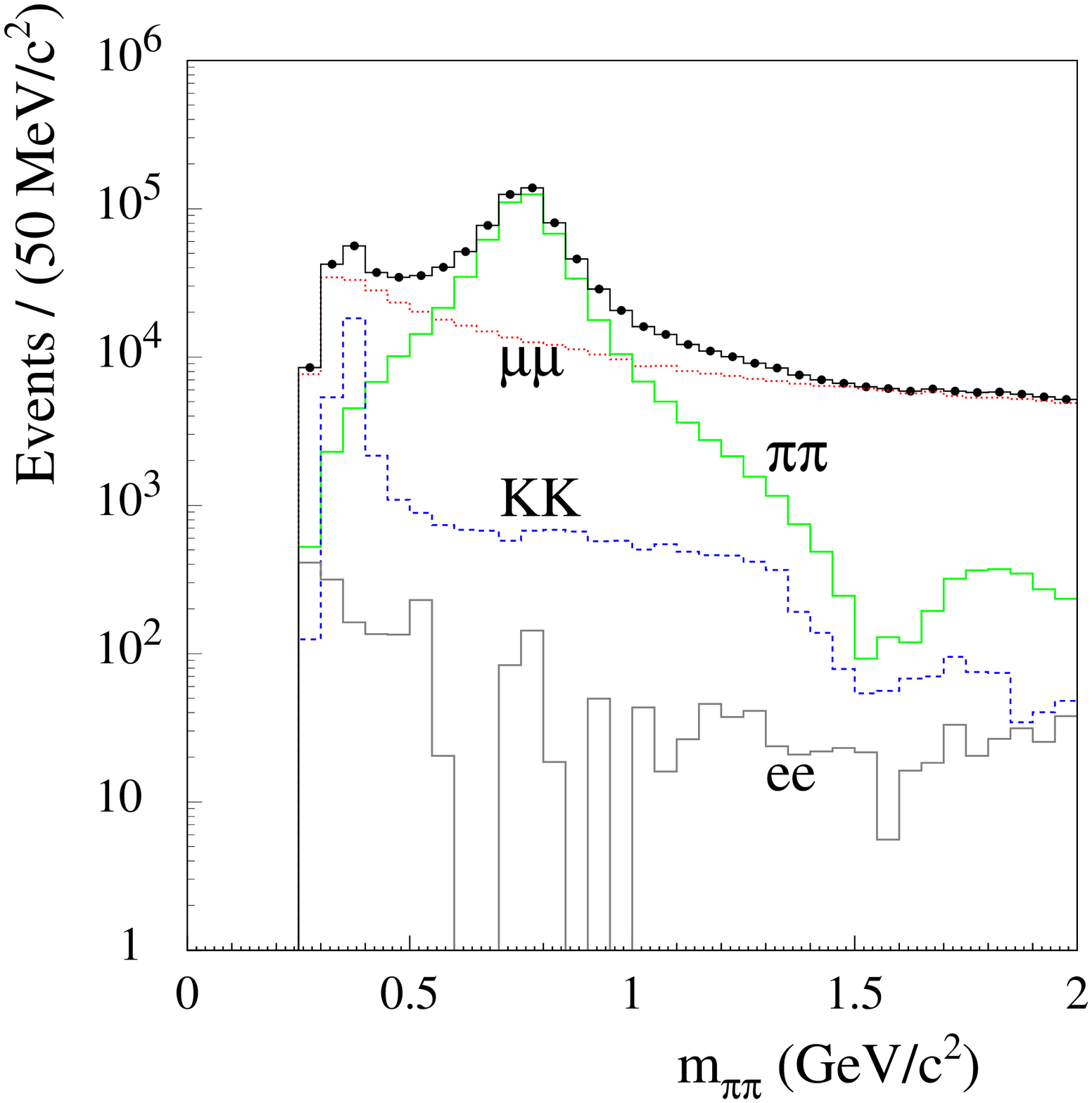}
  \includegraphics[width=7.cm]{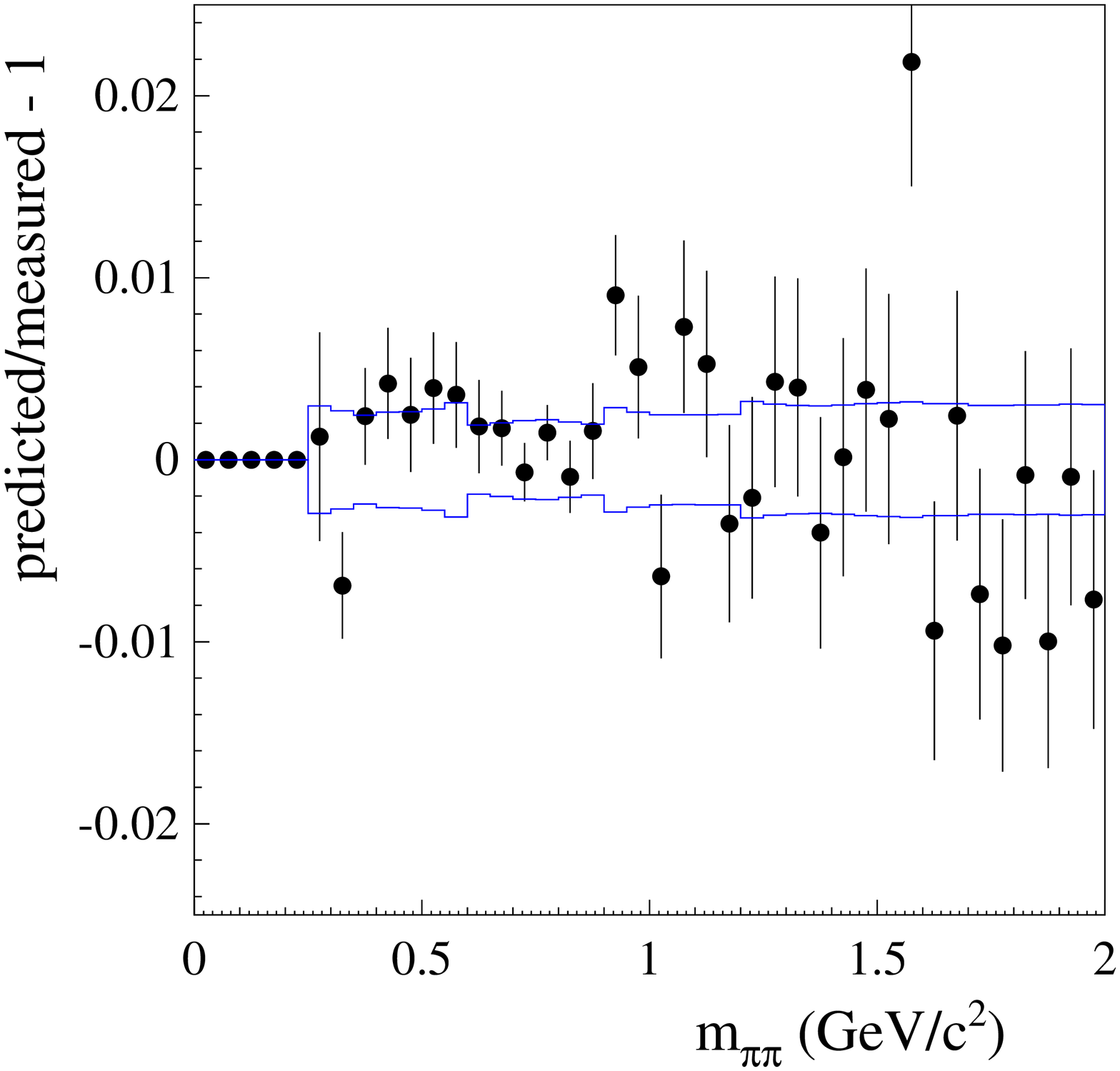}
  \caption{\label{global-all-14} \small(color online).
    Global PID test on data (see text). (top): The $m_{\pi\pi}$ 
spectrum of all $xx\gamma$ events (no PID applied, data points) compared
to  $dN_{xx~\rm pred}/dm_{\pi\pi}$ (histogram). (middle): The different 
components $N_{xx}$ of the histogram in the top plot, with $xx$ labels
indicated, and their sum (top histogram with dots). (bottom): The relative 
difference between the two spectra in the top plot: predicted/measured -1.
The independently-estimated systematic uncertainty is shown by the band.}
  \end{figure}

The $dN_{xx}/dm_{\pi\pi}$ spectrum is compared to
the full inferred one, $dN_{xx~\rm pred}/dm_{\pi\pi}$, obtained by summing the 
$N_{\pi\pi}$, $N_{\mu\mu}$ and $N_{KK}$ components (and the small $ee$ background). 
Figure~\ref{global-all-14}, which shows their relative
difference, contains all the information available in data on the validity
of the ID corrections applied to the different `$ii$' spectra.
The band in Fig.~\ref{global-all-14} represents the limits given by the
quadratic sum of the estimated systematic uncertainties on the $\mu\mu$, 
$\pi\pi$, and $KK$ ID efficiencies. Within the statistical uncertainties of the 
data sample, all deviations are consistent with the band, thus
validating the estimates of the systematic errors. 

\subsubsection{Hard pion specific efficiency}
\label{pipih-eff}

The hard pion identification is required for one of the two tracks when
computing the $\pipi(\gamma)\gamma_{\rm ISR}$ 
cross section off the $\rho$ peak (Sect.~\ref{tails}). The
`$\pi_h$' efficiency and misidentifications determined in simulation 
are controlled in data, and data-MC discrepancies are corrected for.
The efficiency correction of the `$\pi\pi$'$\to`\pi\pi_h$' identification 
can only be determined in the central $\rho$ mass region where backgrounds 
are small in the `$\pi\pi$' sample. Remaining $\mu\mu$ backgrounds are 
subtracted from the `$\pi\pi$' and `$\pi\pi_h$' samples using the
measured $\varepsilon_{\mu\to`\pi\rq}$ misidentification probability and the
likelihood selector $P_{\pi/\mu}$ defined in Section~\ref{pi-hard}, respectively.

\begin{figure}[htp]
  \centering
  \includegraphics[width=7.5cm]{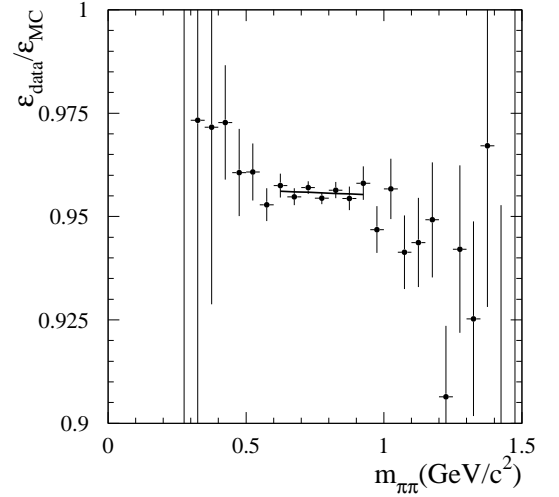}
  \caption{\small
The ratio of the efficiencies for `$\pi\pi_h$' identification 
of $\pi\pi(\gamma)\gamma_{\rm ISR}$ events for data and MC, the line is the linear fit 
result in the $\rho$ region.}
  \label{corr-pipiTOpihpi-for-pipi}
\end{figure}

Compared to the standard `$\pi\pi$' definition, the event ID efficiency is 
reduced by a factor of 0.825 in data and 0.870 in simulation.
The ratio of the efficiencies in data and simulation
is shown in Fig.~\ref{corr-pipiTOpihpi-for-pipi}, exhibiting
no significant mass dependence between 0.4 and 1\gevcc. A decrease is observed above
1\gevcc, which is ascribed to an imperfect representation of the large 
background in this region for the `$\pi\pi$' sample. A linear fit is performed
for $0.6<m_{\pi\pi}<0.9\gevcc$ and extrapolated outside with propagation of 
errors.

\subsubsection{Separation of the different channels using particle identification}
\label{pid-sep}

As shown above for the global PID test on data, the PID efficiencies and misidentification
probabilities 
measured on pure data samples allow one to reliably separate the different two-body ISR 
channels composing the full physics sample before PID assignment.
Equivalently stated, by solving Eqs.~(\ref{sep-pid}) on the total physical sample, 
one obtains the produced spectrum $dN_{\pi\pi}/dm_{\pi\pi}$ free of misidentified
$\mu\mu(\gamma)\gamma_{\rm ISR}$ and $KK(\gamma)\gamma_{\rm ISR}$ backgrounds. 
Likewise, the produced muon spectrum $dN_{\mu\mu}/dm_{\mu\mu}$ is obtained 
free of hadronic backgrounds by solving Eqs.~(\ref{sep-pid}) per bin of $m_{\mu\mu}$. 

The measured `$\pi\pi$' and `$KK$' spectra contain
contributions from multihadronic background, from higher-multiplicity ISR
or $q\overline{q}$ processes, which are subtracted after solving 
Eqs.~(\ref{sep-pid}). There is a 
contribution from the ISR $p\overline{p}\gamma$ process, which appears 
dominantly in the `$\pi\pi$' spectrum. In this procedure it is treated like 
pion pairs and is removed later.
Mistreatment of multihadron events where the final state
involves a $K\pi$ pair is also considered, although they are reduced by 
PID. 

The procedure of solving Eqs.~(\ref{sep-pid}) to separate the two-body ISR channels
is applied to isolate the $\mu\mu(\gamma)\gamma_{\rm ISR}$ channel over the full mass range,
and the $\pi\pi(\gamma)\gamma_{\rm ISR}$ channel in the $\rho$ peak region.
As the method relies on the completeness of PID class assignment of Table~\ref{PID_def}, 
it does not apply when the `$\pi\pi_h$' selection is required off the $\rho$ peak
in the $\pi\pi(\gamma)\gamma_{\rm ISR}$ channel.
In the latter mass regions, the $\mu\mu(\gamma)\gamma_{\rm ISR}$ and $KK(\gamma)\gamma_{\rm ISR}$ 
background subtraction is achieved 
differently as described in Section~\ref{background}.

\section{Efficiency studies (II)}
\label{kin-eff}

Although the ISR and FSR kinematic fits described in Section~\ref{kin-fit} take into account 
potential additional photons, inadequate description of NLO radiation by the simulation
might induce incorrect estimates of the 2D-$\chi^2$ efficiency by MC. Comparisons of
additional radiation in data and MC are performed and data-to-MC corrections of
efficiencies are applied, as detailed below.
 
\subsection{Additional radiation}

\subsubsection{Additional small-angle ISR}

\begin{figure*}[htp]
\begin{center}
  \includegraphics[width=0.45\textwidth]{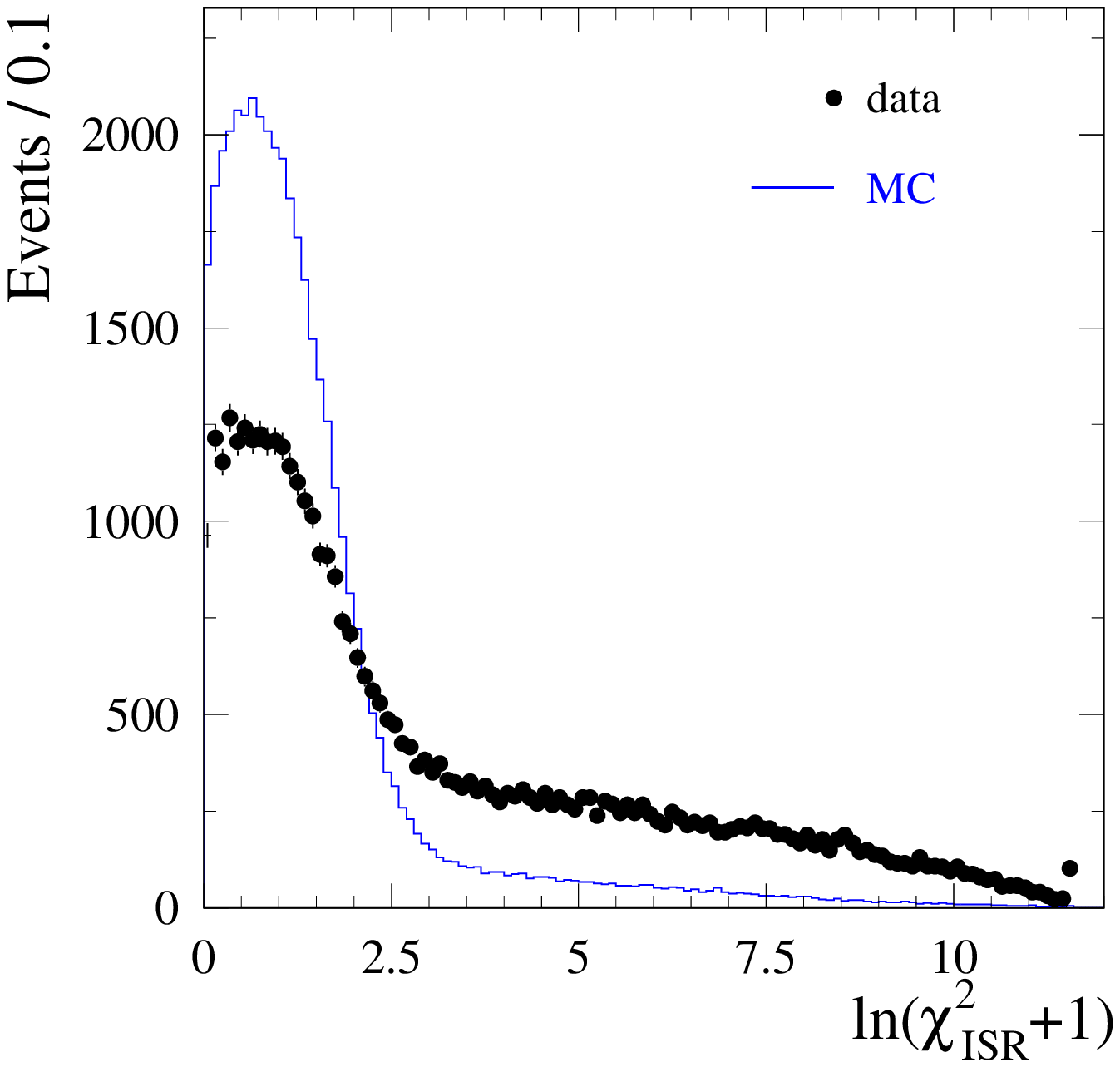}
  \includegraphics[width=0.45\textwidth]{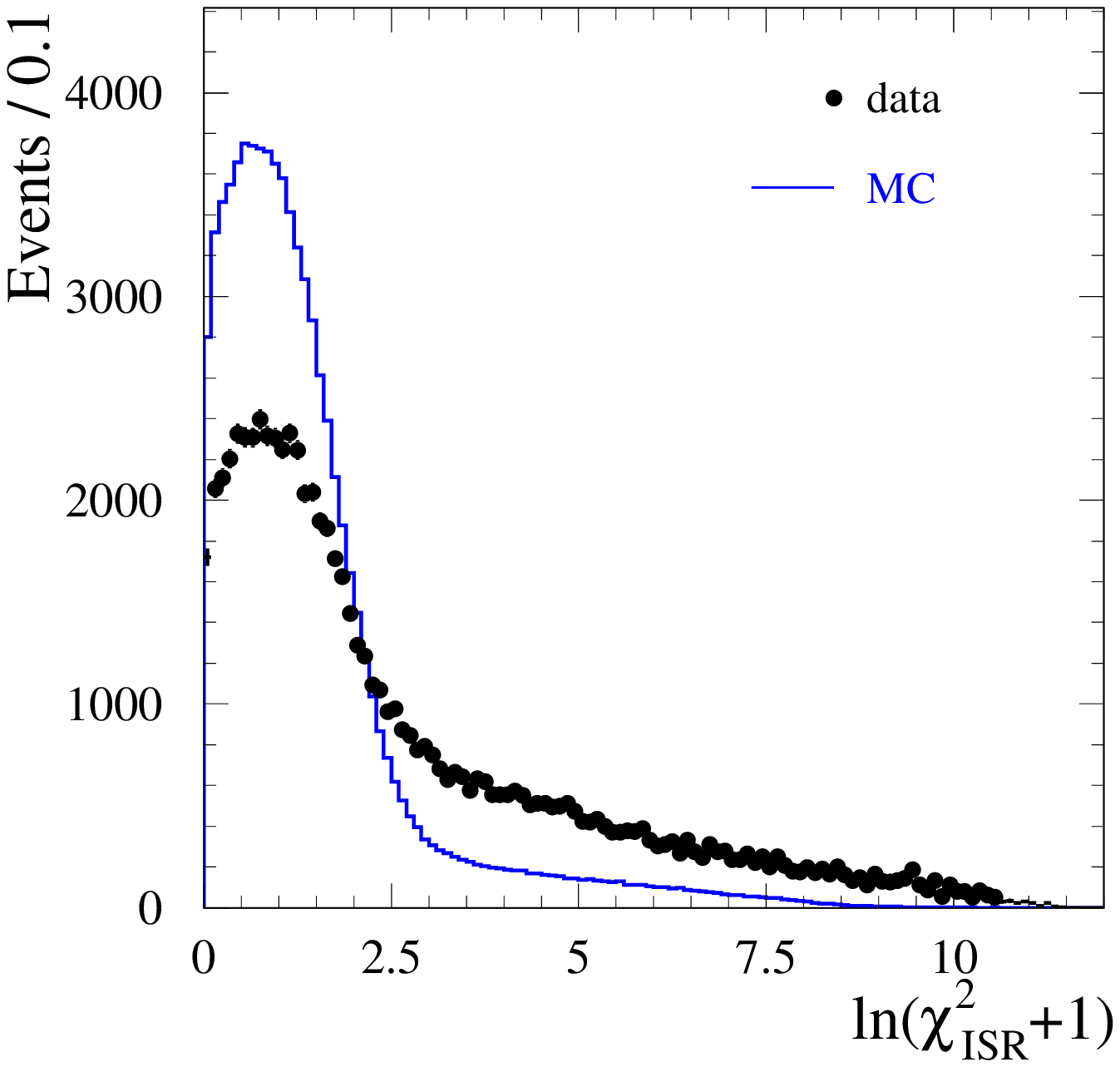}\\
  \caption{\small 
Distributions of $\ln(\chi^2_{\rm ISR}+1)$
for $\mu\mu(\gamma)\gamma_{\rm ISR}$ (left) and 
$\pi\pi(\gamma)\gamma_{\rm ISR}$ (right) events in the 0.2--1\gevcc (0.5--1\gevcc) mass region
for muons (pions), selected with 
$\ln(\chi^2_{\rm FSR}+1)>\ln(\chi^2_{\rm ISR}+1)$ and $E^*_{\gamma~\rm add.ISR}>0.2\gev$ 
for data (black points with errors) and MC (blue histogram). The distributions are normalized to the data 
luminosity.}
  \label{chi2-add-ISR}
\end{center}
\end{figure*}
\begin{figure*}
\begin{center}
  \includegraphics[width=0.45\textwidth]{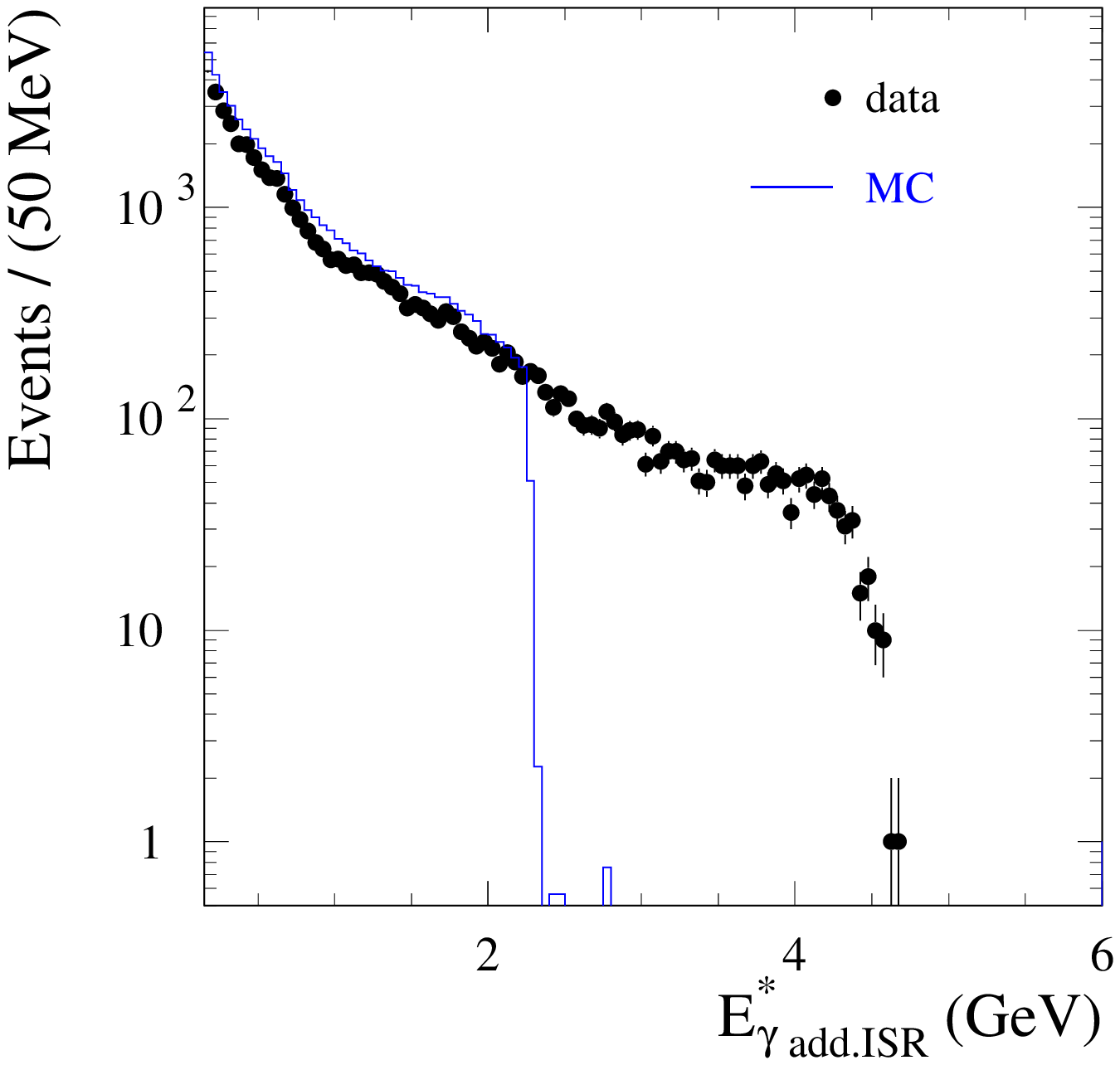}
  \includegraphics[width=0.45\textwidth]{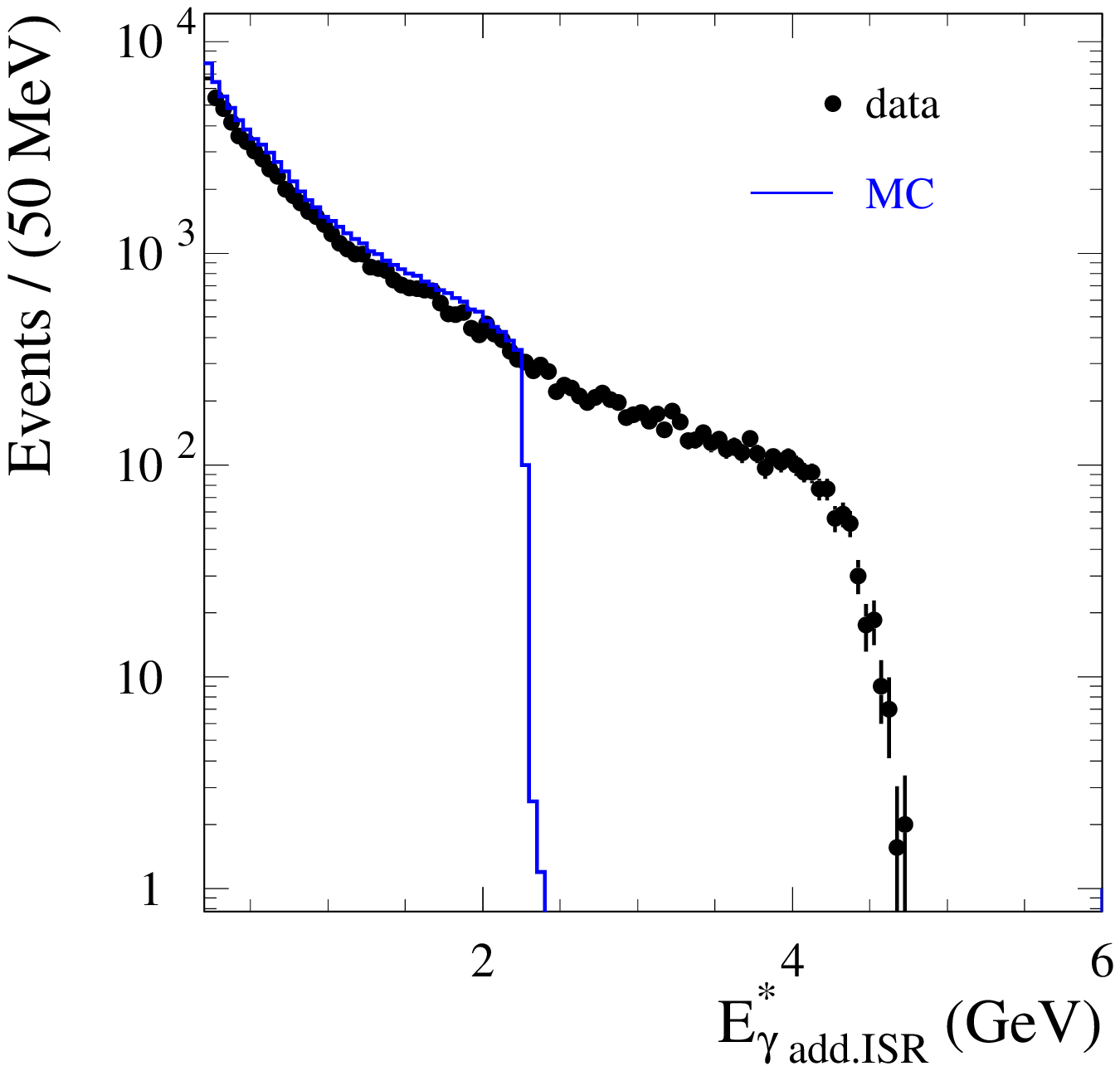}
  \caption{\small 
Distributions of $E^*_{\gamma~\rm add.ISR}$ 
for $\mu\mu(\gamma)\gamma_{\rm ISR}$ (left) and 
$\pi\pi(\gamma)\gamma_{\rm ISR}$ (right) events in the 0.2--1\gevcc (0.5--1\gevcc) mass region
for muons (pions), selected with 
$\ln(\chi^2_{\rm FSR}+1)>\ln(\chi^2_{\rm ISR}+1)$ and $E^*_{\gamma~\rm add.ISR}>0.2\gev$ 
for data (black points with errors) and MC (blue histogram). The distributions are normalized to the data 
luminosity.}
  \label{eg-add-ISR}
\end{center}
\end{figure*}

Additional radiation by the incoming electrons and positrons is evidenced by
the tail seen along the $\chi^2_{\rm FSR}$ vertical axis in Fig.~\ref{pi-2d-chi2} 
for pions and Fig.~\ref{mu-2d-chi2} for muons.
In order to study the $\chi^2_{\rm ISR}$ distributions in data and simulation,
events are selected above the diagonal 
($\ln(\chi^2_{\rm FSR}+1)>\ln(\chi^2_{\rm ISR}+1)$) and the restriction 
$E^*_{\gamma~\rm add.ISR}>0.2\gev$ is applied to ensure a significant level of
extra radiation. The quantity $E^*_{\gamma~\rm add.ISR}$ is obtained from the
2C ISR fit described in Section~\ref{kin-fit}. 

The $\chi^2_{\rm ISR}$ distributions for the selected events are shown in 
Fig.~\ref{chi2-add-ISR}.
As expected for an ISR effect, the situation is found to be identical 
in the muon and pion channels. 
The data distributions are wider than the MC ones. 
This is a consequence of the ISR fit hypothesis that additional ISR photons are
collinear to the beams as assumed in the  AfkQed simulation, as opposed to 
the angular distribution of additional ISR in data.

The corresponding distributions of $E^*_{\gamma~\rm add.ISR}$, normalized to the
data luminosity, are given in Fig.~\ref{eg-add-ISR} for muon and
pion channels. The MC spectra stop at 2.3\gev as a result of the 
$m_{X\gamma_{\rm ISR}}>8\gevcc$ requirement used in AfkQed at generation, 
while the data distributions extend to the kinematic limit.
Below 2.3\gev the data and MC distributions agree well in shape, but MC is
a little higher than data. This is expected since MC includes all additional
ISR photons, while events in data where the additional ISR
photon is at large angle have good $\chi^2_{\rm FSR}$ and therefore are not present 
in the sample considered in this section.

The results found for additional ISR in $\pi\pi(\gamma)\gamma_{\rm ISR}$ and 
$\mu\mu(\gamma)\gamma_{\rm ISR}$ channels are in agreement, as expected from the factorization 
of additional ISR. The lack of angular distribution 
in AfkQed is studied at the 4-vector level using Phokhara,
and acceptance corrections estimated,
but its effect essentially cancels in the $\pi\pi$/$\mu\mu$ ratio 
(Sect.~\ref{add-isr-corr}).

\subsubsection{Additional FSR and large-angle ISR}
\label{add-fsr}

\begin{figure*}[htb]
\begin{center}
  \includegraphics[width=0.45\textwidth]{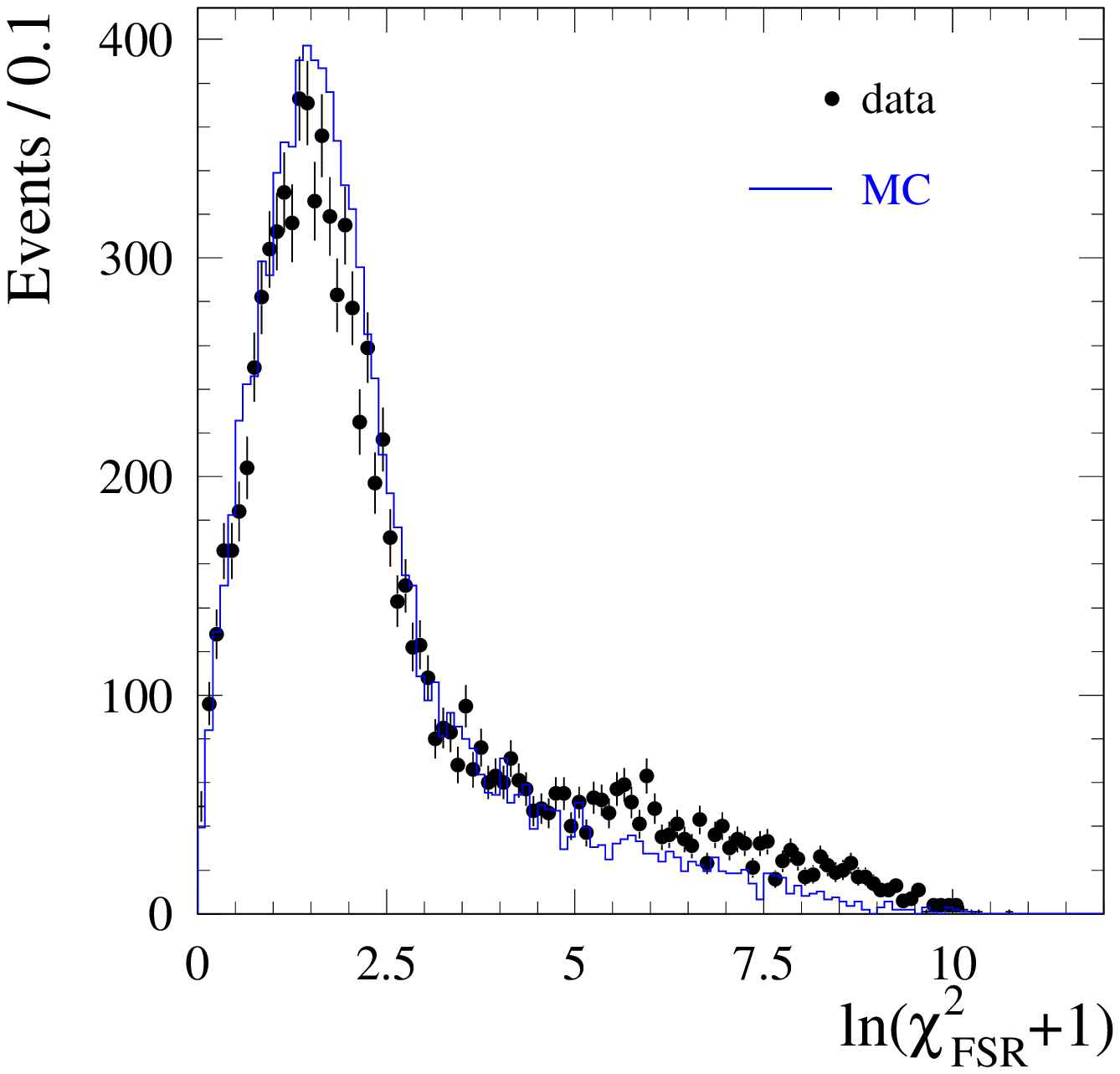}
  \includegraphics[width=0.45\textwidth]{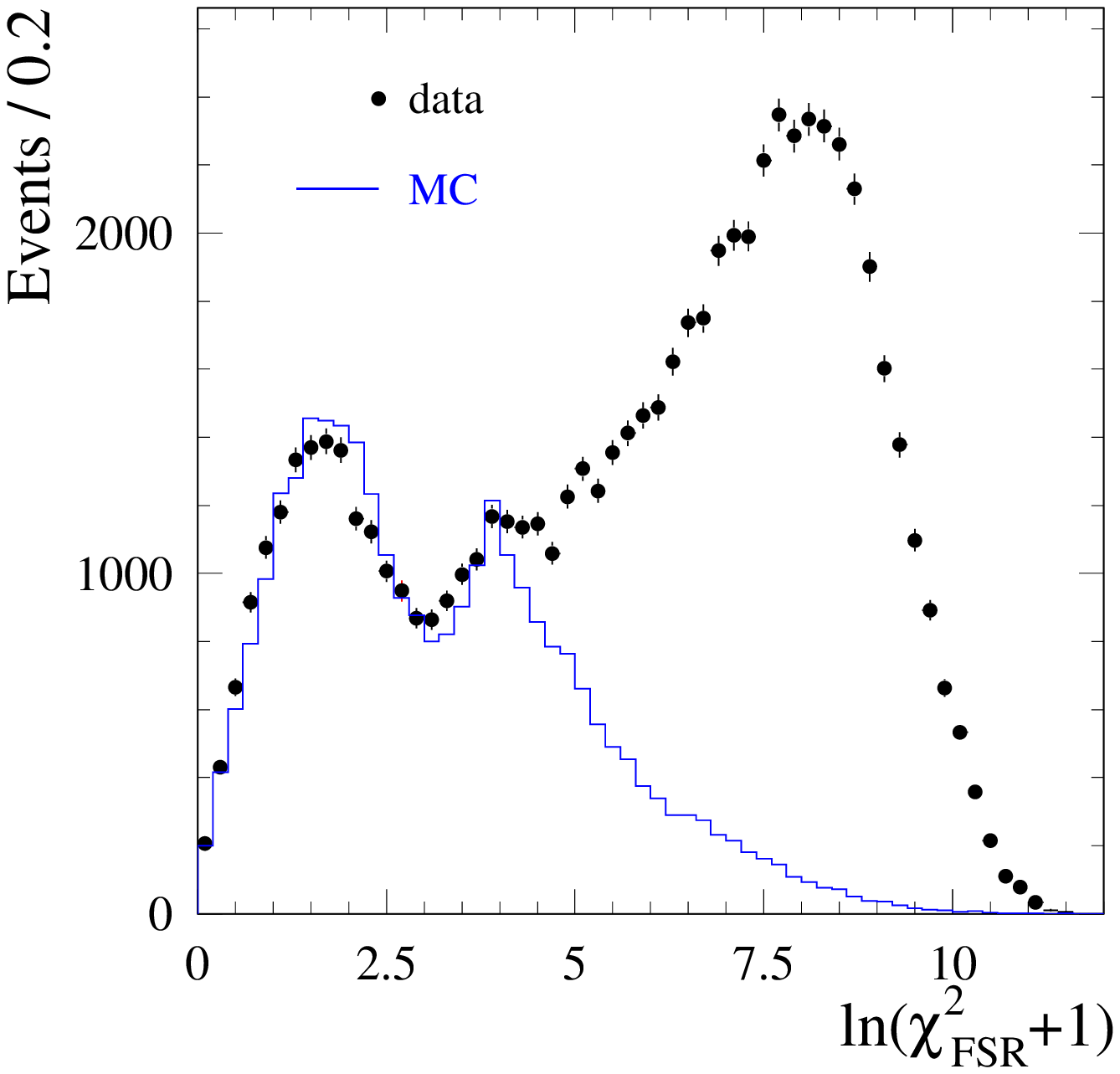}
  \caption{\small 
Distributions of $\ln(\chi^2_{\rm FSR}+1)$ for $\mu\mu\gamma\gamma_{\rm ISR}$ (left) 
and $\pi\pi\gamma\gamma_{\rm ISR}$ (right) events, selected with 
$\ln(\chi^2_{\rm FSR}+1)<\ln(\chi^2_{\rm ISR}+1)$ and $E_{\gamma~\rm add.FSR}>0.2\gev$ 
for data (black points with errors) and MC (blue histogram). The mass regions are chosen to be similar for
muons (0.2-1\gevcc) and pions (0.5-1\gevcc); the MC distribution is 
normalized to the number of events in data for muons and with 
$\ln(\chi^2_{\rm FSR}+1)<2$ for pions.}
  \label{chi2-add-FSR}
\end{center}
\end{figure*}

Similarly, we select a sample of events with an extra photon
in the detector acceptance and compares additional FSR in data and in
the simulation where it is generated using {\small PHOTOS}. 
Events are selected with
$\ln(\chi^2_{\rm FSR}+1)<\ln(\chi^2_{\rm ISR}+1)$ and the fitted energy of the additional 
large-angle photon is restricted to $E_{\gamma~\rm add.FSR}>0.2\gev$  
in the laboratory frame. The request for a large-energy additional FSR photon
effectively restricts the 2D-$\chi^2$ plane to a region 
$\ln(\chi^2_{\rm ISR}+1)\gtrsim3$. 

The corresponding $\chi^2_{\rm FSR}$ distributions for $\mu\mu\gamma\gamma_{\rm ISR}$ 
data and MC shown in Fig.~\ref{chi2-add-FSR} (left) are in reasonable 
agreement. Distributions for the selected $\pi\pi\gamma\gamma_{\rm ISR}$ events 
show a similar agreement,
Fig.~\ref{chi2-add-FSR} (right), except in the large $\chi^2$ tail where 
there are contributions from background in the data. A
contribution from secondary interactions is seen both in data and simulation. 

\begin{figure*}[htp]
\begin{center}
  \includegraphics[width=0.45\textwidth]{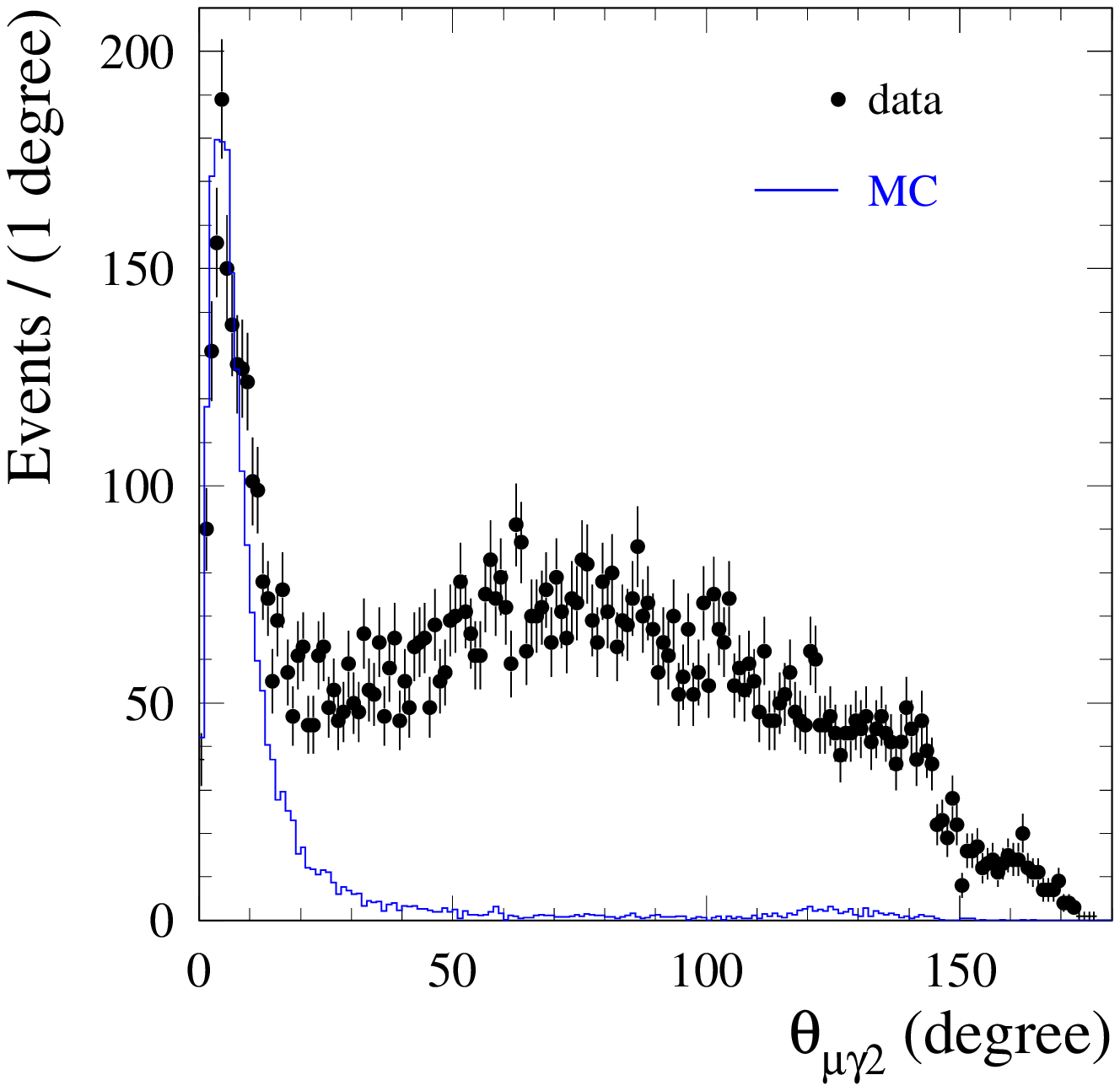}
  \includegraphics[width=0.45\textwidth]{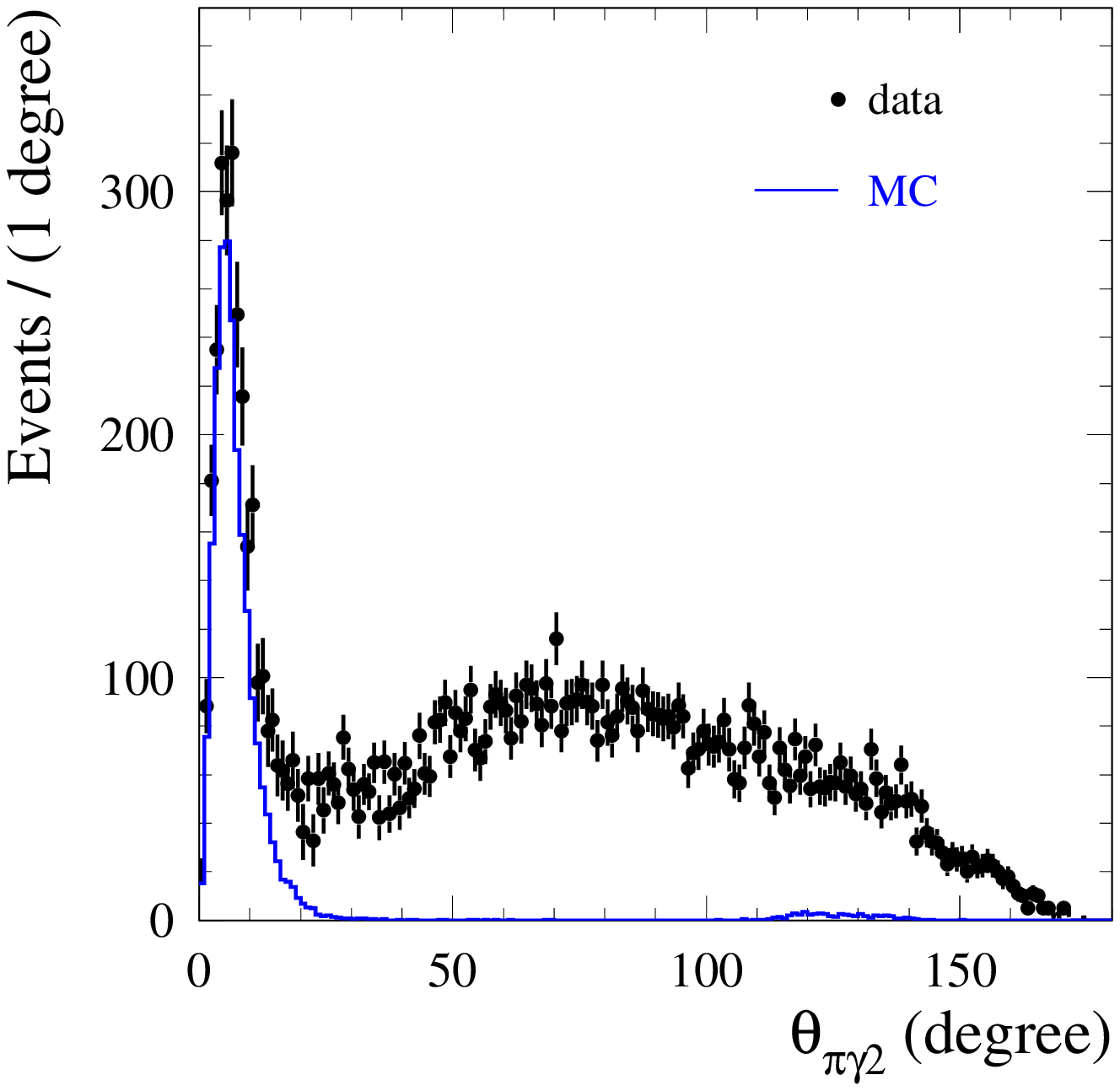}
  \caption{\small 
The additional `FSR' photon angular distribution with
  respect to the closer outgoing muon
  for the $\mu\mu\gamma\gamma_{\rm ISR}$ (left) and background-subtracted
$\pi\pi\gamma\gamma_{\rm ISR}$ (right) events with
  $\ln{(\chi^2_{\rm FSR}+1)}<\ln{(\chi^2_{\rm ISR}+1)}$, $E_{\gamma~\rm add.FSR}>0.2\gev$,
$\ln{(\chi^2_{\rm FSR}+1)}<2.5$,
  and in the mass intervals $0.2<m_{\mu\mu}<1\gevcc$, $0.5<m_{\pi\pi}<1\gevcc$ 
 (data:~black points with errors, MC:~blue histogram). The mass regions are chosen to be similar for
the muon and pion samples and the MC is normalized to data luminosity.}
  \label{FSR-thgam}
\end{center}
\end{figure*}
\begin{figure*}[htp]
\begin{center}
  \includegraphics[width=0.45\textwidth]{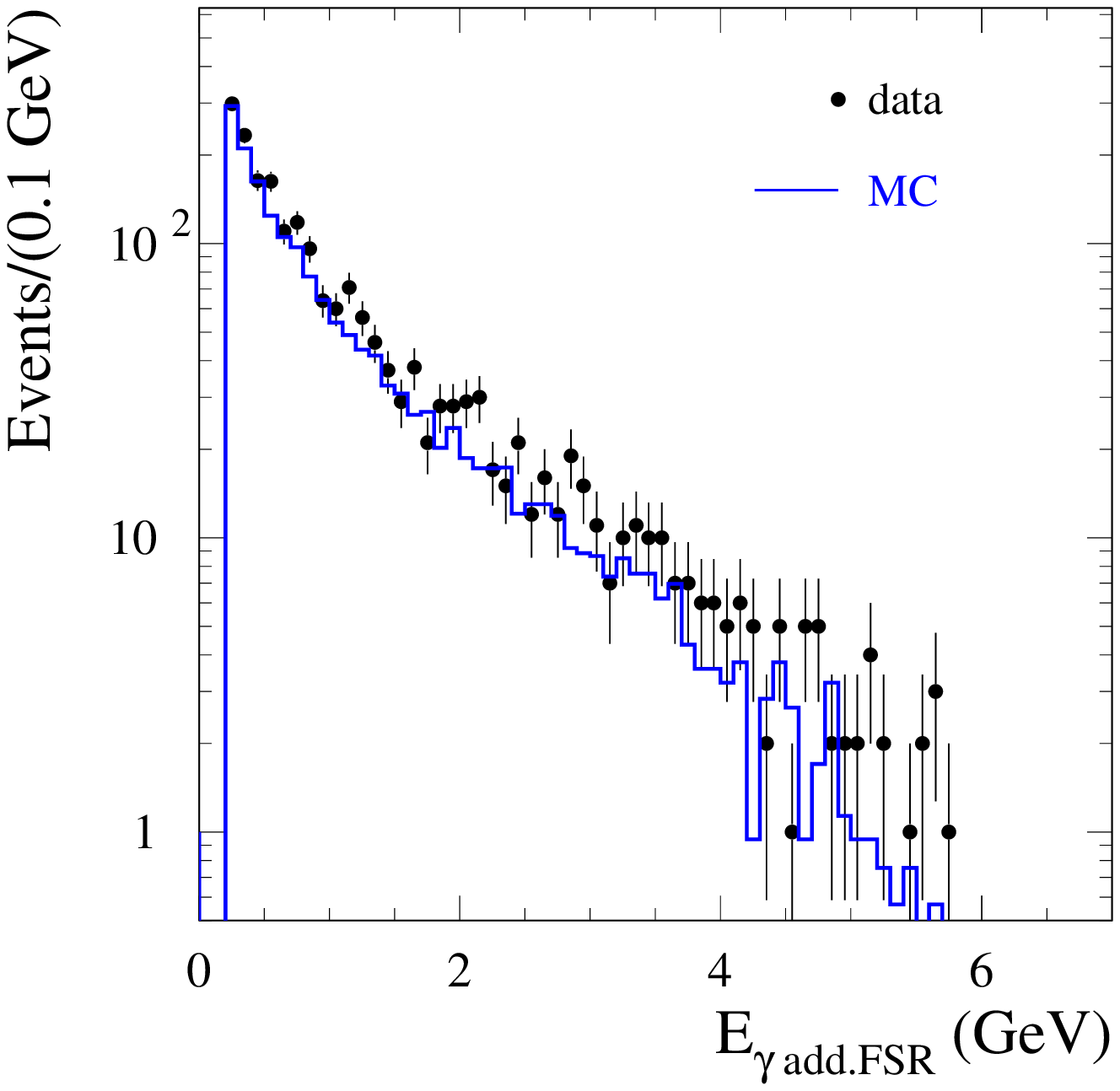}
  \includegraphics[width=0.45\textwidth]{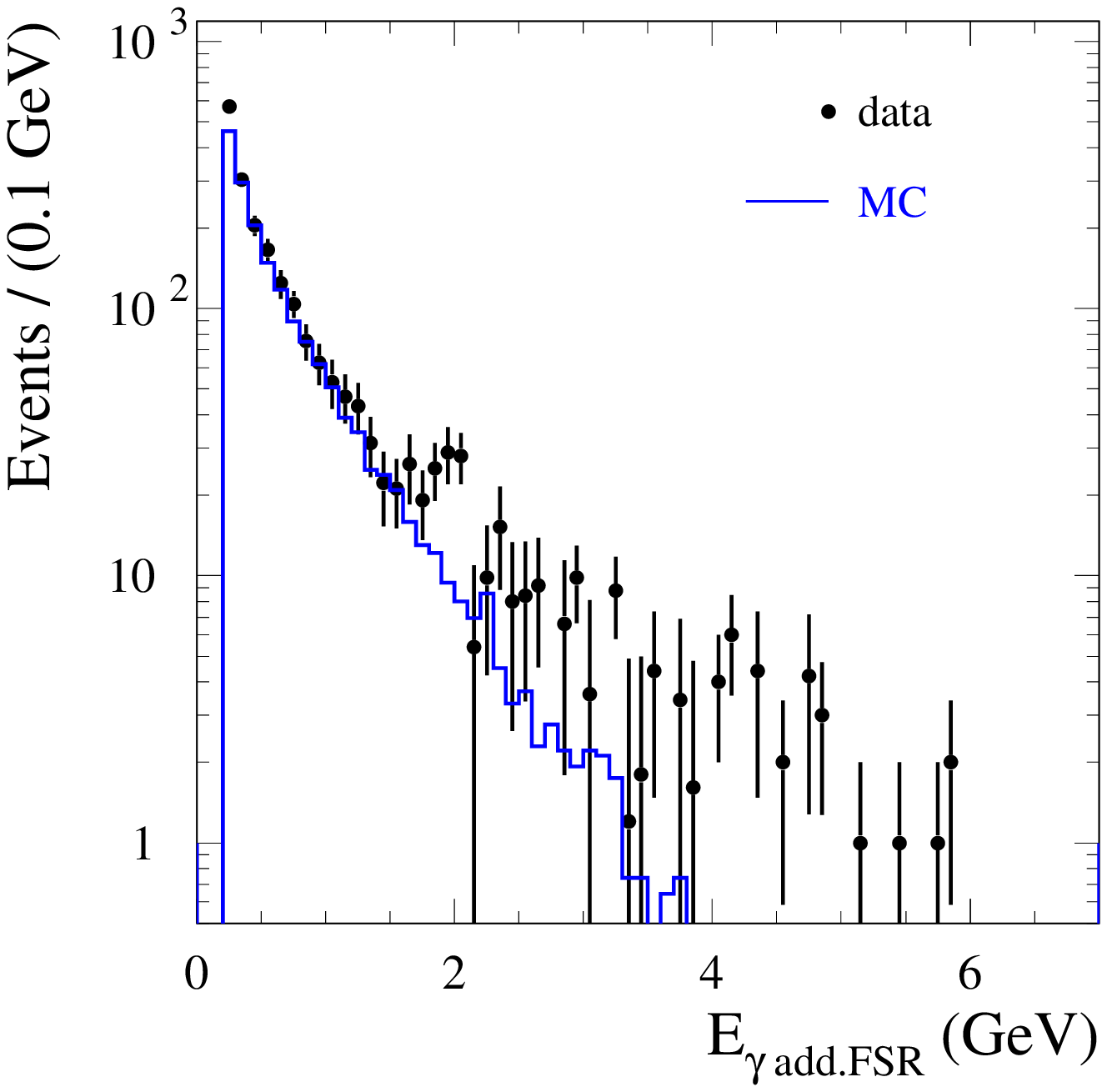}
  \caption{\small 
The additional FSR photon energy distributions
  for the $\mu\mu\gamma\gamma_{\rm ISR}$ (left) and background-subtracted
$\pi\pi\gamma\gamma_{\rm ISR}$ (right) events with
  $\ln{(\chi^2_{\rm FSR}+1)}<\ln{(\chi^2_{\rm ISR}+1)}$, $E_{\gamma~\rm add.FSR}>0.2\gev$,
$\ln{(\chi^2_{\rm FSR}+1)}<2.5$, 
 in the mass intervals $0.2<m_{\mu\mu}<1\gevcc$, $0.5<m_{\pi\pi}<1\gevcc$,
and $\theta_{\mu\gamma_2}<20^\circ$, $\theta_{\pi\gamma_2}<10^\circ$ 
 (data:~black points with errors, MC:~blue histogram). The mass regions are chosen to be similar for
the muon and pion samples and the MC is normalized to data luminosity.}
  \label{FSR-energy}
\end{center}
\end{figure*}

The components of large-angle ISR and
FSR are identified looking at the extra photon angular distribution with
respect to the outgoing charged particles. Figure~\ref{FSR-thgam} (left) shows 
the distribution of $\theta_{\mu\gamma_2}$, the smallest of the two angles 
with either muon  
for $m_{\mu\mu}<~1\gevcc$. A clear peak is observed at small photon-muon angle, 
thus indicating a true FSR signal in data in agreement with the simulation. 
Evidence for large-angle ISR is also seen in data, at variance with AfkQed. 
This major discrepancy is expected, as additional ISR in AfkQed is 
constrained to be collinear with the beams.
The same situation is observed for pions. After subtraction
of a residual background from the ISR $\pi^+\pi^-\pi^0\gamma$ process,
the FSR peaks in data and MC are in fair agreement 
(Fig.~\ref{FSR-thgam} (right)). 

The photon energy distributions for the $\mu\mu\gamma\gamma_{\rm ISR}$ data and MC samples
are given in Fig.~\ref{FSR-energy} (left) for the 
sub-samples satisfying the requirement $\theta_{\mu\gamma_2}<20^\circ$. 
The main component of this sub-sample is from FSR but a contribution
from large-angle ISR is estimated from the $\theta_{\mu\gamma_2}$ distribution
and taken with a 25\% systematic uncertainty.
Absolute rates in data and MC are compared, showing a good agreement up to
$E_{\gamma~\rm add.FSR} \sim 2\gev$, and a small excess in data in the tail above. 
After correction for the remaining ISR contribution below $20^\circ$,
the ratio data/MC for the fraction of 
additional FSR amounts to $0.96\pm0.06$. Therefore the use of {\small PHOTOS} to generate
FSR photons is in good agreement with data and adequate for our precision 
goal, since the uncertainty on the FSR rate represents about $8\times 10^{-4}$ 
of the total $\mu\mu(\gamma)\gamma_{\rm ISR}$ sample.

Figure~\ref{FSR-energy} (right) shows the same distributions for pions after
subtraction of the residual $\pi^+\pi^-\pi^0\gamma$ background.
From the comparison of data and MC events in the
FSR region defined by $\ln(\chi^2_{\rm FSR}+1)<\ln(\chi^2_{\rm ISR}+1)$, 
$E_{\gamma~\rm add.FSR}>0.2\gev$, and $\theta_{\pi\gamma_2}<10^\circ$, 
some excess is observed in $\pi\pi(\gamma)\gamma_{\rm ISR}$ data 
with respect to the {\small PHOTOS} expectation. In the 0.5--1.0\gevcc mass range, the 
excess is $(21\pm5)$\%, taking into account subtraction of background and 
the large-angle ISR contribution. This difference is accounted for in the
determination of the $\chi^2$ selection efficiency, as discussed below.

\subsection{$\chi^2$ selection efficiency}
\label{eff-chi2}
\subsubsection{Overview for $\pi\pi(\gamma)\gamma_{\rm ISR}$ and $\mu\mu(\gamma)\gamma_{\rm ISR}$}

After the hadronic channels are removed using the PID (Sect.~\ref{pid-sep}),
the remaining background contributions to the $\mu\mu(\gamma)\gamma_{\rm ISR}$
sample are essentially from $\tau\tau(\gamma)$ events with two muons in the 
final state, or one muon and a pion misidentified as `muon'. This contribution
is small and well simulated, so it can be handled easily even in the
background (BG) region of the 2D-$\chi^2$ plane (Sect.~\ref{mumu-background}).
Thus the determination of the loose $\chi^2$ selection efficiency is
straightforward in the muon channel.
In contrast, for the pion channel, it is not possible to 
directly measure the efficiency of the 2D-$\chi^2$ selection in data because of 
an overwhelming background in the rejected region (controlling the loose selection), 
even in the intermediate region (controlling the tight selection). 

The rejected signal events with large $\chi^2$ have several sources:
(i) bad input to the kinematic fits, mostly from the direction of the ISR
photon,
(ii) tails of the $\chi^2$ distributions of events with additional ISR or FSR,
(iii) more than one additional photon (mostly ISR), and
(iv) secondary interactions.  
Except for the last type that is specific to pions, the other sources are
common to pions and muons. A very small difference is also expected for the 
tail of the FSR-fit $\chi^2$, as the FSR level is slightly different for 
pions and muons. However the level of additional FSR is measured in data and 
MC and the loss of events due to FSR can be controlled.

The strategy is hence to rely on the $\chi^2$ selection studies performed on muon data
to account for the common losses and to further investigate the losses 
specific to pions. Therefore the $\chi^2$ selection efficiency in data is
derived from the following expression:
\beqn
\label{chi2-mu-pi}
 \varepsilon^{\pi\pi(\gamma)\gamma_{\rm ISR}~\rm data}_{\chi^2} &= &
\varepsilon^{\mu\mu(\gamma)\gamma_{\rm ISR}~\rm data}_{\chi^2} + \delta\varepsilon_{\chi^2}^{\pi/\mu}~,
\eeqn
where the $\mu/\pi$ correction term 
$\delta\varepsilon_{\chi^2}^{\pi/\mu}$ accounts for two effects: 
i) the difference in additional FSR between pions and muons, and ii) pion 
interactions. The contributions of the two components are separated, measured 
in the simulation, and corrected for data/MC discrepancies, 
according to the procedures outlined below.

\subsubsection{Determination of the $\chi^2$ selection efficiency for $\mu\mu(\gamma)\gamma_{\rm ISR}$}

The efficiency of the 2D-$\chi^2$ selection is measured by the rate
of $\mu\mu(\gamma)\gamma_{\rm ISR}$ events in the rejected region. The spectrum in the signal 
region is obtained in data by solving Eqs.~(\ref{sep-pid}) in each mass bin.
The same procedure applied in the BG region yields directly the mass spectrum 
of produced muon events rejected by the loose $\chi^2$ selection. 
A small contribution from $\tau\tau$ must be explicitly
subtracted using the simulation. 
The efficiency is directly deduced from the ratio of the two spectra.

\begin{figure}[htp]
  \centering
  \includegraphics[width=7.5cm]{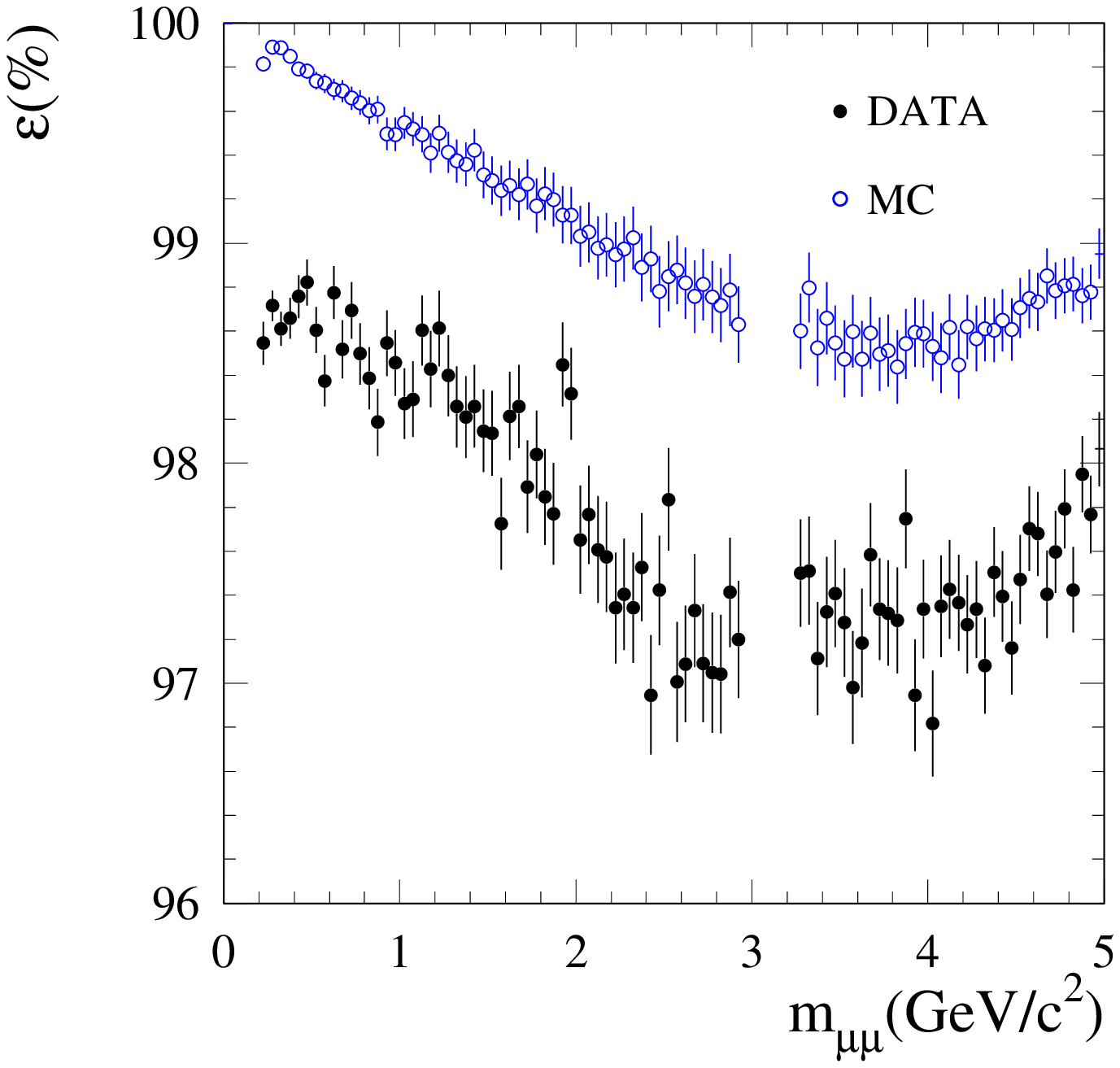}
  \includegraphics[width=7.5cm]{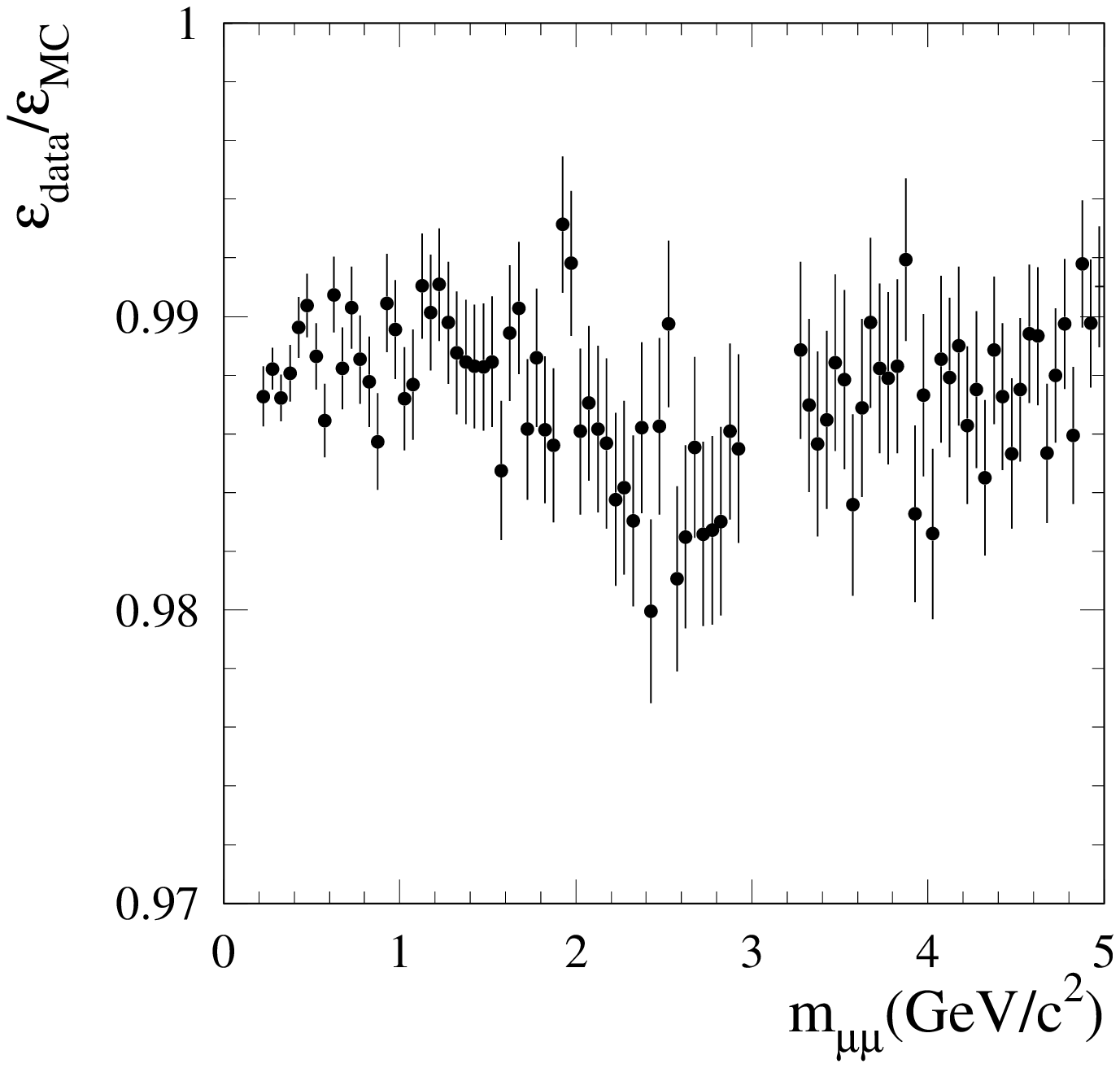}
  \caption{\small The $\chi^2$ efficiency (top) for $\mu\mu(\gamma)\gamma_{\rm ISR}$
  data (after background
  subtraction), MC (AfkQed), and the ratio $C_{\chi^2}$ of data to MC (bottom),
  as a function of $m_{\mu\mu}$. The gap at $3.0-3.2\gevcc$ corresponds to the 
  excluded $J/\psi$ window (Sect.~\ref{mumu-background}).}
  \label{chi2-eff}
\end{figure}

Figure~\ref{chi2-eff} gives the measured $\chi^2$ efficiency as a function 
of $m_{\mu\mu}$. It is lower in data than the prediction
from the simulation by 1.2\%. Most of the discrepancy arises from
the absence of large-angle ISR in AfkQed, which is present in data and 
generates some loss when the large-$\chi^2$ tails are removed, 
as expected from Fig.~\ref{chi2-add-ISR}. Evidence for this loss
appears at $\mu\mu$ threshold where FSR vanishes, while in simulation the 
efficiency deviates from one by only $5\times 10^{-4}$.
The efficiency decrease with $m_{\mu\mu}$ is due to the loss
of large-$\chi^2$ FSR events and reflects the increasing rate of additional
FSR photons with large energy ($E_{\gamma~\rm add.FSR}>0.2\gev$). 
The same behavior is observed in data and
simulation, consistent with the fact that additional FSR is well described
in the simulation.

The systematic uncertainty on the determination of the $\chi^2$ efficiency
comes exclusively from the estimate of the background in the muon channel, 
due to the uncertainty on the $\tau\tau$ cross section and on the PID 
$\varepsilon_{j\to`i\rq}$ efficiencies. These uncertainties amount to a few $\times 10^{-4}$
and are incorporated in the point-to-point errors.

\subsubsection{Effect of additional FSR for pions}
\label{acc-fsr}

The first component in $\delta\varepsilon_{\chi^2}^{\pi/\mu}$ comes from the 
difference in additional FSR between pions and muons. 
The difference in the FSR rate due to the $\pi-\mu$ mass 
difference is observed in simulation at the expected level ($\sim 25$\%) and
the expected loss of efficiency due to additional FSR is consequently lower 
by that amount in $\pi\pi(\gamma)\gamma_{\rm ISR}$ with respect to 
$\mu\mu(\gamma)\gamma_{\rm ISR}$. 

The contribution of additional FSR to the $\chi^2$ inefficiency is the product
of the fraction of FSR events lost by the $\chi^2$ selection times
the rate of FSR events in the full pion data sample.
To estimate this loss, one relies on the $\chi^2_{\rm FSR}$ distributions
of the events with additional FSR as shown in Fig.~\ref{chi2-add-FSR}. The shapes 
are similar in the muon and pion channels up to $\ln(\chi^2_{\rm FSR}+1)=2.5$ 
(which coincides with the lower edge of the `BG' region defined in Fig.~\ref{pi-2d-chi2}),
at which value the pion interaction component turns on. The shapes in data 
agree well with the MC shapes, over the entire distribution in $\mu\mu(\gamma)\gamma_{\rm ISR}$,
and, in $\pi\pi(\gamma)\gamma_{\rm ISR}$, until the distribution
in data is affected by background in addition to interactions. 
The fraction of FSR events lost in the pion channel
is consequently estimated from the fraction of FSR events that fall beyond the 
$\chi^2_{\rm FSR}$ selection boundary in the muon data ($(35\pm5)\%$).
This fraction of lost FSR events is normalized to the rate of additional FSR events 
observed in the full pion MC sample in the $\ln(\chi^2_{\rm FSR}+1)<2.5$ region ($0.64\%$).  
The loss is further corrected to account for the observed $(21\pm5)$\% data/MC difference 
in FSR rates (Sect.~\ref{add-fsr}).

The resulting data/MC correction to the loose $\chi^2$ efficiency due to 
the pion-muon FSR difference is estimated to be $(0.6\pm 0.2)\times 10^{-3}$.  
This correction is slightly overestimated as a fraction of the additional
FSR events is in the signal region of the loose $\chi^2$ criterion. 
The correction to the tight $\chi^2$ selection efficiency is larger, 
$(1.9\pm0.8)\times 10^{-3}$, since all the 
FSR events with photon energy larger than about 0.2\gev are lost with the 
tight criterion ($\ln(\chi^2_{\rm ISR}+1)<3$).

\subsubsection{Effect of pion interactions}
\label{sec-int-study}

The effects of secondary interactions are mostly
seen in the tracking efficiency because of the tight requirements imposed on the 
track pointing to the interaction region. The small residual effect in the 
2D-$\chi^2$ selection efficiency is  
estimated using the simulation, essentially by comparing the behavior
of muon and pion events, and corrected for data-MC difference in interaction rates. 

It is found in simulation that the difference of 2D-$\chi^2$ selection efficiencies,
$\delta\varepsilon_{\chi^2}^{\pi/\mu}$,
between $\pi\pi(\gamma)\gamma_{\rm ISR}$ and $\mu\mu(\gamma)\gamma_{\rm ISR}$ is 
about $-1.2\times 10^{-3}$ at 0.75\gevcc.
As we know that the loss of additional FSR events is smaller for pions,
since the FSR rate is lower,
the smaller efficiency in $\pi\pi(\gamma)\gamma_{\rm ISR}$ is ascribed to pion interactions.
The $\chi^2$ selection efficiency loss from secondary interactions 
estimated this way in simulation is $(2.8\pm0.2)\times 10^{-3}$ for the loose criteria, 
flat with mass, and $(1.4\pm0.1)\times 10^{-2}$ for the tight criteria, with some $\pm20$\% 
relative variation with mass.

Two methods are considered to isolate interacting events in both data and MC,
and data/MC corrections to the $\chi^2$ efficiencies obtained above in simulation
are estimated from the respective rates of pion interactions. 
The corresponding events populate the diagonal region in the
2D-$\chi^2$ plane, extending through the $\chi^2$ selection boundary, and therefore 
affect the $\chi^2$ selection efficiency.

In the first method, interactions are tagged by the
presence of `bad' tracks ({\it i.e.}, tracks not satisfying
the track requirements of the ISR two-body analysis) in addition to the two good
tracks of the selected events, provided a secondary vertex can be found between
a bad track and one of the two good tracks. 
Because of the strict requirements on good tracks, most tagged interactions occur in 
the beam pipe, with further contributions from the first SVT layers. 
The data/MC ratio of interacting events estimated with this method is found to 
be $1.44\pm0.10$ in 
the intermediate $\chi^2$ region, and $1.43\pm0.13$ in the background region. 
However the efficiency of this procedure to tag interacting pions is rather low 
since it keeps about 10\% of the events with secondary 
interactions in the signal (loose) region and 25\% in the background region.

\begin{figure}[htp]
  \centering
  \includegraphics[width=7.5cm]{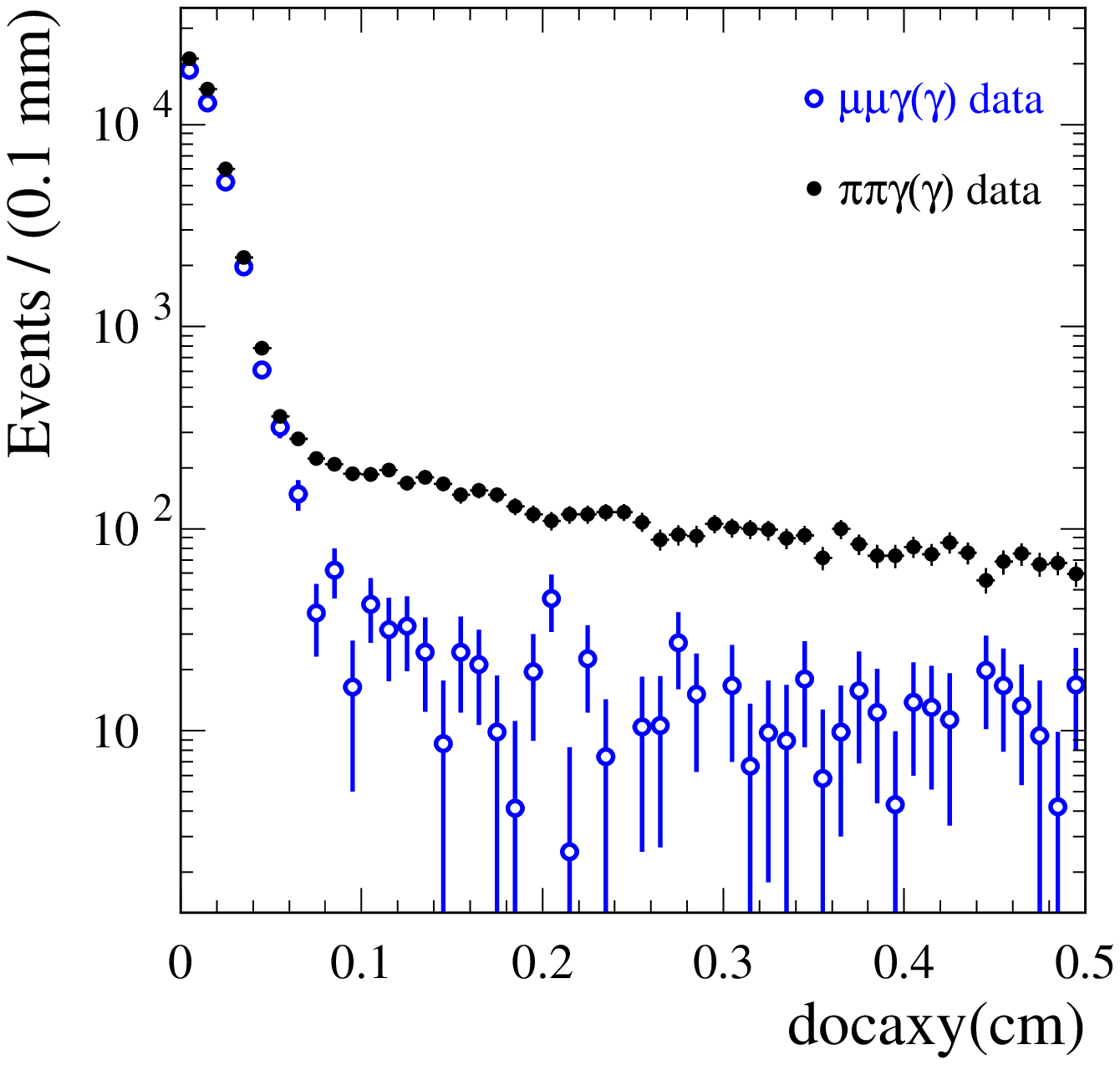}
  \includegraphics[width=7.5cm]{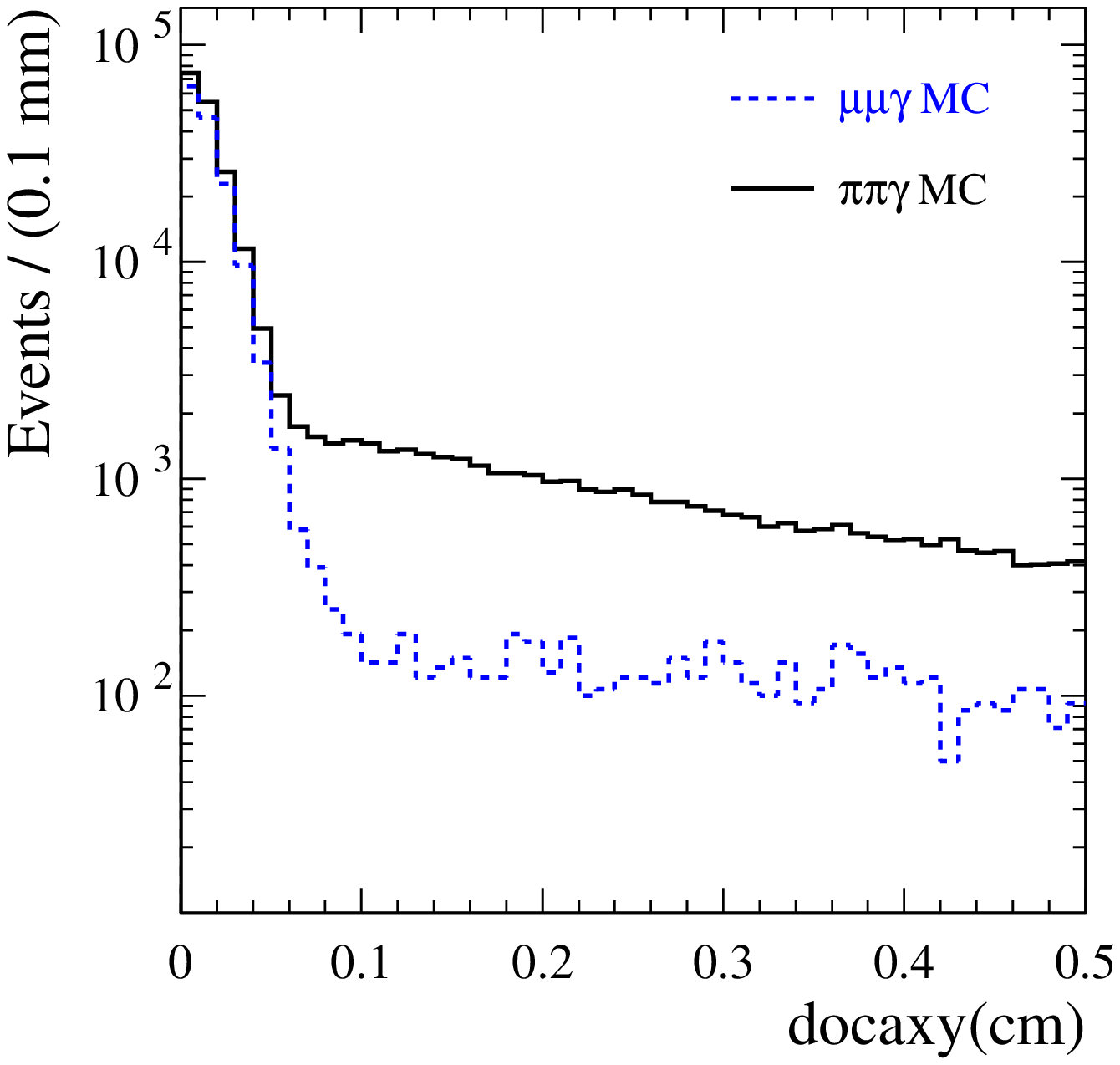}

  \caption{\small 
(top): The distribution of the largest of the two transverse distances of 
closest approach to the interaction point $doca_{\rm xy}^{\rm max}$ for pions and 
muons in data, for the intermediate $\chi^2$ region.
(bottom): Same distributions in simulation.}
  \label{docamax-pi-mu}
\end{figure}

The second method tags a much larger fraction of interacting events. 
The quantity $doca_{\rm xy}^{\rm max}$ is defined to be the largest of
the $doca_{\rm xy}$ for the two tracks in the event, each limited by the requirement
$doca_{\rm xy}<0.5\cm$ used in the good track definition (Sect.~\ref{evt-sel}). 
The sensitivity of this variable
to secondary interactions can be appreciated in Fig.~\ref{docamax-pi-mu}, 
showing a striking difference in the tail of the distributions for pions
and muons. The same behavior is also observed between pions in events 
satisfying the tight or loose $\chi^2$ criteria. 
The selection of events with $doca_{\rm xy}^{\rm max}>0.1\cm$ retains about 50\% 
of interactions, with a background from non-interacting events that is 
estimated from the muon distribution.
Again it is found that the level of secondary interactions is underestimated
in the simulation, with a data/MC ratio of $1.52\pm0.03$ in the
intermediate $\chi^2$ region. A reliable determination cannot be achieved in the
background region with this method, because of multihadronic background. 

The second determination is more accurate and dominates the average data/MC
ratio of interaction rates, which is $1.51\pm0.03$.

\subsubsection{Summary of corrections to the $\chi^2$ selection efficiency for pions}

Adding the two components of the $\chi^2$ selection inefficiency in $\pi\pi(\gamma)\gamma_{\rm ISR}$ 
that are not common to the muon channel, the extra data/MC correction 
to apply to $\delta\varepsilon_{\chi^2}^{\pi/\mu}$ for the 
loose $\chi^2$ selection amounts to 
$(1.4 \pm 0.1)\times 10^{-3}$ for secondary interactions and 
$(0.6 \pm 0.2)\times 10^{-3}$ for FSR. The total correction is 
$(2.0\pm0.3)\times 10^{-3}$.

For the tight $\chi^2$ selection, both corrections are larger:
$(7.1\pm0.4)\times 10^{-3}$ for interactions and
$(1.9\pm0.8)\times 10^{-3}$ for FSR.
The total data/MC correction on the tight $\chi^2$ selection efficiency amounts to 
$(9.0\pm0.9)\times 10^{-3}$.

\section{Backgrounds}
\label{background}

\subsection{Backgrounds in the $\mu\mu(\gamma)\gamma_{\rm ISR}$ channel}
\label{mumu-background}

Contributions from hadrons to the `$\mu\mu$' sample are removed by PID
through solving Eqs.~(\ref{sep-pid}), as described in Section~\ref{pid-sep},
but one must still consider background from 
processes producing real muons.

ISR-produced $J/\psi$ followed by decay to $\mu\mu$ is not a background to the
complete $\mu\mu(\gamma)\gamma_{\rm ISR}$ process, but is a background to the 
purely QED reaction
used for the determination of the ISR luminosity. The $\psi(2S)$
contributes as a
background through its decays to $J/\psi$, either following the
$\pi^0\pi^0J/\psi$ transition or radiative decays through charmonium
states. Both direct and indirect $J/\psi$ production is observed.
These contributions are removed by excluding events where the
measured invariant $\mu\mu$ mass is in the $3.0-3.2\gevcc$ window 
~\footnote{In the final $m_{\mu\mu}$ mass spectrum,
this rejection does not however produce a sharp hole as $m_{\mu\mu}$ is determined
after the $\mu\mu(\gamma)\gamma_{ISR}$ kinematic fits described in Section\ref{kin-fit}.}. 

Another hidden background from $J/\psi$ comes from the radiative decay
$J/\psi\to\mu^+\mu^-\gamma$, which is indeed barely observed in the 
$\mu\mu\gamma$ mass spectrum in data. Its contribution to the $\mu\mu$ 
mass spectrum between 2 and 3\gevcc is of order $10^{-3}$ and neglected. 

The process $e^+e^-\to\tau^+\tau^-(\gamma)$ can contribute to the `$\mu\mu$' 
sample through $\tau\to\mu\nu_{\mu}\nu_{\tau}$ and mis-identified $\tau\to\pi\nu_{\tau}$ decays. 
The contribution is estimated by MC and found to be negligibly small, 
except at masses above 2\gevcc where it reaches a fraction of $10^{-3}$.
It is subtracted using the simulation.

\subsection{Backgrounds in the $\pi\pi(\gamma)\gamma_{\rm ISR}$ channel}
\label{pipi-background}

\subsubsection{Background from $\mu\mu(\gamma)\gamma_{\rm ISR}$}
\label{mubkg-pih}

Separation of each component of the two-prong ISR sample
is achieved through solving Eqs.~(\ref{sep-pid}) in each $m_{\pi\pi}$ mass bin.
This procedure yields the produced spectrum $dN_{\pi\pi}/dm_{\pi\pi}$
and the background contributions to the observed `$\pi\pi$'-identified spectrum. 
The muon background level is less than $4\times 10^{-3}$ at the 
$\rho$ peak but increases rapidly  away from the resonance and reaches a few percent 
at the $\rho$ tail boundaries (Table~\ref{fbg-all-central}).

\begin{figure}[htp]
  \centering
  \includegraphics[width=7.5cm]{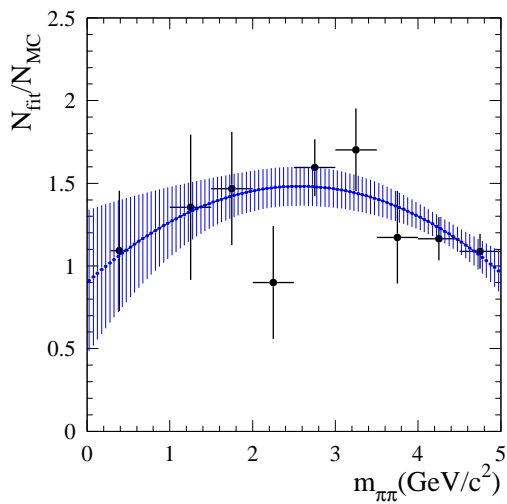}
  \caption{\small The rate of $\mu\mu$ events misidentified
as `$\pi\pi_h$' in data relative to the MC expectation, measured in mass intervals below and
above the $\rho$ region .
The curve with the error band is a second-order polynomial fit to the data
points, used to interpolate through the $\rho$ mass region ($0.6-0.9\gevcc$).}
  \label{mumu-pihpi-fit}
\end{figure}

As the `$\pi\pi_h$' selection applied in mass ranges away from
the $\rho$ precludes using the above procedure, the 
reduced $\mu\mu(\gamma)\gamma_{\rm ISR}$ background contribution is
determined directly from the  `$\pi\pi_h$' sample using the track identified 
as a `$\pi$' with the standard pion identification.
The fit of the distribution, for that track, of the $\pi/\mu$ likelihood 
estimator $P_{\pi/\mu}$ 
(introduced in Section~\ref{pi-hard}) yields the respective true muon and 
pion components of the `$\pi$'-identified tracks, hence of the `$\pi\pi_h$'
sample. Fits are performed in 0.5\gevcc-wide mass bins and, except for the 0.5--1\gevcc 
interval ($\rho$ peak region), the $\mu\mu$ component can be well determined. Above 3\gevcc the muon 
contribution becomes dominant, despite the `$\pi\pi_h$' ID, and the pion 
signal is lost above 4\gevcc. The results of the fits are summarized in 
Fig.~\ref{mumu-pihpi-fit}, which shows, for each mass interval, the 
$\mu\mu\to`\pi\pi_h\rq$ fraction in data relative to the prediction from the simulated 
muon sample after luminosity scaling. A second-order polynomial fit to 
all points allows one to smoothly interpolate between the low and high mass 
regions. The band indicates the error envelope of the fit.

\subsubsection{Background from $KK(\gamma)\gamma_{\rm ISR}$}
\label{Kbkg-pih}

When the standard $\pi$-ID identification is applied to both tracks in the `$\pi\pi$' 
sample, the kaon background is implicitly subtracted through 
solving Eqs.~(\ref{sep-pid}) and stays below the permil level in the $\rho$ peak region
(Table~\ref{fbg-all-central}).
 
\begin{figure}[htp]
  \centering
  \includegraphics[width=7.5cm]{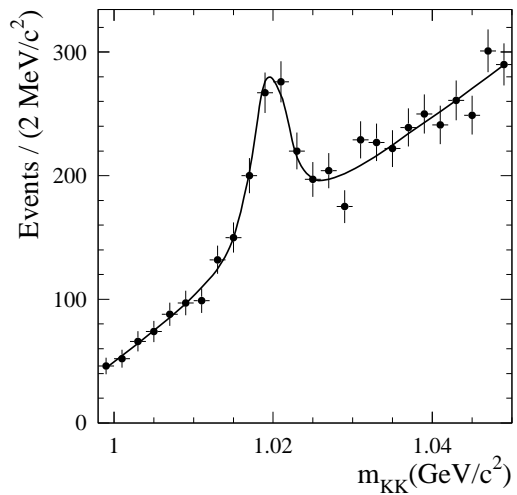}
  \caption{\small The $m_{KK}$ mass distribution in the 
`$\pi\pi_h\gamma$' data sample. The solid line represents the result of
the fit (see text).}
  \label{KK-pipih}
\end{figure}

This background is essentially insensitive to
the further selections applied to the `$\pi\pi_h$' sample.
Since the $KK(\gamma)\gamma_{\rm ISR}$ events are dominated by the narrow $\phi$ resonance, one can
use this feature to determine the $KK$ component directly in data. The 
procedure is illustrated in Fig.~\ref{KK-pipih}, which shows the $m_{KK}$ mass 
distribution of the `$\pi\pi_h$' sample when the $K$ mass is assigned
to both tracks. A $\phi$ signal is fitted with the signal line shape taken 
from simulation and a linear term to describe the dominant true $\pi\pi$ component.
The $\phi$ signal yield provides the normalization of the remaining $KK(\gamma)\gamma_{\rm ISR}$
background contribution to be subtracted from the `$\pi\pi_h$' data. The wide and distorted
shape of the $\phi$ peak reflection in the $m_{\pi\pi}$ mass spectrum 
is taken from simulation. 

\subsubsection{Background from $ee\gamma$ events}
\label{pipi-ee}

Radiative Bhabha events are very strongly suppressed in the event selection
because of the track definition that contains a veto on electrons. 
Remaining events of this type are from distribution tails and various
pathologies. Because of this large selection bias there are very few events 
actually identified as `$ee$'$\gamma$ in the identification process. 
Radiative Bhabha background appears in the `$e\pi$' and `$\pi\pi$' ID topologies. 
This background is identifiable near threshold and at high masses.
Its small contribution cannot be detected in the $\rho$ region and its mass shape and magnitude 
are estimated from a downscaled sample of radiative Bhabha events
normalized near threshold.

\begin{figure}[htp]
  \centering
  \includegraphics[width=8.cm]{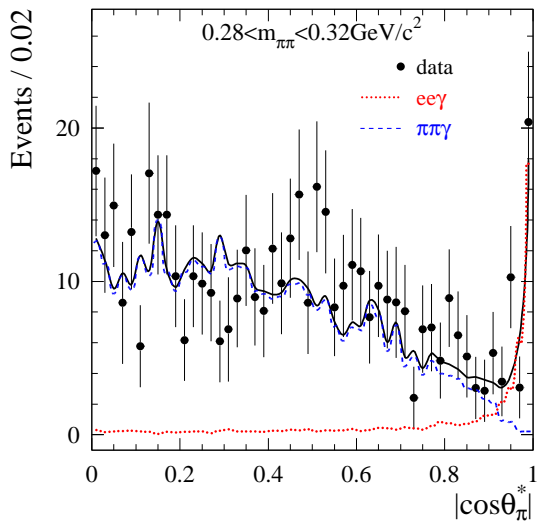}
  \caption{\small(color online). The $|\cos\theta^*_{\pi}|$ distribution of `$\pi\pi_h$' data in
  the 0.28--0.32\gevcc $m_{\pi\pi}$ range, fitted (black curve) to two free 
components: $\pi\pi(\gamma)\gamma_{ISR}$ from MC (blue dashed line) and $ee\gamma$ background from 
downscaled radiative Bhabha events (red dotted line). The small $\mu\mu(\gamma)\gamma_{ISR}$ contribution is 
subtracted out.}
  \label{fit-cosths-pipih}
\end{figure}

The radiative Bhabha background normalization is achieved using the angular 
distribution in the $\pi\pi$ center-of-mass system, assuming the pion mass for 
the particles. The angle $\theta^*_{\pi}$ of the $\pi^+$ is measured relative to the ISR 
photon direction in that frame. 
In the mass range 
$0.28<m_{\pi\pi}<0.32\gevcc$ just above threshold, the remaining  $ee\gamma$
background contribution, still noticeable with the `$\pi\pi_h$' identification,
is obtained by fitting the $|\cos\theta^*_{\pi}|$ distribution.
Backgrounds from $\mu\mu(\gamma)\gamma_{\rm ISR}$ and $KK(\gamma)\gamma_{\rm ISR}$ 
are subtracted before fitting, with shapes 
taken from simulation with correction from the data, and normalized to data luminosity.
The $|\cos\theta^*_{\pi}|$ distribution is fitted with two components: 
$\pi\pi(\gamma)\gamma_{ISR}$, with the shape taken from the simulation, and $ee\gamma$
background with the shape obtained from the downscaled radiative Bhabha sample. The latter
contribution has a characteristic sharp peak near one with a long tail
while the $\pi\pi(\gamma)\gamma_{\rm ISR}$ signal behaves as $\sin^2\theta^*_{\pi}$.
The fit shown in Fig.~\ref{fit-cosths-pipih} on the `$\pi\pi_h$' sample
at threshold provides the normalization factor 
to be applied to the radiative Bhabha sample to describe the $ee\gamma$ 
background.
A similar fit performed at threshold on the `$\pi\pi$' sample yields a consistent
normalization factor. 

The mass dependence of the $ee\gamma$ background is checked using the
sample of events identified as `$e\pi$'. The latter is a rather pure $ee\gamma$ sample
outside the $\rho$ region, and is more representative of 
events mis-identified as `$\pi\pi_h$' than the genuine radiative Bhabha downscaled sample. 
The ratio of the mass spectra of 
`$e\pi$' events to the radiative Bhabha sample is found to be constant within uncertainties, 
for masses away from the $\rho$ peak, {\it i.e.}, just above threshold and in the 
1.5--3\gevcc range.

The $ee\gamma$ background is only noticeable at threshold and near the edges
of the $\rho$ central region: 6\% for $m_{\pi\pi}<0.32\gevcc$ and less than
0.05\% for larger masses for the `$\pi\pi_h$' selection, and 0.63(0.27)\% at 
$m_{\pi\pi}=0.525(0.975)\gevcc$ for the `$\pi\pi$' selection. At the $\rho$ peak 
the fraction drops to only 0.03\%.

A systematic uncertainty of 100\% is assigned to
the $ee$ normalization factor determined at threshold, which is applied up to 
3\gevcc. This precision is adequate in the $\rho$ region as well as in the resonance
tails.

\subsubsection{Conversions and rejection of displaced vertices}
\label{Vxy-ee}

Purely electromagnetic processes may induce backgrounds when one of the
final-state particles interacts with the detector material, allowing the
selection criteria to be satisfied. This is the case at threshold from 
the $e^+e^-\to\gamma\gamma$ process followed by a photon conversion, and 
at large masses from Bhabha scattering where one of final electrons (positrons)
undergoes bremsstrahlung in the beam pipe. In either case, one or both of the
detected tracks can be misidentified as pions. However, as they do not originate 
from the interaction point, this contamination is reduced by requiring
the distance in the transverse plane $V_{\rm xy}$ of the vertex of the two tracks
and the average interaction point to be small. 

The background from conversions is expected to yield a rather wide 
$V_{\rm xy}$ distribution, while prompt particles (from $ee\gamma$ and $\pi\pi(\gamma)\gamma_{ISR}$)
produce a peak at zero. The requirement $V_{\rm xy}<0.5\cm$ is applied in the $\rho$ tails, 
which are the mass regions affected by the background from conversions.
The conversions are reduced to a negligible level ($<5\times 10^{-4}$) by the selection.
The efficiency of the $V_{\rm xy}$ requirement for $\pi\pi(\gamma)\gamma_{ISR}$ events is controlled
by the two-track vertexing and pion secondary interactions. The former effect
is studied in data and simulation using the $\mu\mu(\gamma)\gamma_{ISR}$ sample.
The effect of the pion secondary interactions
is studied in the $\rho$ region, both in data and MC. The overall correction
to the MC efficiency is $(1.1\pm0.1)$\% at 0.4\gevcc and smaller for larger masses.

\subsubsection{$p\overline{p}\gamma_{\rm ISR}$ process}

Proton ID is not considered in the particle identification process, since
the process $p\overline{p}\gamma_{ISR}$ contributes at a very small level. 
With the chosen ID classes protons are classified as pions, and antiprotons
sometimes as electrons. The cross section for the $p\overline{p}\gamma_{\rm ISR}$ process 
has been measured
by \babar~\cite{ppb} and the results are used to reweight the MC prediction.
The overall contamination is taken from the reweighted simulation and 
subtracted statistically. It amounts to less than 0.5\% in the $\rho$ central 
region (Tab.~\ref{fbg-all-central}) and exceeds the percent level at large masses only
($m_{\pi\pi}> 1.1\gevcc$, Tab.~\ref{fbg-all-tails}).

\subsubsection{Multihadrons from the $q\overline{q}$ process}
\label{bkg-had}

Hadronic processes, either direct or ISR-produced, introduce a background
in the pion sample that is considerably reduced by the requirement of only 
two good tracks and the $\chi^2$ selection of the kinematic fits.
This contribution is estimated using simulated samples of the
$e^+e^-\to q\overline{q}$ process.
However, the {\small JETSET} prediction for $q\overline{q}$ fragmentation into 
low-multiplicity final states is not necessarily reliable, so the MC rate is 
normalized using data. 

In backgrounds from the $q\overline{q}$ process, the ISR photon candidate 
actually originates from the decay of an energetic $\pi^0$.  
Such a signature is searched for, both in data and in MC, by pairing the ISR 
photon candidate with all detected additional photons and the MC normalization
is obtained from the observed $\pi^0$ rates. The pair with $\gamma\gamma$ mass 
closest to the nominal $\pi^0$ mass is retained.
Fits to $\gamma\gamma$ mass distributions are performed
in data and MC assuming a Gaussian 
shape for the $\pi^0$ signal and taking into account `background' from 
$\pi\pi(\gamma)\gamma_{\rm ISR}$ events and contributions from other processes, like 
$e^+e^-\to\tau\tau(\gamma)$ events, both taken from the simulation.

\begin{figure}[htp]
  \centering
  \includegraphics[width=7.5cm]{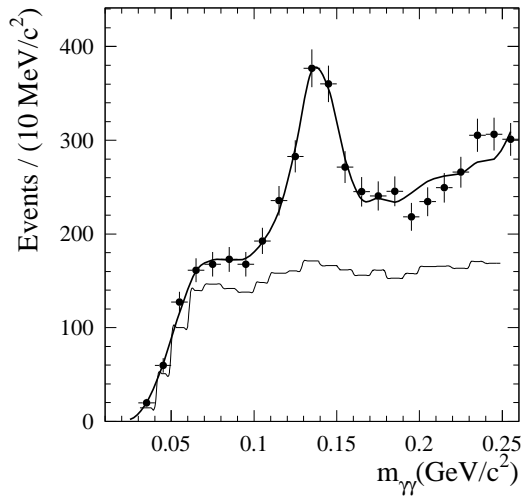}
  \caption{ \small 
The $\gamma_{\rm ISR}\gamma$ mass distribution for data events in a background 
enriched region.
The $\pi^0$ signal is fitted with a Gaussian while additional
contributions are represented by a linear term.
The histogram is the $\pi\pi(\gamma)\gamma_{\rm ISR}$ MC distribution. 
Contribution from $\tau\tau$ events has been subtracted.}
  \label{pi0-fit-sleeve}
\end{figure}

The $\pi^0$ fits are carried out in wide $\pi\pi$ mass bins (0.5\gevcc) between 
threshold and 3\gevcc, covering the practical range for the analysis.  
Fits are also performed in background enriched regions
to check the sensitivity to the final-state multiplicity. 
An example of fits is shown in Fig.~\ref{pi0-fit-sleeve}.
{\small JETSET} is found to overestimate the background 
contributions by a factor of 1.3, almost independent 
of the $\pi\pi$ mass and whether it is determined in the signal or background 
enriched regions.

We check that the $\pi^0$ finding efficiency does not depend on the final state
produced by fragmentation in {\small JETSET}. The results are consistent within 
5\% for $\pi^+\pi^-\pi^0$, $\pi^+\pi^-2\pi^0$, and the full $q\overline{q}$ 
contribution. The final state with lowest multiplicity, $\pi^+\pi^-\pi^0$, 
which is topologically and kinematically identical to the signal, is further
controlled on data. As the $\pi\pi$ mass distribution for these events in 
{\small JETSET} peaks between 1 and 2\gevcc, well above the $\rho$ peak, we 
search for a $\pi^0$ signal in that mass range where the 
$e^+e^-\to\pi\pi(\gamma)\gamma_{\rm ISR}$ contribution is small. The weak 
signal observed in data is consistent with the
negligible {\small JETSET} expectation within a 50\% uncertainty.

After renormalization, the $q\overline{q}$ background fraction is below 1.8\% 
in the loose $\chi^2$
region (Tab.~\ref{fbg-all-central}) and much smaller when the tight $\chi^2$ 
criterion is applied, except 
above 1.3\gevcc\ where it reaches a few percent (Tab.~\ref{fbg-all-tails}).

The MC statistical errors are included in the subtracted $q\overline{q}$ 
background spectra, while the uncertainty in the normalization 
(from $\pi^0$ fits), typically 10\%, is taken as a systematic error. 

\subsubsection{Multihadronic ISR processes}
\label{bkg-isr}

The background is estimated using simulated processes $e^+e^-\to X\gamma_{\rm ISR}$
where $X$ stands for the final states: $\pi^+\pi^-\pi^0$, $\pi^+\pi^-2\pi^0$,
$2\pi^+2\pi^-$, $2\pi^+2\pi^-\pi^0$, $\eta\pi^+\pi^-$, and $K_SK_L$. 
The dominant contributions are from
$e^+e^-\to \pi^+\pi^-\pi^0\gamma$ and
$e^+e^-\to \pi^+\pi^-2\pi^0\gamma$. They sum to about 10\%
at the lower edge of the $\rho$ peak (Tab.~\ref{fbg-all-central}) 
but are strongly reduced by the tight $\chi^2$ criterion.

An approach similar to $q\overline{q}$ comparing data and MC is followed 
for the $\pi^+\pi^-\pi^0$ ISR normalization. This
process is dominated by the production of the $\omega$ and $\phi$
narrow resonances, which are used as calibration signals. 
A kinematic fit to the $\pi^+\pi^-3\gamma$ final state is performed using 
a $\pi^0$ constraint, and the $\pi^+\pi^-\pi^0$ mass distribution is fitted.
The ratio of the contributions in data and MC is found to be $0.99\pm0.04$. 
The error on the normalization factor is taken as a systematic uncertainty.

The remaining ISR processes are 
higher-multiplicity $\pi^0$ hadronic states such as $2\pi 3\pi^0\gamma$. 
These cross sections have not yet been measured by \babar\, but we estimate
that the contributions of these channels to the total background do not 
exceed the $10^{-3}$ level.
It is estimated from MC alone, assuming a normalization uncertainty of 10\%.

\subsubsection{Background from other processes}
\label{bkg-other}

The $e^+e^-\to \tau^+\tau^-(\gamma)$ process contributes significantly only 
at $\pi\pi$ masses higher than the range of interest for this analysis.
Although at a very small level, this background is subtracted using simulation.
Two-photon processes with hard radiation such as 
$e^+e^-\to (e^+ e^-) \pi^+\pi^-\gamma$ and the similar reaction with muons
have been specifically looked for in kinematic regions where they are expected
to contribute, but without finding a significant effect.

A summary of backgrounds and related errors are given in Tables~\ref{fbg-all-central} 
and ~\ref{fbg-all-tails}.

\subsubsection{Overall test of the multihadronic background}
\label{bkg-test}

The multihadron background fraction estimated above reaches sizeable values near
the boundaries of the central $\rho$ region, but with a quite small 
uncertainty, $4.8\times 10^{-3}$ at 0.5\gevcc and $3.0\times 10^{-3}$
at 1\gevcc, the value at the $\rho$ peak being negligible. In the $\rho$ tails,
the estimated systematic errors due to multihadron backgrounds, which are 
strongly reduced by the tight $\chi^2$ criterion, do not exceed a few permil.

\begin{figure}[thp]
  \centering
  \includegraphics[width=7.5cm]{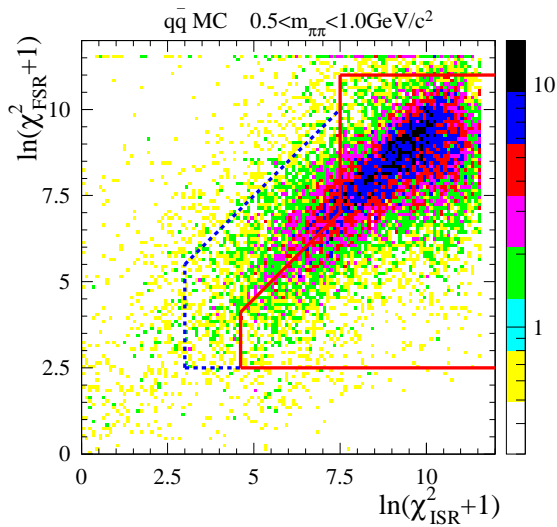}
  \caption{ \small(color online).
2D-$\chi^2$ distribution of $q\overline{q}$ MC events normalized
to the data luminosity for $0.5<m_{\pi\pi}<1\gevcc$. The solid broken line
indicates the loose $\chi^2$ criterion, while the dashed line defines
a `sleeve' in the signal region where most of the background is concentrated.}
  \label{chi2-sleeve}
\end{figure}

\begin{figure}[htp]
  \centering
  \includegraphics[width=7.5cm]{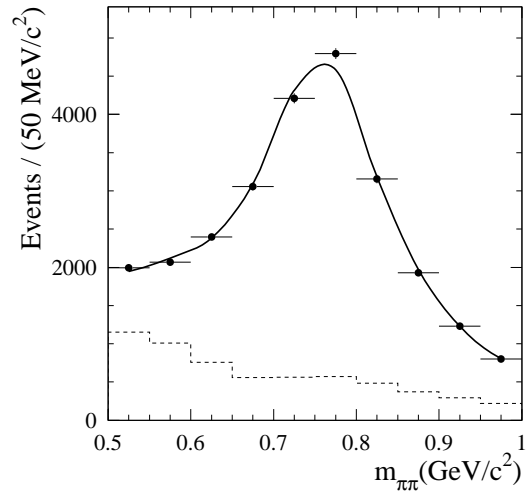}
  \caption{\small 
The $m_{\pi\pi}$ distribution in the background-rich `sleeve' region. 
The solid line represents a fit to the data with $\pi\pi$ signal 
and multihadron background components, with their shapes both taken from 
simulation.}
  \label{sleeve-mass}
\end{figure}

We assess both the rate and the mass distribution of the multihadron background
in data,  
in the 2D-$\chi^2$ region where it is the largest, {\it i.e.}, in the 
`sleeve' outlined in Fig.~\ref{chi2-sleeve}. We fit the $\pi\pi$ mass distribution 
in this region (Fig.~\ref{sleeve-mass}) to background and signal components,
with shapes taken from MC. The ratio of the fitted background to the
one estimated in the above sections is found to be $0.968\pm0.037$, consistent with unity. 
This is translated into an uncertainty of $4.5\times 10^{-3}$ at 0.5\gevcc and 
$1.5\times 10^{-3}$ at 1\gevcc on the background fraction in the full $\pi\pi$ sample. 
These values are below the quoted uncertainties, which validates the multihadron 
background estimate and confirms that the mass distribution 
of the background from the simulation is appropriate.  


\section{Mass spectra determination}
\label{mass-spectrum}
The spectra of $\mu\mu(\gamma)\gamma_{\rm ISR}$ and $\pi\pi(\gamma)\gamma_{\rm ISR}$ events 
after event selection are obtained as functions of the two-track mass 
$m_{\mu\mu}$ ($m_{\pi\pi}$) given by the best $\chi^2_{\mu\mu}$ ($\chi^2_{\pi\pi}$) 
fit. These spectra are background-subtracted, and mass-dependent corrections
for data/MC efficiency differences are applied as described above. 
To account for FSR effects and resolution 
smearing due to the detector response, unfolding is required to
obtain the $dN/d\sqrt{s'}$ spectra as functions of the final-state mass 
including FSR, which are used to measure the cross sections through 
Eq.~(\ref{def-lumi}).

\subsection{Unfolding of the mass spectra}
\label{unfold}

\subsubsection{The unfolding method}

The unfolding technique used in the present analysis is a simplified version 
of a method developed for more complex unfolding problems~\cite{bogdan}. 
The folding probability $P_{ij}$ of an event produced in a true ($\sqrt{s'}$) 
bin $j$ to be reconstructed in a ($m_{xx}$) bin $i$ is computed directly in 
simulation from the transfer matrix $A_{ij}$ (the number of events produced 
in a true bin $j$ that are reconstructed in bin $i$)~\footnote{The  matrix of folding probabilities is related to the 
transfer matrix $A_{ij}$ by $P_{ij} = A_{ij}/\sum_{k=1}^{N}{A_{kj}}$
while the matrix of unfolding probabilities is
$P'_{ij} = A_{ij}/\sum_{k=1}^{N}{A_{ik}}$, where $A_{ij}$ is the number of 
events produced in a true bin $j$ that are reconstructed in bin $i$.}. 
Conversely, the matrix of unfolding probabilities $P'_{ij}$ indicates 
the probability for an event reconstructed in a bin $i$ to originate from the 
true bin $j$, and is also computed from the transfer matrix.
$A_{ij}$ and $P'_{ij}$ depend on the assumed true spectrum
while $P_{ij}$, which describes detector and FSR effects, does not.
The method used to unfold the $m_{xx}$ spectra
is based on the idea that if the MC describes well enough the true spectrum in 
data and if the folding probabilities are well simulated, the matrix of unfolding 
probabilities determined in simulation can be applied to data.

If the first condition is not fulfilled, that is if the data spectrum after unfolding
differs significantly from the true MC spectrum, several steps are
considered where the transfer matrix is improved by re-weighting the true MC, 
keeping the folding probabilities unchanged.
Differences between data and folded (`reconstructed') MC spectra are ascribed to
differences in the unfolded (`true') spectra. At each step of the iterative 
re-weighting process, the data-MC differences of reconstructed spectra are unfolded 
and added to the true MC spectrum. 
Such iterative procedures can result in a significant bias to the final results
if statistical fluctuations are mis-interpreted as true differences between data and 
MC distributions.
The stability of the method is provided in this analysis by the use of 
a regularization function to avoid unfolding large fluctuations in the data,
due for example to a large background subtraction.  
Details on the method are given in Ref.~\cite{bogdan}.

\subsubsection{Procedure}

Unfolding is applied to the reconstructed $m_{xx}$ 
spectrum, after background subtraction and data/MC corrections for 
efficiencies, obtained as described in the previous sections. The unfolding procedure
handles detector resolution and distortion effects, and corrects for FSR.
Thus the process delivers the `true' distribution of events in the detector 
acceptance as a function of $\sqrt{s'}$,  and a covariance matrix 
containing the statistical uncertainties and their bin-to-bin correlations.
The covariance matrix is obtained from pseudo-experiments (toys), where both 
the spectrum and the transfer matrix are statistically fluctuated.

For the $\pi\pi(\gamma)\gamma_{\rm ISR}$ analysis, the same energy range 0--3\gev is chosen for 
data and the MC transfer matrix.
The spectra obtained under the central $\rho$ region conditions (loose $\chi^2$ criterion) 
and under the $\rho$ tails conditions (tight $\chi^2$ criterion) are unfolded separately 
over the full mass range.  
The unfolded spectra are combined afterwards, each being used in its 
respective mass region. Different bin sizes are used: 10\mevcc for the tails 
and 2\mevcc for the central part. 

Since the loose condition retains events in the intermediate 2D-$\chi^2$
region, 
resolutions in data and simulation are compared in specific ranges across this region. 
It is noteworthy that, although the large-angle additional-ISR events, absent in MC,
populate the intermediate region, they do not contribute to resolution tails 
in data as the mass in that case is given by the (good) FSR fit.
Mass spectra shapes are found to be well simulated, but the rate of events 
in the degraded mass resolution regions is underestimated by MC. A reweighting
of the MC sample in the corresponding $\chi^2$ regions is applied, thus modifying the
MC transfer matrix.  The relative distortion of the mass spectrum due to 
reweighting is less than one percent (see Fig.~\ref{loose-tight-resol}), hence 
the systematic uncertainty from the imperfect knowledge of the transfer matrix 
is estimated to be a fraction of $10^{-3}$. 

The unfolding of the monotonic featureless mass distribution in the 
$\mu\mu(\gamma)\gamma_{ISR}$ sample is much less sensitive to resolution effects and the
transfer of events involves mostly FSR. Here a larger mass range (0--6\gevcc) is
considered in 50\mevcc intervals (120$\times$120 matrix), although only the 
first half is needed for luminosity purposes.

\begin{figure*}[htp]
\centering
\includegraphics[width=10cm]{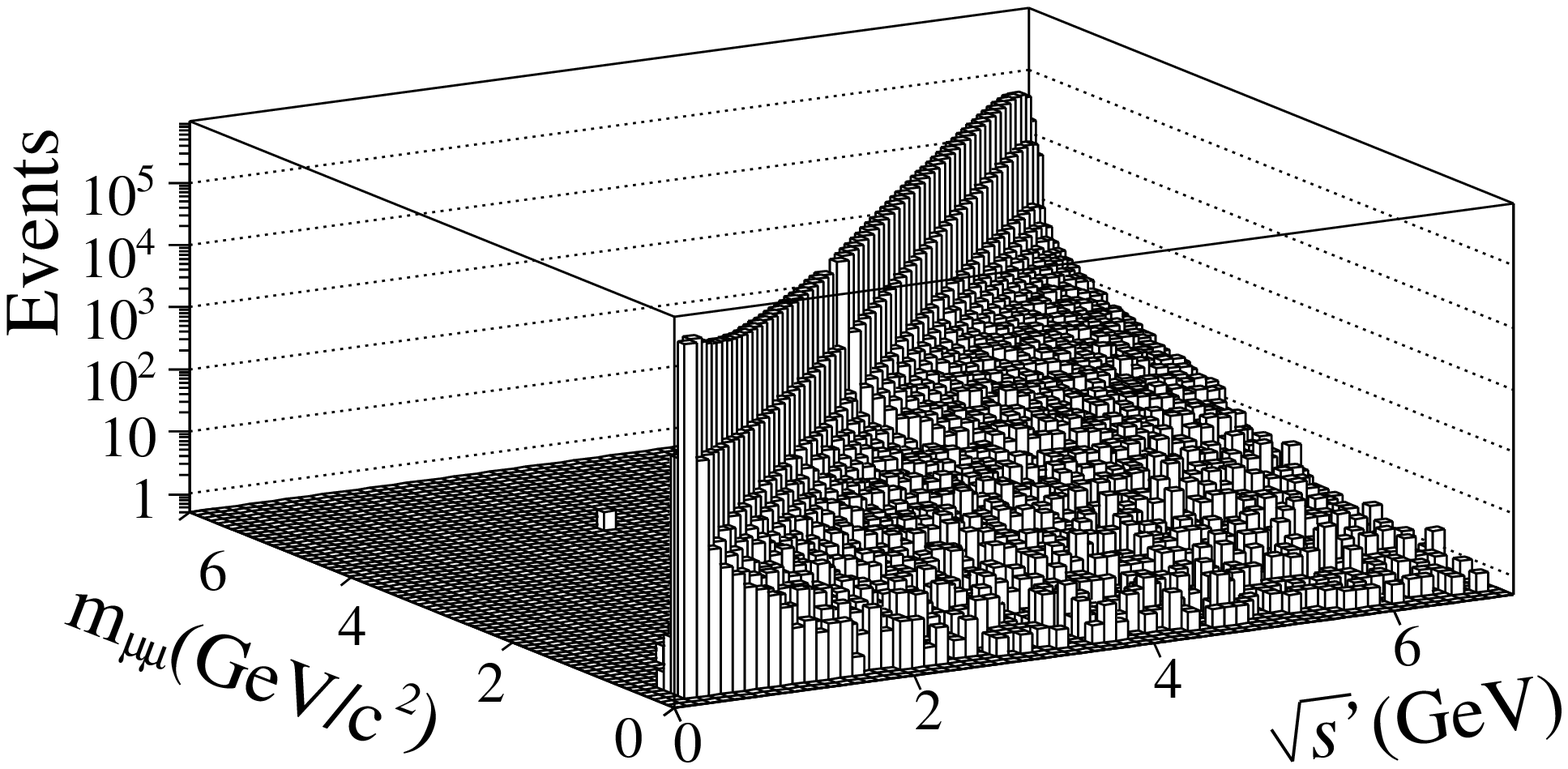}
\includegraphics[width=10cm]{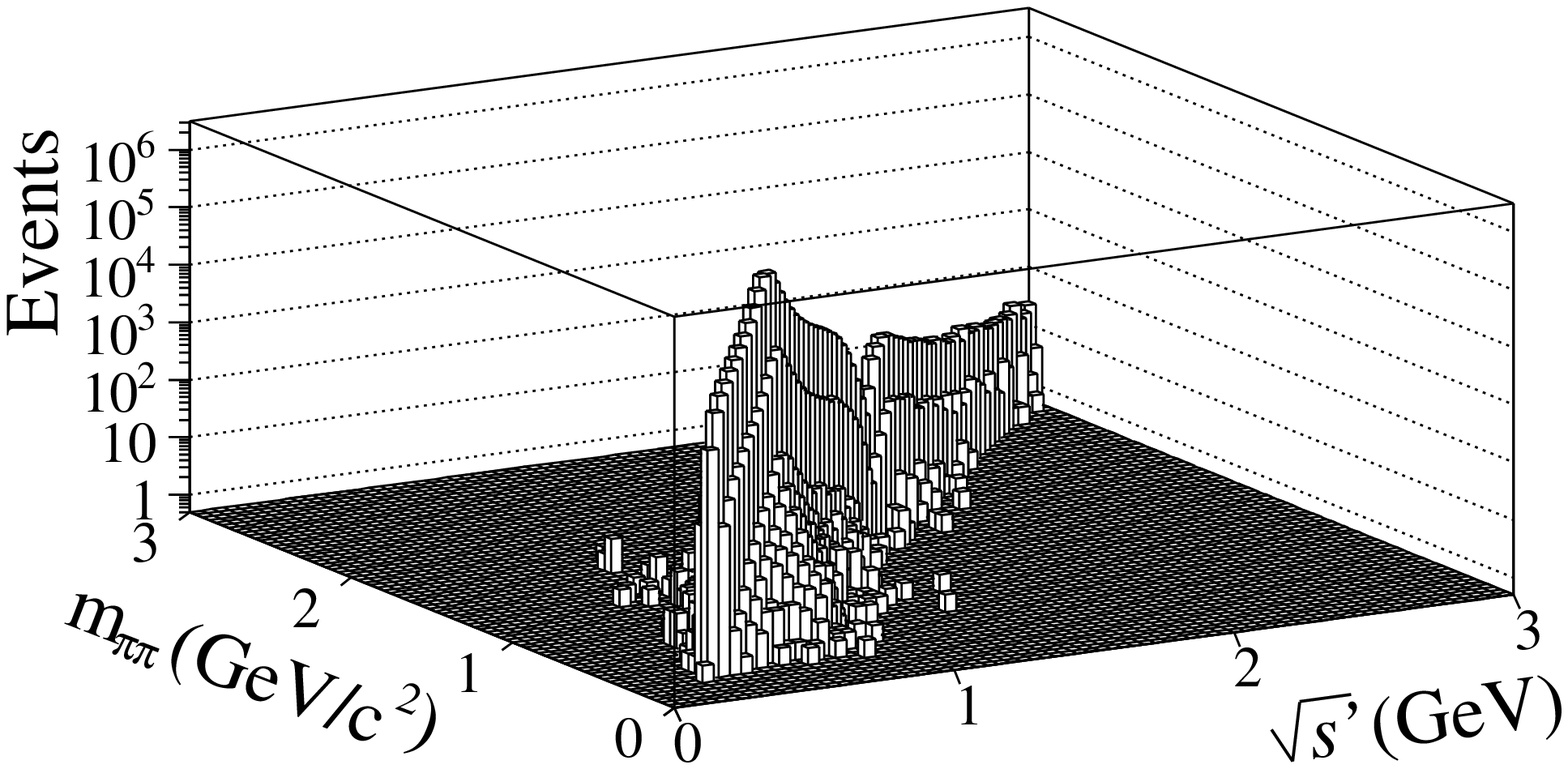}
\includegraphics[width=10cm]{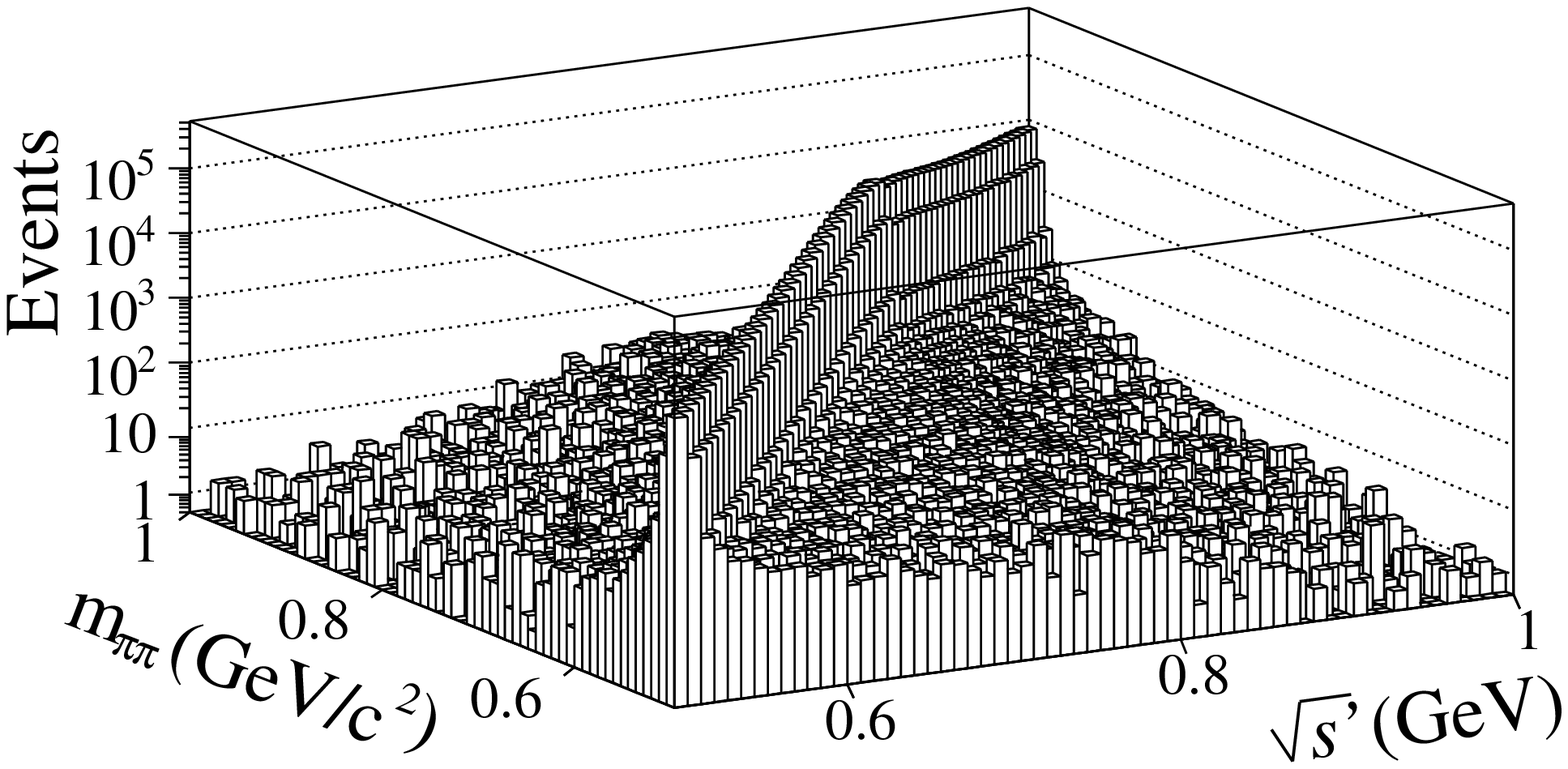}
\caption{\small
The initial mass-transfer matrix $A_{ij}$ from the simulation giving the
number of events generated with a (true) mass $\sqrt{s'}$ in a bin $j$ and
reconstructed with a (measured) mass $m_{xx}$ in a bin $i$: 
$\mu\mu(\gamma)\gamma_{ISR}$ (top), $\pi\pi(\gamma)\gamma_{\rm ISR}$ with tight $\chi^2$ criterion (middle)
and with loose $\chi^2$ criterion (bottom). For the latter case only the relevant range
0.5--1.0\gevcc is shown. The $\sqrt{s'}$ dependence comes from QED and a model 
of the pion form factor used in the AfkQed generator, respectively.} 
\label{mass-matrix}
\end{figure*}

The initial mass-transfer matrices for $\mu\mu(\gamma)\gamma_{\rm ISR}$ events, and 
$\pi\pi(\gamma)\gamma_{\rm ISR}$ under the
loose and tight conditions, are
shown in Fig.~\ref{mass-matrix}. The large diagonal component corresponds to
a mass resolution~\footnote{For illustration here we use an effective mass resolution
obtained by taking the weighted-average of the standard deviations from a 
two-Gaussian fit of the resolution function in simulation.} of 3.2\mevcc at $m_{\pi\pi}=0.78\gevcc$ (loose 
$\chi^2$ criterion), and 4.6\mevcc at $m_{\pi\pi}=1.5\gevcc$ (tight $\chi^2$ criterion). 

\begin{figure}[htp]
\centering
\includegraphics[width=9cm]{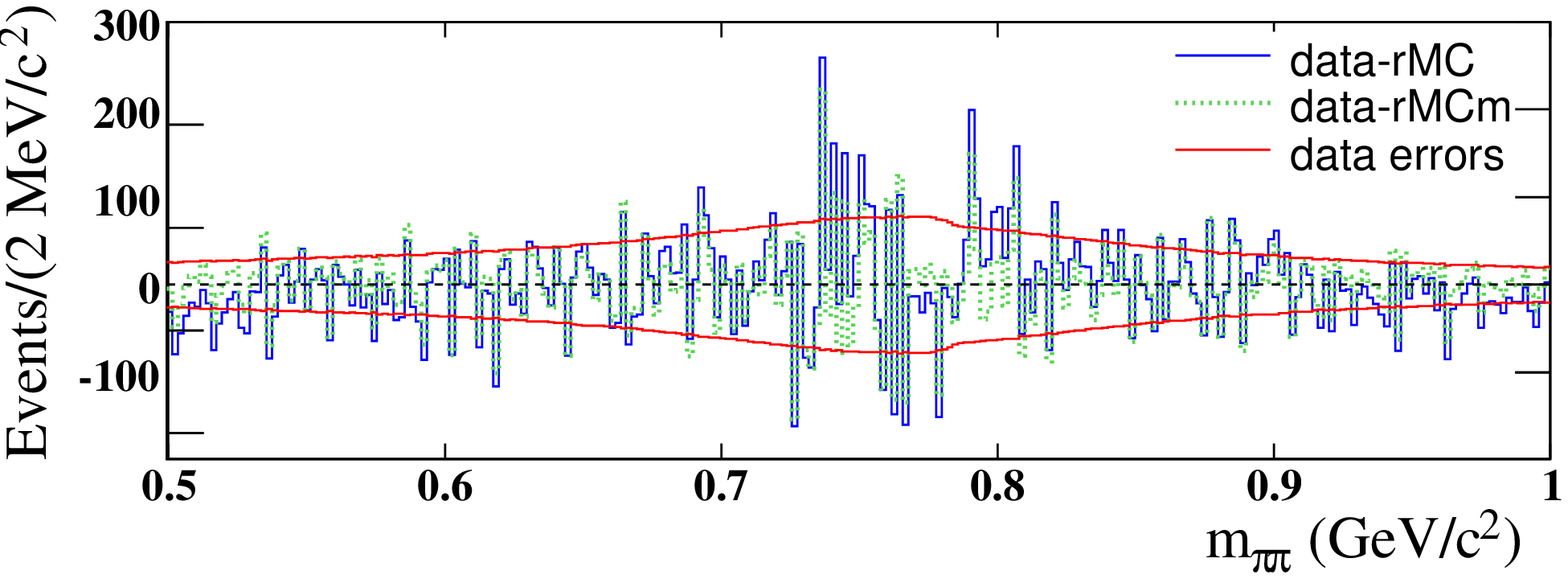}
\includegraphics[width=9cm]{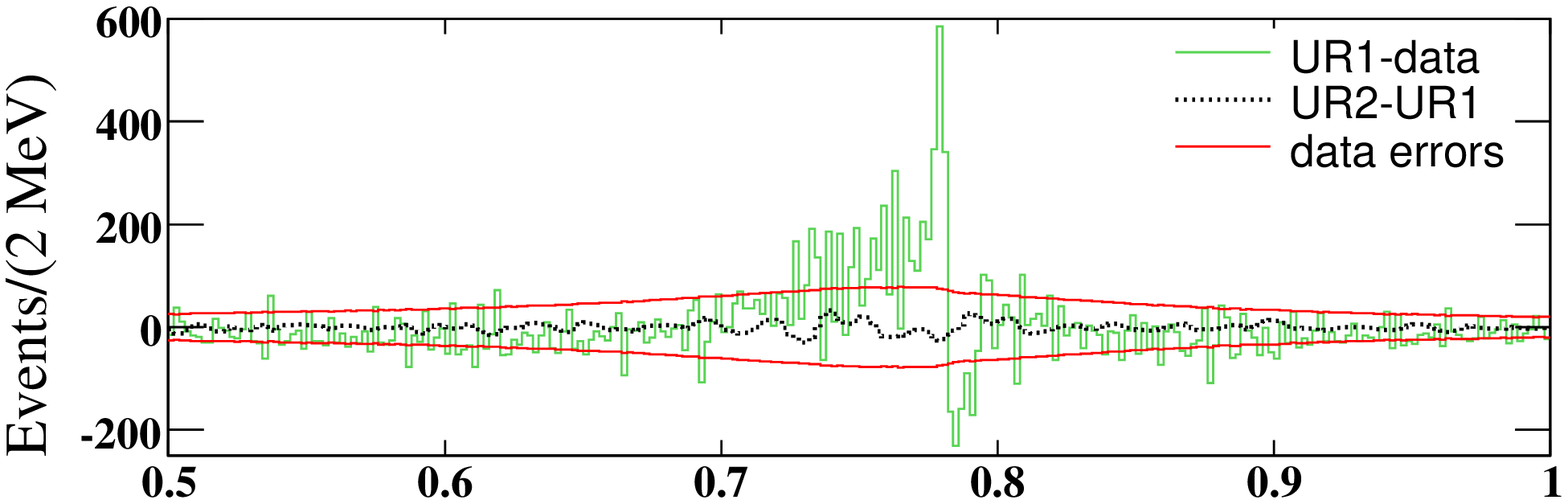}
\put(-60,0){$\sqrt{s'}$ (\gev)}
\caption{\small(color online). 
(top): 
The difference between the $\pi^+\pi^-$ mass distributions of data and 
reconstructed MC before unfolding (data$-$rMC) and after one iteration (data$-$rMCm),
for the loose $\chi^2$ selection used in the central $\rho$ region.
The data statistical errors ($\pm 1\sigma$) are shown for comparison.
The correction to the initial MC distribution is small, but significant in the
peak and tail regions.
(bottom): 
The difference between the result of the first unfolding (UR1) 
and the initial data for the same loose $\chi^2$ criterion. 
It exceeds the data statistical error (band) only in the $\rho-\omega$ 
interference region. 
No significant improvement is observed between the first (UR1) and second 
(UR2) unfolding results.} 
\label{pi-unfold}
\end{figure}

The most significant difference between data  
and reconstructed MC in relative terms occurs in the region 
1.7--2\gevcc, where the pion form factor is not well simulated. 
Smaller differences, not exceeding the statistical errors for 2-MeV bins, 
are observed in the $\rho$ lineshape, in the tails and in the peak region 
with the $\rho-\omega$ interference (Fig.\ref{pi-unfold} (top)). 
These differences are assigned to the generated mass distribution in the MC,
as resolution effects between data and MC are studied separately: 
the resolution broadening in the intermediate $\chi^2$ region was discussed
above and the resolution for the tight $\chi^2$ condition is presented in 
Section~\ref{mass-resol}. 
The differences are corrected for in the iterative way described above, 
but it is observed that already after the first step of the re-weighting 
procedure, they are reduced to a negligible level.
The residual systematic differences have indeed very little effect on the 
result of the unfolding. The first unfolding
result is very close to the initial data (well within the statistical error), 
except in the $\rho-\omega$ interference region, as expected since the 
mass resolution is not small compared to the $\omega$ width (8.5\mev). 
Adding one iteration in the unfolding does not result in further improvement,
as shown in Fig.~\ref{pi-unfold} (bottom).

\subsubsection{Tests of the unfolding method}

Tests of the unfolding procedure are performed, in the $\mu\mu$ and $\pi\pi$ channels, investigating 
potential systematic biases introduced by the method.
The test uses toy distributions of true spectra and their corresponding 
reconstructed distributions obtained by folding using the nominal transfer 
matrix $A$. The reconstructed toy spectrum is then unfolded with a transfer 
matrix ($\tilde{A}$) obtained after statistically fluctuating $A$. 
Finally the unfolded result is compared to the true toy spectrum.

The true toy distribution is constructed from the true MC with a bias added.
In order to build a test as close as possible to the real situation, the bias 
is taken as the difference between data and the normalized initial 
reconstructed MC. Two variations of the test are considered, where the 
reconstructed spectrum is additionally fluctuated statistically or not. The
first situation is closer to the real unfolding operation and could reveal
spurious effects due to the limited statistics in the data (and MC). 
The second test allows one to search for potential systematic 
effects of the method itself.

It is found that the systematic bias on the $\sqrt{s'}$ spectrum from the unfolding technique is 
negligible in the $\mu\mu$ channel. In the $\pi\pi$ channel, it is below the 
$10^{-3}$ level, except in the ranges 0.5--0.6\gev ($1.9\times 10^{-3}$) and 
0.9--1.0\gev ($1.2\times 10^{-3}$). The latter two values are anti-correlated with 
the rest of the spectrum, hence the systematic uncertainty that affects 
the dispersion integral used in the $a_{\mu}$ calculation remains smaller than $10^{-3}$.

\subsubsection{Distortion of the mass spectrum due to excess FSR}

The small excess of events with additional FSR in data compared to the 
simulation produces a distortion of the mass spectrum not taken into account
in the mass-transfer matrix. By appropriately reweighting the energy 
distribution of FSR photons by the energy-dependent excess fraction one 
obtains the resulting systematic uncertainty on the mass distribution.
The maximum deviation in the $\rho$ region
occurs at 0.5--0.6\gevcc at the $2\times 10^{-3}$ level, while it decreases to
$-0.8\times 10^{-3}$ at the $\rho$ peak and $-0.5\times 10^{-3}$ at 1\gevcc. These values are 
taken as systematic uncertainties on the cross section. Because of the 
anti-correlation occuring below and above the peak, this effect produces
a systematic uncertainty on the dispersion integral well below $10^{-3}$.

\subsection{Mass scale calibration}
\label{mass-calib}

The absolute $\pi\pi$ mass scale depends on the momenta and angular measurements
and the kinematic fit. Unlike at threshold where the mass scale is governed by the
angular measurements, the uncertainty from the momentum scale is dominant at the 
$\rho$ mass and above. Therefore systematic effects are studied using
ISR-produced $J/\psi\to\mu\mu$ events, which are treated in the same way as
the di-pion sample.

The $\mu\mu$ mass distribution is fitted in the
3.0--3.2\gevcc range across the $J/\psi$ peak with a linear term for the QED 
background and a signal shape
obtained by convoluting the sum of the natural $J/\psi$ Breit-Wigner and
the QED-$J/\psi$ interference with a Gaussian resolution shape. The free
parameters are the amplitude of the signal, the $J/\psi$ mass $m_{J/\psi}$,
the resolution $\sigma_m$ and the two constants of the linear background term. 
Three bins (1--3; 3--5; 5--8\gevc) are defined for the two track momenta $p_{\rm min}$ 
and $p_{\rm max}$ ($p_{\rm min}<p_{\rm max}$) and fits are performed in 6 boxes in the 
($p_{\rm min},p_{\rm max}$) plane, not 
distinguishing charges.

\begin{figure}[htp]
  \centering
  \includegraphics[width=7.5cm]{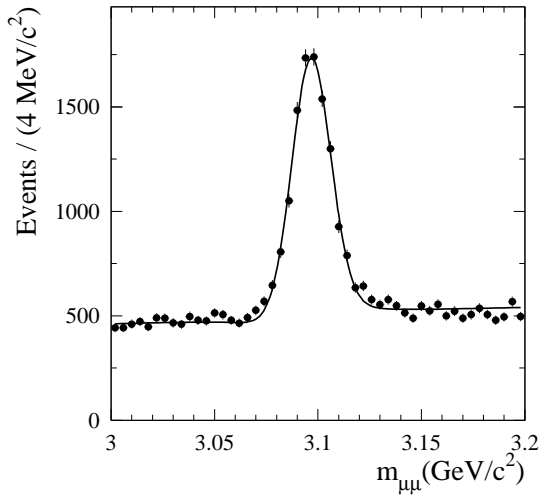}
  \caption{\small 
Fit of the `$\mu\mu$' mass distribution in the $J/\psi$ region including
the QED-$J/\psi$ interference as a momentum calibration test.}
  \label{fit-Jpsi}
\end{figure}

Whereas $\sigma_m$ increases for larger momenta as expected, the fitted values
for $m_{J/\psi}$ are consistent for all boxes, showing no evidence for a
momentum-dependent calibration change. Therefore the whole sample is
considered, the corresponding fit being shown in Fig.~\ref{fit-Jpsi}.
The small symmetric excess observed in the tails does not affect 
significantly the central value. The result
\beqn
\label{res-fit-Jpsi}
 m_{J/\psi} &=& (3096.30\pm0.13)\mevcc \\
\label{resol-Jpsi}
 \sigma_m   &=& (9.38\pm0.04)\mevcc~,
\eeqn
is compared with the world-average value~\cite{pdg}, $(3096.92\pm0.01)\mevcc$.
The difference, $(-0.62\pm0.13)\mevcc$, is interpreted as a momentum scale shift
of $(-2.00\pm0.04)\times 10^{-4}$.

This momentum-calibration scale factor translates into a shift for the
$\rho$ mass of $(-0.16\pm0.16)\mevcc$ where the full correction is taken 
as a systematic uncertainty.

\subsection{Mass resolution}
\label{mass-resol}

Since detector resolution effects are corrected for through the unfolding
procedure using the mass transfer matrix from simulation, we
check that simulation reproduces data in this respect. This problem is not
crucial since the $\pi^+\pi^-$ annihilation cross section is dominated by
the wide $\rho$ resonance, except in the $\rho-\omega$ interference region
because of the small $\omega$ width (8.5\mev compared to the 7.6\mev FWHM
resolution).

In the $\rho$ region the mass resolution is dominated by momentum, rather
than angular measurements. Thus the $J/\psi$ study described in 
Section~\ref{mass-calib} is again relevant. The mass resolution found in data
(Eq.~(\ref{resol-Jpsi})) is slightly better than the result from the simulation
of continuum $\mu\mu(\gamma)\gamma_{ISR}$ events in the 3.0--3.2\gevcc range (no $J/\psi$
contribution is generated in AfkQed), which is found to be $(10.0\pm0.1)\mevcc$.

The contribution of the decay angle measurement to the mass resolution is
obtained for data and simulation from a study of the decays 
$K^0_S\rightarrow \pi^+\pi^-$, from a sample of ISR-produced $\phi$ mesons
decaying into $K^0_SK^0_L$. In this case the angular measurement plays the
dominant role compared to momentum, thus this is a situation complementary to the
$J/\psi$ one. Taking into account the smaller contribution from momentum taken
from the $J/\psi$ case, the study yields the average resolutions on the decay 
opening angle of $(1.65\pm0.03)\mrad$ in data and $(1.59\pm0.03)\mrad$ in
simulation.

Combining the momentum and angular contributions one obtains the full mass
resolutions of $(3.03\pm0.03)\mevcc$ and $(3.20\pm0.03)\mevcc$ for data and
simulation, respectively. This resolution difference results in a bias on 
the measured resonance widths after unfolding of the mass spectrum, given by
\beqn
 \Delta \Gamma_\rho    &=& (+0.016 \pm 0.004)\mev,\\
 \Delta \Gamma_\omega  &=& (+0.27 \pm 0.07)\mev.
\eeqn
As for the mass scale calibration the full biases are taken conservatively 
as the corresponding systematic uncertainties on the measured $\rho$ and
$\omega$ widths.

\section{Results on $e^+e^-\to \mu^+\mu^-(\gamma)\gamma_{\rm ISR}$ cross section 
and ISR Luminosity}

Simultaneous measurement of the $e^+e^-\to \mu^+\mu^-(\gamma)\gamma_{\rm ISR}$ and
$e^+e^-\to \pi^+\pi^-(\gamma)\gamma_{\rm ISR}$ channels is a major feature of
this analysis. In this section, we report the results on the absolute 
$e^+e^-\to \mu^+\mu^-(\gamma)\gamma_{\rm ISR}$ cross section measurement and
the comparison to QED. The measured $m_{\mu\mu}$ mass spectrum is corrected for 
all efficiencies described in the preceding sections and unfolded. Further
corrections are specific to the absolute cross section measurement, which
necessitate dedicated studies described in this section. We then express the 
results on the $e^+e^-\to \mu^+\mu^-(\gamma)\gamma_{\rm ISR}$ spectrum in terms
of the effective ISR luminosity used in the $e^+e^-\to \pi^+\pi^-(\gamma)$ 
cross section measurement.

\subsection{Acceptance effects specific to the $\mu\mu(\gamma)\gamma_{ISR}$ analysis}
\label{mu-accept}

\subsubsection{Relevance of these studies}

As stressed in the Introduction, the measurement of the $e^+e^-\to \pi\pi(\gamma)$ cross section relies
on the measurement of the $\pi\pi/\mu\mu$ ratio $R_{\rm exp}(\sqrt{s'})$. 
A major advantage is that the effect from additional ISR essentially cancels 
in the ratio, leaving only second-order effects that are studied specifically. 

However, for the QED test, we use the absolute measurement of the $\mu\mu(\gamma)\gamma_{\rm ISR}$ 
cross section in order to perform a direct comparison to QED at NLO. This is 
a stringent check of the full understanding of all involved systematic effects. 
Dedicated studies are conducted in order to assess the importance of
NLO effects in the MC generator.
The QED test, if successful, demonstrates that these effects are properly taken into account 
and that their residual impact on the $\pi\pi/\mu\mu$ ratio measurement, in which they largely 
cancel, is indeed very small.

\subsubsection{Extra radiation in the MC generators}

We use AfkQed as the $\mu\mu(\gamma)\gamma_{\rm ISR}$ event generator to produce
a large sample (5.3 times the data) of fully-simulated events. 
The radiator function $dW/d\sqrt{s'}$ and vacuum polarization correction 
$\alpha(s')/\alpha(0)$ entering Eq.~(\ref{lumi-eff}) are included in the 
generator. However, although AfkQed
describes correctly the lowest-order process, it has some shortcomings in the 
generation of extra radiation: (i) additional ISR photons are generated with the 
structure function method in the collinear approximation, with a photon energy 
cut-off near 2.3\gev in the $e^+e^-$ c.m.\ (coming from the requirement  
$m_{X\gamma_{\rm ISR}}>8\gevcc$ applied at generation), (ii) generation of 
additional FSR photons follows the {\small PHOTOS} algorithm.

The effects of these limitations are studied with the Phokhara 4.0 
generator~\cite{phokhara}. The advantage
of Phokhara is that it uses the almost-exact QED NLO calculation (without
ISR-FSR interference). However, contrary to AfkQed, it does not include the 
contribution from two FSR photons nor higher-order ISR emission. 
Both effects are expected to be at a very small level. The contribution
of two FSR photons is suppressed by the smallness of LO FSR (about 1\% at 
1\gev and 15\% at 3\gev) and NLO FSR ($<1$\% at 1\gev and 2.7\% at 3\gev) corrections for 
photon energies $E_{\gamma~\rm add.FSR}>0.2\gev$. 
Even at 3\gev 
the expected contribution of $4\times 10^{-3}$ has a negligible effect
on the acceptance. The contribution of higher-order ISR emission is
relevant only if the third photon has a significant energy. From the
acceptance change between Phokhara and AfkQed, and the fraction of NLO ISR 
above photon energies of 1\gev in the $e^+e^-$ c.m., one estimates a maximum
acceptance bias of $2\times 10^{-3}$ at threshold and $10^{-3}$ at 1\gev.

\subsubsection{Fast simulation studies with Phokhara and AfkQed}
Since any number of additional photons are accepted at event selection, an imperfect 
simulation of NLO ISR does not affect the event topology selection but 
alters the event acceptance through kinematic effects.

The main criteria affecting the geometrical acceptance are: both muon tracks 
in the polar angle range $0.4<\theta_\mu<2.45\rad$, with momenta larger 
than 1\gevc; the most energetic photon in the c.m.\ (ISR candidate) with 
$E^*_\gamma>3\gev$ and in the polar angle range 
$0.35<\theta_\gamma<2.4\rad$.
The full acceptance involves all the event selection criteria.

As calculated using the AfkQed generator and the full 
simulation, the acceptance needs to be corrected for the effects resulting from 
the imperfect description of NLO ISR. For this correction Phokhara and AfkQed 
are compared at the generator level. 
Since the effect of the NLO differences is to give
different longitudinal boosts to the events, one expects deviations in the
geometrical and momentum acceptance. This justifies the use of the generators
at 4-vector level. To improve on this, track and photon parameters are smeared 
using resolution functions obtained from data. The acceptance is defined at
this level by the polar angle ranges for the ISR photon and the two muons,
and the $p>1\gevc$ requirement on the muons. 

\begin{figure}\centering
  \includegraphics[width=7.5cm]{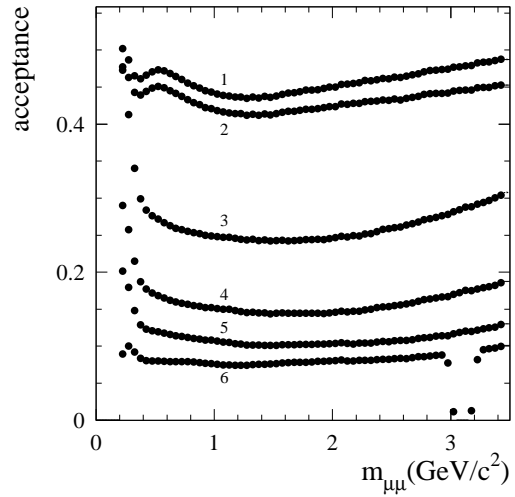}\\
  \caption{\small 
Breakdown of the full simulation acceptance with respect to the generated 
events in the ISR photon angular range in the $e^+e^-$ center-of-mass 
$20^0-160^0$ for $\mu\mu(\gamma)\gamma_{\rm ISR}$ events. The numbers refer to the 
sequential application of the selection requirements:
(1) trigger + acceptance selection for reconstructed ISR photon and tracks;
(2) preselection of ISR events + $E^*_\gamma>3\gev$;
(3) $p>1\gevc$ for both tracks;
(4) tracks in IFR active area;
(5) tracks in DIRC active area;
(6) `$\mu\mu$'-ID + $\chi^2$ selection + $J/\psi$ rejection + minor selections.}
  \label{acc_fast-full}
\end{figure}

We test the sensitivity of the results to using only fast 
simulation. Smearing generates a relative shift of
$1.0\times 10^{-3}$ for the acceptance correction. So any inadequacy of the resolution 
functions is expected to be at a lower level. 
Some effects are not included in the fast simulation while they enter 
the full MC. The main components of the overall efficiency for the full simulation 
are shown in Fig.~\ref{acc_fast-full}. While the
loss of acceptance is estimated to be 92\% near the $\rho$ mass,
the fast simulation accounts for 78\% only. The major contribution to the
difference between full and fast simulation comes from the DIRC crack removal 
and the IFR active area (cracks and bad areas). These azimuthal effects are 
essentially insensitive to the longitudinal boosts from additional ISR photons. 
Fast simulation is consequently adequate to describe the event acceptance 
changes generated by the additional ISR photon kinematics.  

\subsubsection{Effect of collinear additional ISR in AfkQed}
\label{add-isr-corr}
The angular distribution of hard additional ISR photons can
produce a significant transverse momentum that affects the event 
acceptance and preselection efficiency. The change of acceptance for 
collinear and non-collinear additional 
ISR is investigated with Phokhara as it provides an MC sample with 
additional ISR following the QED angular distribution. 
A significant decrease of the acceptance is observed 
as a function of the polar angle of the additional hard ($>0.2\gev$) 
ISR photon. The difference between Phokhara and AfkQed is aggravated by the
$m_{X\gamma_{\rm ISR}}>8\gevcc$ requirement used at generation in AfkQed, which 
suppresses hard additional ISR. Both of these effects are kinematic in nature,
and are well studied at the 4-vector level.

\begin{figure}\centering
  \includegraphics[width=8cm]{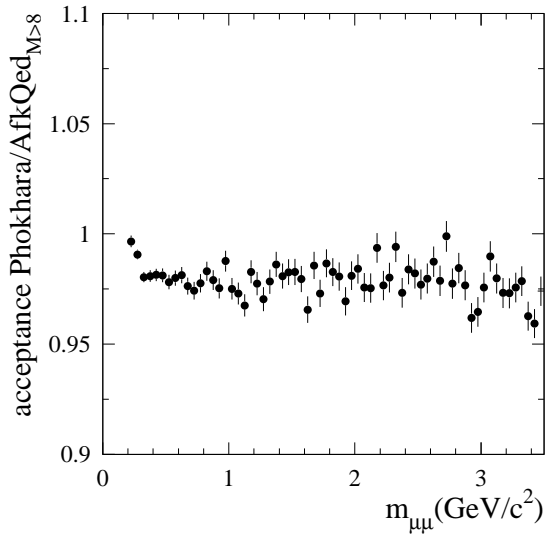}
  \caption{\small The ratio of the $\mu\mu(\gamma)\gamma_{ISR}$ acceptances determined 
  in Phokhara and AfkQed at the generator level with fast simulation.}
  \label{pho-afk-fs}
\end{figure}

The observed differences between data and AfkQed for the angular and energy 
distributions for NLO ISR (Sect.~\ref{add-fsr}) are overcome in Phokhara, which 
provides a much better description of the data.
The geometrical acceptances computed with the smeared 4-vectors in Phokhara 
and AfkQed differ by about 2\% in most of the mass range, Phokhara leading 
understandably to a lower acceptance (Fig.~\ref{pho-afk-fs}). 
The global efficiency $\epsilon_{\mu\mu(\gamma)\gamma_{ISR}}$ obtained with AfkQed is corrected 
by this factor when computing the $\mu\mu(\gamma)\gamma_{\rm ISR}$ cross section.

\subsection{ISR photon efficiency correction}

A coarse $\gamma_{\rm ISR}$ detection efficiency map in the ($E^*_\gamma,\theta_\gamma$) plane is derived
from $\mu^+\mu^-\gamma_{\rm ISR}$ events selected on the basis of the muons only in data
and full MC simulation.
The data/MC correction for the ISR photon efficiency is obtained as a function
of $m_{\mu\mu}$ by sampling the efficiency maps using the simulated sample.
The efficiency is found to be lower in data by ($1.5\pm0.1$)\% below 2\gevcc, 
with a difference slightly smaller above.
A systematic uncertainty of $3\times 10^{-3}$ is assigned to cover the effects
originating from the limited map granularity.

\subsection{Distributions of kinematic variables}
\label{mu-kinvar}

The comparison of distributions of relevant kinematic variables (polar angle
$\theta_{\gamma}$ of the ISR photon, angular and momentum distributions 
of the muons) observed in data and 
simulation is an important cross-check of the analysis, as the true 
distributions are predicted by QED. Not all detailed corrections that are 
applied to the full simulation as a function of the $\mu\mu$ mass are available 
for these variables, and we only consider corrections from PID for this test.  
Knowing some deficiencies of AfkQed for additional ISR, the comparison is made 
for events without excessive extra radiation, requiring the $\chi^2_{\mu\mu}$ 
of the 1C fit that uses only the two tracks to be less than 15.

\begin{figure}[htp]
\centering
  \includegraphics[width=7.5cm]{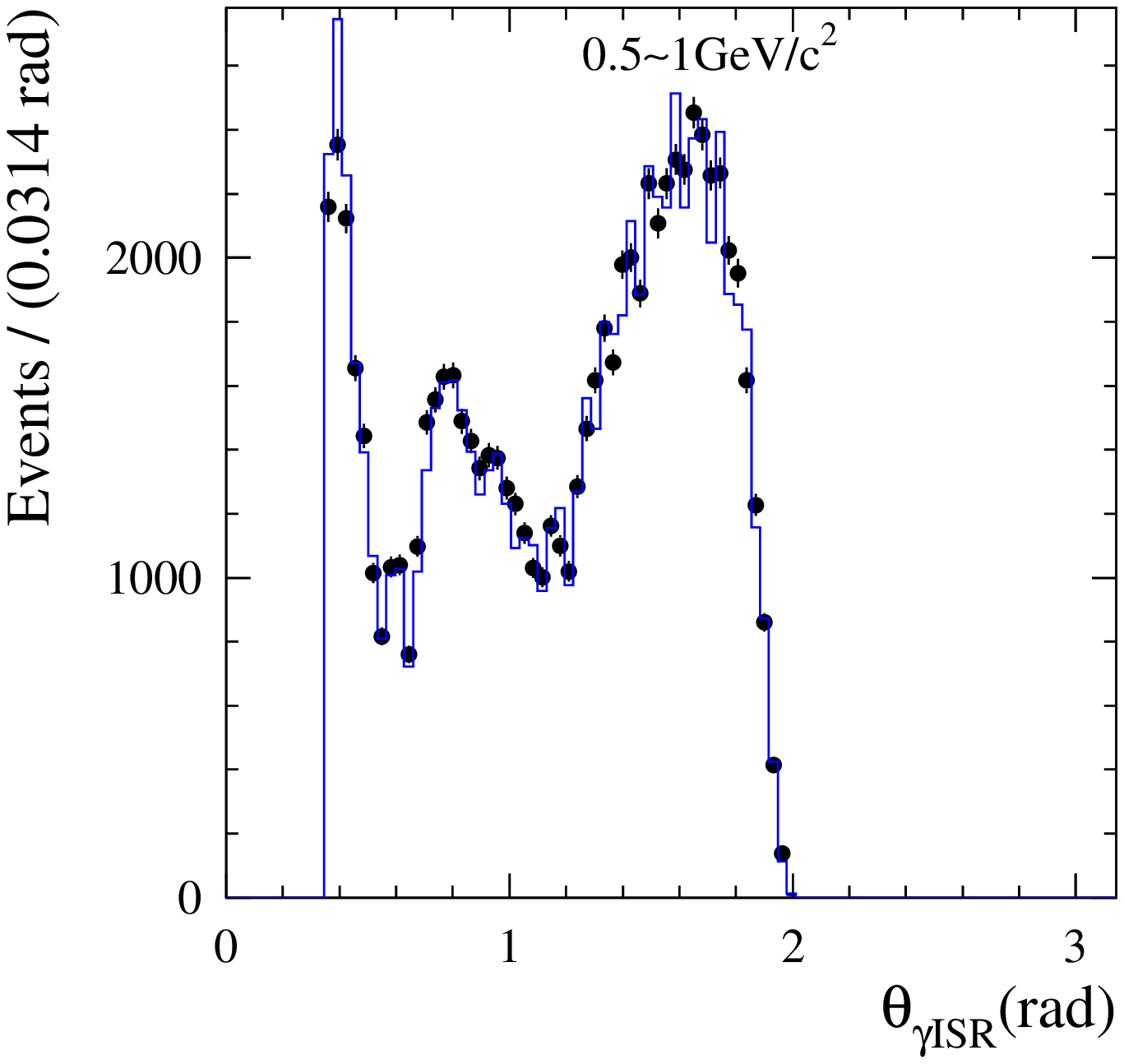}
  \includegraphics[width=7.5cm]{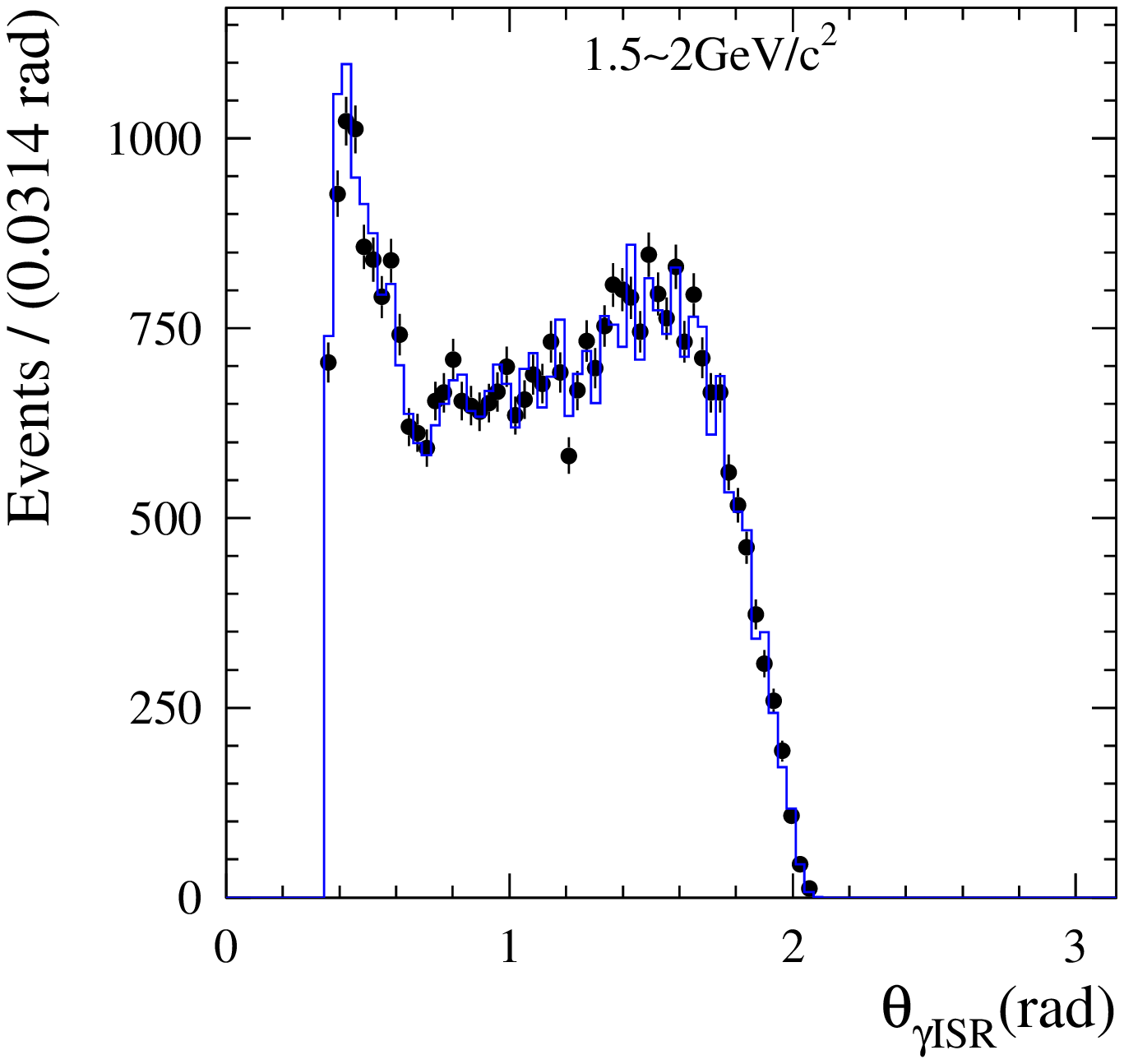}
  \includegraphics[width=7.5cm]{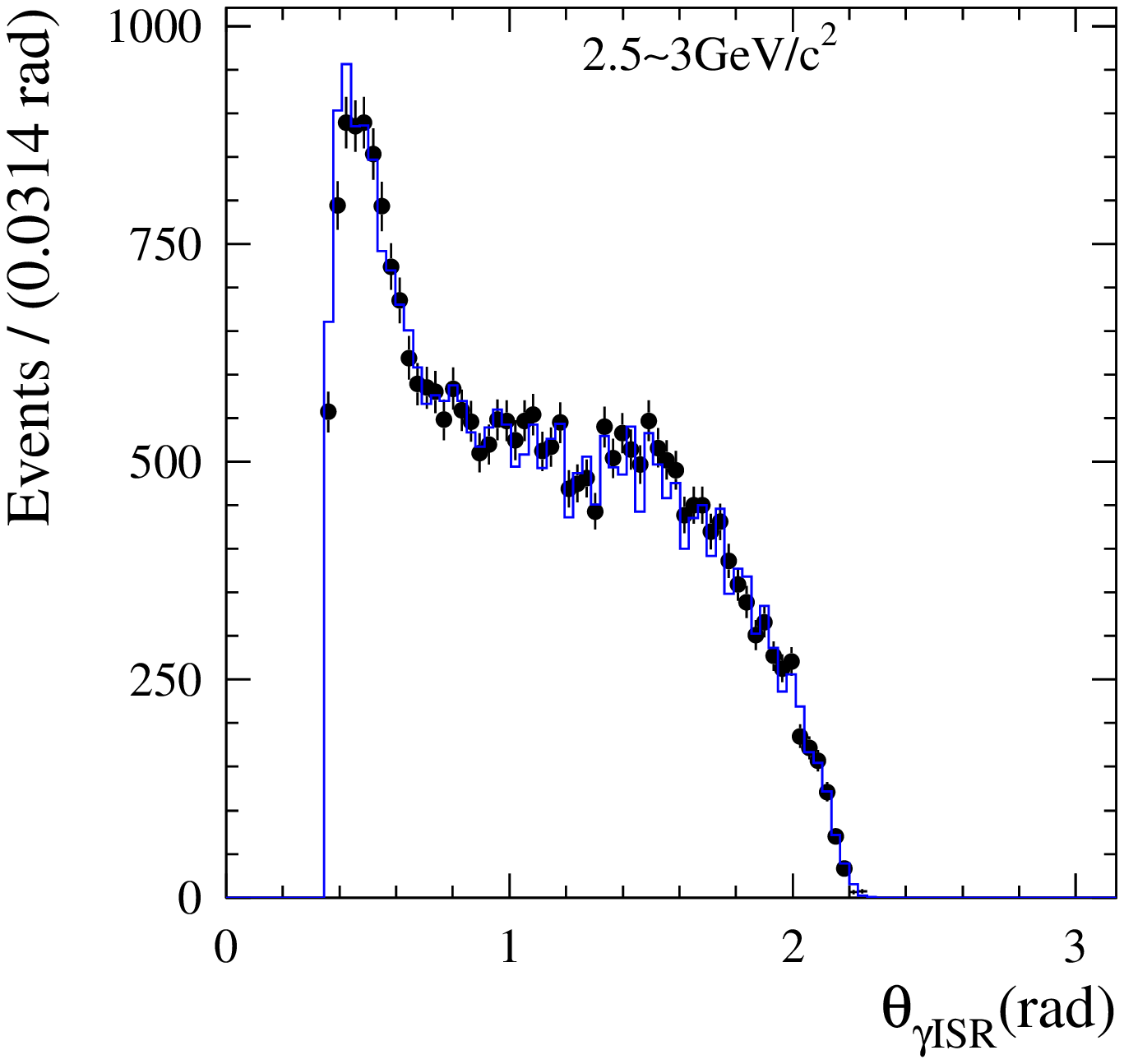}
  \caption{\small 
The comparison between the distributions of data (points with errors) and
simulation  corrected for data/MC differences in PID (blue histogram), 
for $\theta_\gamma$ in radians in the $m_{\mu\mu}$
intervals $0.5-1\gevcc$ (top), $1.5-2\gevcc$ (middle), $2.5-3\gevcc$ (bottom).}
  \label{mumu-thetag}
\end{figure}

\begin{figure}[htp]
\centering
  \includegraphics[width=7.5cm]{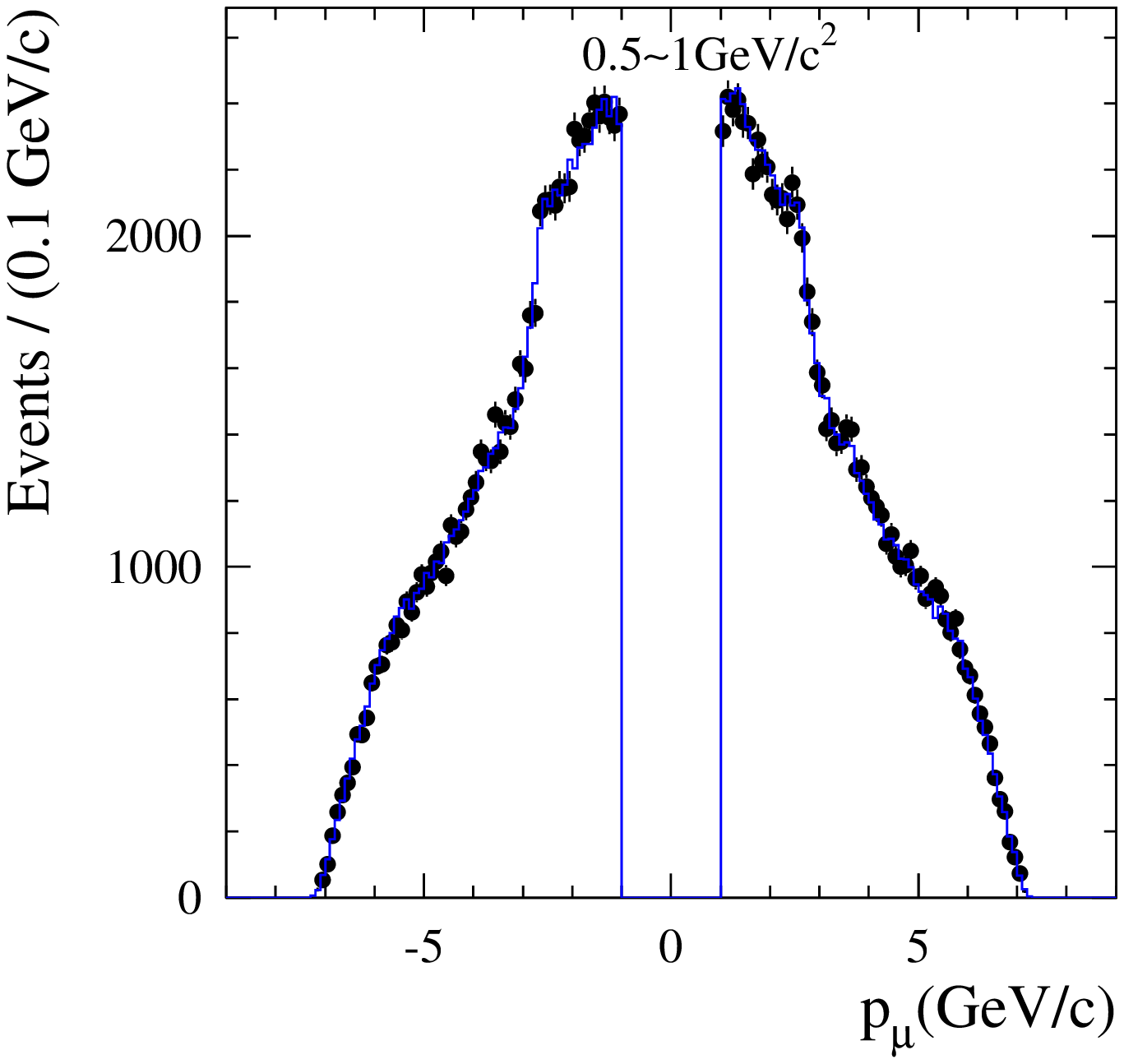}
  \includegraphics[width=7.5cm]{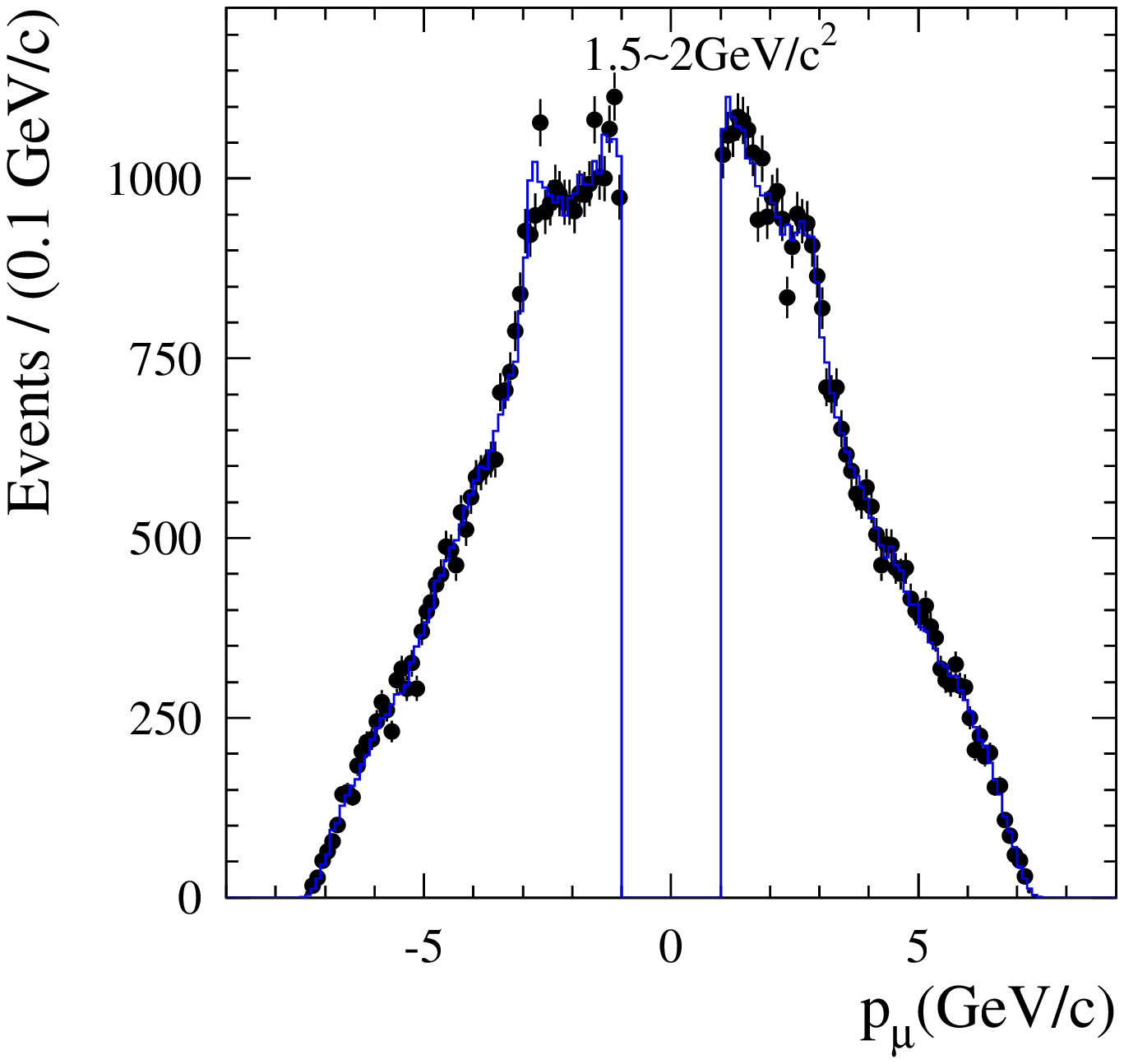}
  \includegraphics[width=7.5cm]{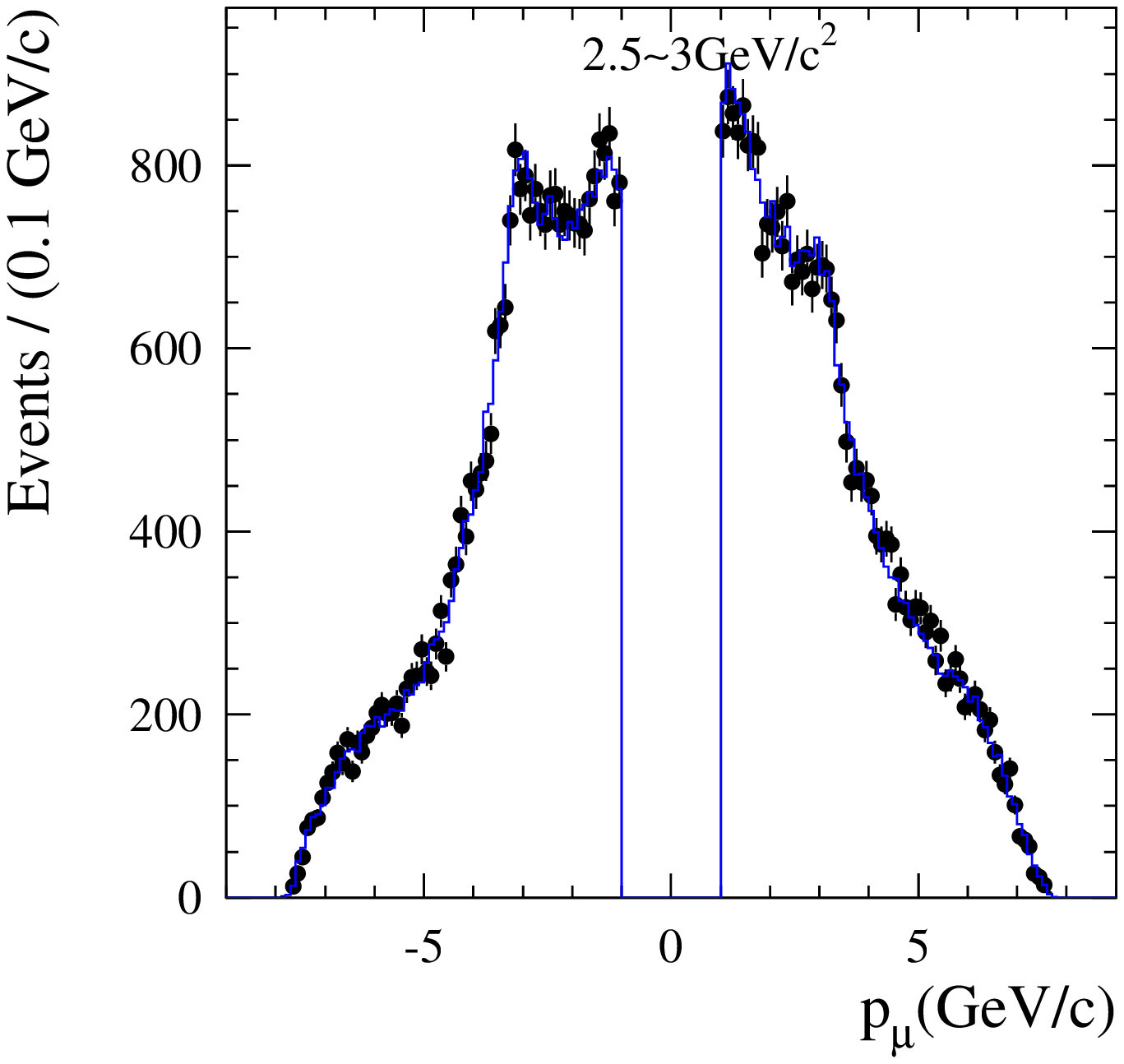}
  \caption{\small 
The comparison between the distributions of data (points with errors) and
simulation corrected for data/MC differences in PID (blue histogram), 
for $\pm p_{\mu^\pm}$ in $\gevc$ in the $m_{\mu\mu}$
intervals $0.5-1\gevcc$ (top), $1.5-2\gevcc$ (middle), $2.5-3\gevcc$ (bottom).}
  \label{mumu-pmu}
\end{figure}

Figures~\ref{mumu-thetag} and \ref{mumu-pmu}
show the distributions of $\theta_{\gamma}$ and $p_{\mu}$ in three mass intervals. 
In each case, MC is normalized to data as we are interested in
testing the shapes. The agreement between data and the simulation is good, except for
the ISR photon distribution at small angles where the data lies below the
simulation. This effect, which cancels in the $\pi\pi/\mu\mu$ ratio,
is imputed to a data/MC difference for the photon 
efficiency at small angles. 

The angular distribution ($\theta_{\mu}^*$) of the muons in the $\mu\mu$  
c.m.\ with respect to the ISR photon direction in this frame is 
of particular interest since it is predicted by QED to behave as
\beqn
\label{mumu-angular}
\frac {dN}{d\cos\theta_{\mu}^*}\sim 1+\cos^2\theta_{\mu}^* + (1-\beta^2)\sin^2\theta_{\mu}^*~,
\eeqn
for pure ISR production, with the muon velocity $\beta=\sqrt{1-4m_\mu^2/s'}$. So we expect
the distribution to be flat at threshold 
and to follow a $1+\cos^2\theta_{\mu}^*$ distribution at intermediate mass. 
At higher masses a larger fraction of the `ISR'-selected photon comes in fact 
from FSR, increasingly modifying the $\cos{\theta_{\mu}^*}$ distribution. 

\begin{figure}[htp]
\centering
  \includegraphics[width=7.5cm]{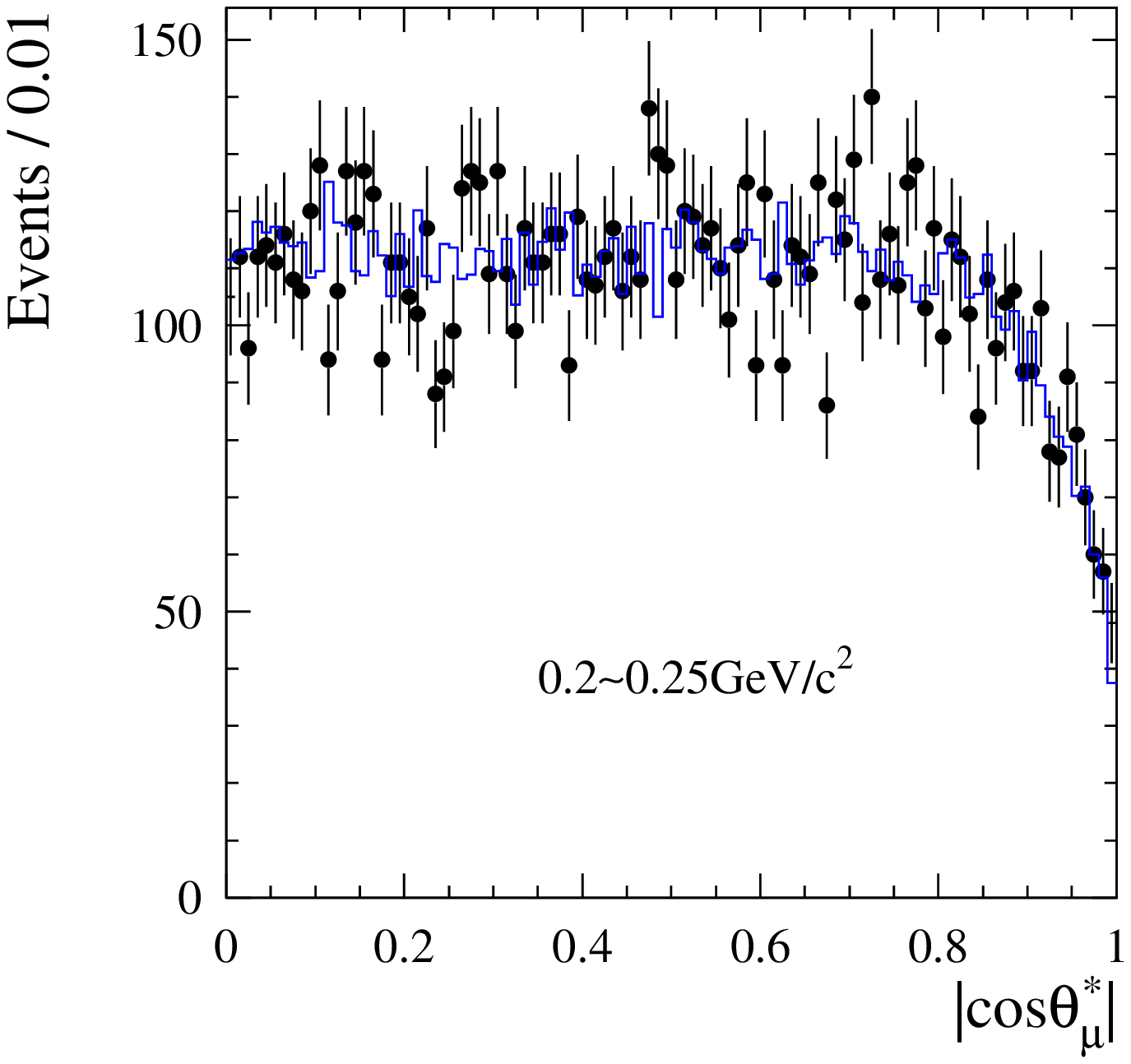}
  \includegraphics[width=7.5cm]{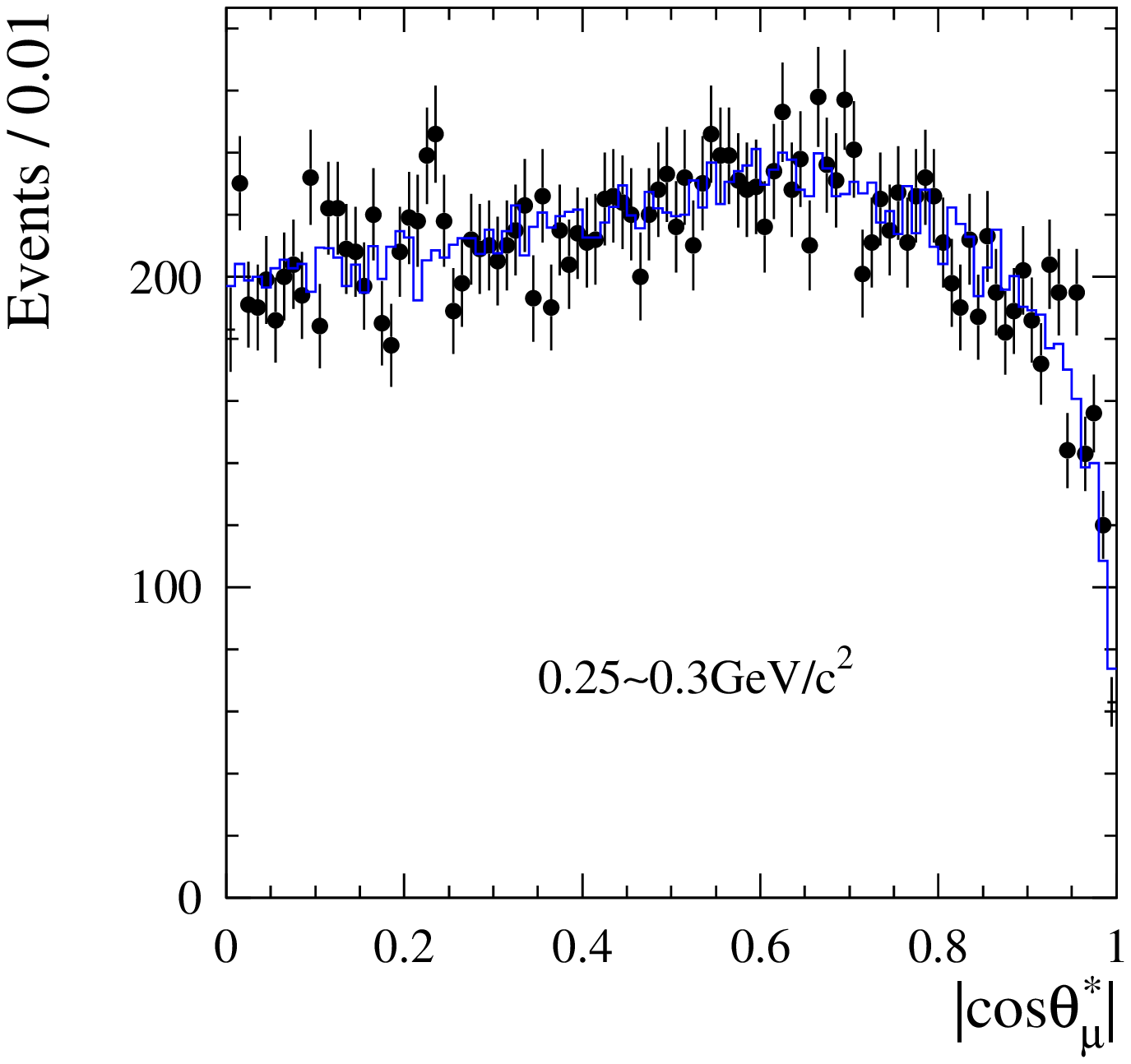}
  \includegraphics[width=7.5cm]{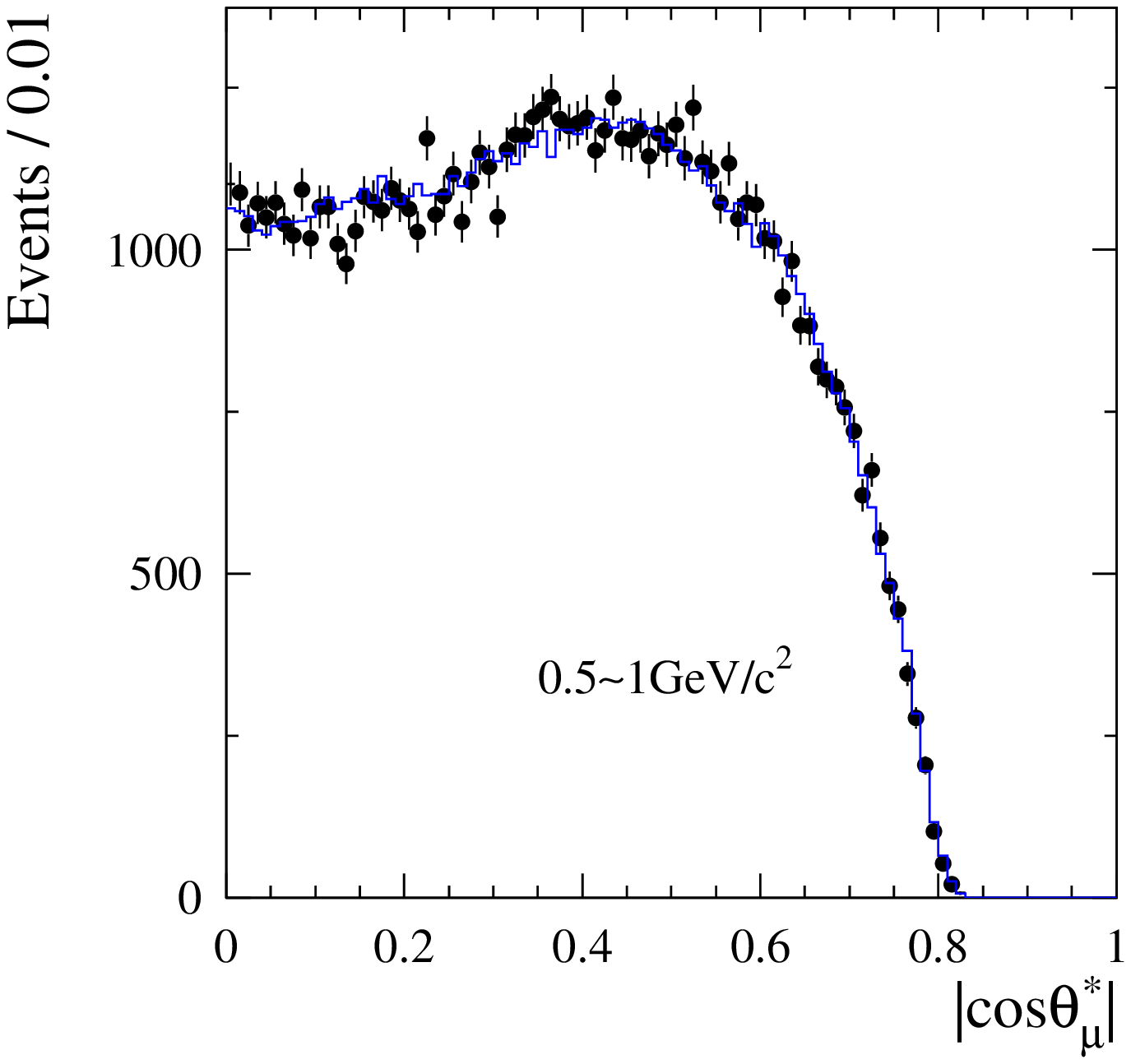}
  \caption{\small 
The comparison between the distributions of data (points with errors) and
simulation corrected for data/MC differences in PID (blue histogram), 
for $|\cos\theta_{\mu}^*|$ in the $m_{\mu\mu}$ intervals:
(from top to bottom) $0.20-0.25\gevcc$, $0.25-0.30\gevcc$, and $0.5-1\gevcc$.}
  \label{costheta-star-low}
\end{figure}

The distributions of $|\cos{\theta_{\mu}^*}|$ for different mass intervals agree
well with expectation as seen in Fig.~\ref{costheta-star-low}.
Although they are strongly biased by the $p>1\gevc$ requirement, which 
depletes the region near one, distributions in the threshold region 
indeed show the behavior expected from Eq.~(\ref{mumu-angular}), in
agreement with the MC.

Thus the distributions of the main kinematic variables of the selected 
$\mu\mu(\gamma)\gamma_{ISR}$ sample are in good agreement with expectations
from QED as implemented in the MC generator.

\subsection{Systematic uncertainties on the absolute 
$\mu\mu(\gamma)\gamma_{\rm ISR}$ cross section}

The statistical errors of the measured efficiencies are included with the main
statistical uncertainty on the $\mu\mu$ mass spectrum. However, in some cases,
remaining systematic uncertainties are attached to the efficiency measurement 
process. Estimated systematic 
uncertainties on the measured cross section are summarized in 
Table~\ref{mu-syst-err} for the mass range from threshold to 2\gevcc. Above 2\gevcc
the uncertainties are smaller, essentially because of the more straightforward
determination of the muon-ID efficiencies. In some cases no systematic error
is quoted when all uncertainties proceed from measurements and are already 
included in the point-to-point statistical errors. 

We find that Phokhara at the fast simulation level
is adequate to correct the AfkQed generator and that the systematic uncertainty
on the acceptance resulting from the fast simulation is $10^{-3}$. 
The effect of the momentum calibration uncertainty is only at the $10^{-4}$
level. 

The absolute $\mu\mu(\gamma)\gamma_{\rm ISR}$ cross section makes use of the
effective luminosity function defined by Eq.~(\ref{lumi-eff}) that 
includes the \babar\ luminosity $L_{ee}$.
The latter is obtained for all the analyzed data using
measurements of $ee\to ee$, $\mu\mu$ and $\gamma\gamma$, and amounts to
230.8 fb$^{-1}$. The corresponding systematic uncertainty is 0.94\%.

The uncertainty assigned to the QED cross section comes from the neglect of 
the NNLO contribution. The latter is estimated from the NLO fraction as given 
by {\small PHOKHARA}, equal to $(4.33\pm0.11)$\% within the selection used in 
this analysis. Assuming a geometric growth of the coefficients of the expansion
in $\alpha$, the NNLO fraction is estimated to be $2\times 10^{-3}$, which is 
taken as the systematic uncertainty.

Summarizing, the overall systematic uncertainty on the absolute 
$\mu\mu(\gamma)\gamma_{\rm ISR}$
cross section is 1.1\%, dominated by the \babar\ luminosity error.

\begin{table} [t] \centering 
\caption{ \label{mu-syst-err} \small 
Systematic uncertainties (in $10^{-3}$) on the absolute 
$\mu\mu(\gamma_{\rm FSR})$ cross section from the determination 
of the various efficiencies in the $\mu\mu$ 
mass range up to 2\gevcc. The statistical part of the efficiency measurements 
is included in the total statistical error in each mass bin. For those
contributions marked `-' all the relevant uncertainties come from measurements
and are already counted in the statistical errors.}
\vspace{0.5cm}
\setlength{\extrarowheight}{1.5pt}
\begin{tabular}{lc} \hline\hline\noalign{\vskip1pt}
 Sources & systematic errors $(10^{-3})$          \\ \hline\noalign{\vskip1pt}
 triggers and background filter          &  0.3   \\    
 tracking                                &  1.3   \\
 muon ID                                 &  3.3   \\
 $\pi\pi$ and $KK$ backgrounds           &  -     \\
 multihadronic background                &  -     \\
 $\chi^2$ cut efficiency                 &  -     \\
 angle and momentum acceptance           &  1.0   \\
 ISR photon efficiency                   &  3.4   \\ 
 $e^+e^-$ luminosity                     &  9.4   \\
 NNLO corrections to $\sigma_{\rm QED}$  &  2.0   \\  \hline\noalign{\vskip1pt}
 sum                                     & 10.9   \\  \hline\hline
\end{tabular}
\end{table}

\subsection{QED test with the $\mu\mu(\gamma)\gamma_{\rm ISR}$ events}

The comparison of the $\mu\mu(\gamma)\gamma_{\rm ISR}$ cross section with QED is made through 
the ratio of the distribution in the data as a function of $m_{\mu\mu}$
to the same distribution of the simulation. Specifically, the
distribution of the data is background-subtracted, and the distribution of
the AfkQed-based full simulation, normalized to the data luminosity, is
corrected for all data/MC detector and reconstruction 
effects and for the generator NLO limitations using the Phokhara/AfkQed
comparison with fast simulation. 
Because of the latter adjustments, discussed in detail in Section~\ref{mu-accept},
the corrected ratio of $m_{\mu\mu}$ spectra is equivalent to a direct comparison 
of data with the NLO QED cross section.

The QED prediction for the $m=m_{\mu\mu}$ distribution is obtained in the 
following way:
\beqn \nonumber
 \frac {dN_{\rm QED}}{dm} &=& L_{ee}~\sigma_{\rm Phokhara}^{\rm NLO}~\left( \frac {1}{N_0}
\frac {dN}{dm}\right)_{\rm full sim}^{\rm AfkQed,M>8}~\\  
& & \times
\frac {\left(\frac {1}{N_0} \frac {dN}{dm}\right)_{\rm fast sim}^{\rm Phokhara}}
      {\left(\frac {1}{N_0} \frac {dN}{dm}\right)_{\rm fast sim}^{\rm AfkQed,M>8}}
\times C_{\rm data/MC}~,
\eeqn
where for each case $N_0$ is the generated number of events, $dN/dm$ the mass
spectrum of events satisfying all criteria. The ratio of spectra at the generator
level with fast simulation are labeled `$\rm fast sim$', while `$\rm full sim$' 
denotes the spectrum of events with full detector simulation. The notation
`$\rm M>8$' recalls that AfkQed was run with a requirement limiting hard additional ISR, 
$m_{X\gamma_{\rm ISR}}>8\gevcc$. Finally, the $C_{\rm data/MC}$
factor incorporates all data/MC corrections for detector
efficiencies, such as trigger, tracking, muon ID, $\chi^2$ selection, and ISR photon 
efficiency.

Because the PID efficiency varied with time due to the degradation of the
IFR detector, the ratio data/QED is determined separately splitting the running period in two. 
Both distributions are flat from threshold to 3.5\gevcc and consistent with 
unity within errors with satisfactory $\chi^2$ values. 
The difference
of the ratios for the two data sets is $(6.0\pm4.0\pm3.5\pm4.4)\times 10^{-3}$, 
where the first error is statistical, the second from non-common systematics
(uncorrelated parts of the $\mu$-ID systematic uncertainties), and the third 
from the \babar\ luminosity. 
Since the two samples correspond
to different performances of the IFR detector, this test provides a 
confirmation that the muon-ID efficiency has been handled adequately.

\begin{figure}
\centering
  \includegraphics[width=8.cm]{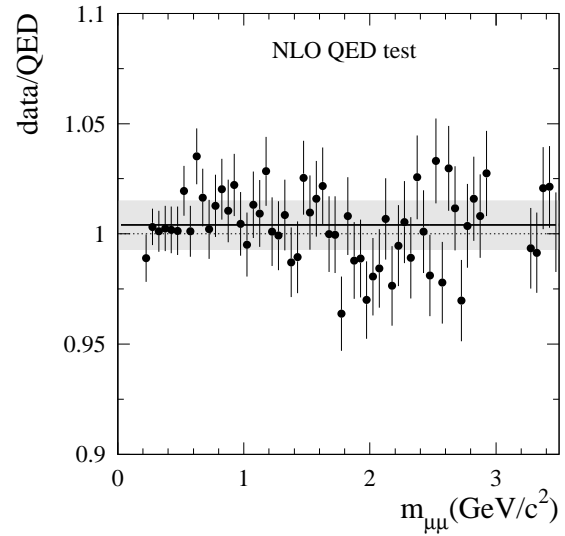}\\
  \caption{\small The ratio of the $\mu\mu$ mass spectrum in data over the
absolute prediction from QED using the \babar\ luminosity. The
NLO QED prediction is obtained from the data-corrected (for detector 
simulation) and Phokhara-corrected (for NLO effects) AfkQed mass spectrum.
The band is drawn around the fit of the 0.2--3.5\gevcc region to a free constant, with
a width given by $\pm$ the total expected systematic uncertainty.}
  \label{data-qed-14}
\end{figure}

The fit to the full data set is shown in Fig.~\ref{data-qed-14} and yields
\beqn
\!\!\frac {\sigma_{\mu\mu(\gamma)\gamma_{\rm ISR}}^{\rm data}} {\sigma_{\mu\mu(\gamma)\gamma_{\rm ISR}}^{\rm NLO~QED}} 
&=& 1+(4.0\pm1.9\pm5.5\pm9.4)\!\times\!10^{-3},   
\eeqn
where the first error is statistical, the second from systematics,
and the third from the \babar\ luminosity. 
The value found for the ratio is consistent with unity over the full mass range
explored in this analysis.
We conclude that the measurement of the $e^+e^-\to\mu^+\mu^-(\gamma)\gamma_{\rm ISR}$ cross
section using the \babar\ luminosity agrees with NLO QED in the $\mu\mu$ mass 
range from threshold to 3.5\gevcc within an overall accuracy of 1.1\%.

\subsection{Determination of ISR luminosity}
\label{isr-lumi}

In this section we express the results obtained for the $\mu\mu(\gamma)\gamma_{\rm ISR}$
sample in terms of the effective ISR luminosity, following Eq.~(\ref{def-lumi}).
As discussed in Section~\ref{mu-accept} the $\mu\mu(\gamma)\gamma_{\rm ISR}$ event 
acceptance, $\epsilon_{\mu\mu(\gamma)\gamma_{ISR}}$, appearing in Eq.~(\ref{def-lumi}) is 
obtained from the large simulated sample generated with AfkQed with corrections 
for detector and reconstruction effects. Corrections specific to the QED test,
{\it i.e.}, NLO and ISR photon efficiency corrections that cancel in the 
$\pi\pi/\mu\mu$ ratio, are not applied.

Several factors need to be considered in addition: (i) the LO FSR correction, 
(ii) unfolding of the data from $m_{\mu\mu}$ to $\sqrt{s'}$ to 
include the possible emission of an additional FSR photon, and (iii) the 
QED cross section $\sigma^0(e^+e^-\to\mu^+\mu^-(\gamma_{\rm FSR}))~(s')$ at the Born level 
concerning ISR, but including FSR. Except for (ii), which has been discussed in
Section~\ref{unfold}, we address these points in turn 
before giving the final result on the ISR luminosity.

\subsubsection{Lowest-order FSR correction}

The most energetic detected photon is assumed to be emitted by the initial
state. This is largely true at low mass, but there is an increasing probability
at larger $s'$ values that this photon originates from muon radiation. Thus
the observed $\mu\mu$ mass spectrum has to be corrected in order to
keep only ISR production, since for all practical purposes at \babar\, where 
$\sqrt{s}\sim 10.58\gev$ and $\sqrt{s'}<5\gev$, LO FSR production 
is negligible for hadronic processes.

\begin{figure}\centering
  \includegraphics[width=7.5cm]{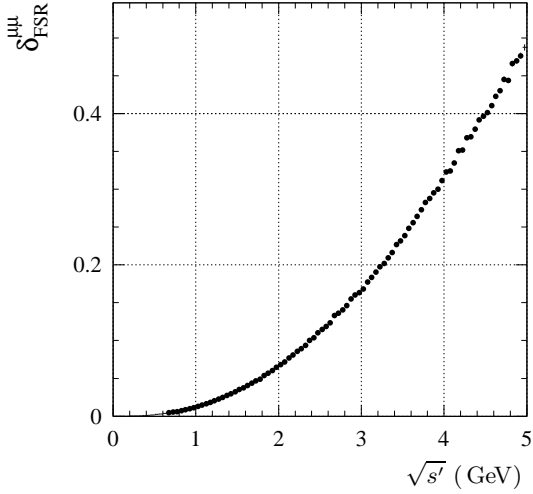}
  \put(-60,6){$\sqrt{s'}$ (\gev)}
  \caption{\small The FSR correction 
$\delta_{\rm FSR}^{\mu\mu}$
obtained with AfkQed.}
  \label{fsr-frac}
\end{figure}

Figure~\ref{fsr-frac} shows the $\delta_{\rm FSR}^{\mu\mu}$ correction to the cross section,  
defined as
\beqn
\delta_{\rm FSR}^{\mu\mu}= \frac {|{\cal A_{\rm FSR} + \cal A_{\rm add.ISR,add.FSR}}|^2}{|{\cal A_{\rm ISR} +\cal A_{\rm add.ISR,add.FSR}}|^2}~,
\label{fsr-delta}
\eeqn
as a function of $\sqrt{s'}$, where $\cal A_{\rm FSR}$ ($\cal A_{\rm ISR}$) is the 
LO FSR (ISR) amplitude and
$\cal A_{\rm add.ISR,add.FSR}$ is the NLO contribution.
$\delta_{\rm FSR}^{\mu\mu}$ is obtained with AfkQed at the generator level. 
It would be preferable to use Phokhara
instead, as we know additional ISR generation is approximate in AfkQed, but by construction
the FSR or ISR origin of photons is not available in Phokhara, hence $s'$
is not accessible on an event-by-event basis. However the difference in 
$\delta_{\rm FSR}^{\mu\mu}$
is expected to be at a negligible level, about $10^{-4}$ and $2\times 10^{-3}$ 
at 1\gev and 3\gev, respectively.

\subsubsection{Born QED cross section with additional FSR}
\label{born-fsr}

The cross section for $e^+e^-\to\mu^+\mu^-(\gamma_{\rm FSR})$, at the Born level
for the initial state and without vacuum polarization, can be calculated
exactly in QED at NLO. It has the form:
\beqn
 \sigma^0_{\mu\mu(\gamma)}(s')=\sigma_{\rm pt}(s')~(1+\delta^{\mu\mu}_{\rm add.FSR})~,
  \label{born-fsr-sig}
\eeqn
with
\beqn
 \sigma_{\rm pt}(s')=\frac {4\pi\alpha^2(0)}{3s'}~\frac {\beta(3-\beta^2)}{2} \\
 \delta^{\mu\mu}_{\rm add.FSR}=\frac {\alpha(0)}{\pi}\eta(s')\\
 \eta(s')=\eta_h(s')+\eta_s(s')+\eta_v(s')~,
\eeqn
where $\beta$ is the muon velocity and $\eta_{h,s,v}$ are the $O(\alpha)$ 
contributions to the final state 
from hard and soft bremsstrahlung, and the one-loop/Born interference
(`virtual' contribution), respectively. 

The sum of $\eta_v$ and $\eta_s$ is infrared-finite, while the total sum is 
independent of the choice of the energy used to separate soft and hard
photons (within reasonable limits). Expressions for all three components can be
found in many papers, for example in Refs.~\cite{fsr-kuraev,fsr-kuehn}.
By virtue of the Kinoshita-Lee-Nauenberg (KLN) theorem~\cite{kln}, the dominant 
logarithmic terms cancel
in the sum of the (soft+virtual) and hard contributions. Although the 
two terms reach a level of a few percent, they have opposite signs and
the sum $\delta^{\mu\mu}_{\rm add.FSR}$ stays in the few $\times 10^{-3}$ range. 
This explains why a sizeable hard 
additional-FSR signal is seen in data, despite the fact that the total 
additional-FSR contribution is very small.

\subsubsection{The effective ISR luminosity for the $\pi\pi(\gamma)\gamma_{\rm ISR}$ analysis}
\label{eff-lumi}

For the $\pi\pi(\gamma)\gamma_{\rm ISR}$ analysis, the luminosity $dL_{\rm ISR}^{\rm eff}$
integrates all configurations up to two ISR photons, where at least one photon has 
$E_\gamma^*>3\gev$ and is in the angular range $(\theta_{\rm min}^*,\theta_{\rm max}^*)$
in the $\mee$ c.m.\ with
$\theta_{\rm min}^*=180^\circ-\theta_{\rm max}^*=20^\circ$.

\begin{figure}\centering
  \includegraphics[width=7.5cm]{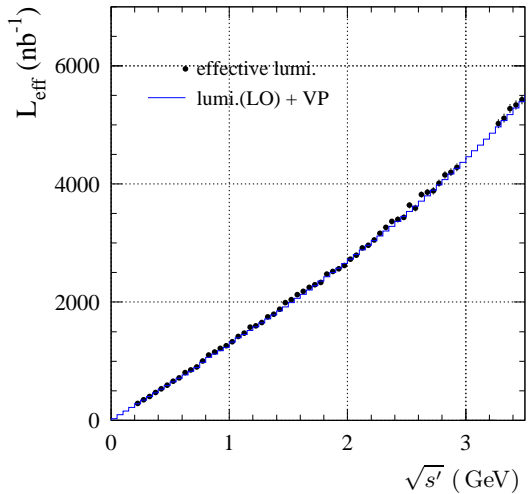}
  \put(-60,6){$\sqrt{s'}$ (\gev)}
  \caption{\small 
The effective ISR luminosity for the $\pi\pi$ analysis: 
the data points give $\L_{\rm ISR}^{\rm eff}$ in $\Delta \sqrt{s'}=50\mev$ bins. 
The conditions for the detected/identified ISR 
photon are $E_\gamma^*>3\gev$ and $20^\circ<\theta_\gamma^*<160^\circ$ in the 
$e^+e^-$ c.m.\ frame, while one additional ISR photon is allowed without any 
restriction. The superimposed histogram is the lowest-order ISR prediction 
following Eq.~(\ref{fsrlo}). The $J/\psi$ mass region is removed for the
luminosity determination.
}
  \label{ISRL-full}
\end{figure}

The full effective ISR luminosity $dL_{\rm ISR}^{\rm eff}/d\sqrt{s'}$ is derived 
from the measured $\mu\mu(\gamma)\gamma_{\rm ISR}$ spectrum
according to Eq.~(\ref{def-lumi}). The event 
acceptance is taken from AfkQed, with corrections for detector and reconstruction effects.
Unfolding of the background-subtracted $\mu\mu$ mass spectrum is performed as explained in
Section~\ref{unfold}.
The result is shown in Fig.~\ref{ISRL-full}.
The effective luminosity derived this way implicitly includes
the VP factor since the $\mu\mu(\gamma)\gamma_{\rm ISR}$ data include vacuum polarization effects,
while the bare cross section $\sigma^0_{\mu\mu(\gamma)}(\sqrt{s'})$ entering
Eq.~(\ref{def-lumi}) does not.  
 
The measured effective luminosity is compared to the standard estimate of
Eq.~(\ref{lumi-eff}) using LO QED, given by
\beqn
\label{fsrlo}
 \frac {dL_{\rm LO}}{d\sqrt{s'}}= L_{ee}~\frac {dW_{\rm LO}}{d\sqrt{s'}}~
  \left(\frac {\alpha(s')}{\alpha(0)}\right)^2~,
\eeqn
where the LO radiator function is~\cite{isr3,isr4}
\beqn
\nonumber
\frac {dW_{\rm LO}}{d\sqrt{s'}}=\frac {\alpha(0)}{\pi x}~\left[ (2-2x+x^2)
  \ln \frac {1+c}{1-c}~-x^2c\right]~\frac {2\sqrt{s'}}{s}~,
\eeqn
with $x=1-s'/s$ and $c=\cos{\theta_{\rm min}^*}$. We insert the VP term $\alpha(s')/\alpha(0)$
in Eq.~(\ref{fsrlo}) for a convenient comparison with the effective luminosity. 
The computed VP factor includes both leptonic and
hadronic contributions. The hadronic contribution is taken from the
parametrization used in AfkQed and is found to agree well
with an independent determination using the tools of Ref.~\cite{dhmz09}.

The LO+VP prediction is superimposed on the measured effective luminosity in 
Fig.~\ref{ISRL-full}.
The measured luminosity is found to be about 2\% larger than the
LO+VP QED result. This difference 
varies slowly with mass and includes 
systematic effects on the \babar\ luminosity determination, the effect of the 
NLO contribution in data, the difference between the ISR 
photon efficiency in data and MC and any residual effect in the detection 
efficiency. The latter contribution  
is small, in accordance with the successful QED test performed with the
$\mu\mu(\gamma)\gamma_{\rm ISR}$ cross section.

The effective luminosity shown in Fig.~\ref{ISRL-full} is measured in 50\mev bins. 
This interval size is too wide near narrow resonances ($\omega$ 
and $\phi$) because of the rapid variation of hadronic vacuum polarization. 
Therefore, we compute the local variation inside each 50\mev interval as the product 
of the lowest-order QED radiator function times the vacuum polarization
factor. The result is normalized to the effective luminosity in the
interval. In this way the detailed local features are 
described, while preserving the measured effective luminosity 
as a function of mass. 

The statistical error of the $\pi\pi$ cross section is limited
in the $\rho$ resonance region by the number of events available to determine 
the ISR luminosity. Bin-to-bin
statistical fluctuations are reduced by a suitable averaging of the ratio of measured 
to LO ISR luminosities. The ratio distribution in 50\mev bins is 
smoothed by averaging five consecutive bins in a sliding way. This value 
is chosen as a compromise between smoothing and the validity of the assumption 
of slow variation. This method does not improve in principle the ISR 
luminosity statistical error because the reduced local error is compensated 
by the correlation between neighbouring bins. A slight improvement 
in the dispersion integral is however observed due to the weighting in different 
mass regions.

The statistical errors on the ISR effective luminosity 
from the measurement of efficiencies are included in 
the statistical covariance matrix, while the systematic errors 
from the different procedures are accounted for separately. 
These errors are $0.3\times 10^{-3}$ for the trigger, $1.3\times 10^{-3}$ for tracking, $2.9\times 10^{-3}$ for 
$\mu$-ID, and $1.0\times 10^{-3}$ for acceptance, for a total systematic uncertainty
of $3.4\times 10^{-3}$. The uncertainty  from the correlated loss
of $\mu$-ID for both tracks is not included here, since it is anticorrelated
with the pion rate. It is counted in the systematic errors on the $\pi\pi$
cross section.

\section{Measurement of the $e^+e^-\to \pi^+\pi^-(\gamma)$ cross section}

The $e^+e^-\to \pi^+\pi^-(\gamma)$ bare cross section is measured from the 
$\sqrt{s'}$ distribution of produced $\pi^+\pi^-(\gamma)\gamma_{\rm ISR}$ events 
divided by the effective ISR luminosity obtained from $\mu^+\mu^-(\gamma)\gamma_{\rm ISR}$
on the same data. The $\sqrt{s'}$ distribution is obtained from the 
observed $m_{\pi\pi}$ mass spectrum, after background subtraction, corrections
for data/MC efficiency differences, unfolding, and MC acceptance corrections, as
described in detail in the preceding sections. Because the cross section
spans several orders of magnitude over the energy region considered, from
threshold to 3\gev, the analysis strategy depends on the mass region.     
Event selection and background subtraction are optimized separately for the 
$\rho$ resonance central region or for the resonance tails, with corresponding 
efficiency corrections and unfolding matrix. To facilitate comparison with other experiments,
the results are also shown in terms of the pion form factor fitted with a 
vector-dominance model (VDM).

\subsection{The central $\rho$ region ($0.5<m_{\pi\pi}<1\gevcc$)}
\label{central}

\subsubsection{Strategy}
\label{chi2-loose-cut}

The mass region between 0.5 and 1\gevcc, dominated by the $\rho$ resonance,
provides the dominant contribution to the vacuum-polarization dispersion integrals
and their errors. The need for small systematic uncertainties, congruous with the 
small statistical errors, together with the low background level in that region, lead 
to an event selection with the largest efficiency.  
Therefore the loose $\chi^2$ criterion, the same as for the $\mu\mu(\gamma)\gamma_{ISR}$ analysis, and 
standard $\pi$-ID for both tracks are used.

\subsubsection{Summary of backgrounds}

The backgrounds are obtained as described in Section~\ref{background}. 
The dominant contribution is from multihadronic
processes, mostly ISR ($\pi^+\pi^-\pi^0\gamma$, $\pi^+\pi^-2\pi^0\gamma$)
and $q\overline{q}$, with a fraction amounting to $8.4\times 10^{-3}$ at the 
$\rho$ peak. The $p\overline{p}\gamma$ contribution is much
smaller ($<10^{-3}$). 

\begin{table} [thb] 
\centering 
\caption{ \label{fbg-all-central} \small 
Estimated background fractions (in \%) in the `$\pi\pi$' sample for 
$m_{\pi\pi}$=0.525, 0.775, 0.975\gevcc. The quoted errors include both 
statistical and systematic uncertainties.}
\vspace{0.5cm}
\setlength{\extrarowheight}{1.5pt}
\setlength{\tabcolsep}{2pt}
\begin{tabular}{lccc} \hline\hline\noalign{\vskip1pt}
 Process &  0.525\gevcc & 0.775\gevcc & 0.975\gevcc  \\ \hline
 $\mu\mu$                & $3.48\pm0.36$ & $0.37\pm0.23$ & $2.71\pm0.31$  \\
 $KK$                    & $0.08\pm0.01$ & $0.01\pm0.01$ & $0.08\pm0.01$  \\ 
\hline  
 $\gamma 2\pi\pi^0$      & $8.04\pm0.41$ & $0.39\pm0.05$ & $0.88\pm0.19$  \\
 $q\overline{q}$         & $1.11\pm0.17$ & $0.26\pm0.03$ & $1.81\pm0.19$  \\
 $\gamma 2\pi 2\pi^0$    & $1.29\pm0.16$ & $0.06\pm0.01$ & $0.46\pm0.09$  \\
 $\gamma 4\pi$           & $0.20\pm0.04$ & $0.09\pm0.01$ & $0.24\pm0.06$  \\
 $\gamma p\overline{p}$  & $0.22\pm0.02$ & $0.04\pm0.01$ & $0.52\pm0.06$  \\
 $\gamma \eta 2\pi$      & $0.02\pm0.01$ & $0.03\pm0.01$ & $0.09\pm0.01$  \\
 $\gamma K_SK_L$         & $0.18\pm0.03$ & $0.01\pm0.01$ & $0.10\pm0.02$  \\
 $\gamma 4\pi 2\pi^0$    & $<0.01      $ & $<0.01      $ & $<0.01      $  \\
 $\tau\tau$              & $0.17\pm0.03$ & $0.04\pm0.01$ & $0.31\pm0.05$  \\
 $\gamma ee$             & $0.63\pm0.63$ & $0.03\pm0.03$ & $0.27\pm0.27$  \\
\hline
 total                  & $15.38\pm0.87$ & $1.31\pm0.24$ & $7.37\pm0.51$  \\
\hline\hline
\end{tabular}
\end{table}

The fractions of all the considered backgrounds 
are given in Table~\ref{fbg-all-central} at three mass values.
For convenience, we also show the level of the $\mu\mu$ and $KK$
background contributions in the `$\pi\pi$'-identified sample, although they
are implicitly subtracted when solving Eqs.~(\ref{sep-pid}) for
the produced spectrum $N_{\pi\pi}$.

\begin{figure}[htp]
  \centering
  \includegraphics[width=7.5cm]{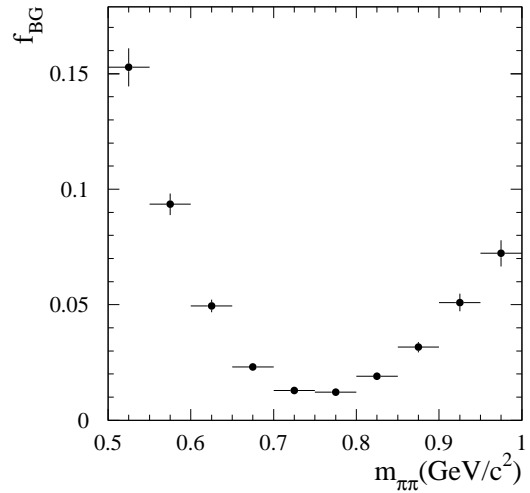}
  \caption{\small 
The total background fraction for the $\pi\pi(\gamma)\gamma_{\rm ISR}$ sample in the
central $\rho$ region (loose $\chi^2$ selection).}
  \label{bkg-central}
\end{figure}

The total background fraction as a function of $m_{\pi\pi}$ is shown in
Fig.~\ref{bkg-central}. It is 1.3\% at the $\rho$ peak, but reaches $\sim$15\% 
at 0.5\gevcc and $\sim$7\% at 1\gevcc. These sharp increases justify the limits 
chosen to define the `central region'. At the worst place, near 0.5\gevcc, 
the total uncertainty from the estimated non-$\mu\mu$/$KK$ 
background fraction is 0.8\%, which is still tolerable. 
At the peak the uncertainty is less than 0.1\%. 

\subsubsection{Background-subtracted $m_{\pi\pi}$ mass distribution}

\begin{figure}[thp]
\centering
  \includegraphics[width=7.5cm]{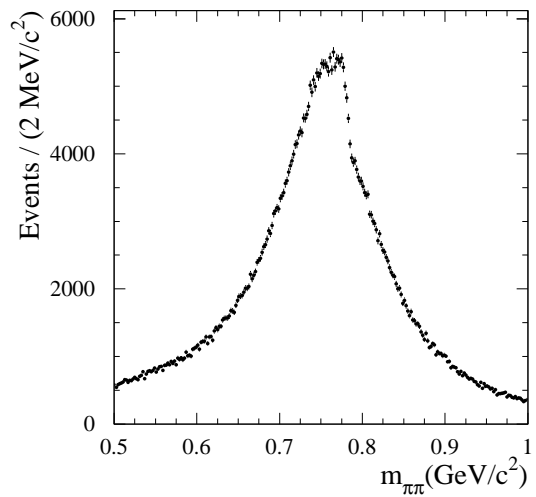}\\
 \caption{\small The $m_{\pi\pi}$ spectrum of $\pi\pi(\gamma)\gamma_{\rm ISR}$
events in the $\rho$ region, in 2\mevcc bins.}
  \label{pipi-central}
\end{figure}

The background-subtracted $m_{\pi\pi}$ spectrum obtained in the $\rho$ region 
before unfolding, 
with loose $\chi^2$ criterion and `$\pi$'-ID for both pions, is shown in 
Fig.~\ref{pipi-central}. Only the statistical errors 
in the 2\mevcc mass 
intervals are given, amounting to 1.4\% on peak and 4.4\% near the boundaries. 
Apart from the $\rho$ resonance shape, a clear $\rho-\omega$ interference 
pattern is observed.

\subsection{The $\rho$ tail regions ($m_{\pi\pi}<0.5$, $m_{\pi\pi}>1\gevcc$)}
\label{tails}

\subsubsection{Strategy}

The pion cross section decreases very rapidly away from the $\rho$ resonance, while the
backgrounds from $\mu\mu(\gamma)\gamma_{ISR}$, $KK(\gamma)\gamma_{ISR}$, $p\bar{p}\gamma_{ISR}$, and
multihadron events show a smooth variation with the $\pi\pi$ mass. To keep the background
levels under control in the $\rho$ tail regions,  
the selection of $\pi\pi(\gamma)\gamma_{ISR}$ events is tightened with respect 
to the criteria used in the $\rho$ region.
Two handles are simultaneously used: (i) the tight $\chi^2$ criterion 
$\ln(\chi^2_{\rm ISR}+1)<3$ is chosen to reduce multihadronic backgrounds,
and (ii) the pion-ID is strengthened to improve muon (and also electron) 
rejection. In addition, the $V_{\rm xy}$ requirement described in Section~\ref{Vxy-ee}
is applied to remove backgrounds from photon conversions and bremsstrahlung in
the beam pipe. 

\subsubsection{Summary of backgrounds}

For two-body ISR and $ee$ backgrounds, the
tighter $\chi^2$ criterion is not useful, so the harder $\pi$ identification, 
`$\pi_{h}$' (Sect.~\ref{pi-hard}) is required for at least one of the 
two `$\pi$'-identified tracks, giving a further rejection of $\mu$ and $e$. 
The downside is that, because the `$\pi_h$' identification breaks the completeness of PID classes, 
the $\mu\mu$ and $KK$ backgrounds cannot be subtracted anymore from the 
`$\pi\pi_h$' sample by solving the Eqs.~(\ref{sep-pid}) system. 

The $\mu\mu(\gamma)\gamma_{ISR}$ background is estimated according to Section~\ref{mubkg-pih}.
With the `$\pi\pi_h$' selection, it is reduced by a factor $\sim7$ with 
respect to the background remaining after the standard pion-ID.
However the corresponding factor for the background fraction in the
final pion sample is smaller because the `$\pi\pi_h$' efficiency also 
reduces the signal (Sect.~\ref{pipih-eff}). 

Neither the tight $\chi^2$ criterion nor the `$\pi\pi_h$' ID brings significant reduction 
of the $KK(\gamma)\gamma_{ISR}$ background compared to the selection used in the $\rho$
peak region. Its small contribution is estimated from data by using the 
procedure described in Section~\ref{Kbkg-pih} and subtracted.

The backgrounds from $q\overline{q}$ and multihadronic ISR events are estimated
as discussed in Sections~\ref{bkg-had} and \ref{bkg-isr}, respectively. They are much reduced 
compared to the central $\rho$ region because of the tight $\chi^2$ condition.

\begin{figure*}
\begin{minipage}[ht]{0.45\textwidth}
  \centering
  \includegraphics[width=7.5cm]{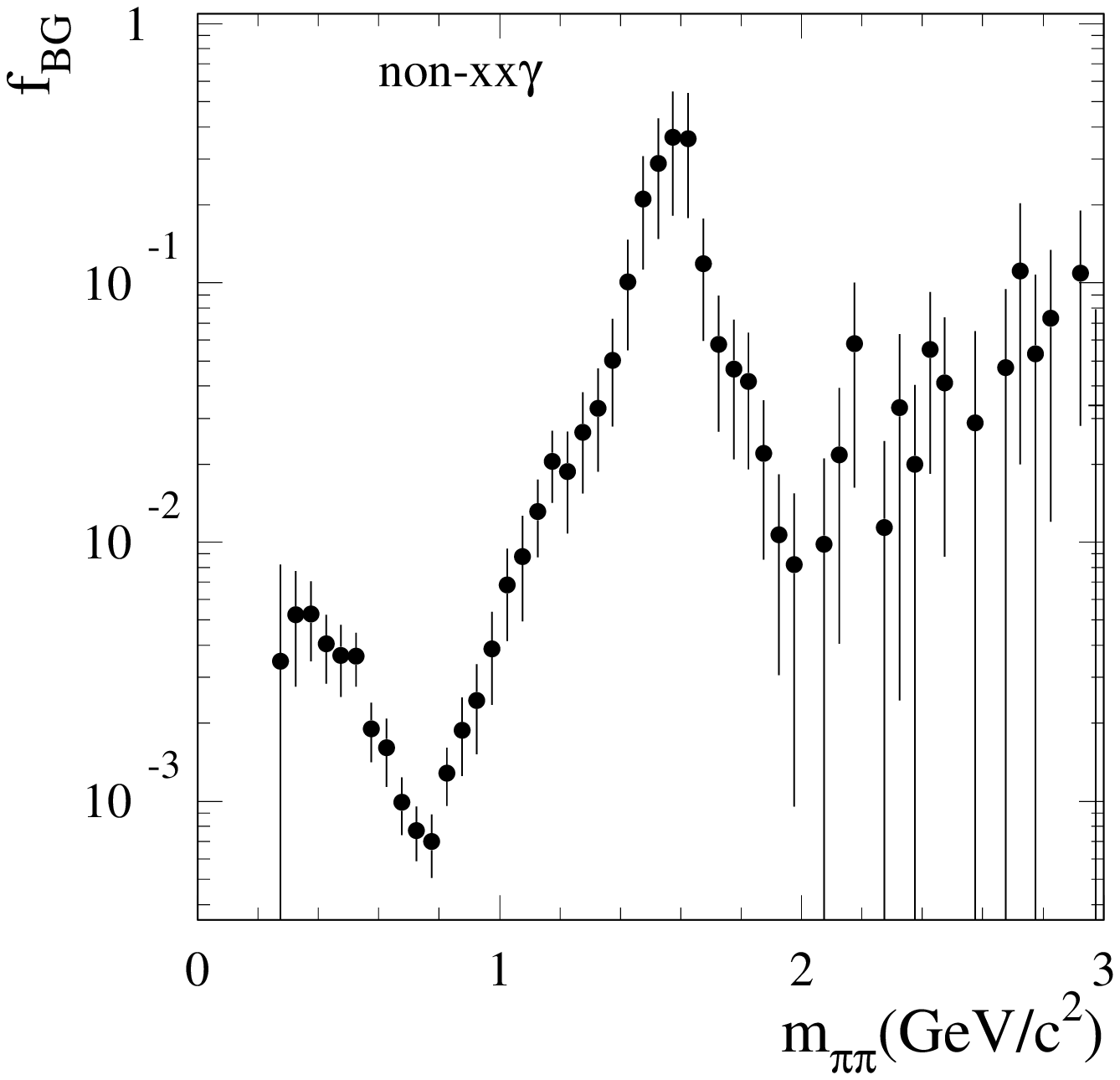}
  \includegraphics[width=7.5cm]{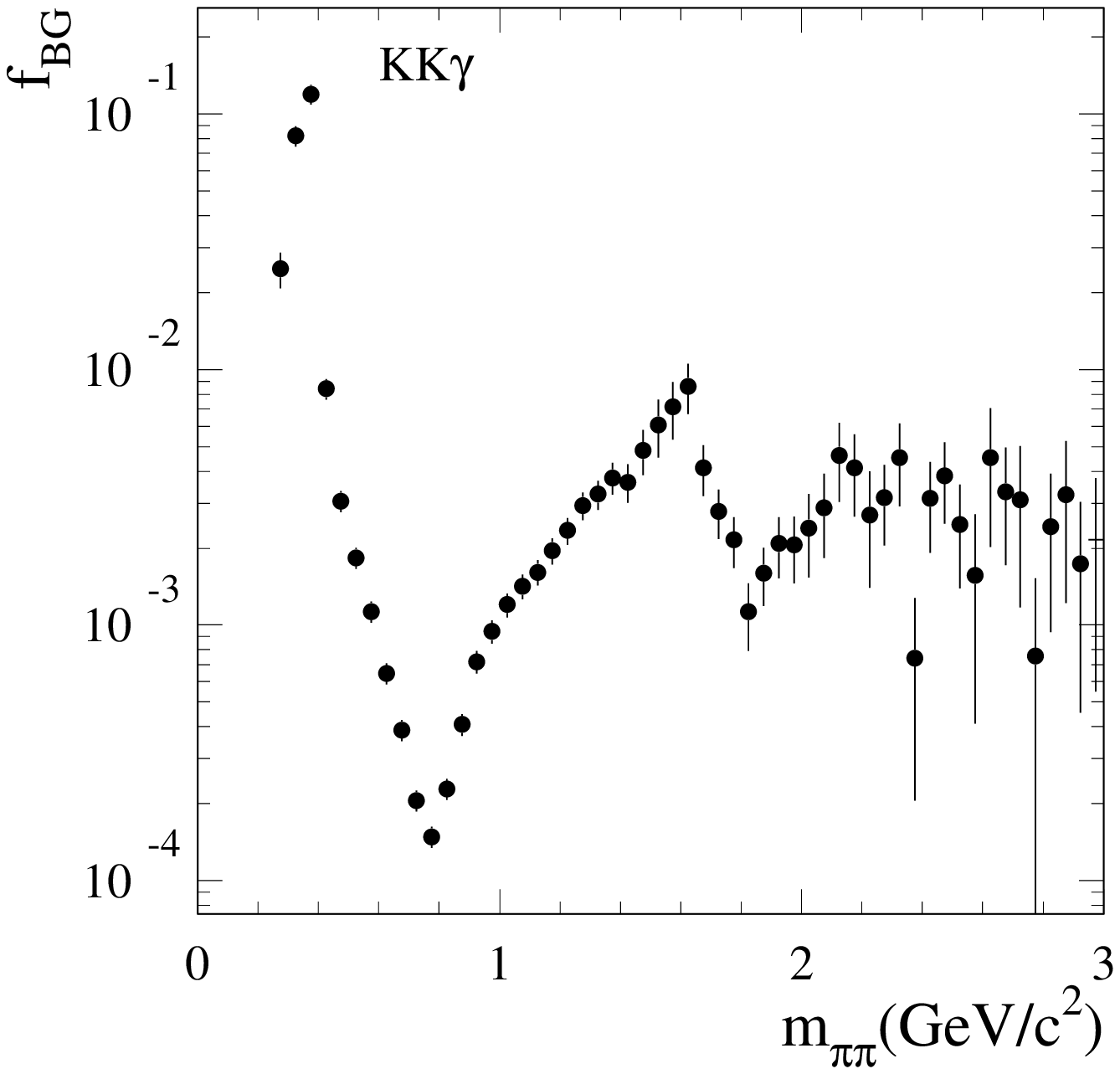}
  \end{minipage}\hfill
  \begin{minipage}[ht]{0.45\textwidth}
  \centering
  \includegraphics[width=7.5cm]{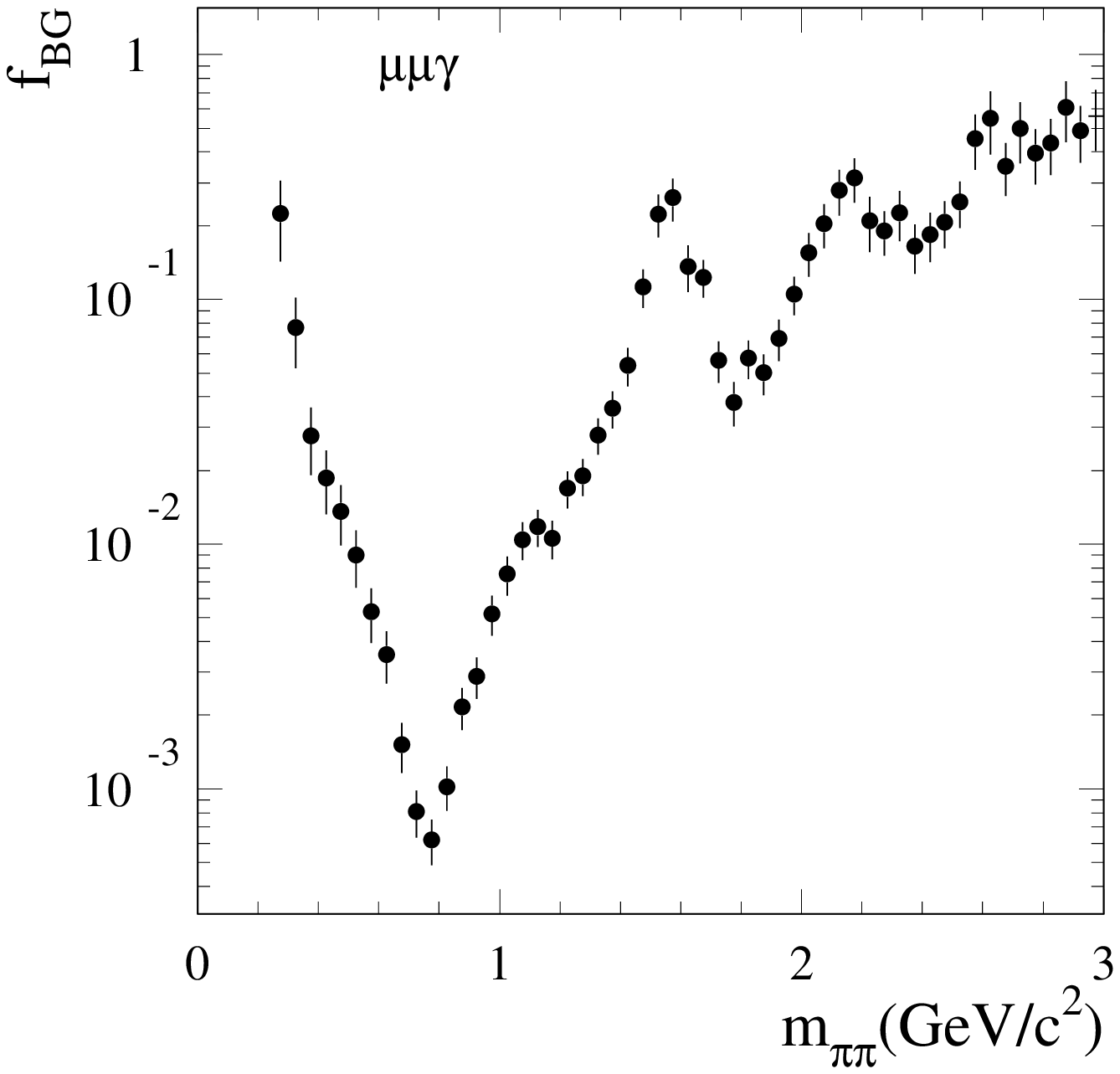}\\
  \includegraphics[width=7.5cm]{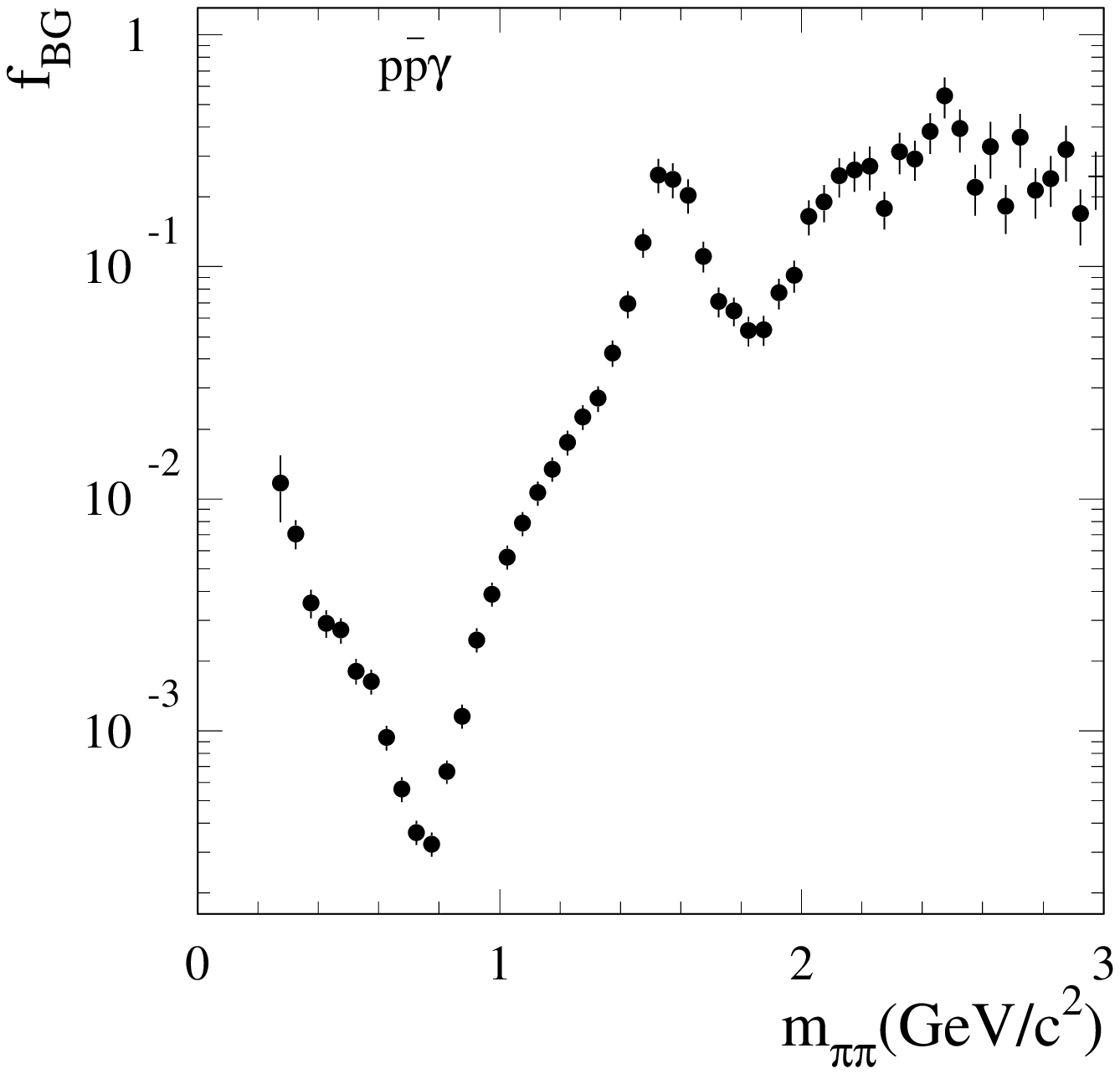}
\end{minipage}
\caption{\small The fractions of different backgrounds in the physical 
sample with the tight 2D-$\chi^2$ criterion and strengthened `$\pi\pi_h$'-ID 
(as used in the $\rho$ tails region) as a function of the $\pi\pi$ mass.
(top left): Multihadrons, including $\tau\tau$.
(top right): $\mu\mu(\gamma)\gamma_{\rm ISR}$ (data + measured mis-ID). (bottom left):
$KK(\gamma)\gamma_{\rm ISR}$ (data + measured mis-ID). (bottom right): $p\overline{p}\gamma_{\rm ISR}$ (MC).}
\label{fbg-tailreg}
\end{figure*}

The different fractions of background in the region of the $\rho$ tails, with 
`$\pi\pi_h$' ID and $\ln(\chi^2_{\rm ISR}+1)<3$, are given in Fig.~\ref{fbg-tailreg}. 
Fractions at specified masses are listed in Table~\ref{fbg-all-tails}. 
The total background contribution is obtained by summing all the 
individual contributions obtained above.

\begin{table*} 
\begin{minipage}[ht]{1.\textwidth}
  \centering
\caption{ \label{fbg-all-tails} \small 
Estimated background fractions (in \%) in the `$\pi\pi_h$' sample for 
$m_{\pi\pi}$=0.325, 0.475, 0.975, 1.375, 1.975, and 2.975\gevcc. The entries
marked as `$-$' correspond to a negligible fraction. Processes with
fractions less than 0.05\% in all intervals are not listed. The quoted errors
include both statistical and systematic uncertainties.}
\vspace{0.5cm}
\setlength{\extrarowheight}{1.5pt}
\setlength{\tabcolsep}{2pt}
\begin{tabular}{lcccccc} \hline\hline\noalign{\vskip1pt}
 Process &  0.325\gevcc & 0.475\gevcc & 0.975\gevcc & 1.375\gevcc & 1.975\gevcc & 2.975\gevcc \\ \hline
 $\mu\mu$                & $7.7\pm2.5$ & $1.4\pm0.4$ & $0.5\pm0.1$ & 
                           $3.6\pm0.6$ & $10.5\pm1.9$ & $56.2\pm15.8$  \\
 $KK$                    & $8.2\pm0.7$ & $0.3\pm0.1$ & $0.1\pm0.1$ &
                           $0.4\pm0.1$ & $0.2\pm0.1$ & $0.2\pm0.2$  \\ 
\hline  
 $\gamma 2\pi\pi^0$      & $0.4\pm0.2$ & $0.3\pm0.1$ & $0.1\pm0.1$ &
                           $-$         & $-$         & $-$    \\
 $q\overline{q}$         & $0.1\pm0.1$ & $0.1\pm0.1$ & $0.3\pm0.2$ &
                           $5.0\pm2.2$ & $0.8\pm0.7$ & $3.4\pm4.5$  \\
 $\gamma p\overline{p}$  & $0.7\pm0.1$ & $0.3\pm0.1$ & $0.4\pm0.1$ &
                           $4.3\pm0.5$ & $9.2\pm1.4$ & $24.5\pm6.8$  \\
\hline
 total                   & $17.1\pm2.6$ & $2.3\pm0.4$ & $1.4\pm0.2$ &
                           $13.2\pm2.3$ & $20.7\pm2.5$ & $84.3\pm17.8$  \\
\hline\hline
\end{tabular}
\end{minipage}
\end{table*}

\subsubsection{Background-subtracted $m_{\pi\pi}$ mass distribution}

\begin{figure}[tph]
  \centering
  \includegraphics[width=7.5cm]{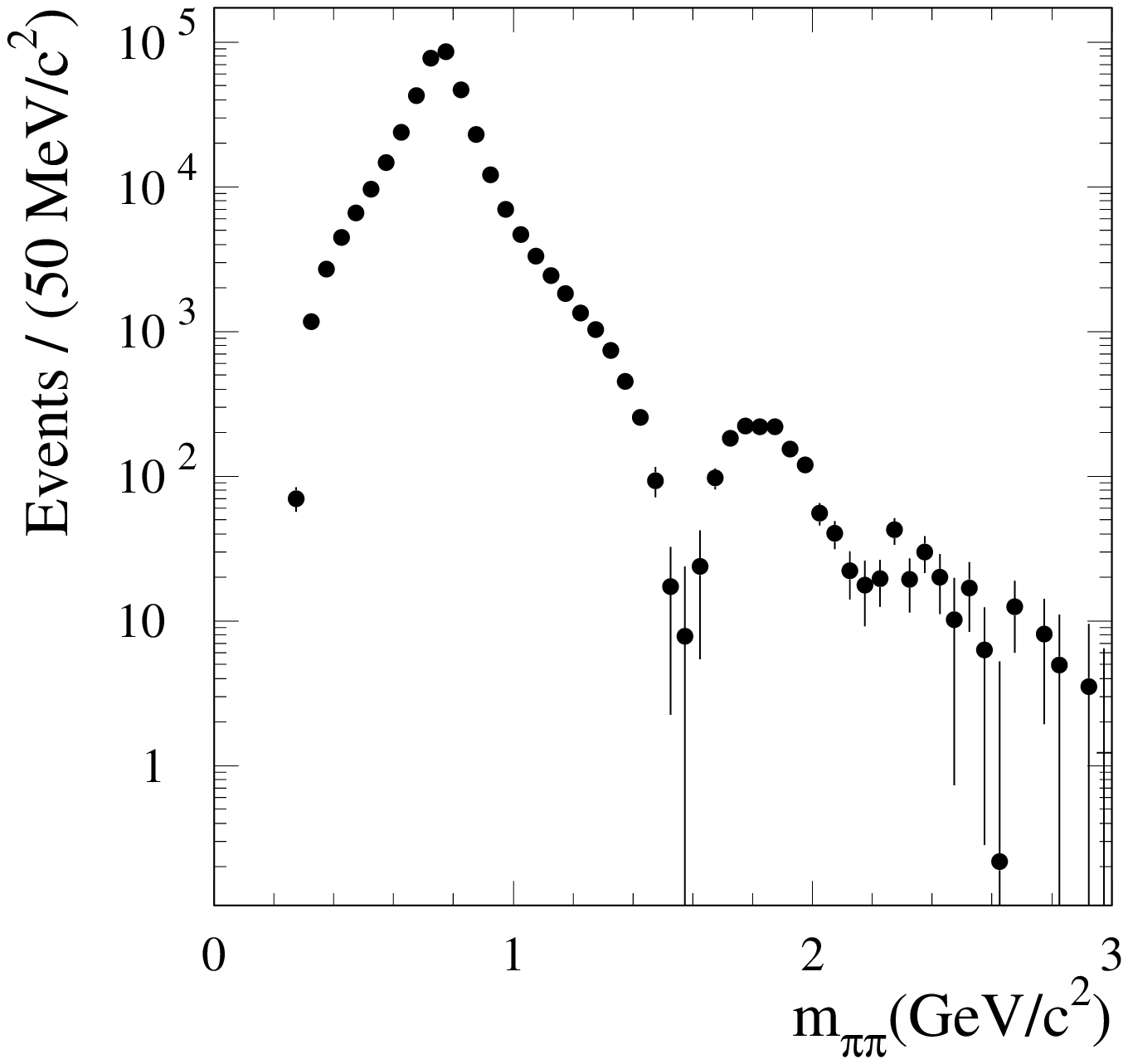}
  \caption{\small The $m_{\pi\pi}$ spectrum of $\pi\pi(\gamma)\gamma_{\rm ISR}$ events 
selected
with `$\pi\pi_h$' identification and the tight $\ln(\chi^2_{\rm ISR}+1)<3$ criterion,
from threshold to 3\gevcc in 50\mevcc mass intervals.}
  \label{pipih-bkg-sub}
\end{figure}

The background-subtracted $m_{\pi\pi}$ distribution of $\pi\pi(\gamma)\gamma_{\rm ISR}$ events 
before unfolding,
using `$\pi\pi_h$' identification and $\ln(\chi^2_{\rm FSR}+1)<3$ is plotted 
from threshold to 3\gevcc in 50\mevcc mass intervals in Fig.~\ref{pipih-bkg-sub}. 
A dynamic range of $10^3$--$10^4$ is observed
between the $\rho$ peak and either the first bin above threshold or at 3\gevcc.
The dip structure at 1.6\gevcc seen by the DM2 experiment~\cite{dm2} is confirmed
with high statistics and a new structure shows up near 2.2\gevcc. 

%
\subsection{Angular distribution in the $\pi\pi$ center-of-mass}
\label{theta*-central}

\begin{figure}[t!]
  \centering
  \includegraphics[width=7.5cm]{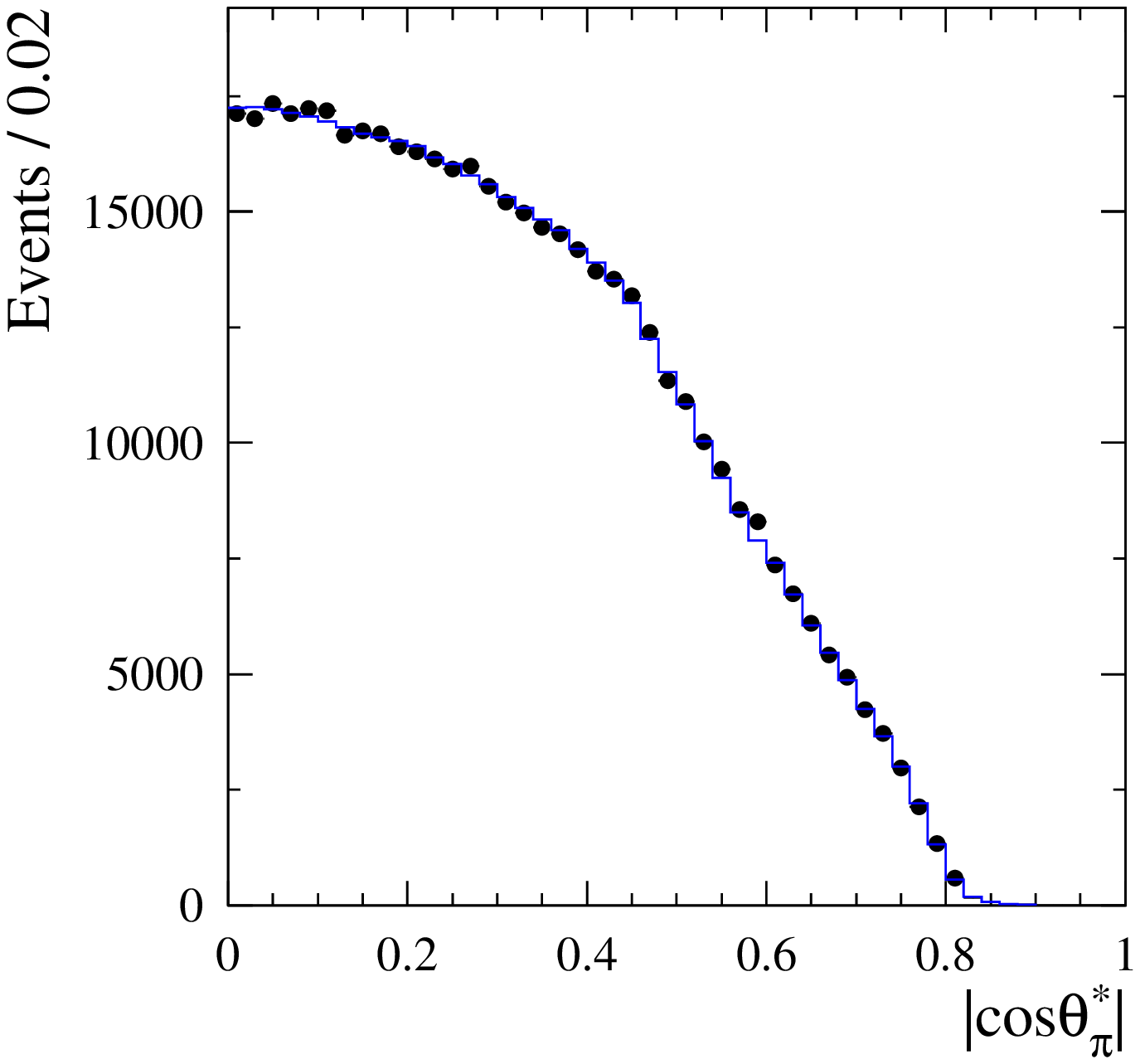}
  \caption{\small 
The angular pion distribution in the $\pi\pi$ system with respect to the ISR
photon direction as function of $|\cos\theta_{\pi}^*|$ for background-subtracted
$\pi\pi(\gamma)\gamma_{\rm ISR}$ data (points) in the $\rho$ central region
($0.5<m_{\pi\pi}<1\gevcc$). The blue histogram is the shape
obtained in the simulation, normalized to the data.}
  \label{cos-theta*-central}
\end{figure}

The distributions of kinematic variables such as the ISR photon polar angle and 
the pion momenta and angles depend on the hadronic structure we seek to 
measure. Thus the detailed comparisons between data and MC distributions expected
from theory, which are performed in the $\mu\mu(\gamma)\gamma_{\rm ISR}$ channel with QED 
(Sect.~\ref{mu-kinvar}), are meaningless in the pion channel. However, one 
distribution, namely the pion angular distribution 
in the $\pi\pi$ center-of-mass, with respect to the ISR photon direction in 
that frame, is model-independent. The $\cos\theta_{\pi}^*$ distribution 
is consistent with $\sin^2\theta_{\pi}^*$ as expected in the
$e^+e^-\rightarrow \pi^+ \pi^-$ process,
but it is strongly distorted at $|\cos\theta_{\pi}^*|$ values near one by the
$p>1\gevc$ requirement on the tracks.

The $|\cos\theta_\pi^*|$ distributions for background-subtracted data 
and MC are compared in
Fig.~\ref{cos-theta*-central} for the 0.5--1\gevcc mass range: they agree with
each other within the statistical errors, as expected for a pure pion sample.

\subsection{Acceptance and corrections}

The overall efficiency $\eps_{\pi\pi(\gamma)\gamma_{\rm ISR}}$ entering Eq.~(\ref{def-lumi}) 
for the pion channel is 
calculated using the AfkQed generator and full simulation in the same way as 
for the muon channel. It is corrected for differences
in efficiencies between data and MC, which are introduced 
as mass-dependent corrections applied to the event spectrum (Eq.~(\ref{eff-corr})). 

As discussed in Section~\ref{mu-accept}, NLO approximations are made in simulation,
which affect the acceptance. 
While the FSR prescription by {\small PHOTOS}
is found to agree reasonably well with data, this is not the case for additional ISR
as simulated in AfkQed. The problems have been studied in detail for muons, 
since they affect the absolute 
measurement of the $\mu\mu(\gamma)\gamma_{\rm ISR}$ cross section and the comparison with QED.
However here we deal with acceptance corrections that apply to the 
pion cross section measured from the ratio of the ${\pi\pi(\gamma)\gamma_{\rm ISR}}$ spectrum 
to the effective luminosity. 
As the additional-ISR issues are common to the $\pi\pi(\gamma)\gamma_{\rm ISR}$ and 
$\mu\mu(\gamma)\gamma_{\rm ISR}$ channels, they cancel in the 
$R_{\rm exp}(\sqrt{s'})$ ratio, except for second-order effects addressed below. 
Thus the $\pi\pi$ measurement does not rely on the accurate description of NLO 
effects by the MC generator, a fact that is a strength of this analysis method.

\begin{figure}[t!]
\centering
  \includegraphics[width=7.5cm]{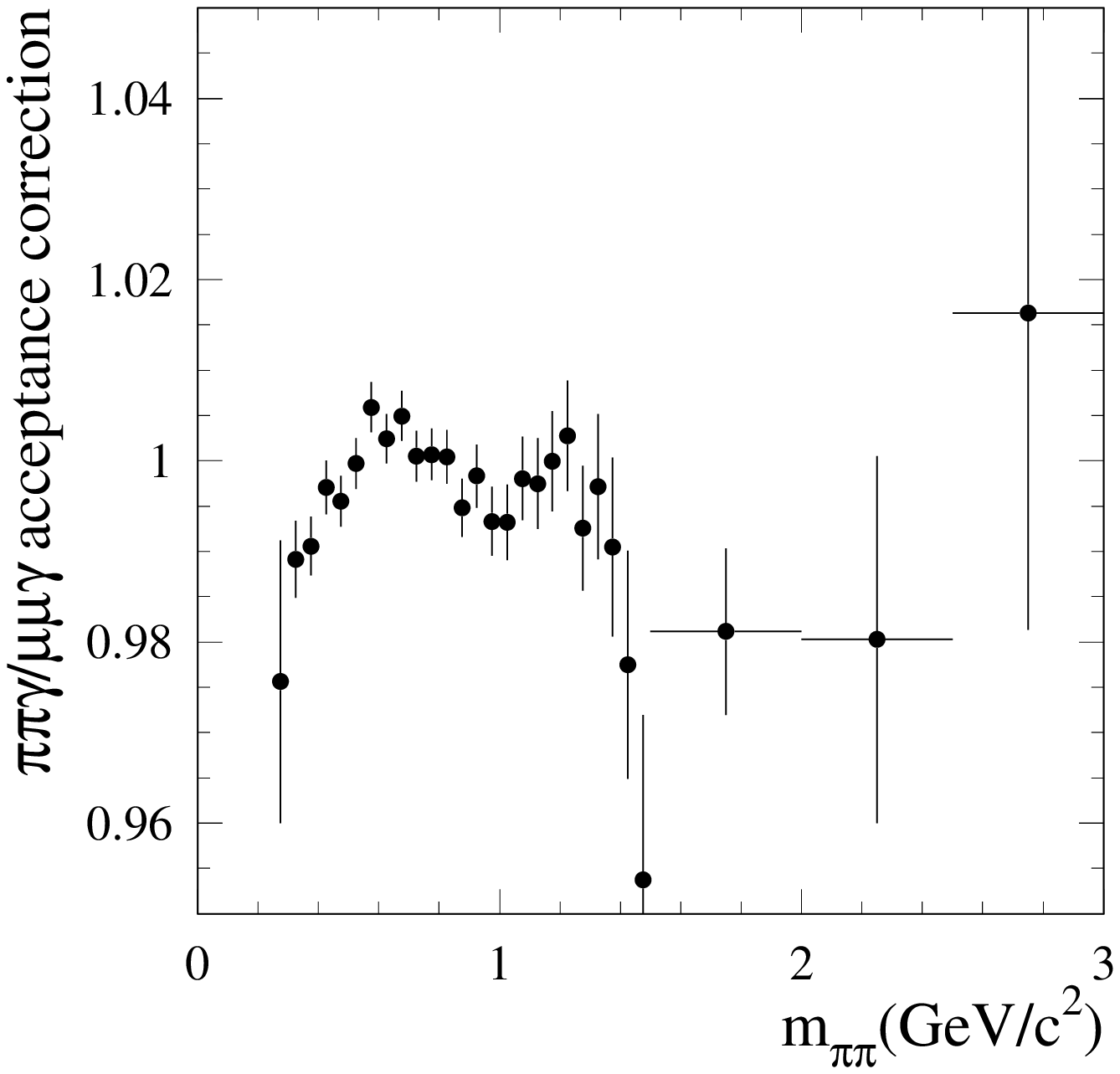}
  \caption{\small 
The full correction to the $\eps_{\pi\pi(\gamma)\gamma_{ISR}}/\eps_{\mu\mu(\gamma)\gamma_{ISR}}$ acceptance ratio 
for non-canceling effects (see text).}
  \label{pipi-mumu-corr}
\end{figure}

Acceptance, including preselection efficiency, is mainly affected by kinematics, {\it i.e.},
the angular and energy distributions of the hard additional-ISR photon. We study 
these effects for the pion channel as we have done for the muon channel, using
AfkQed and Phokhara at the generator level and fast simulation with parametrized 
efficiencies and resolutions. 
The resulting data/MC correction on acceptance for pions is consistent with 
the correction obtained for muons. Second-order corrections
induced by pion secondary interactions are investigated, using
the full simulation.
The total acceptance correction amounts to a few $\times 10^{-3}$ in the $\rho$
mass region. It is shown as a function of the $\pi\pi$ mass in 
Fig.~\ref{pipi-mumu-corr}. As for the other corrections, the statistical 
uncertainties are included in the final cross section errors. Since part of the
correction is derived using the fast simulation at the generator level, a significant 
fraction of the correction ($\sim$25\%) is taken as a systematic uncertainty, 
$10^{-3}$ in the 0.6--0.9\gevcc region and larger outside.

\subsection{Summary of the treatment of statistical uncertainties}
\label{summary-stat}

The statistical covariance matrix of the cross section includes the 
bin-to-bin correlations affecting the $\pi\pi$ spectrum and the luminosity.

The statistical covariance matrix of the $\pi\pi$ spectrum is not diagonal,
due firstly to correlations introduced by the transfers of events in the 
unfolding process. In addition, the data/MC efficiency corrections and subtracted background 
spectra are initially computed in 50\mevcc bins, but applied to 
spectra with 2\mevcc bins (in the central $\rho$ region) or 10\mevcc bins (in the $\rho$ 
tail regions) using splines. The resulting covariance matrix 
is obtained from a large series of toy experiments.

The ratio of the measured luminosity to the LO luminosity including vacuum 
polarisation is initially computed in (almost) uncorrelated bins of 50\mev. 
The procedure of sliding bins, used for smoothing this distribution,
introduces correlations between the final values (Sect.~\ref{eff-lumi}).
The luminosity errors for the final cross section (2 or 10\mev) bins are 
$100\%$ correlated within a 50\mev bin, whereas additional correlations occur
between the 50\mev bins because of the bin-sliding procedure.
Finally, the correlation effect from unfolding the $\mu\mu$ spectrum is 
rather weak, but it is however propagated to the final correlation
matrix.

\subsection{Systematic errors}
\label{summary-syst}
Systematic uncertainties affecting the $\pi\pi$ sample in different mass 
regions are now summarized.
The statistical errors of the measured efficiencies are included with the main
statistical uncertainty on the $\pi\pi$ mass spectrum. However, in some cases,
remaining systematic uncertainties are attached to the efficiency measurement 
process and quoted as such. Details have been given for each efficiency
study in Sections~\ref{eff} and \ref{eff-chi2}. The results for all systematic 
uncertainties are listed in Table~\ref{pi-syst-err}. 

The overall relative systematic uncertainty on the $\pi\pi(\gamma_{\rm FSR})$ 
cross section is $5.0\times 10^{-3}$ in the 0.6--0.9\gev range,
but significantly larger below and above the central region.
For comparison, the statistical error of the measured efficiency corrections
amounts to $4.7\times 10^{-3}$ at the $\rho$ peak, while the statistical error of 
the raw spectrum is $1.35\%$ at that mass.

A full treatment of the systematic uncertainties is implemented,
using a covariance matrix. To achieve this we consider the individual 
systematic errors (for each source, as given in Table~\ref{pi-syst-err}) to be 
100\% correlated in all the mass bins. Then
the total systematic covariance matrix is built as the sum of the covariance 
matrices corresponding to each individual systematic source.

\begin{table*} 
\begin{minipage}[ht]{1.\textwidth}
\centering 
\caption{ \label{pi-syst-err} \small 
Systematic uncertainties (in $10^{-3}$) on the cross section for 
$e^+e^-\rightarrow\pi\pi(\gamma_{\rm FSR})$ from the determination of the various 
efficiencies in different $\pi\pi$ mass ranges (in\gevcc).
The statistical part of the efficiency measurements is included in 
the total statistical error in each mass bin. The last line gives the 
total systematic uncertainty on the $\pi\pi$ cross section, including the
systematic error on the ISR luminosity from muons.}
\vspace{0.5cm}
\setlength{\extrarowheight}{1.5pt}
\setlength{\tabcolsep}{2pt}
\begin{tabular}{lcccccccc} \hline\hline\noalign{\vskip1pt}
 Sources &  0.3-0.4 & 0.4-0.5 & 0.5-0.6 & 0.6-0.9 & 0.9-1.2 & 1.2-1.4 & 1.4-2.0 & 2.0-3.0 \\ \hline
 trigger/ filter            & 5.3 & 2.7 & 1.9 & 1.0 & 0.7 & 0.6 & 0.4 & 0.4\\ 
 tracking                   & 3.8 & 2.1 & 2.1 & 1.1 & 1.7 & 3.1 & 3.1 & 3.1\\ 
 $\pi$-ID                   &10.1 & 2.5 & 6.2 & 2.4 & 4.2 &10.1 &10.1 &10.1\\
 background                 & 3.5 & 4.3 & 5.2 & 1.0 & 3.0 & 7.0 &12.0 &50.0\\
 acceptance                 & 1.6 & 1.6 & 1.0 & 1.0 & 1.6 & 1.6 & 1.6 & 1.6\\
 kinematic fit ($\chi^2$)   & 0.9 & 0.9 & 0.3 & 0.3 & 0.9 & 0.9 & 0.9 & 0.9\\
 correl $\mu\mu$ ID loss    & 3.0 & 2.0 & 3.0 & 1.3 & 2.0 & 3.0 &10.0 &10.0\\
 $\pi\pi/\mu\mu$ non-cancel.& 2.7 & 1.4 & 1.6 & 1.1 & 1.3 & 2.7 & 5.1 & 5.1\\
 unfolding                  & 1.0 & 2.7 & 2.7 & 1.0 & 1.3 & 1.0 & 1.0 & 1.0\\
 ISR luminosity             & 3.4 & 3.4 & 3.4 & 3.4 & 3.4 & 3.4 & 3.4 & 3.4\\
\hline
 sum (cross section)        &13.8 & 8.1 &10.2 & 5.0 & 6.5 &13.9 &19.8 &52.4\\
\hline\hline
\end{tabular}
\end{minipage}
\end{table*}

\subsection{Consistency check with tight and loose $\chi^2$ selection}

The loose $\chi^2$ criterion is used in the $\rho$ central region, while the tight
one is used in the tails where backgrounds are larger. However it is possible
to compare the results obtained with the two selections in the central region. This
provides a test of the $\chi^2$ selection efficiency and of the multihadronic
background. The test is also sensitive to unfolding, as mass resolutions
are different in different 2D-$\chi^2$ regions. 
For this test, events are selected with the `$\rho$ central' conditions, and with
either the tight or loose $\chi^2$ criterion.

\begin{figure}[t!]
  \centering
  \includegraphics[width=8.0cm]{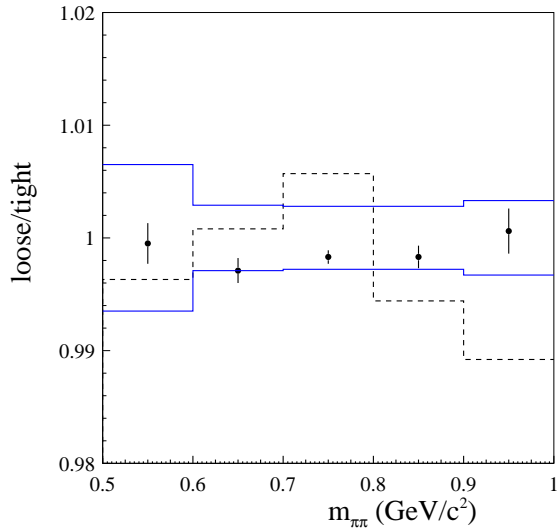}
  \caption{\small
    The ratio of the corrected and unfolded mass spectra (data points) for loose 
    over tight 2D-$\chi^2$ selection in the central $\rho$ region fitted in 100\mevcc 
    bins, compared to the band of independently estimated 
    uncertainties (solid lines). The MC mass-matrix resolution 
    correction is shown as the dashed histogram.}
  \label{loose-tight-resol}
\end{figure}

The result of the test is expressed as the ratio of the efficiency-corrected 
and unfolded spectra for the loose over the tight $\chi^2$ selections. The fitted value 
of this ratio over the full central range (0.5--1.0\gevcc) is found to be 
consistent with unity within errors, $0.9983 \pm 0.0049$ with a $\chi^2/{\rm DF}$ 
of 53.6/49 for 10\mevcc bins. Fits in 100\mevcc intervals, 
given in Fig.~\ref{loose-tight-resol}, do not show any significant trend for 
a resolution mismatch between data and corrected MC.
Deviations from unity are at a much smaller level than the resolution correction applied to 
the MC in the intermediate region (Sect.~\ref{unfold}, shown by the dashed histogram).
They are also within the range of estimated uncertainties between the two
$\chi^2$ conditions (background and $\chi^2$ selection efficiencies).
We thus conclude that the procedure used for correcting the MC mass-transfer
matrix is consistent within the quoted systematic uncertainties.

\subsection{Cross section results}

\subsubsection{Estimates of LO FSR in the $\pi\pi(\gamma)\gamma$ process}
\label{lofsr-pi}

In Eq.~(\ref{eq-3}) it is assumed that the contribution of LO FSR to the
$\pi\pi(\gamma)\gamma$ cross section is negligible. This approximation is 
supported by calculations using specific models. 

The simplest model uses an extrapolation of the pion form factor to the large
value of  $s\sim 112\gev^2$, and assumes point-like pions to compute LO FSR
photon emission, as for additional FSR.  This procedure, questionable for large
energies $E_\gamma^* > 3\gev$, yields a very small relative contribution from
$|{\cal A_{\rm FSR}}|^2$, $\delta^{\pi\pi}_{\rm FSR} \sim 10^{-7}$.  
A more realistic
model~\cite{lu-schmidt} considers radiation from quarks and recombination into a
pion pair, with the parametrization of the produced even-spin states based on
Ref.~\cite{DGP}.  In this case the contribution is at a few $\times 10^{-4}$ level for
masses below $1\gev$.  However the FSR rate could be enhanced on specific
resonances that are not explicitly taken into account in the model. 
Estimates~\cite{chernyak} using a $\gamma^* \gamma
f_2(1270)$ transition form factor evaluated in the asymptotic regime by
perturbative QCD indicate a FSR contribution of about 0.9\% on the
$f_2(1270)$ resonance. Contributions from $f_0(980)$ and $f_0(1370)$ are
expected to be lower.

Finally, a direct test with \babar\ data has been performed with a measurement of
charge asymmetry, which is proportional to the interference between LO ISR and 
FSR amplitudes. This work in progress, which will be published separately, 
yields results that do not exceed the estimates above.

The estimated LO FSR contributions are at levels smaller than the quoted systematic
uncertainties on the $\pi\pi(\gamma)\gamma$ cross section, much smaller actually for the 
$\rho$ region. 
No subtraction has been applied to the measured cross section. 

\subsubsection{Results on the bare cross section with FSR}
\label{pi-xsec}

\begin{figure*}
\begin{minipage}[htp]{0.8\textwidth}
  \centering
  \includegraphics[width=0.8\textwidth]{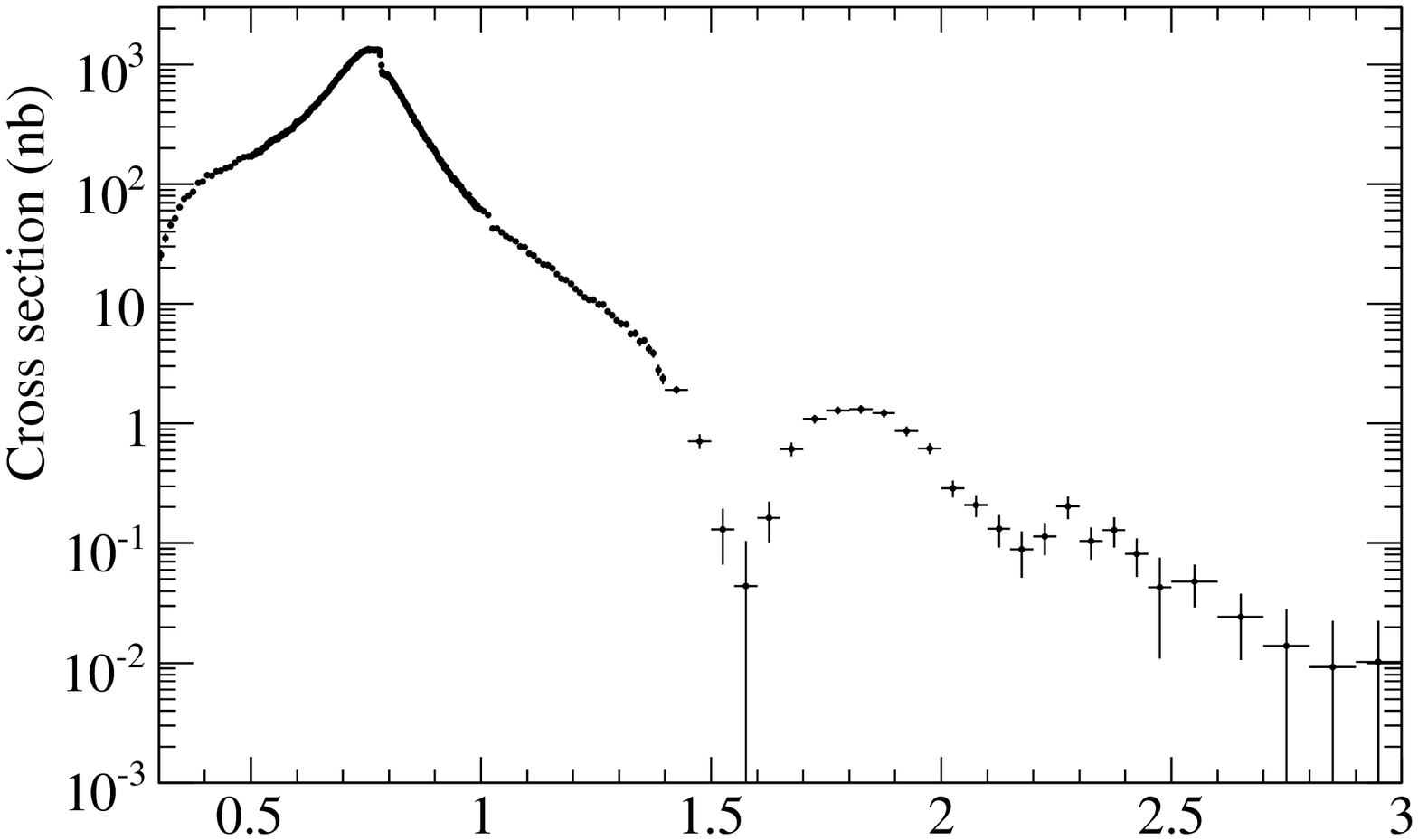}
  \put(-60,6){\large {$\sqrt{s'}$ (\gev)}}
  \caption{\small
 The measured cross section for $e^+e^-\to\pi^+\pi^-(\gamma)$
over the full mass range. Systematic and statistical uncertainties are shown,
but based only on the diagonal elements of the covariance matrix (see text).}
  \label{babar-log}
  \end{minipage}\hfill
  \begin{minipage}[htp]{0.8\textwidth}
  \centering
  \includegraphics[width=0.8\textwidth]{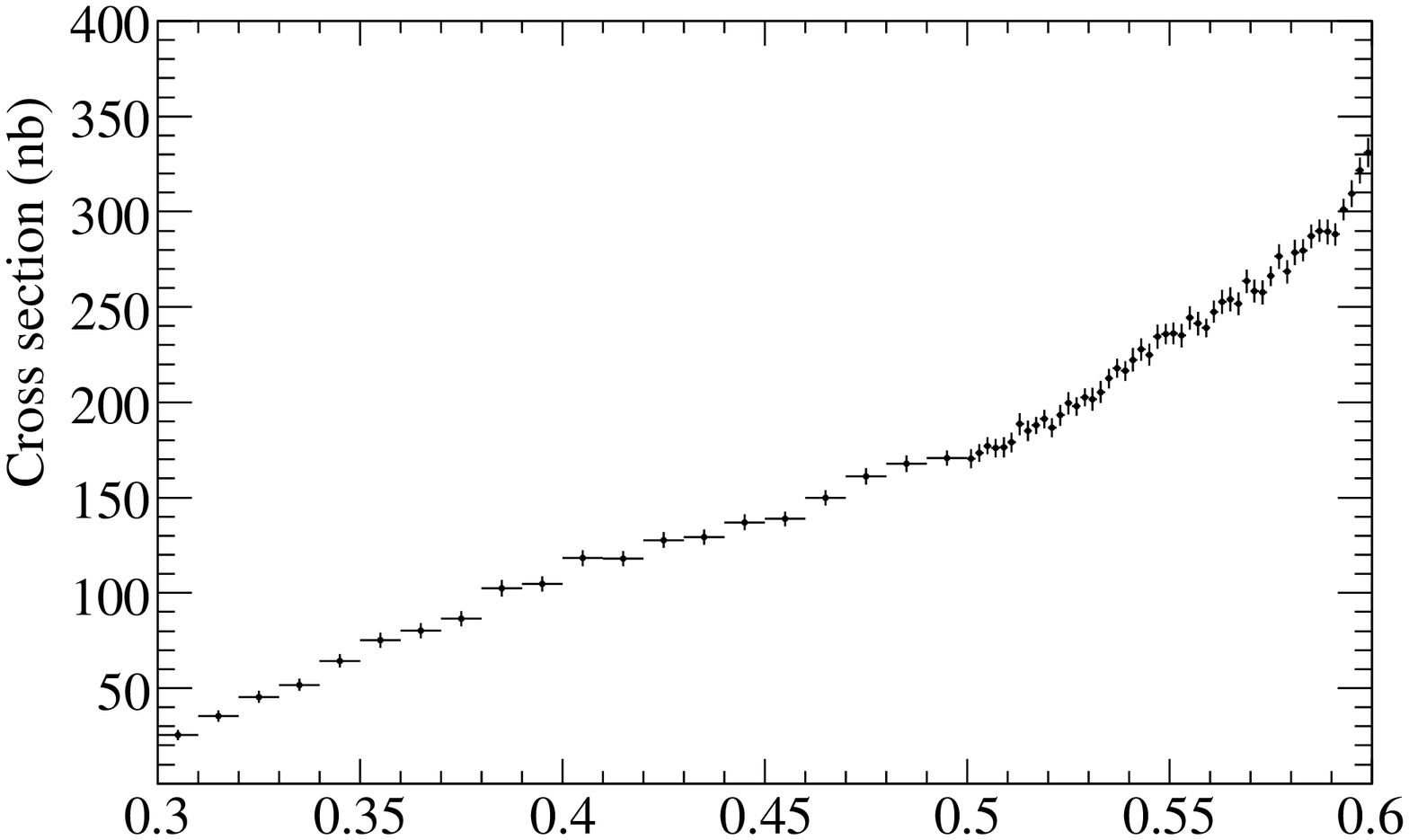}
  \put(-60,6){\large {$\sqrt{s'}$ (\gev)}}
  \caption{\small
 The measured cross section for $e^+e^-\to\pi^+\pi^-(\gamma)$
in the lower mass range.  Systematic and statistical uncertainties are shown,
but based only on the diagonal elements of the covariance matrix (see text).}
  \label{babar-low}
\end{minipage}
\end{figure*}

\begin{figure*}
\begin{minipage}[htp]{0.8\textwidth}
  \centering
  \includegraphics[width=0.8\textwidth]{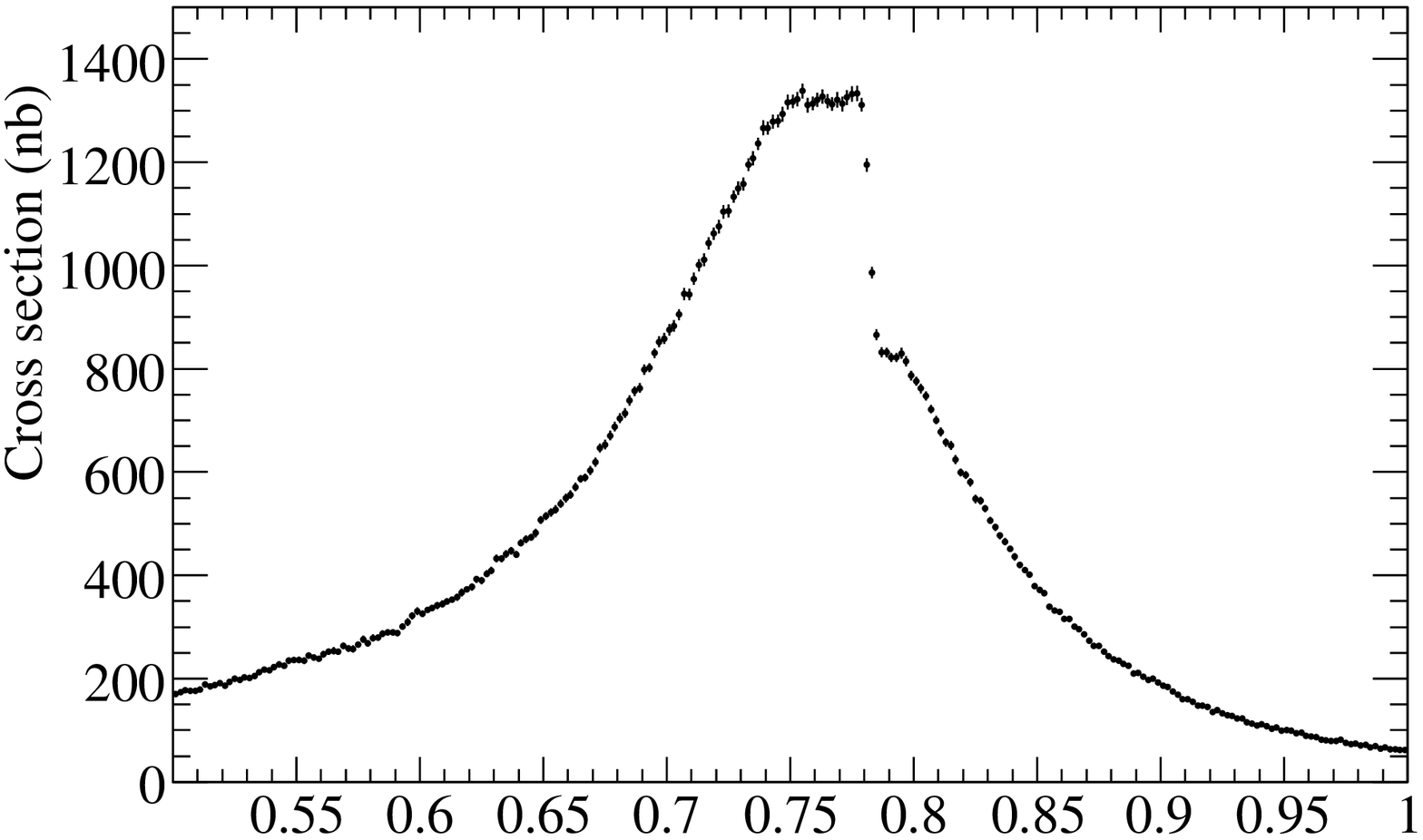}
  \put(-60,6){\large {$\sqrt{s'}$ (\gev)}}
  \caption{\small
 The measured cross section for $e^+e^-\to\pi^+\pi^-(\gamma)$
in the central $\rho$ region.  Systematic and statistical uncertainties are 
shown, but based only on the diagonal elements of the covariance matrix (see text).}
  \label{babar-central}
\end{minipage}
\end{figure*}

The results for the $e^+e^-\to\pi^+\pi^-(\gamma_{\rm FSR})$ bare cross section 
including FSR, $\sigma^0_{\pi\pi(\gamma_{\rm FSR})}$, are given in
Figs.~\ref{babar-log},~\ref{babar-low}, and~\ref{babar-central} as a function of $\sqrt{s'}$. 
The cross section is dominated
by the wide $\rho$ resonance, with structures at larger masses.
The dip region near 1.6\gev, usually
interpreted as resulting from interference between the $\rho'$ and $\rho''$
amplitudes, is mapped with a much increased precision compared to previous
experiments. 
There is also an indication for a structure in the 2.2--2.25\gev region,
which could be due to a still higher-mass $\rho'''$ vector meson.

Files containing the cross section data and their covariance 
matrices are provided in the EPAPS repository~\cite{epaps}.

\subsection{Pion form factor fits}
\label{vdm-ff}

The square of the pion form factor is defined as usual by the ratio of the 
dressed cross section without FSR, divided by the lowest-order cross section 
for point-like spin 0 charged particles. Thus,
\beqn
  |F_\pi|^2(s') = \frac {3s'}{\pi\alpha^2(0)\beta_\pi^3}\,\sigma_{\pi\pi}(s')~,
\eeqn
with 
\beqn
  \sigma_{\pi\pi}(s') = \frac {\sigma^0_{\pi\pi(\gamma)}(s')}
                              {1+\delta^{\pi\pi}_{\rm add.FSR}}~
                   \left(\frac {\alpha(s')}{\alpha(0)}\right)^2~,
\eeqn
and $\beta_\pi$ the pion velocity.
The FSR correction~\cite{fsr-kuraev,fsr-kuehn}, 
$\delta^{\pi\pi}_{\rm add.FSR}=\alpha(0)/\pi~\eta (s')$, decreases slowly
with $s'$ and amounts to $8.0\times 10^{-3}$ at the $\rho$ mass. 

A vector-dominance model (VDM) is used to fit the \babar\ pion 
form factor. It is a way to interpret the observed structures beyond 
the $\rho$ resonance in terms of higher-mass isovector vector mesons. The fit
also provides a convenient means to interpolate through the \babar\ data points 
in order to facilitate the comparison to other experiments. 

The VDM parameterization, including $\rho-\omega$ interference, is given by:
\begin{widetext}
\beqn
F_\pi(s) = \frac{BW^{\rm GS}_\rho (s, m_\rho, \Gamma_\rho)
\frac{1+c_\omega BW^{\rm KS}_\omega(s, m_\omega, \Gamma_\omega)}{1+c_\omega} 
+ c_{\rho'}   BW^{\rm GS}_{\rho'}   (s, m_{\rho'},  \Gamma_{\rho'}) 
+ c_{\rho''}  BW^{\rm GS}_{\rho''}  (s, m_{\rho''}, \Gamma_{\rho'''})
+ c_{\rho'''} BW^{\rm GS}_{\rho'''} (s, m_{\rho'''},\Gamma_{\rho'''}) }{1+c_{\rho'}+c_{\rho''}+c_{\rho'''}},~~
\eeqn
\end{widetext}
which satisfies $F_\pi(0)=1$.
The amplitudes of the Breit-Wigner (BW) functions are complex: 
$c_\omega = |c_\omega| e^{i\phi_\omega}$, 
$c_{\rho'} = |c_{\rho'}| e^{i\phi_{\rho'}}$,
$c_{\rho''} = |c_{\rho''}| e^{i\phi_{\rho''}}$ and  
$c_{\rho'''} = |c_{\rho'''}| e^{i\phi_{\rho'''}}$.
The BW of the $\omega$ is taken as:
\beqn
BW_\omega^{\rm KS}(s, m, \Gamma) = \frac{m^2}{m^2 - s - i  m \Gamma}~.
\eeqn
The wide $\rho$, $\rho'$, $\rho''$ and $\rho'''$ resonances are described by the 
Gounaris-Sakurai(GS) model~\cite{gounaris}, which takes into account the variation 
of their width with energy:
\beqn  
BW^{\rm GS}(s, m, \Gamma) = \frac{m^2 (1 + d(m) \Gamma/m) }{m^2 - s + f(s, m, \Gamma) - i m \Gamma (s, m, \Gamma)}~,~~
\eeqn   
with
\beqn  
\Gamma (s, m, \Gamma) = \Gamma  \frac{s}{m^2} \left( \frac{\beta_\pi (s) }{ \beta_\pi (m^2) } \right) ^3~,
\eeqn   
where $\beta_\pi (s) = \sqrt{1 - 4m_\pi^2/s}$. In principle this energy dependence 
is justified only below 1\gev, as 4-pion final states dominate at larger 
energies, but it is used for simplicity. Detailed 
studies of the high mass states cannot be performed only on the basis of pion form factor 
fits, and require complex coupled-channel analyses. Such studies are beyond 
the scope of this paper, but the present $2\pi$ data constitute a very 
useful ingredient for them.

The auxiliary functions used in the GS model are:
\begin{widetext}
\beqn  
d(m) = \frac{3}{\pi} \frac{m_\pi^2}{k^2(m^2)} \ln \left( \frac{m+2 k(m^2)}{2 m_\pi} \right) 
   + \frac{m}{2\pi  k(m^2)}
   - \frac{m_\pi^2  m}{\pi k^3(m^2)}~,
\eeqn   
\beqn  
f(s, m, \Gamma) = \frac{\Gamma  m^2}{k^3(m^2)} \left[ k^2(s) (h(s)-h(m^2)) + (m^2-s) k^2(m^2)  h'(m^2)\right]~,
\eeqn   
\end{widetext}
where 
\beqn 
k(s)& =& \frac{1}{2} \sqrt{s}  \beta_\pi (s)~,\\ 
h(s)& =& \frac{2}{\pi}  \frac{k(s)}{\sqrt{s}}  \ln \left( \frac{\sqrt{s}+2 k(s)}{2 m_\pi} \right).
\eeqn   
and $h'(s)$ is the derivative of $h(s)$.

\begin{table}[h] \centering 
\caption[.]{\label{ParFitGS} \small
Parameters obtained for the VDM fit (described in the text) to the 
\babar\ pion form factor data. The errors include both statistical and
systematic uncertainties. The errors shown in parentheses for the $\rho$ and
$\omega$ parameters stem from the mass calibration and resolution 
uncertainties (see text).}
\vspace{0.5cm}
\setlength{\extrarowheight}{1.5pt}
\setlength{\tabcolsep}{2pt}
\begin{tabular*}{0.45\textwidth}{@{\extracolsep{\fill}}lc@{\hskip10pt}} 
\hline\hline\noalign{\vskip2pt}
Parameter   & Value  $\pm$ Error   \\
\hline\noalign{\vskip1pt}
$m_\rho$~(\mevcc)         & $775.02  ~\pm~ 0.31 ~~~(\pm~ 0.16)$ \\
$\Gamma_\rho$~(\mev)      & $149.59   ~\pm~ 0.67 ~~~(\pm~ 0.02)$\\
$m_\omega$~(\mevcc)       & $781.91  ~\pm~ 0.18 ~~~(\pm~ 0.16)$ \\
$\Gamma_\omega$~(\mev)    & $8.13    ~\pm~ 0.36 ~~~(\pm~ 0.27)$ \\
$|c_\omega|$              & $(1.644  ~\pm~ 0.061)\times 10^{-3}$\\
$\phi_\omega$~(\rad)      & $-0.011   ~\pm~ 0.037$              \\
$m_{\rho'}$~(\mevcc)      & $1493    ~\pm~ 15$                  \\
$\Gamma_{\rho'}$~(\mev)   & $427     ~\pm~ 31$                  \\
$|c_{\rho'}| $            & $0.158   ~\pm~ 0.018$               \\
$\phi_{\rho'}$~(\rad)     & $3.76    ~\pm~ 0.10$                \\
$m_{\rho''}$~(\mevcc)     & $1861    ~\pm~ 17$                  \\
$\Gamma_{\rho''}$~(\mev)  & $316     ~\pm~ 26$                  \\
$|c_{\rho''}|$            & $0.068  ~\pm~ 0.009$                \\
$\phi_{\rho''}$~(\rad)    & $1.39    ~\pm~ 0.20$                \\
$m_{\rho'''}$~(\mevcc)    & $2254    ~\pm~ 22$                  \\
$\Gamma_{\rho'''}$~(\mev) & $109     ~\pm~ 76$                  \\
$|c_{\rho'''}|$           & $0.0051  ~^{+0.0034}_{-0.0019}$     \\
$\phi_{\rho'''}$~(\rad)   & $0.70    ~\pm~ 0.51$                \\
\noalign{\vskip1pt}\hline\hline
\end{tabular*}
\end{table}

\begin{figure*}
\begin{minipage}[thp]{0.7\textwidth}
  \centering
  \includegraphics[width=0.8\textwidth]{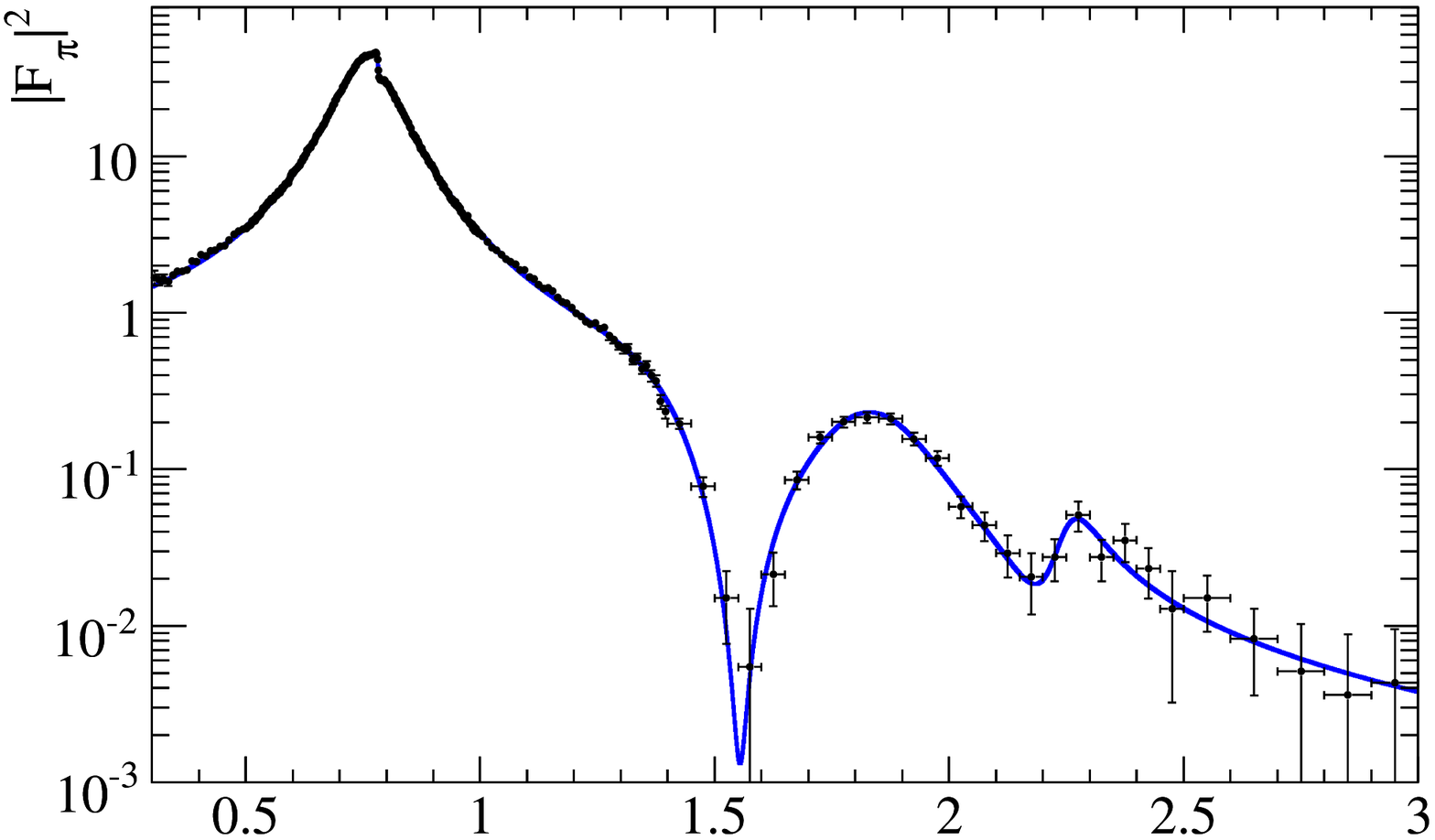}
  \put(-60,6){\large {$\sqrt{s'}$ (\gev)}}
  \caption{\small
 The pion form factor-squared measured by \babar\ as a function of 
$\sqrt{s'}$ from 0.3 to 3\gev and the VDM fit described in the text.}
  \label{ff-fit-log}
  \end{minipage}\hfill
  \begin{minipage}[ht]{0.7\textwidth}
  \centering
  \includegraphics[width=0.8\textwidth]{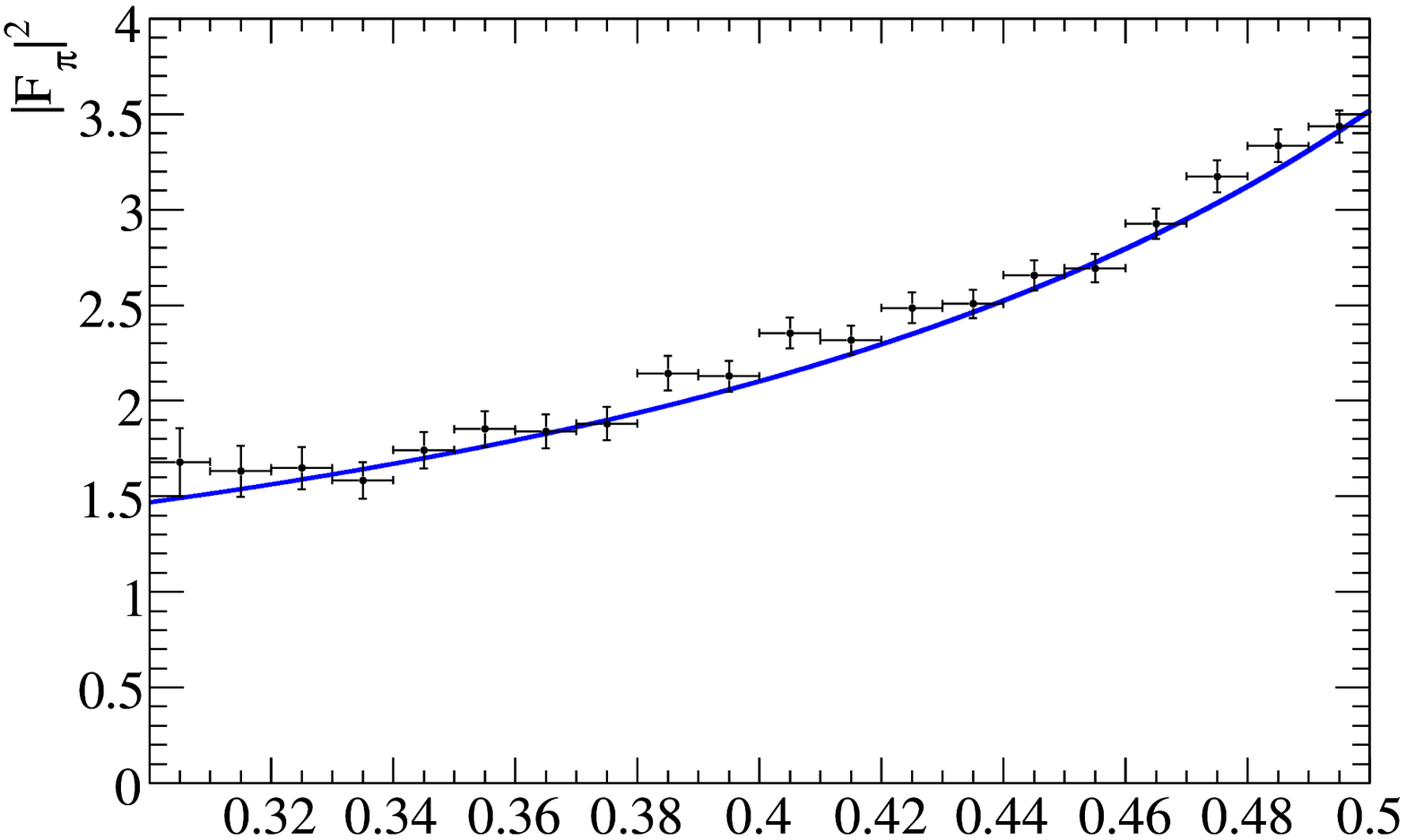}
  \put(-60,6){\large {$\sqrt{s'}$ (\gev)}}\\
  \includegraphics[width=0.8\textwidth]{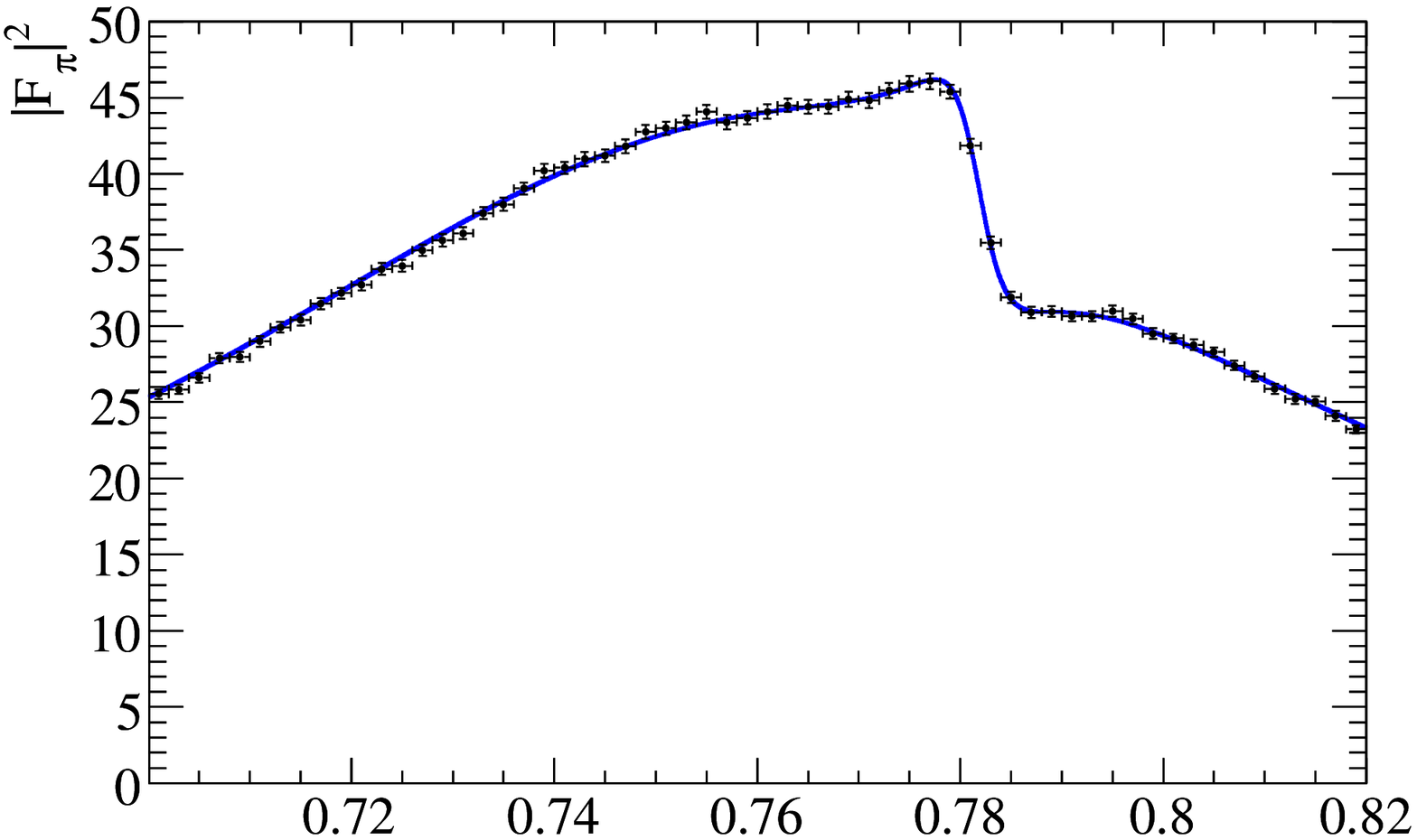}
  \put(-60,6){\large {$\sqrt{s'}$ (\gev)}}
  \caption{\small
 The pion form factor-squared measured by \babar\ as a function of 
$\sqrt{s'}$ and the VDM fit from 0.3 to 3\gev described in the text.
(top): Low-mass region (0.3--0.5\gev). (bottom): $\rho$ peak region with
$\rho-\omega$ interference (0.70--0.82\gev).}
  \label{ff-fit-zoom}
\end{minipage}
\end{figure*}

\begin{figure*}
\begin{minipage}[htp]{0.70\textwidth}  
 \centering
  \includegraphics[width=0.8\textwidth]{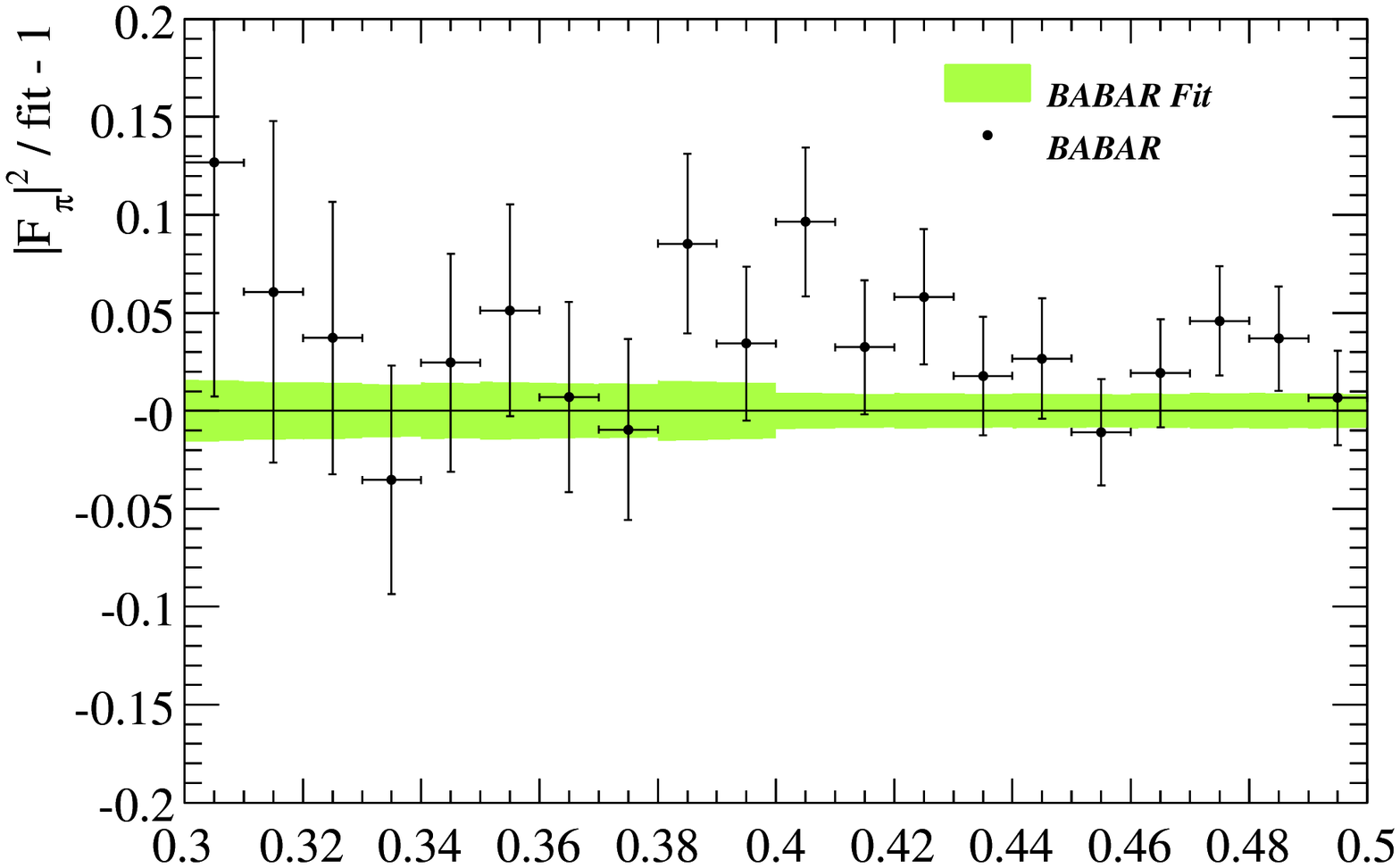}
  \put(-60,6){ {$\sqrt{s'}$ (\gev)}}\\
  \includegraphics[width=0.8\textwidth]{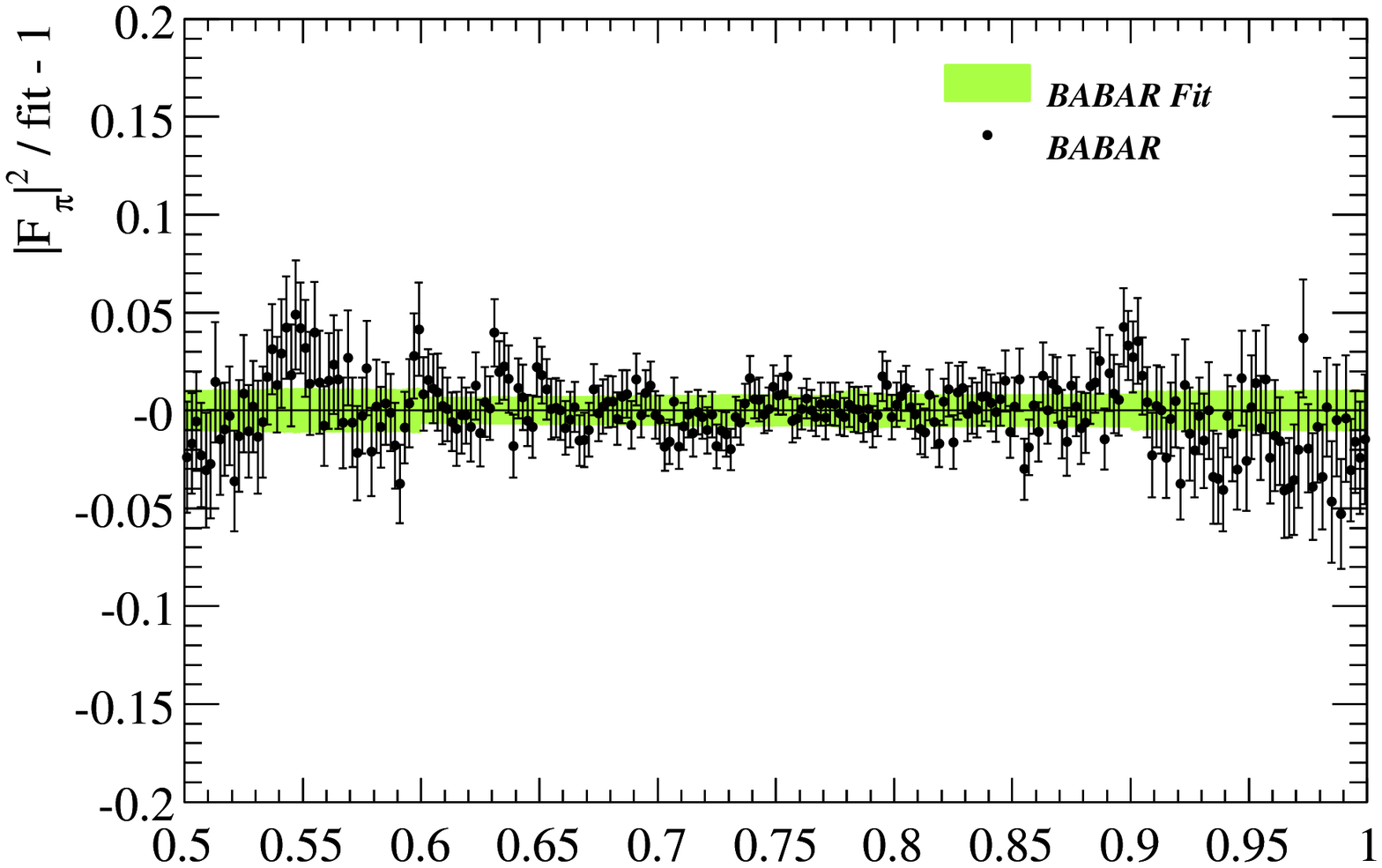}
  \put(-60,6){ {$\sqrt{s'}$ (\gev)}}\\
  \includegraphics[width=0.8\textwidth]{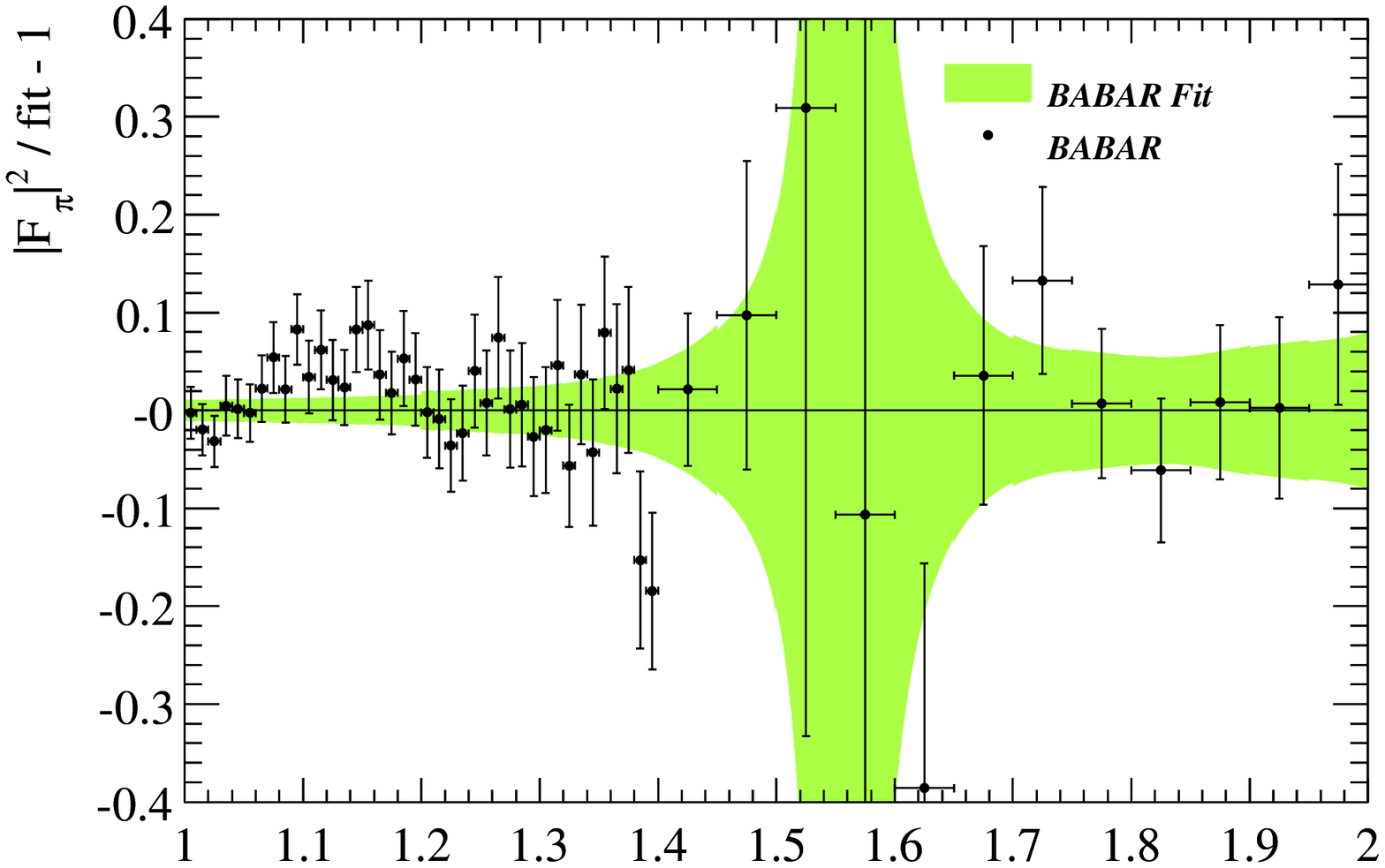}
  \put(-60,6){ {$\sqrt{s'}$ (\gev)}}\\
  \caption{\small
 The relative difference between the pion form factor-squared from \babar\ 
data and the 18-parameter phenomenological fit in three mass regions.  
Systematic and statistical uncertainties are included for data (diagonal 
errors). The width of the band shows the propagation of statistical errors
from the fit and the quoted systematic uncertainties, added quadratically.}
  \label{babar-babar}
\end{minipage}
\end{figure*}

The form factor data is fitted in the full energy range, from 0.3 to 3.0\gev, 
involving 18 free parameters: the mass and width of the $\rho$, and for 
each other resonance ($\omega$, $\rho'$, $\rho''$, $\rho'''$) the amplitude (modulus and 
phase) with respect to the $\rho$, and mass and width. 
According to a well-known effect~\cite{dagostini}, 
the $\chi^2$ minimization returns fitted values that are systematically shifted with
respect to the data points when the full covariance matrix 
is used in the fit, while the fit using diagonal errors is verified to be bias-free. 
This feature is due to correlations, which here arise from 
both statistical and systematic origins, but mostly from the ISR-luminosity 50\mev sliding bins 
(Sect.~\ref{eff-lumi}) and systematic errors. To circumvent the problem, we 
fit the data with only diagonal errors to obtain the central values of the fitted parameters.
The error on each parameter is taken as the largest error obtained from the 
fit either with the full covariance matrix or with only diagonal errors.
The biases on the mass scale calibration and the resolution obtained in
Sections~\ref{mass-calib} and \ref{mass-resol} are included in the fit results 
on the $\rho$ and $\omega$ resonance parameters in Table~\ref{ParFitGS}, with 
the corresponding systematic uncertainties indicated. 

As shown in Fig.~\ref{ff-fit-log}, the VDM fit provides an adequate description of 
the \babar\ data over the full 0.3--3\gev range ($\chi^2/\rm DF=351/319$).
The goodness of the fit shows that the GS
parametrization of the dominant $\rho$ resonance describes the data in a reasonable manner,
as well as the contributions from the higher $\rho'$, $\rho''$ and $\rho'''$
resonances. In particular the strong interference dip near 1.6\gev is well
reproduced. Beyond 2\gev, the $\rho'''$ is required in order to reproduce the 
structure seen in the data.
The quality of the fit is shown in more detail in Fig.~\ref{ff-fit-zoom} 
in the low-mass range and in the $\rho$ peak region with the $\rho-\omega$ 
interference. 

The relative ratio $(|F_\pi|^2_{\rm data}/|F_\pi|^2_{\rm VDM} -1)$ 
is shown in Fig.~\ref{babar-babar} over the full energy range. Some deviation is observed
in the low-mass region where the fit underestimates the data. Some oscillation 
is also observed between 0.9 and 1.2\gev.
This shows that the GS function, the parameters of 
which are mainly determined in the $\rho$ peak region, together with the 
constraint at $s'=0$, does not accurately describe the resonance tails.
At higher masses, the validity of the VDM description that involves the parametrization 
of very broad resonances with large inelasticity, is somewhat arguable. 
However the overall agreement is satisfactory, notably in the 0.5--1.0\gev region.

We compare the results for the resonance parameters (Table~\ref{ParFitGS}) 
to those obtained by other experiments, noting that the comparison  
can be biased if the mass range or the parametrizations are different.
The fitted $\rho$ parameters are compared to the results from 
CMD-2~\cite{cmd-2} and SND~\cite{snd}: for the mass $m_\rho$, these two experiments obtain
$(776.0\pm0.8)\mevcc$ and $(774.6\pm0.6)\mevcc$, respectively, while for the
width $\Gamma_\rho$, they obtain $(146.0\pm0.9)\mev$ and $(146.1\pm1.7)\mev$. 
The fitted value of the phase $\phi_\omega$ of the $\rho-\omega$ interference is not in 
good agreement with the CMD-2 value
($0.182\pm0.067\rad$); SND uses a different parametrization. However in the CMD-2 
fit, the $\omega$ mass is fixed to the world-average value $m_\omega=782.65\mevcc$
~\cite{pdg}. If we fix $m_\omega$ to this value in the
\babar\ fit, the phase comes out to be $(0.137\pm0.023)\rad$,
in agreement with CMD-2. In fact in the 18-parameter fit the fitted values
for $m_\omega$, $\phi_\omega$, and $c_\omega$ are strongly correlated (80\%).
The fitted $\omega$ width $\Gamma_\omega$ is found to be consistent with 
the world-average value $(8.49\pm0.08)\mev$ obtained from the dominant 
$\pi^+\pi^-\pi^0$ decay mode~\cite{pdg}.

As the CMD-2 and SND experiments at Novosibirsk are well calibrated in energy with the
resonant depolarization method, one can use the VDM fit to check the mass 
calibration by leaving the $\omega$ mass free, and using the CMD-2 result 
for the $\rho-\omega$ phase. One obtains
\beqn  
 m_\omega = (782.68\pm0.12\pm0.27)\mevcc~,
\eeqn   
where the first error is from the fit to the data and the second from the
uncertainty on the CMD-2 value for $\phi_\omega$. The absolute difference
with the world average $\omega$ mass is
\beqn  
 m_\omega^{\rm fit}-m_\omega^{\rm PDG} = (0.03\pm0.29)\mevcc~,
\eeqn   
consistent with the calibration from the $J/\psi$ study reported in 
Section~\ref{mass-calib}, $(-0.16\pm0.16)\mevcc$.

\subsection{Comparison to other experiments}

\subsubsection{Pion form factor from  $e^+e^-\to\pipi$ cross section}

The measured form factor $F_\pi({s'})$ is compared to published data from the 
CMD-2~\cite{cmd-2}, SND~\cite{snd}, and KLOE experiments. While the Novosibirsk
results are obtained in the scan mode at fixed energy points, KLOE, like \babar\, uses the
ISR method, albeit at a much smaller energy ($\sqrt s=1.02\gev$).
The KLOE~\cite{kloe} data are obtained without direct detection of the ISR photon.
More recently, KLOE has performed
a new analysis~\cite{kloe10} where the ISR photon is detected at large 
angles, allowing them to collect data down to the threshold region.

\begin{figure*}
\begin{minipage}[h!]{0.5\textwidth}  
  \centering
  \includegraphics[width=0.9\textwidth]{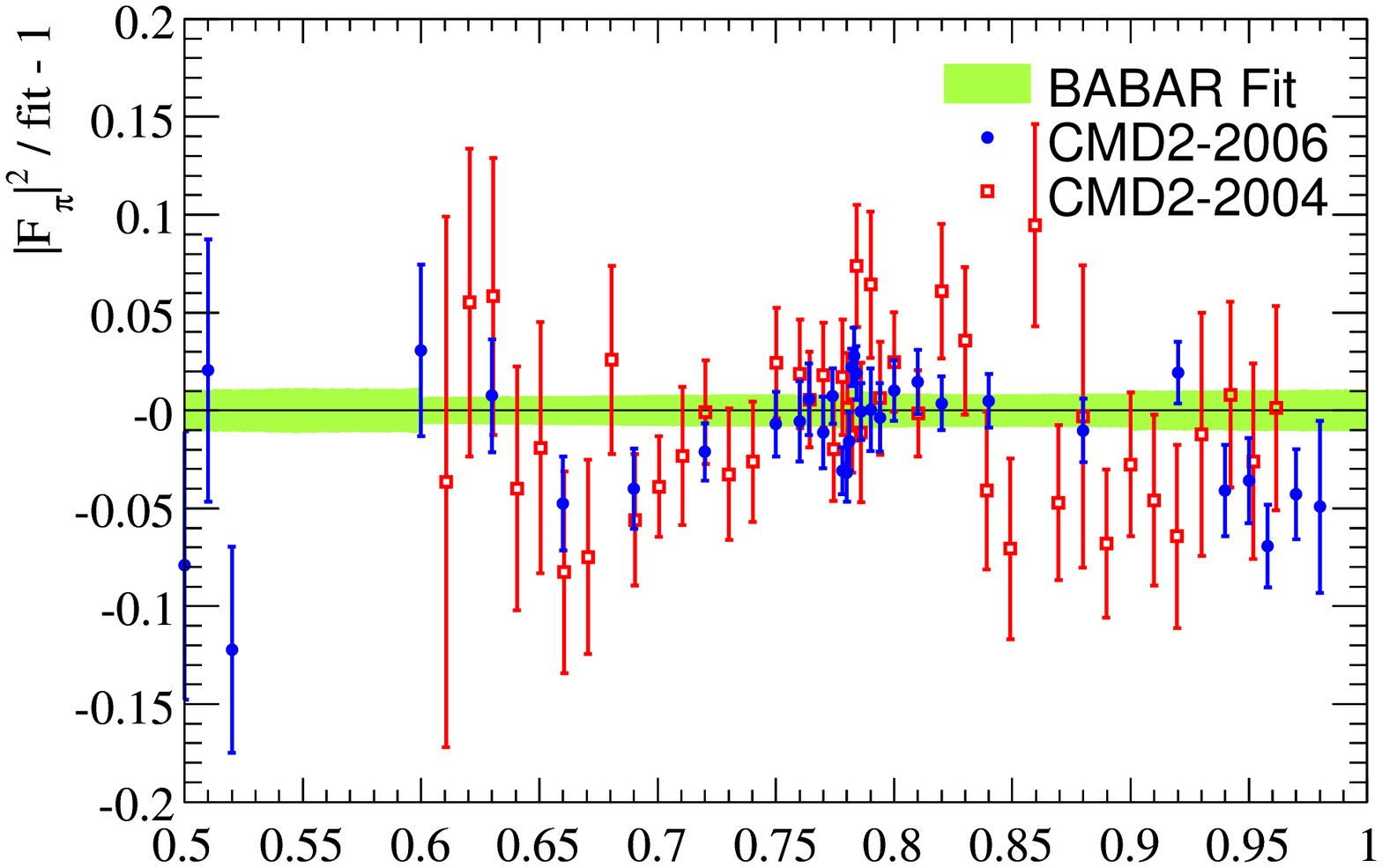}
  \put(-60,0){ {$\sqrt{s'}$ (\gev)}}\\
\end{minipage}\hfill
\begin{minipage}[htp]{0.5\textwidth}  
  \centering
  \includegraphics[width=0.9\textwidth]{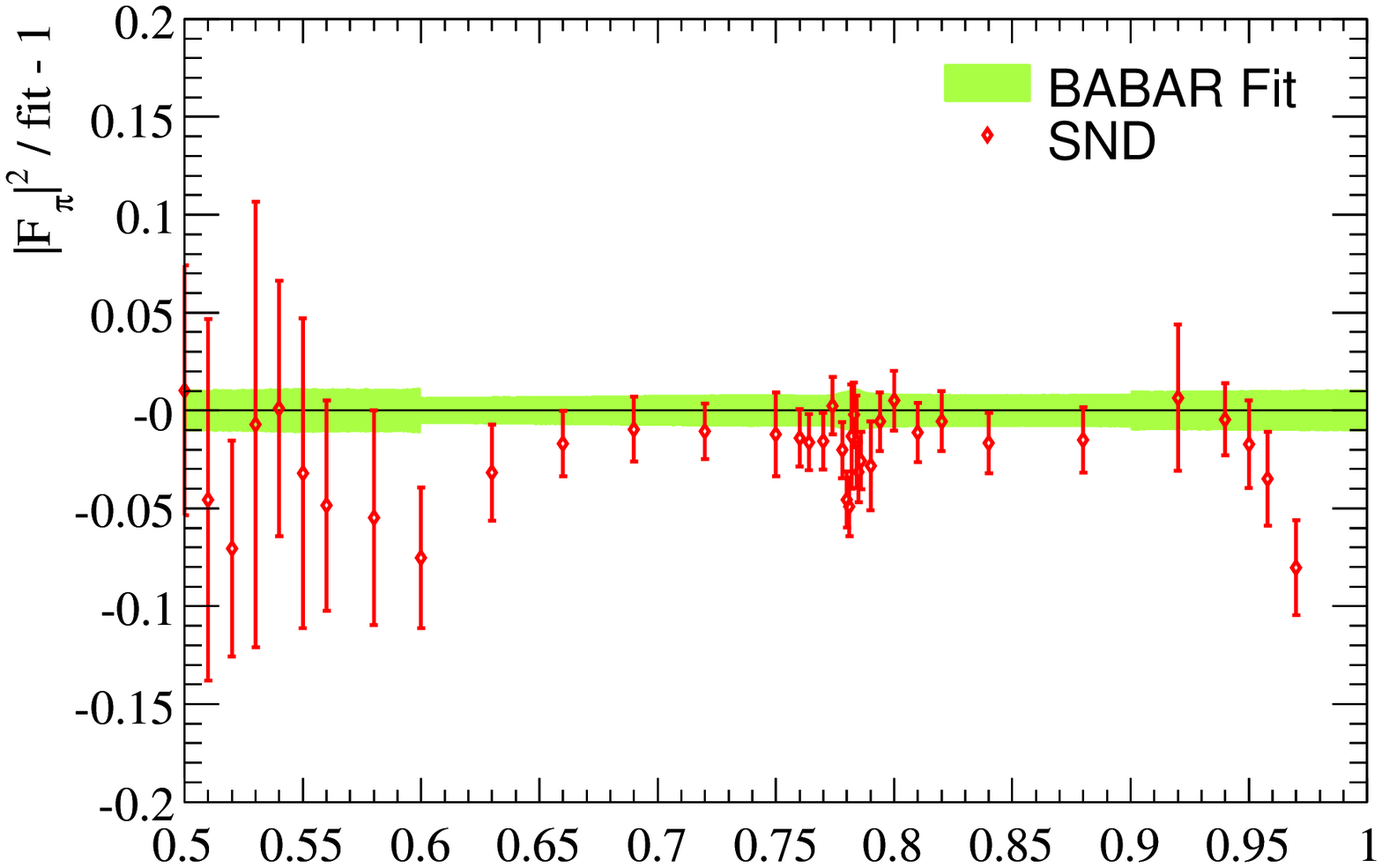}
  \put(-60,0){ {$\sqrt{s'}$ (\gev)}}\\
\end{minipage}
\caption{\small
 The relative difference of pion form factor-squared from the
\babar\ fit in the 0.5--1\gev region with CMD-2 (left) and with SND (right). 
Systematic and statistical uncertainties are included in the data points. 
The width of the \babar\ band shows the propagation of statistical errors
from the fit and the quoted systematic uncertainties, added quadratically.} 
  \label{babar-cmd2-snd}
\end{figure*}

\begin{figure*}
\begin{minipage}[htp]{0.5\textwidth}  
  \centering
  \includegraphics[width=0.9\textwidth]{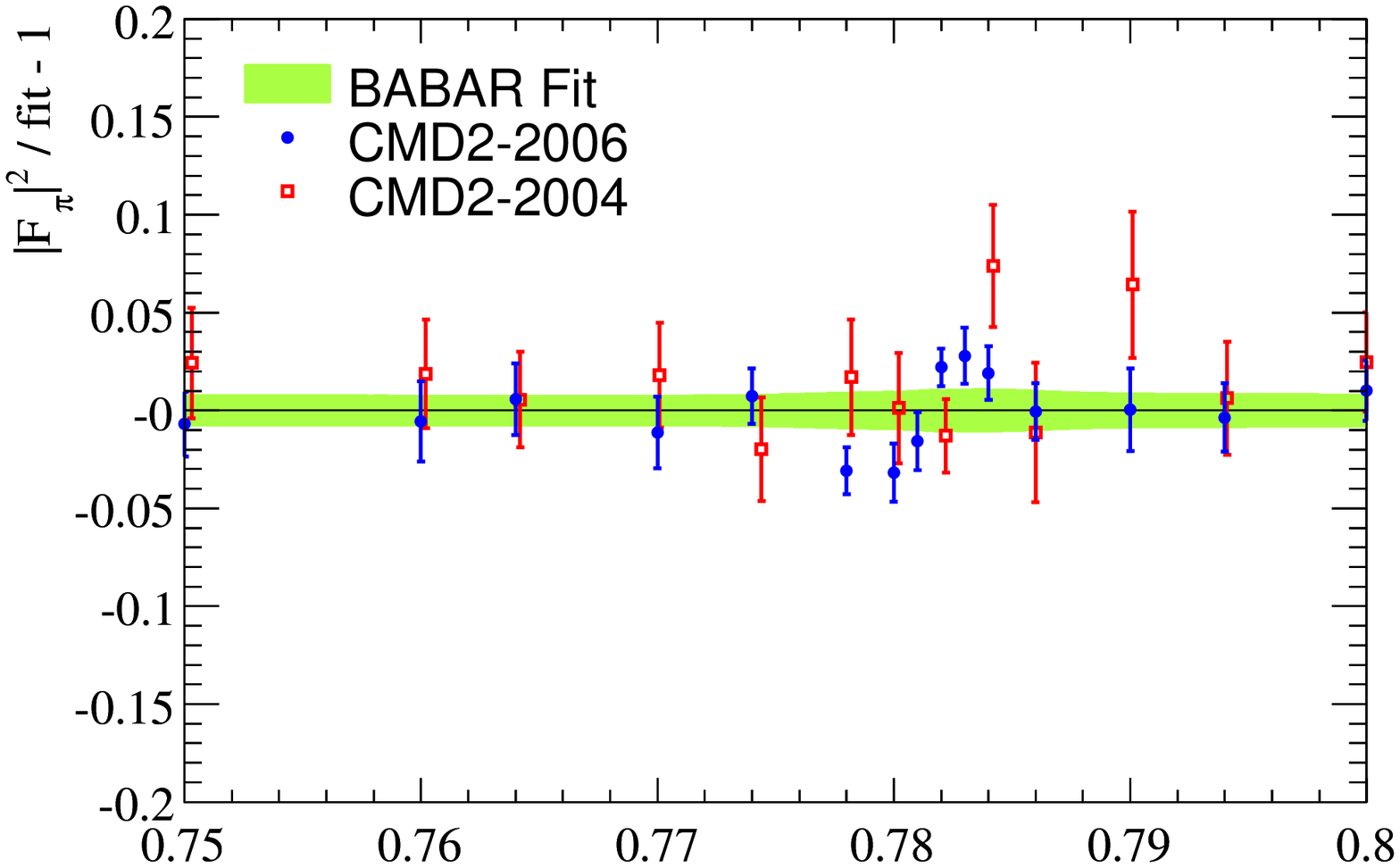}
  \put(-60,0){ {$\sqrt{s'}$ (\gev)}}\\
\end{minipage}\hfill
\begin{minipage}[htp]{0.5\textwidth}  
  \centering
  \includegraphics[width=0.9\textwidth]{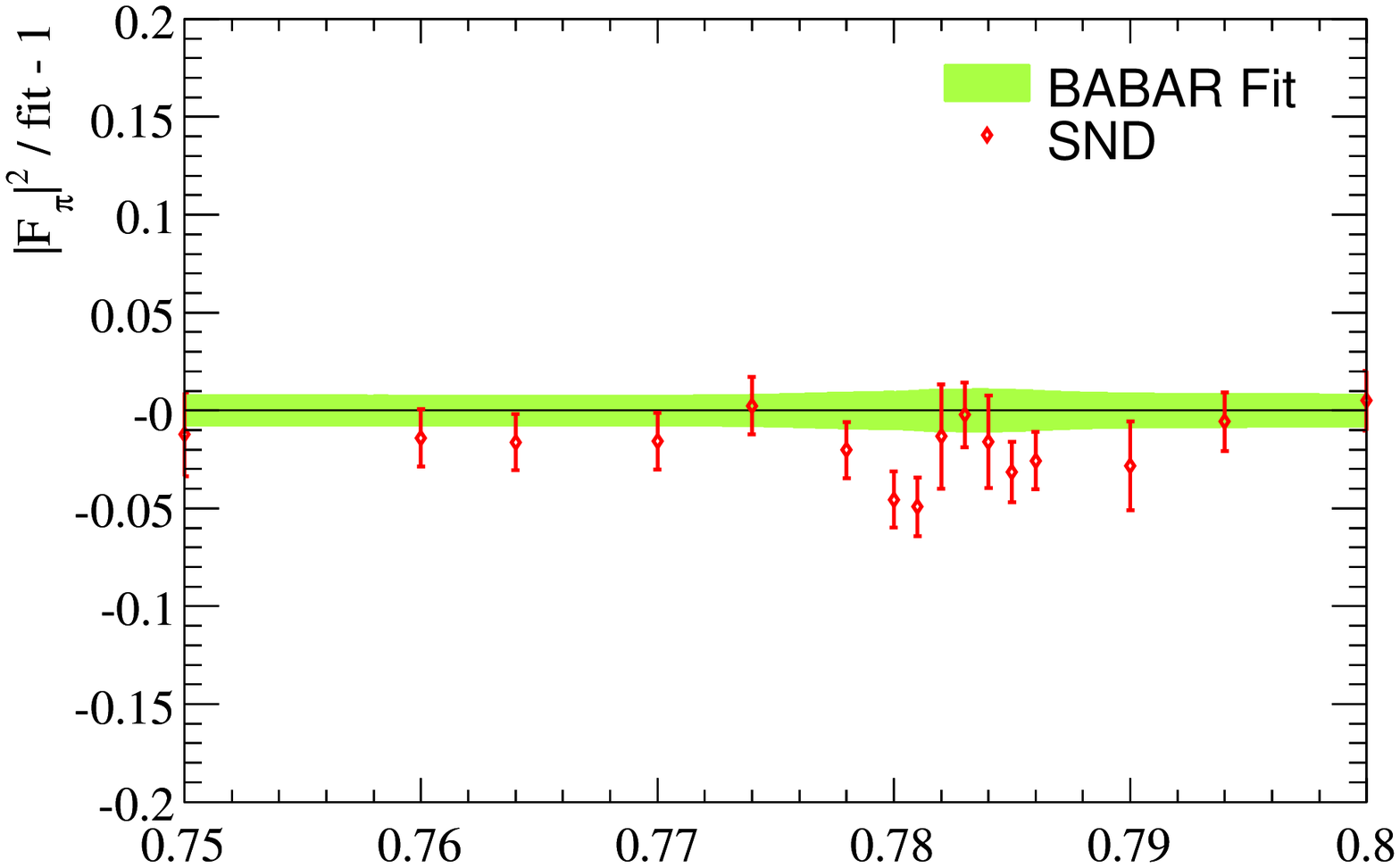}
  \put(-60,0){ {$\sqrt{s'}$ (\gev)}}\\
\end{minipage}
  \caption{\small
 The relative difference of pion form factor-squared from the
\babar\ fit in the $\rho-\omega$ mass region with CMD-2 (left) and with SND (right). 
Systematic and statistical uncertainties are included in the data points. 
The width of the \babar\ band shows the propagation of statistical errors
from the fit and the quoted systematic uncertainties, added quadratically.}   
  \label{babar-cmd2-snd-rhom}
\begin{minipage}[htp]{0.5\textwidth}  
  \centering
  \includegraphics[width=0.9\textwidth]{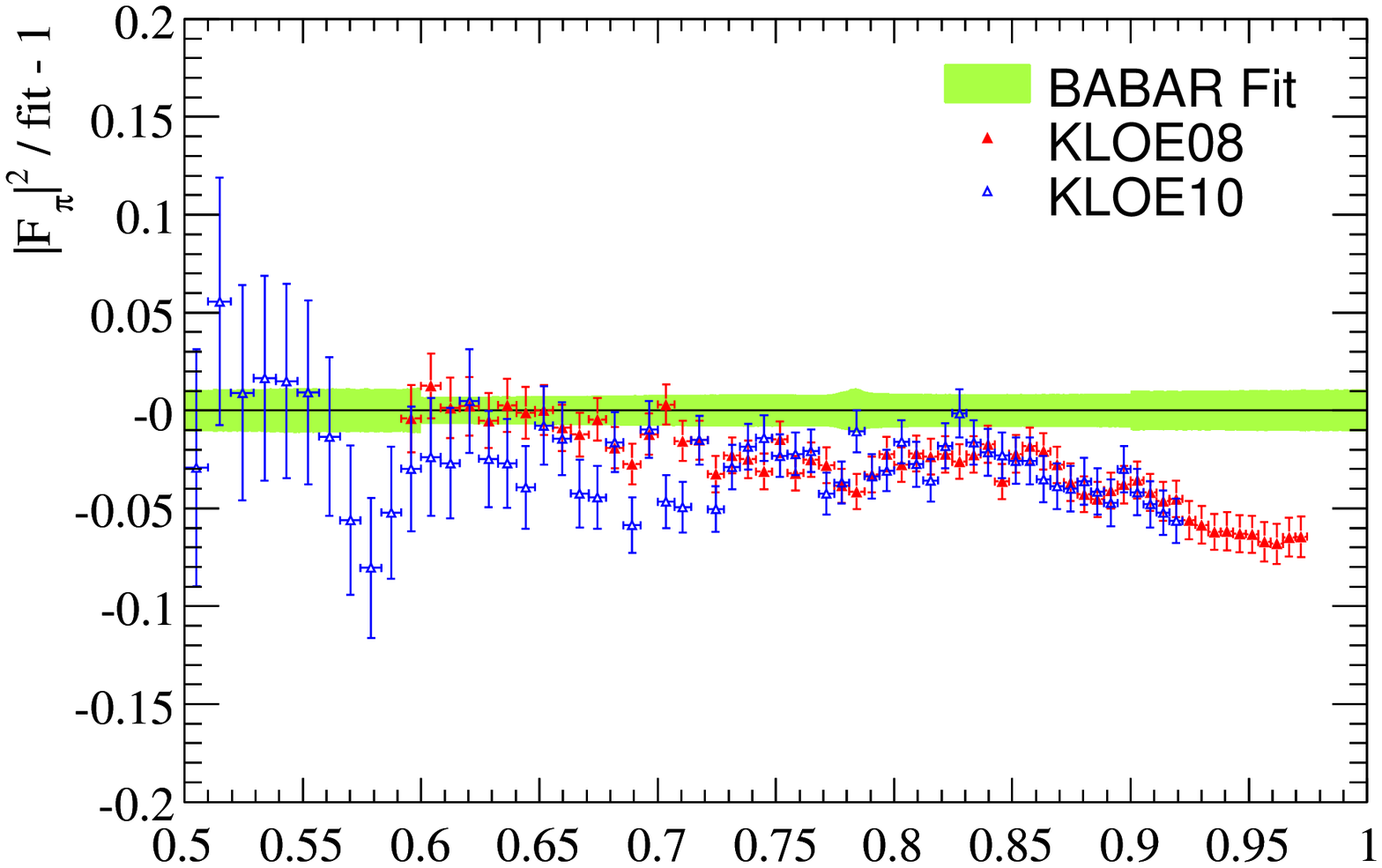}
  \put(-60,0){ {$\sqrt{s'}$ (\gev)}}\\
  \caption{\small
 The relative difference of pion form factor-squared from KLOE and the
\babar\ fit in the 0.5--1\gev region. 
Systematic and statistical uncertainties are included in the data points. 
The width of the \babar\ band shows the propagation of statistical errors
from the fit and the quoted systematic uncertainties, added quadratically.}  
  \label{babar-kloe}
\end{minipage}
\end{figure*}

\begin{figure*}
\begin{minipage}[htp]{0.49\textwidth}  
  \centering
  \includegraphics[width=1.0\textwidth]{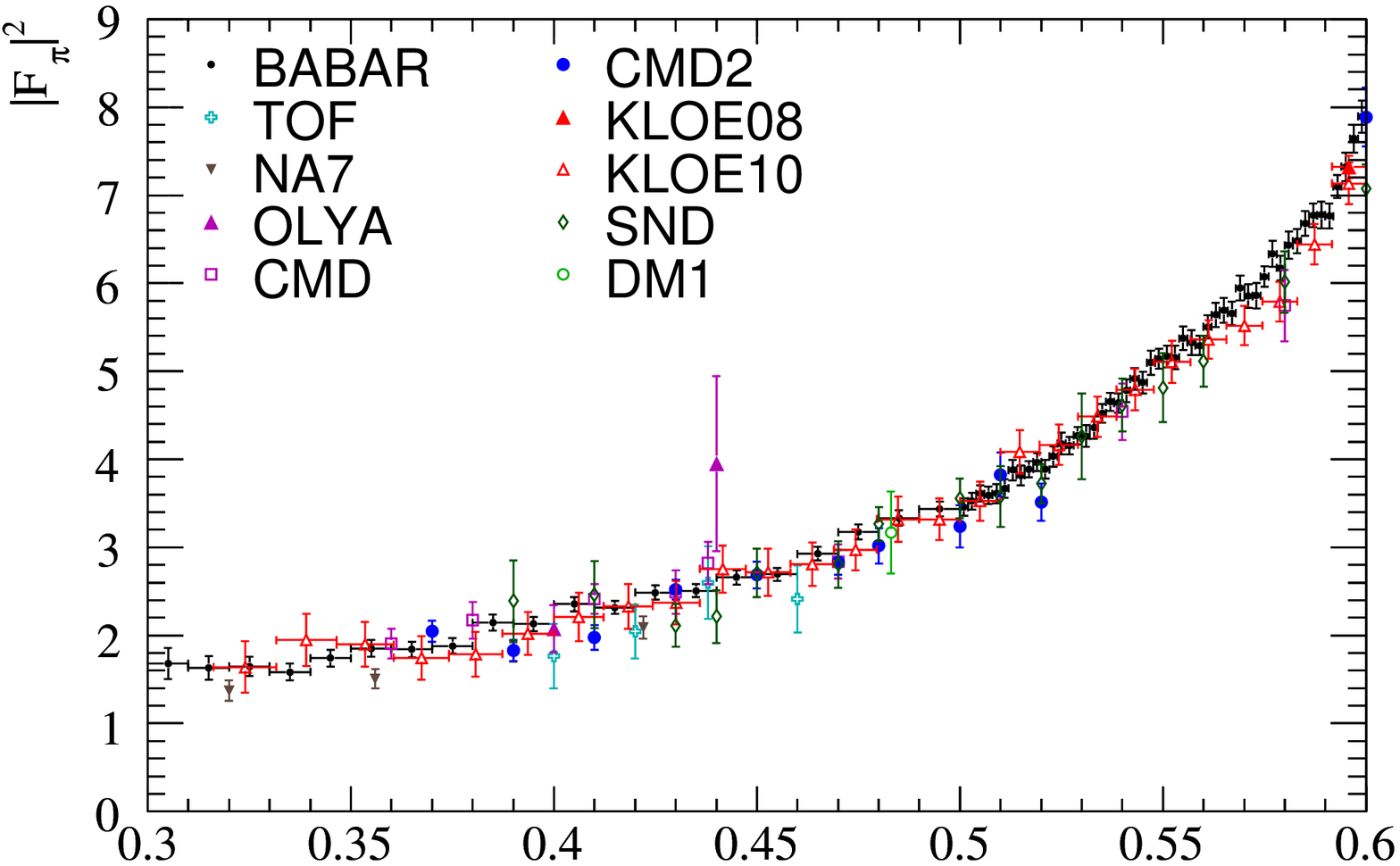}
  \put(-60,0){ {$\sqrt{s'}$ (\gev)}}\\
  \caption{\small(color online).
 The measured pion form factor-squared  compared to
published results from other experiments. 
Systematic and statistical uncertainties are shown for all results,
with the diagonal elements of the \babar\ covariance matrix.}
  \label{babar-low-others}
\end{minipage}\hfill
\begin{minipage}[htp]{0.49\textwidth}  
  \centering
  \includegraphics[width=1.0\textwidth]{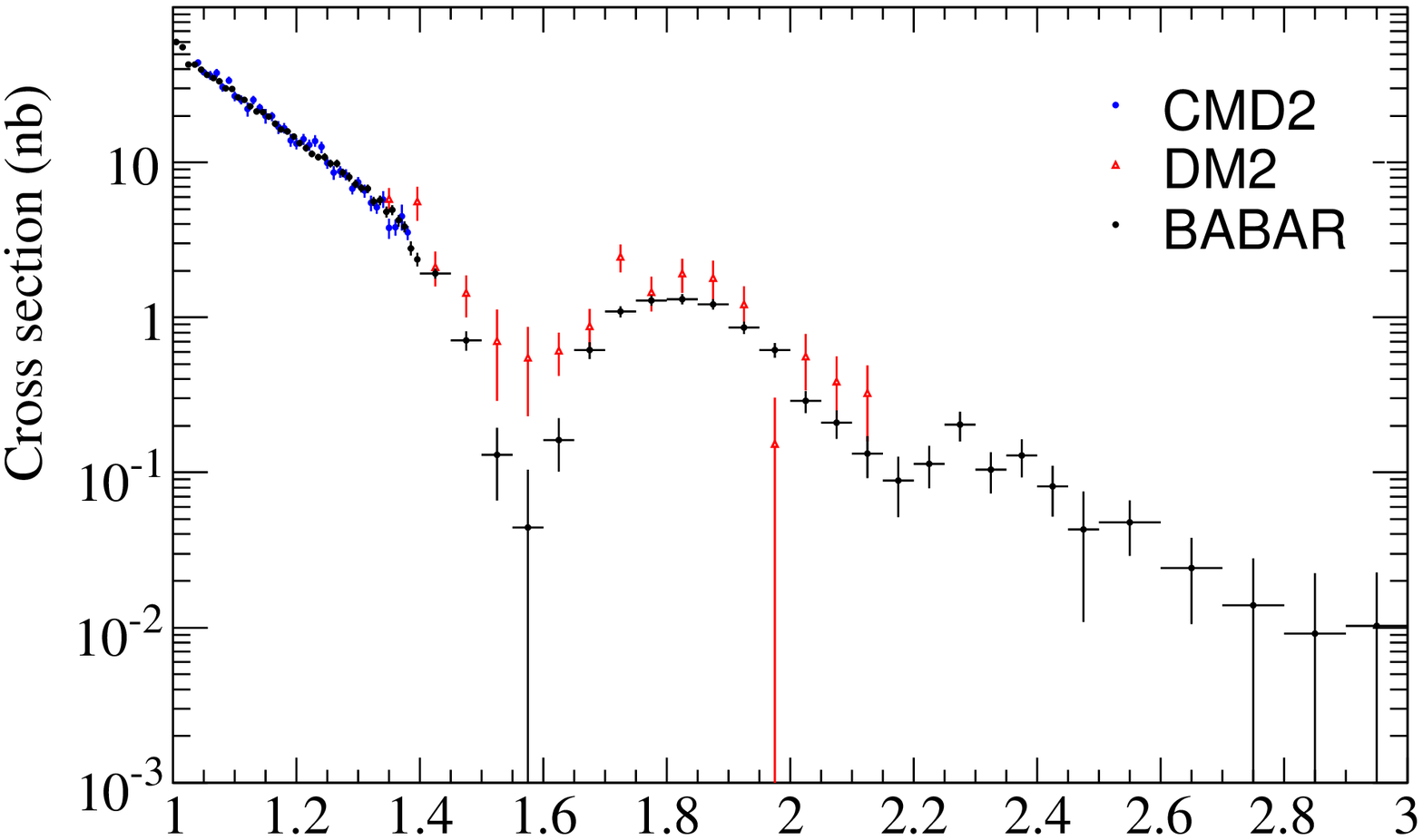}
  \put(-60,0){ {$\sqrt{s'}$ (\gev)}}\\
  \caption{\small(color online).
 The measured cross section for $e^+e^-\to\pi^+\pi^-(\gamma)$ compared to
published results from CMD-2 up to 1.4\gev and DM2 above. 
Systematic and statistical uncertainties are shown for all results,
with the diagonal elements of the \babar\ covariance matrix.}
  \label{babar-high-others}
\end{minipage}
\end{figure*}

The data of the other experiments are compared with
the result from the \babar\ form factor fit, which was shown in the previous 
section to describe well the \babar\ data itself. Each plot shows the relative 
difference between the form factor-squared of the other experiment and \babar\
as data points, while the width of the band around zero is the result of the
propagation of statistical errors from the \babar\ fit with systematic 
uncertainties in each mass region (Table~\ref{pi-syst-err}) added quadratically.

The comparisons with other experiments are shown in Figs.~\ref{babar-cmd2-snd}
and \ref{babar-kloe}, where the errors on the data points include both statistical and
systematic uncertainties.
The agreement looks rather reasonable with the CMD-2 and SND measurements 
within systematic errors, the \babar\ results lying generally above on the lower
side of the $\rho$ resonance. The discrepancy is larger with KLOE on and 
above the $\rho$ peak.

The region of the $\rho-\omega$ interference is examined in more detail in
Fig.~\ref{babar-cmd2-snd-rhom}. No evidence is found 
for a significant variation in the steep part of the interference pattern
around the $\omega$ mass, showing that the \babar\ mass calibration
is not shifted with respect to CMD-2 and SND by more than 0.3\mevcc.

\begin{figure*}
\begin{minipage}[htp]{0.49\textwidth}  
  \centering
  \includegraphics[width=1.0\textwidth]{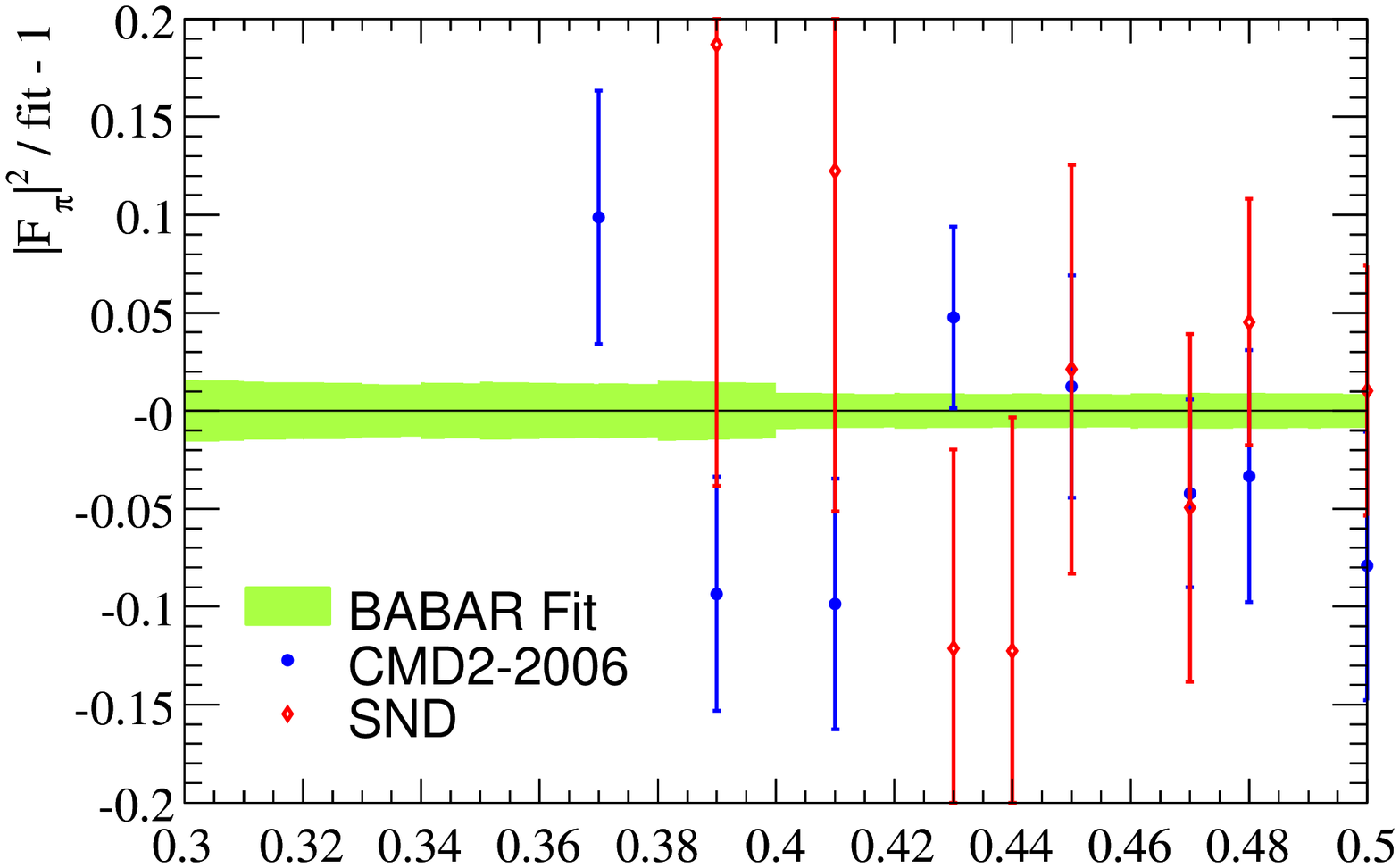}
  \put(-60,0){ {$\sqrt{s'}$ (\gev)}}\\
  \caption{\small
 The relative difference of pion form factor-squared from CMD-2 and SND 
and the \babar\ fit in the region below 0.5\gev. 
Systematic and statistical uncertainties are included in the data points. 
The width of the \babar\ band shows the propagation of statistical errors
from the fit and the quoted systematic uncertainties, added quadratically.}  
  \label{babar-others-low}
\end{minipage}\hfill
\begin{minipage}[htp]{0.49\textwidth}  
  \centering
  \includegraphics[width=1.0\textwidth]{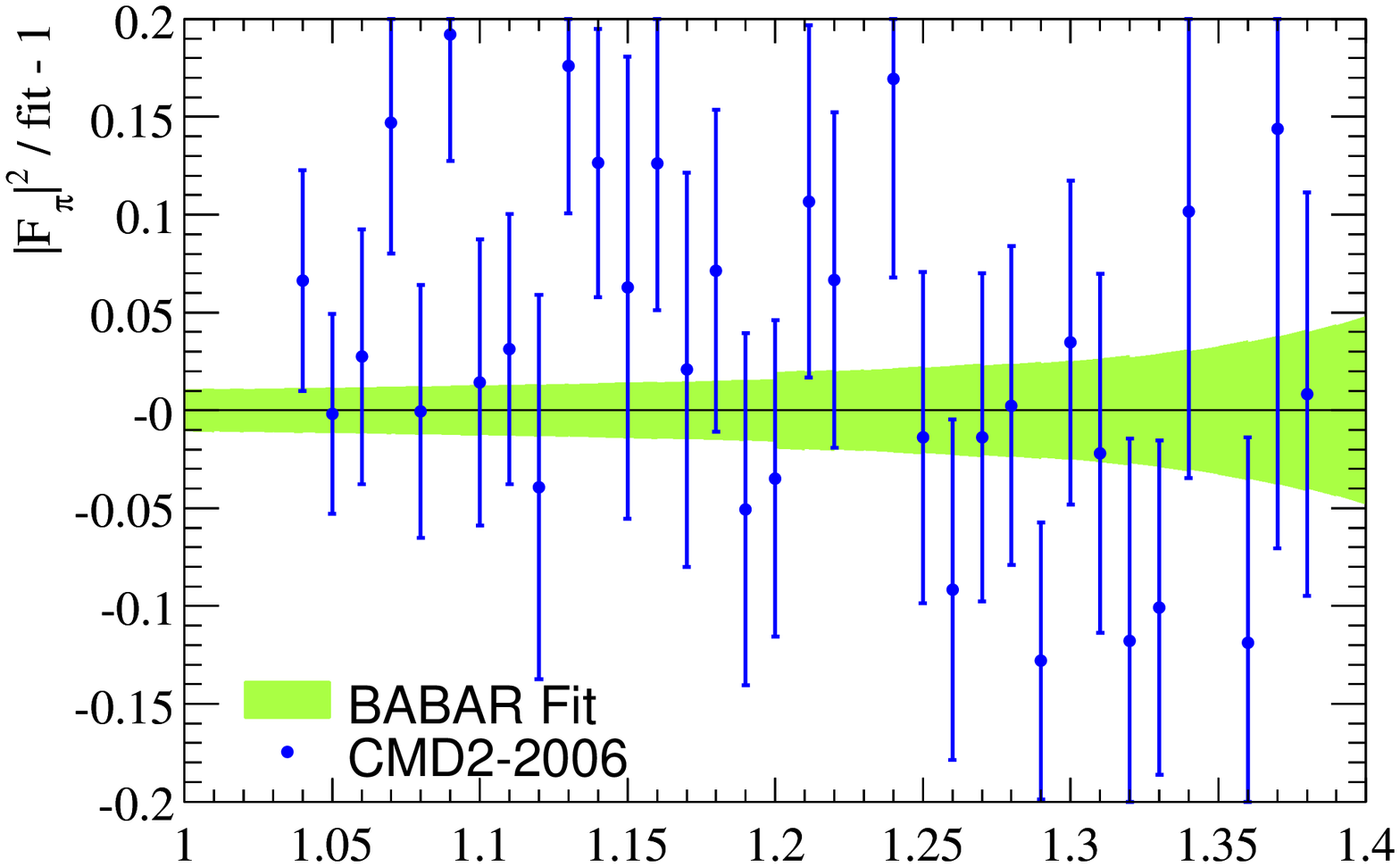}
  \put(-60,0){ {$\sqrt{s'}$ (\gev)}}\\
  \caption{\small
 The relative difference of pion form factor-squared from CMD-2 and the
\babar\ fit in the region above 1\gev. 
Systematic and statistical uncertainties are included in the data points. 
The width of the \babar\ band shows the propagation of statistical errors
from the fit and the quoted systematic uncertainties, added quadratically.}  
  \label{babar-others-high}
\end{minipage}
\end{figure*}

The comparison of the form factor-squared in the low-mass region is made in 
Fig.~\ref{babar-low-others}. The agreement is reasonable, 
with some systematic departure with respect to the NA7 experiment at CERN.
A direct cross section comparison is made in the large mass region in 
Fig.~\ref{babar-high-others}. The \babar\ results agree with CMD-2 
up to 1.4\gev, while the DM2 cross section~\cite{dm2} appears to be larger 
by about 30-40\%.
  
The comparison in relative terms of \babar\ to other experiments is presented
in Figs.~\ref{babar-others-low} and \ref{babar-others-high} for masses lower
than 0.5\gev and between 1.0 and 1.4\gev, respectively. 
The small discrepancy noticed between the \babar\ fit and CMD-2 is in fact also
observed in Fig.~\ref{babar-babar} where \babar\ data are compared to the fit. 
So it points to a problem in the VDM parametrization rather than in the data.

\subsubsection{$\tau$ spectral functions}
It is also appropriate to compare the present results to the 
$\tau\to\nu_\tau\pi\pi^0$ 
spectral function. Taking isospin-breaking (IB) into account, the 
conserved vector current (CVC) relation 
between the $e^+e^-\to\pipi(\gamma)$ bare cross section with FSR 
$\sigma_{\pipi(\gamma)}^0$ and the 
normalized hadronic invariant mass distribution in $\tau\to\nu_\tau\pi\pi^0$ 
decays is modified~\cite{cen,lopez} as follows:
\begin{widetext}
\beqn  
\sigma_{\pi^+\pi^-(\gamma)}^0= \frac {1}{D(s)} \frac {B_{\pi\pi}}{B_e} 
   \left(\frac {1}{N_{\pi\pi}} \frac {dN_{\pi\pi}}{ds}\right)
   \frac {R_{\rm IB}}{S_{\rm EW}} \left(1+\frac {\alpha(0)}{\pi}\eta (s)\right)~, 
\eeqn   
\end{widetext}
where
\beqn  
  D(s)= \frac {3|V_{ud}|^2~s}{2\pi\alpha(0)^2m_\tau^2}
      \left(1-\frac {s}{m_\tau^2}\right)^2\left(1+\frac {2s}{m_\tau^2}\right)~,
\eeqn   
and
\beqn  
 R_{\rm IB}(s)=\frac {1}{G_{\rm EM}(s)} \left(\frac {\beta_0}{\beta_-}\right)^3
      \frac {|F_0(s)|^2}{|F_-(s)|^2}~.
\eeqn   
$B_{\pi\pi}$ and $B_e$ are the branching fractions for $\tau$ decay into the
$\nu_\tau\pi\pi^0$ and $\nu_\tau e \overline{\nu}_e$ final states. $G_{\rm EM}(s)$
is the long-distance QED radiative correction and $S_{\rm EW}$ the short-distance
electroweak radiative correction. $F_0(s)$ and $F_-(s)$ are the electromagnetic
and weak form factors, while $\beta_0$ and $\beta_-$ are the pion velocities
in the $\pi^+\pi^-$ and $\pi\pi^0$ center-of-mass systems, respectively.

Isospin-breaking (IB) corrections have been recently re-evaluated~\cite{newtau}, 
and a new $\tau$
analysis for the muon $g-2$ presented, taking advantage of the Belle data. 
It updates the previous analysis~\cite{dehz03}. The $G_{\rm EM}$
factor takes also into account $\rho-\omega$ interference, and the charged and
neutral $\rho$ mass difference, a charged and neutral $\rho$ width difference 
from radiative decays, and the $m_{\pi^\pm}-m_{\pi^0}$ mass difference in the 
form factor~\cite{rho-width}. Other recent approaches to IB breaking have been 
considered, often based on specific models~\cite{benayoun,jeger-szafron}. 

\begin{figure*}[th]
  \centering
  \includegraphics[width=8cm]{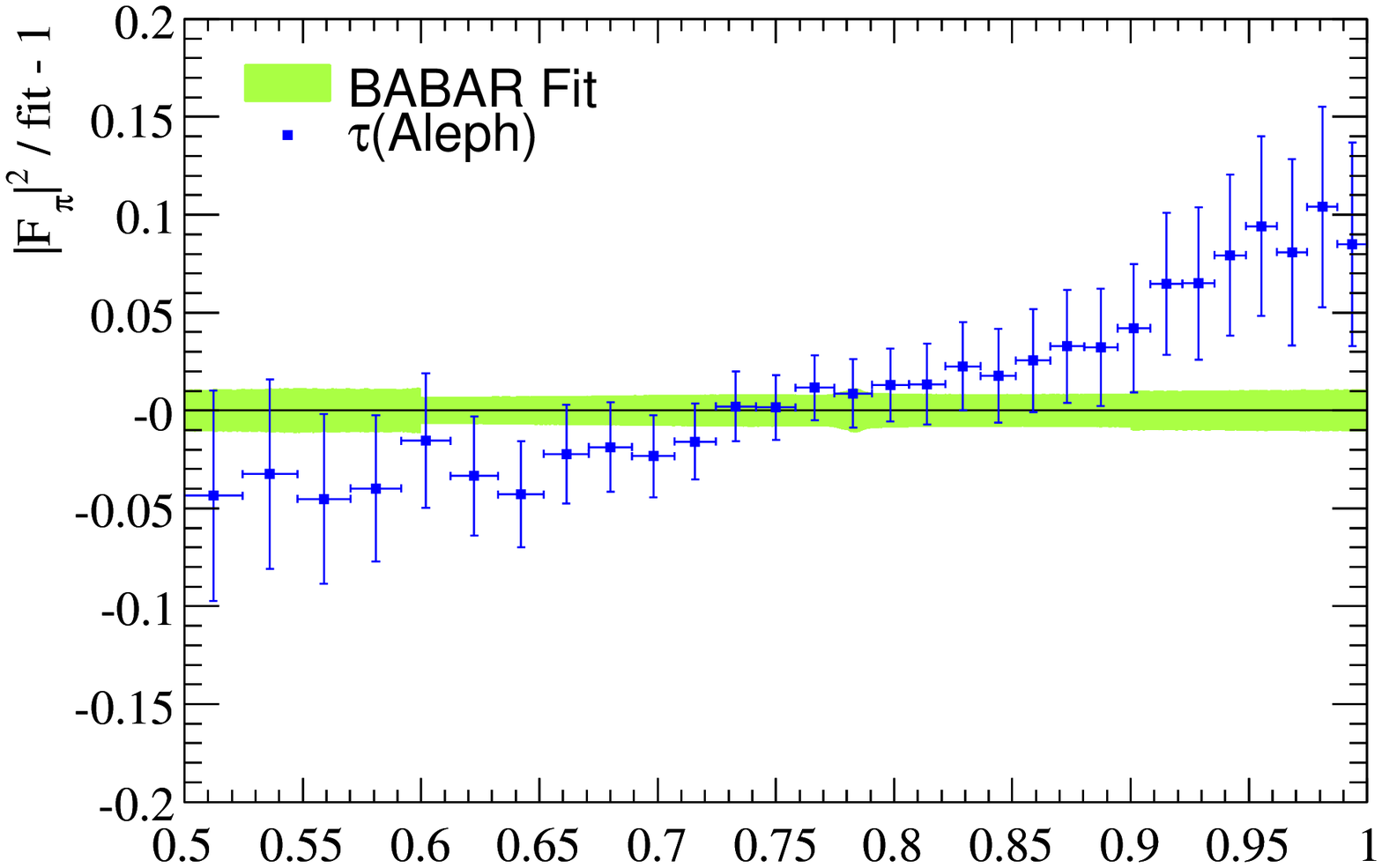}
  \put(-60,0){ {$\sqrt{s'}$ (\gev)}}\\
  \includegraphics[width=8cm]{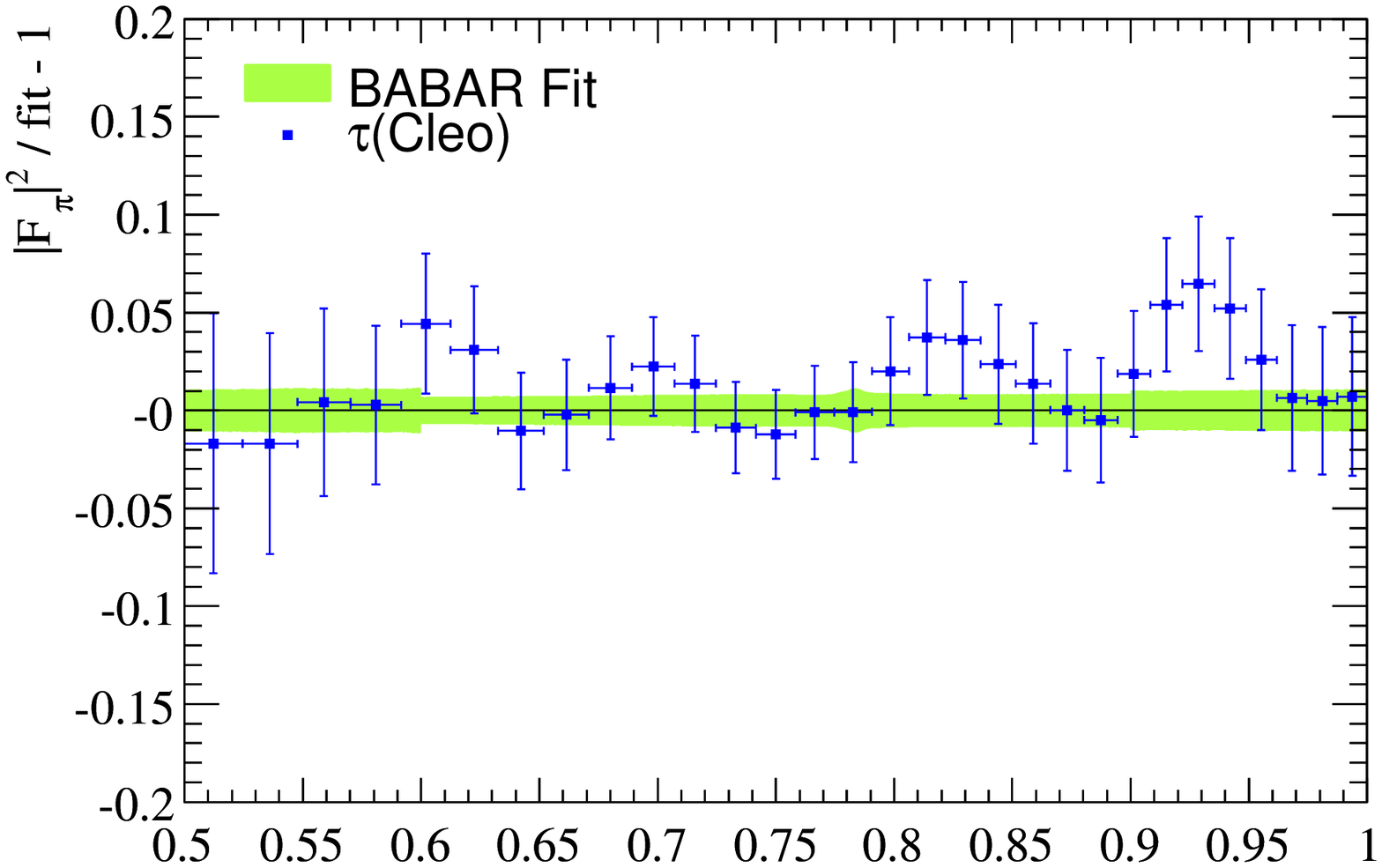}
  \put(-60,0){ {$\sqrt{s'}$ (\gev)}}\\
  \includegraphics[width=8cm]{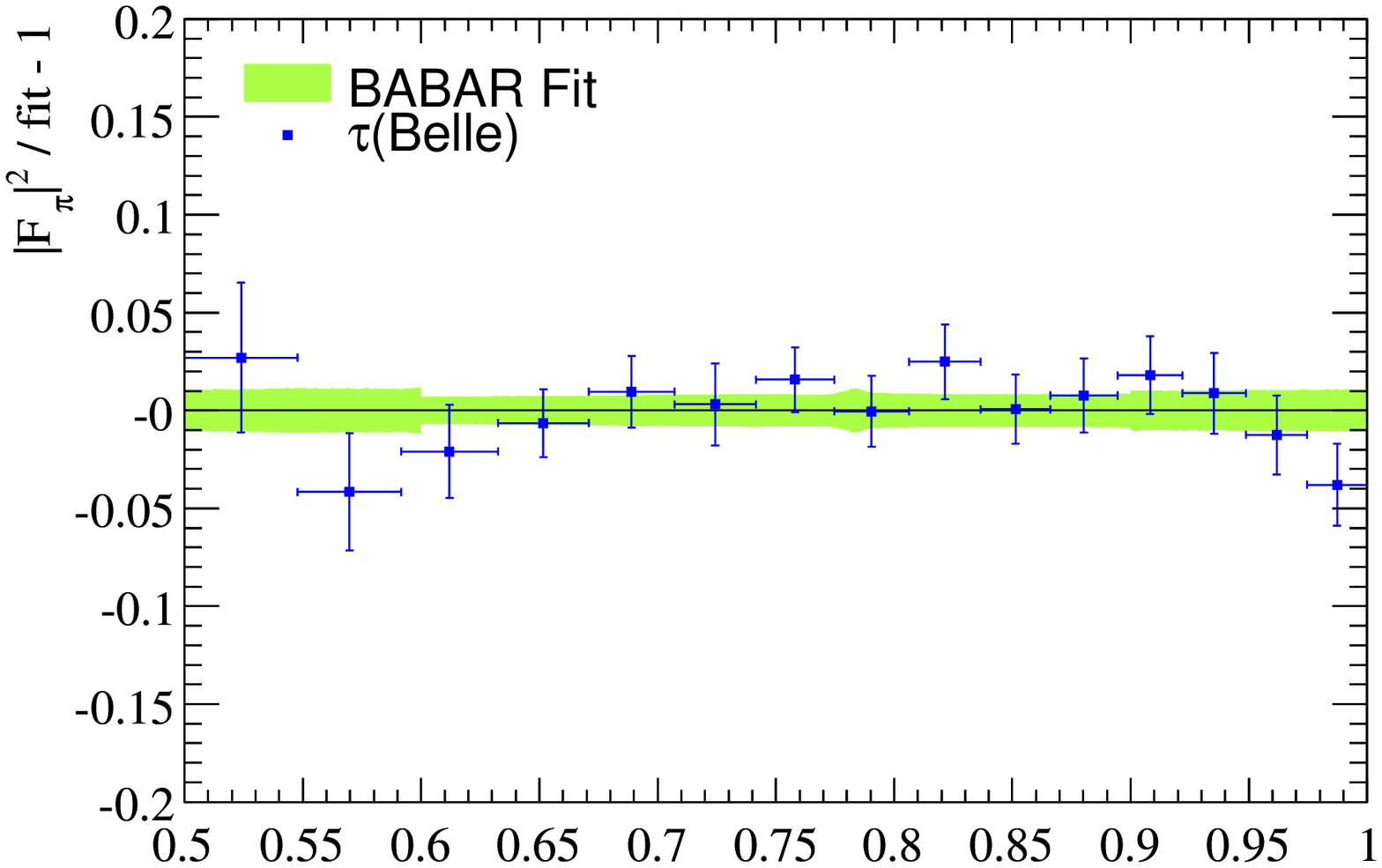}
  \put(-60,0){ {$\sqrt{s'}$ (\gev)}}\\
  \caption{\small
The relative difference of the form factor-squared from the $\tau$ data 
of ALEPH (top), CLEO (middle) and Belle (bottom) with respect to the 
$e^+e^-\to\pipi$ \babar\ measurements in the 0.5--1\gev region. 
Systematic and statistical uncertainties are included in the data points. 
The width of the \babar\ band shows the propagation of statistical errors
from the fit and the quoted systematic uncertainties,
added quadratically. The $\tau$ data are normalized to the value of $B_{\pi\pi}$
measured by each experiment independently.}
  \label{babar-aleph-cleo-belle}
\end{figure*}

Using the results from Ref.~\cite{newtau} the corrected $\tau$ and \babar\
data can be compared directly. 
This is achieved in Figs.~\ref{babar-aleph-cleo-belle}, 
for the ALEPH~\cite{aleph}, CLEO~\cite{cleo}, and Belle~\cite{belle} 
experiments, in a manner similar to the $e^+e^-$ comparisons. Here there is 
another uncertainty resulting from the IB theoretical corrections, 
corresponding roughly to a scale uncertainty of 0.3\%. For this comparison 
the spectral functions are normalized by the 
$B_{\pi\pi} \equiv B(\tau\to\pi\pi^0\nu_\tau)$ value measured by 
each experiment, rather than using the world average as usually done. 
In this way the spectral functions are really independent. The errors 
on the $\tau$ data points include all sources of statistical
and systematic uncertainties, $B_{\pi\pi}$ and IB corrections.

The comparison with ALEPH shows consistency within the systematic uncertainties
up to the $\rho$ peak and some 
slope above, with the remark that the ALEPH points are strongly correlated. 
Agreement is also observed within errors with the results of CLEO and Belle,
CLEO being somewhat in between ALEPH and Belle above $0.8\gev$.

\section{The $\pi\pi$ contribution to the anomalous muon magnetic moment}

\subsection{The \babar\ Result}

The lowest-order loop contribution of the $\pi\pi(\gamma)$ intermediate state
to the muon magnetic anomaly is given by~\cite{bouchiat}
\beqn  
\label{eq:int_amu}
    a_\mu^{\pi\pi(\gamma),\rm LO} \:=\: 
        \frac{1}{4\pi^3}\!\!
        \intl_{4m_\pi^2}^\infty\!\!ds\,K(s)\,\sigma^{0}_{\pi\pi(\gamma)}(s)~,
\eeqn   
where $K(s)$ is the QED kernel~\cite{brodsky},
\begin{widetext}
\beqn  
      K(s) \:=\: x^2\left(1-\frac{x^2}{2}\right) \,+\,
                 (1+x)^2\left(1+\frac{1}{x^2}\right)
                      \left[{\ln}(1+x)-x+\frac{x^2}{2}\right] \,+\,
                 x^2\,{\ln}x\frac{1+x}{1-x}~,
\eeqn   
\end{widetext}
with $x=(1-\beta_\mu)/(1+\beta_\mu)$ and $\beta_\mu$ the muon velocity.

The integration is carried out numerically over the measured cross section per
mass bins. The statistical and systematic errors are computed using the corresponding
covariance matrices described in Sections~\ref{summary-stat} and
\ref{summary-syst}. 

Several tests are performed.
\begin{itemize}
\item When the integral is performed with the original 50\mev bins of ISR luminosity  
the result is $(514.40\pm2.54\pm3.11)\times 10^{-10}$ in the range 0.3--1.8\gev, while the value
$(513.54\pm2.22\pm3.11)\times 10^{-10}$ is obtained with the chosen sliding-bin method. The
difference is consistent with the statistical fluctuations of the luminosity 
in the 50\mev bins (Fig.~\ref{ISRL-full}) and the $K(s)$ kernel weighting 
effect in the $a_\mu$ integral.
\item In the 0.5--1.0\gev range one compares the results obtained with the
`$\rho$ central' and the `$\rho$ tails' conditions. 
The main difference is the $\chi^2$ selection, which affects the background level,
the $\chi^2$ efficiency, and the mass resolution, hence the performance of the
unfolding. For the 0.5--1.0\gev range, the result of the integration with the 
`central' conditions 
is $445.94\times 10^{-10}$ in 2\mev bins, and $446.56\times 10^{-10}$
with the `tails' conditions in 10\mev bins. Thus the effect of 
different resolution and efficiencies has little effect on the integral. 
The difference of $0.62\times 10^{-10}$ between the two analyses is consistent 
with their estimated non-common systematic errors and non-common statistical 
errors, which induce an uncertainty on the integral of $1.8\times 10^{-10}$. 
\end{itemize}

The evaluation of the integral in the threshold region was performed in previous 
estimates~\cite{dehz03} using a polynomial expansion in $s'$ for the pion form factor, 
incorporating constraints on the normalization that $F_{\pi}(0)=1$ and that the 
derivative of the form factor at $s'=0$ be given by the known quadratic charge radius 
of the pion. This
procedure also compensated for the relatively poorer quality of data in this 
region. The \babar\ continuous low-mass data permit a direct 
evaluation, consistent with the constrained method. The very small contribution
$(0.55\pm0.01)\times 10^{-10}$ between the $2\pi$ threshold and 0.3\gev is 
evaluated using the extrapolation of the constrained fit to the data between 0.3-0.5\gev.

\begin{table} [htb] \centering 
\caption{ \label{amu-babar} \small 
Evaluation of $a_\mu^{\pi\pi(\gamma),\rm LO}$ using the \babar\ data 
in different mass regions (see text for details). 
The first error is statistical and the second systematic. }
\vspace{0.5cm}
{
\setlength{\extrarowheight}{1.5pt}
\setlength{\tabcolsep}{5pt}
\begin{tabular}{cr} \hline\hline\noalign{\vskip2pt}
$m_{\pi\pi}$ range (\gev)&$a_\mu^{\pi\pi(\gamma),\rm LO} (\times 10^{-10})$\\\hline
0.28$-$0.30 &   $0.55\pm0.01\pm0.01$ \\
0.30$-$0.50 &  $57.62\pm0.63\pm0.55$ \\
0.50$-$1.00 & $445.94\pm2.10\pm2.51$ \\
1.00$-$1.80 &   $9.97\pm0.10\pm0.09$ \\ \hline
0.28$-$1.80 & $514.09\pm2.22\pm3.11$ \\ 
\hline\hline
\end{tabular}
}
\end{table}

The \babar\ results are given in Table~\ref{amu-babar} in different 
mass ranges. The upper boundary (1.8\gev) is chosen in accordance with previous 
evaluations~\cite{dehz03}, in which the contribution of the higher energy region
was computed using QCD. The contribution in the 1.8--3\gev range, obtained with 
the present \babar\ data, is indeed only $(0.21\pm0.01)\times 10^{-10}$, 
thus negligible with respect to the uncertainty in the main region.
The contribution from threshold to 1.8\gev is obtained for the first
time from a single experiment:
\beqn  
 a_\mu^{\pi\pi(\gamma),\rm LO} ~=~ (514.09\pm2.22\pm3.11)\times 10^{-10}~,
\eeqn   
where the first error is statistical and the second systematic. The total
uncertainty is $3.82\times 10^{-10}$, so that the precision of the measurement is 0.74\%.

\subsection{Comparison to other determinations using $e^+e^-$ and $\tau$ data}

Direct comparison to the results from other experiments is complicated by two 
facts: (i) $e^+e^-$ scan experiments provide cross section measurements at 
discrete and unequally spaced energy values, while the ISR method provides 
a continuous spectrum, (ii) unlike \babar, other experiments do not all cover the
complete mass spectrum from threshold up to energies where the contributions
become negligible. The latter problem is alleviated by appropriately combining 
different sets of measurements performed by the same experiment.
Where gaps remain, they are filled by using the
weighted-average cross section values from the other experiments.
This approach has been thoroughly treated in Ref.~\cite{dhmz09}, from which 
we extract the relevant integrals. The fraction of contributions to 
the integrals estimated in this way ranges from 3\% for KLOE to 11\%
for CMD-2 (7\% for SND), engendering some correlations between the total 
values (given in the energy range from $2m_\pi$ to 1.8\gev).

\begin{figure}[thp]
  \centering
  \includegraphics[width=7.5cm]{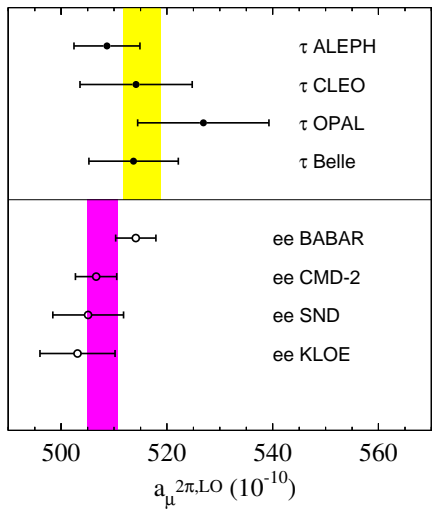}
  \caption{\small
The LO hadronic VP $2\pi$ contributions to the muon magnetic anomaly, evaluated
in the $2m_\pi-1.8$\gev range from the present analysis and other
analyses using $e^+e^-$ data~\cite{dhmz09} and $\tau$ data~\cite{newtau}. The
vertical bands show the values obtained for each data set by combining the
respective spectral functions.}
  \label{comp-all}
\end{figure}

Correlations between systematic uncertainties have also been taken into 
account in Ref.~\cite{dhmz09}, particularly for radiative corrections, when 
combining the results from all experiments. The combination is achieved locally
at the cross section level, taking into account possible disagreements leading
to an increased uncertainty of the resulting average. The results
are summarized in Table~\ref{res-DHMZ} and Fig.~\ref{comp-all} and allow a 
direct comparison of all the
determinations. They are indeed consistent within the errors, \babar\ and CMD-2
being almost a factor of two more precise than SND and KLOE.  
Internal discrepancies between the measurements increase the final
uncertainty to $3.2\times 10^{-10}$, whereas the ideal value would be 
$2.4\times 10^{-10}$ for fully consistent experiments.

\begin{table} [tb] \centering 
\caption{ \label{res-DHMZ} \small 
Evaluation of LO hadronic VP $2\pi$ contributions to the muon magnetic anomaly 
in the energy range $[2m_\pi,1.8\gev]$
from \babar , other $e^+e^-$ experiments~\cite{dhmz09}, 
and $\tau$ experiments~\cite{newtau} (see text for details). 
The errors are from both statistical and systematic sources. For the
values derived from $\tau$ decays, a common systematic error of 1.9 is included to account for 
uncertainties in the isospin-breaking corrections. Note that the 
combined results are not the weighted average of the different values, but
originate from a proper local combination of the respective spectral 
functions~\cite{dhmz09,newtau}. }
\vspace{0.5cm}
{
\setlength{\extrarowheight}{1.5pt}
\setlength{\tabcolsep}{5pt}
\begin{tabular}{cll} \hline\hline\noalign{\vskip2pt}
Experiment && $a_\mu^{\pi\pi(\gamma),\rm LO} (\times 10^{-10})$ \\ \hline\noalign{\vskip2pt}
\babar\ && $514.1\pm3.8$  \\
CMD-2   && $506.6\pm3.9$  \\
SND     && $505.1\pm6.7$  \\
KLOE    && $503.1\pm7.1$  \\ \hline
combined $e^+e^-$ && $507.8\pm3.2$  \\ \hline \hline
ALEPH   && $508.7\pm5.9\pm1.9$  \\
CLEO    && $514.2\pm10.4\pm1.9$ \\
OPAL    && $526.9\pm12.3\pm1.9$ \\
Belle   && $513.7\pm8.2\pm1.9$  \\ \hline
combined $\tau$   && $515.2\pm3.0\pm1.9$  \\
\hline\hline
\end{tabular}
}
\end{table}

Similarly, the \babar\ result is compared to determinations using $\tau$ decays
with IB corrections~\cite{newtau} in Table~\ref{res-DHMZ} and Fig.~\ref{comp-all}. 
The agreement is 
found to be satisfactory, so the \babar\ data reduces the previous tension
between the $e^+e^-$ and $\tau$'s values~\cite{dehz03}. Regarding the consistency of 
all results, it is notable that the four inputs (CMD-2/SND, KLOE, \babar, $\tau$) 
have completely independent systematic uncertainties. 

\subsection{Impact of this result on the comparison of the Standard Model prediction 
and the direct  measurement of the muon magnetic anomaly}

Even though the $2\pi$ contribution is the dominant part of the hadronic LO VP 
component in the Standard Model prediction for the muon magnetic anomaly, all the
contributions must be evaluated. The \babar\ experiment has measured most of
the relevant cross sections by the ISR method. 
Except for a few channels still unmeasured, \babar\ results 
bring a new level of precision and dominate the picture. 
We follow here the recent analysis of Ref.~\cite{dhmz10} that uses all these
measurements and those from other experiments. More recently, similar 
results have been obtained~\cite{hlmnt2011}.

Adding all contributions (QED, electroweak, hadronic LO VP other than $2\pi$,
hadronic higher-order VP, hadronic light-by-light~\cite{lbl})
to the present $2\pi$ result alone, one obtains the predicted value of the muon magnetic anomaly:
\beqn  
 a_\mu ~=~ (11~659~186.5\pm5.4)\times 10^{-10}~,
\eeqn   
to be compared to the direct measurement~\cite{bnl}, slightly 
updated~\cite{pdg-HM}:
\beqn  
 a_\mu^{\rm exp} ~=~ (11~659~208.9\pm6.3)\times 10^{-10}~.
\eeqn   

The experimental value exceeds the prediction by $(22.4\pm8.3)\times 10^{-10}$, 
{\it i.e.}, 2.7 standard deviations. When the present \babar\ cross section 
is combined with all available $2\pi$ data from
$e^+e^-$ experiments~\cite{dhmz10}, the deviation increases to 
$(28.7\pm8.0)\times 10^{-10}$, {\it i.e.}, 3.6 standard deviations. 
Although the deviation is not significant enough to claim a departure from 
the Standard Model, it confirms the trend of earlier results using previous 
$e^+e^-$ data~\cite{dehz03,md-pisa,hmnt2007}.

\section{Conclusion}

The cross sections for the processes 
$e^+e^- \rightarrow \mu^+\mu^-(\gamma)\gamma_{\rm ISR}$ and 
$e^+e^- \rightarrow \pi^+\pi^-(\gamma)\gamma_{\rm ISR}$ have been measured by the
\babar\ experiment, where the additional photon may be produced either by 
FSR or ISR. 
Thanks to the properties of the ISR method, the 
corresponding $e^+e^- \rightarrow \mu^+\mu^-(\gamma)$ and
$e^+e^- \rightarrow \pi^+\pi^-(\gamma)$ cross sections have been determined from their 
thresholds to 3\gev, thus covering completely the interesting region for
calculating hadronic vacuum polarization in the $\pi^+\pi^-$ channel.

For $e^+e^- \rightarrow \mu^+\mu^-(\gamma)$ the cross section is measured
using the $e^+e^-$ luminosity, while the cross section for
$e^+e^- \rightarrow \pi^+\pi^-(\gamma)$, for which the highest precision is
required, is obtained from the ratio of the radiative $\pi^+\pi^-(\gamma)\gamma_{\rm ISR}$ to
$\mu^+\mu^-(\gamma)\gamma_{\rm ISR}$ mass spectra.
In this way the pion results are independent of the $e^+e^-$ luminosity and 
important systematic effects cancel. As a major asset of the method,
the pion-pair cross section is not sensitive to the model of radiative 
corrections in the generator used for MC simulation.

The measured absolute $e^+e^- \rightarrow \mu^+\mu^-(\gamma)\gamma_{\rm ISR}$
cross section is found to agree with QED at NLO
from threshold to 3\gev, with a precision of 1.1\% dominated by the $e^+e^-$
luminosity determination.

The cross section for $e^+e^- \rightarrow \pi^+\pi^-(\gamma)$ is obtained for
the first time continuously in the energy range from threshold to 3\gev. Its
precision exceeds that of previous experiments in most of this range. The
achieved systematic uncertainty is 0.5\% in the dominant $\rho$ region from
0.6 to 0.9\gev.

Fits of the pion form factor have been performed using a sum of contributions 
from isovector vector mesons: besides the dominant $\rho$ resonance and
isospin-violating $\rho-\omega$ interference, three higher states are needed
to reproduce the structures observed in the measured spectrum.

The results are in fair agreement with previous data from CMD-2 and SND, but
some discrepancies are observed when compared to the KLOE data, particularly
on the $\rho$ peak (3\%) and above (up to 6\% at 0.95\gev). These differences
exceed the uncertainties quoted by either experiment. The \babar\ results are in
agreement with the spectral functions derived from 
$\tau\rightarrow \nu_\tau \pi^-\pi^0$ data, although some local deviations 
are seen in the lineshape at the 2\% level with Belle.

Finally, the \babar\ results are used as input to the dispersion integral 
yielding the $\pi^+\pi^-(\gamma)$ vacuum polarization contribution at LO to
the muon magnetic anomaly. This contribution amounts to
$(514.1 \pm2.2_{stat} \pm3.1_{syst})\times 10^{-10}$, the most precise result yet 
from a single experiment. This result brings the contribution estimated
from all $e^+e^- \rightarrow \pi^+\pi^-(\gamma)$ data combined in better
agreement with the $\tau$ estimate. When adding all other Standard Model 
contributions to the present $2\pi$ result, in particular using all available 
\babar\ data on multihadronic processes, the predicted muon magnetic anomaly
is found to be $(11~659~186.5\pm5.4)\times 10^{-10}$, which is smaller than the 
direct measurement at BNL by 2.7$\sigma$. Adding all previous $2\pi$ data
increases the deviation to 3.6$\sigma$.

A claim for a breakdown of the Standard Model requires even more
precise data, both for the direct anomaly measurement and the hadronic cross
sections. But since a deviation of the observed size could be mediated by new 
physics at a scale of a few 100\gev, the present direct exploration for new
phenomena performed at the Tevatron and the LHC will certainly bring valuable 
and complementary information.

We are grateful for the 
extraordinary contributions of our \pep2\ colleagues in
achieving the excellent luminosity and machine conditions
that have made this work possible.
The success of this project also relies critically on the 
expertise and dedication of the computing organizations that 
support \babar.
The collaborating institutions wish to thank 
SLAC for its support and the kind hospitality extended to them. 
This work is supported by the
US Department of Energy
and National Science Foundation, the
Natural Sciences and Engineering Research Council (Canada),
the Commissariat \`a l'Energie Atomique and
Institut National de Physique Nucl\'eaire et de Physique des Particules
(France), the
Bundesministerium f\"ur Bildung und Forschung and
Deutsche Forschungsgemeinschaft
(Germany), the
Istituto Nazionale di Fisica Nucleare (Italy),
the Foundation for Fundamental Research on Matter (The Netherlands),
the Research Council of Norway, the
Ministry of Education and Science of the Russian Federation, 
Ministerio de Ciencia e Innovaci\'on (Spain), and the
Science and Technology Facilities Council (United Kingdom).
Individuals have received support from 
the Marie-Curie IEF program (European Union) and the A. P. Sloan Foundation (USA).


\end{document}